\PassOptionsToPackage{table,xcdraw}{xcolor}
\PassOptionsToPackage{bookmarksnumbered=true}{hyperref}
\documentclass[review=false]{jfp-epi}
\usepackage[utf8]{inputenc}
\usepackage{hyperref}
\usepackage{graphicx}
\usepackage{amsmath}
\usepackage{amssymb}
\usepackage{algorithm}
\usepackage{multirow}
\usepackage{algpseudocode}
\usepackage{booktabs}
\usepackage{mathtools}
\usepackage{proof}
\usepackage{siunitx}
\usepackage{tabularx,booktabs,caption}
\usepackage{varwidth}
\usepackage{verbatim}
\usepackage{caption}
\usepackage{subcaption}
\usepackage{graphics}
\usepackage{tikz}
\usepackage{url}
\usepackage{color}
\usepackage{moreverb}
\usepackage{enumitem}
\usepackage{stmaryrd}
\usepackage{float}
\usepackage{newfloat}
\usepackage{comment}
\usepackage{datetime2}
\usepackage{siunitx}

\DeclareTextCommandDefault{\nobreakspace}{\leavevmode\nobreak\ }

\usepackage{epstopdf}

\usepackage{natbib}
\usepackage{xcolor}
\usepackage{listings}
\usepackage[altpo]{backnaur}
\usepackage{multicol}

\definecolor{mGreen}{rgb}{0,0.6,0}
\definecolor{mGray}{rgb}{0.5,0.5,0.5}
\definecolor{mPurple}{rgb}{0.58,0,0.82}
\definecolor{backgroundColour}{rgb}{0.95,0.95,0.92}

\lstdefinestyle{CStyle}{
    commentstyle=\color{mGreen},
    keywordstyle=\color{magenta},
    numberstyle=\tiny\color{mGray},
    stringstyle=\color{mPurple},
    basicstyle=\ttfamily\footnotesize,
    breakatwhitespace=false,
    breaklines=true,
    captionpos=b,
    keepspaces=true,
    numbers=left,
    numbersep=5pt,
    showspaces=false,
    showstringspaces=false,
    showtabs=false,
    tabsize=2,
    language=C
}

\usepackage{mdframed}
\usepackage[nameinlink]{cleveref}

\newenvironment{nicebox}[1]
{\begin{mdframed}[backgroundcolor=#1,linewidth=0pt,roundcorner=10pt]}
{\end{mdframed}}

\definecolor{ColorBlue}{rgb}{0.8,0.95,1} 
\definecolor{DefinitionColor}{rgb}{1.0,0.9,0.92} 

\newfloat{deffloat}{tb}{deff}
\floatname{deffloat}{Lesson}
\floatstyle{plain}
\restylefloat{deffloat}
\crefname{deffloat}{Lesson}{Lesson}
\newenvironment{lesson}[2][tb]{
  \begin{deffloat}[#1]
    \begin{nicebox}{ColorBlue}
      \refstepcounter{deffloat}
      \noindent\textbf{Lesson \thedeffloat:} #2\par\vspace{0.5em}}%
{\end{nicebox}\end{deffloat}}

\newcolumntype{P}[1]{>{\centering\arraybackslash}p{#1}}

\usepackage{amsmath}
\usepackage{amssymb}
\usepackage{latexsym}
\usepackage{relsize}
\usepackage{color}
\usepackage{mathtools}
\usepackage{listings}
\usepackage{xspace}

\definecolor{identifierColor}{rgb}{0.65,0.16,0.16}
\definecolor{commentColor}{rgb}{0,0.6,0}

\def\coqinline{\lstinline[language=Coq, mathescape=true]}
\def\cinline{\lstinline[language=C, mathescape=true]}

\lstdefinestyle{dynwinlistingstyle}
{columns=fullflexible,rulecolor=\color{green}}

\newcommand*{\dynwinlisting}[2][]{\lstinputlisting[style=dynwinlistingstyle,#1]{#2}}

\newcommand{\fCompose}{\mathbin{\circ}}

\newcommand{\hoperator}[1]{{\mathtt{#1}}}

\DeclareMathOperator{\mr}{\mathtt{M\mkern-3mu R}}

\newcommand{\iUnion}[5]{{\mr_{#1,#2,#3}(\lambda #4. (#5))}}

\newcommand{\ifunc}[3]{{#1}^{#2\rightarrow #3}}
\newcommand{\interval}{\mathbb{I}}
\newcommand{\finNat}[1]{\interval_{#1}}

\newcommand{\bnfd}[1]{$\langle$#1$\rangle$}

\newcommand{\OLL}{\emph{OL}\xspace} 
\newcommand{\SOL}{\texorpdfstring{$\Sigma$-\emph{OL}}{Sigma-OL}\xspace}

\newcommand{\coqlang}{Coq\xspace}
\newcommand{\llvm}{LLVM\xspace}
\newcommand{\spiral}{SPIRAL\xspace}
\newcommand{\helix}{HELIX\xspace}
\newcommand{\HCOL}{\emph{HCOL}\xspace}
\newcommand{\SHCOL}{\texorpdfstring{$\Sigma$-\emph{HCOL}}{Sigma-HCOL}\xspace}
\newcommand{\MSHCOL}{\emph{MSHCOL}\xspace}
\newcommand{\DHCOL}{\emph{DHCOL}\xspace}
\newcommand{\RHCOL}{\emph{RHCOL}\xspace}
\newcommand{\FHCOL}{\emph{FHCOL}\xspace}
\newcommand{\VIR}{\emph{VIR}\xspace}

\newcommand{\N}{\mathbb{N}}
\newcommand{\Z}{\mathbb{Z}}
\newcommand{\R}{\mathbb{R}}
\newcommand{\Complex}{\mathbb{C}}
\newcommand{\Q}{\mathbb{Q}}
\newcommand{\B}{\mathbb{B}}

\newcommand{\ctype}{\mathcal{R}}
\newcommand{\rtheta}{\mathcal{R}_\theta}
\newcommand{\rstheta}{\mathcal{R}_{\theta'}}
\newcommand{\actype}{\mathcal{R}_{\_}}

\makeatletter
\newcommand*{\pmzerodot}{%
  \nfss@text{%
    \sbox0{$\vcenter{}$}
    \sbox2{0}%
    \sbox4{0\/}%
    \ooalign{%
      0\cr
      \hidewidth
      \kern\dimexpr\wd4-\wd2\relax 
      \raise\dimexpr(\ht2-\dp2)/2-\ht0\relax\hbox{%
        \if b\expandafter\@car\f@series\@nil\relax
          \mathversion{bold}%
        \fi
        $\cdot\m@th$%
      }%
      \hidewidth
      \cr
      \vphantom{0}
    }%
  }%
}

\newcommand{\TODO}[1]{}
\newcommand{\CalvinTODO} [1]{}
\newcommand{\IliaTODO}   [1]{}
\newcommand{\IreneTODO}  [1]{}
\newcommand{\SteveTODO}  [1]{}
\newcommand{\VadimTODO}  [1]{}
\newcommand{\YannickTODO}[1]{}
\newcommand{\yz}[1]{}
\newcommand{\vz}[1]{}
\newcommand{\sz}[1]{}
\newcommand{\iy}[1]{}
\newcommand{\calvin}[1]{}
\newcommand{\proposecut}[1]{}

\usepackage{color}
\usepackage{listings}

\definecolor{ltblue}{rgb}{0,0.4,0.4}
\definecolor{dkblue}{rgb}{0,0.1,0.6}
\definecolor{dkgreen}{rgb}{0,0.35,0}
\definecolor{dkviolet}{rgb}{0.3,0,0.5}
\definecolor{dkred}{rgb}{0.5,0,0}

\lstdefinelanguage{Coq}{ 
mathescape=true,
texcl=false, 
morekeywords=[1]{Section, Module, End, Require, Import, Export,
  Variable, Variables, Parameter, Parameters, Axiom, Hypothesis,
  Hypotheses, Notation, Local, Tactic, Reserved, Scope, Open, Close,
  Bind, Delimit, Definition, Let, Ltac, Fixpoint, CoFixpoint, Add,
  Morphism, Relation, Implicit, Arguments, Unset, Contextual,
  Strict, Prenex, Implicits, Inductive, CoInductive, Record,
  Structure, Canonical, Coercion, Context, Class, Global, Instance,
  Program, Infix, Theorem, Lemma, Corollary, Proposition, Fact,
  Remark, Example, Proof, Goal, Save, Qed, Defined, Hint, Resolve,
  Rewrite, View, Search, Show, Print, Printing, All, Eval, Check,
  Projections, inside, outside, Def},
morekeywords=[2]{forall, exists, exists2, fun, fix, cofix, struct,
  match, with, end, as, in, return, let, if, is, then, else, for, of,
  nosimpl, when},
morekeywords=[3]{Type, Prop, Set, true, false, option},
morekeywords=[4]{pose, set, move, case, elim, apply, clear, hnf,
  intro, intros, generalize, rename, pattern, after, destruct,
  induction, using, refine, inversion, injection, rewrite, setoid_rewrite, congr,
  unlock, compute, ring, field, fourier, replace, setoid_replace, fold, unfold,
  change, cutrewrite, simpl, have, suff, wlog, suffices, without,
  loss, nat_norm, assert, cut, trivial, revert, bool_congr, nat_congr,
  symmetry, transitivity, auto, split, left, right, autorewrite},
morekeywords=[5]{by, done, exact, reflexivity, tauto, romega, omega,
  assumption, solve, contradiction, discriminate},
morekeywords=[6]{do, last, first, try, idtac, repeat},
morecomment=[s]{(*}{*)},
showstringspaces=false,
morestring=[b]",
tabsize=3,
extendedchars=false,
sensitive=true,
breaklines=false,
basicstyle=\ttfamily\small,
captionpos=b,
basewidth={2em, 0.5em},
columns=flexible,
identifierstyle={\ttfamily\color{black}},
keywordstyle=[1]{\ttfamily\bfseries\color{dkviolet}},
keywordstyle=[2]{\ttfamily\bfseries\color{dkgreen}},
keywordstyle=[3]{\ttfamily\bfseries\color{ltblue}},
keywordstyle=[4]{\ttfamily\color{dkblue}},
keywordstyle=[5]{\ttfamily\color{dkred}},
stringstyle=\ttfamily,
commentstyle={\ttfamily\itshape\color{dkgreen}},
moredelim=**[is][\ttfamily\color{red}]{/&}{&/},
literate=
    {fun}{{\color{dkgreen}{$\lambda\;$}}}1
    {bool}{{$\mathbb{B}$}}1
    {nat}{{$\mathbb{N}$}}1
    {Vforall2}{Vforall2}1 
    {nat\_equiv}{nat\_equiv}1 
    {forall}{{\color{dkgreen}{$\forall\;$}}}1
    {exists}{{$\exists\;$}}1
    {<-}{{$\leftarrow\;\;$}}1
    {=>}{{$\Rightarrow\;\;$}}1
    {==}{{\texttt{==}\;}}1
    {==>}{{$\Longrightarrow\;\;$}}1
    {->}{{$\rightarrow\;\;$}}1
    {<-->}{{$\longleftrightarrow\;\;$}}1
    {<->}{{$\leftrightarrow\;\;$}}1
    {<==}{{$\leq\;\;$}}1
    {\#}{{$^\star$}}1 
    {\\o}{{$\circ\;$}}1 
    {\/\\}{{$\wedge\;$}}1
    {\\\/}{{$\vee\;$}}1
    {++}{{\texttt{++}}}1
    {~}{{$\sim$}}1
    {\@\@}{{$@$}}1
    {\\mapsto}{{$\mapsto\;$}}1
    {\\hline}{{\rule{\linewidth}{0.5pt}}}1
}[keywords,comments,strings]

\lstnewenvironment{coq}{\lstset{language=Coq}}{}

\def\coqe{\lstinline[language=Coq, breaklines, basicstyle=\ttfamily\small]}

\usepackage{color}
\usepackage{listings}

\definecolor{ltblue}{rgb}{0,0.4,0.4}
\definecolor{dkblue}{rgb}{0,0.1,0.6}
\definecolor{dkgreen}{rgb}{0,0.35,0}
\definecolor{dkviolet}{rgb}{0.3,0,0.5}
\definecolor{dkred}{rgb}{0.5,0,0}

\lstdefinelanguage{Gappa}{ 
mathescape=true,
texcl=false, 
morekeywords=[1]{in},
morecomment     = [l]{\#},    
showstringspaces=false,
morestring=[b]",
tabsize=3,
extendedchars=false,
sensitive=true,
breaklines=false,
basicstyle=\ttfamily\small,
captionpos=b,
basewidth={2em, 0.5em},
columns=flexible,
identifierstyle={\ttfamily\color{black}},
keywordstyle=[1]{\ttfamily\bfseries\color{dkviolet}},
keywordstyle=[2]{\ttfamily\bfseries\color{dkgreen}},
keywordstyle=[3]{\ttfamily\bfseries\color{ltblue}},
keywordstyle=[4]{\ttfamily\color{dkblue}},
keywordstyle=[5]{\ttfamily\color{dkred}},
stringstyle=\ttfamily,
commentstyle={\ttfamily\itshape\color{dkgreen}},
literate=
    {<-}{{$\leftarrow\;\;$}}1
    {->}{{$\rightarrow\;\;$}}1
    {<->}{{$\leftrightarrow\;\;$}}1
    {\/\\}{{$\wedge\;$}}1
    {\\\/}{{$\vee\;$}}1
}[keywords,comments,strings]

\lstnewenvironment{gappa}{\lstset{language=Gappa}}{}

\allowdisplaybreaks

\title{HELIX: Verified compilation of cyber-physical control systems to LLVM IR}

\jfpJournal{JFP}
\jfpDOI{10.1017/xxxxx}
\jfpYear{2025}

\author{Vadim Zaliva}
\orcid{0000-0002-9145-3288}
\affiliation{%
  \institution{University of Cambridge}
  \country{UK}
  \authoremail{vz231@cl.cam.ac.uk}
}

\author{Yannick Zakowski}
\orcid{0000-0003-4585-6470}
\affiliation{%
  \institution{INRIA}
  \country{France}
  \authoremail{yannick.zakowski@inria.fr}
}

\author{Ilia Zaichuk}
\orcid{0000-0003-1617-3259}
\affiliation{%
  \institution{Taras Shevchenko National University}
  \country{Ukraine}
}
\affiliation{%
  \institution{\ Digamma.ai}
  \country{USA}
  \authoremail{zoickx@ztld.org}
}

\author{Valerii Huhnin}
\orcid{0009-0006-0572-7809}
\affiliation{%
  \institution{Université Paris Cité}
  \country{France}
}
\affiliation{%
  \institution{\ Digamma.ai}
  \country{USA}
  \authoremail{valerii.huhnin@etu.u-paris.fr}
}

\author{Calvin Beck}
\orcid{0000-0002-3469-7219}
\affiliation{%
  \institution{University of Pennsylvania}
  \country{USA}
  \authoremail{hobbes@seas.upenn.edu}
}

\author{Irene Yoon}
\orcid{0000-0003-3388-1257}
\affiliation{%
  \institution{University of Pennsylvania}
  \country{USA}
  \authoremail{inbox@ireneyoon.com}
}

\author{Steve Zdancewic}
\orcid{0000-0002-3516-1512}
\affiliation{%
  \institution{University of Pennsylvania}
  \country{USA}
  \authoremail{stevez@seas.upenn.edu}
}

\begin{document}

\begin{abstract}
    This paper presents the design of HELIX, an end-to-end verified
    code generation system with a focus on the intersection of
    high-performance and high-assurance numerical computing. The code
    generation can be fine-tuned to generate efficient code for a
    broad set of computer architectures while providing formal
    guarantees of the correctness of such generated code.

    Using a real-life example of a cyber-physical robot system, this
    paper demonstrates how, by using HELIX, one can start from a
    high-level mathematical formulation of the problem, apply a series
    of algebraic transformations that target intermediate languages,
    and generate an efficient imperative implementation. This is done
    while formally verifying semantic preservation from the original
    formulation down to LLVM IR.

    The method we used for high-performance code compilation is the
    algebraic transformation of vector and matrix computations into a
    dataflow optimised for parallel or vectorised processing on target
    hardware. The abstraction used to formalise and verify this
    technique is an operator language and accompanying
    semantics-preserving term rewriting. We use sparse vector
    abstraction to represent partial computations, enabling us to use
    algebraic reasoning to prove parallel decomposition properties.

    HELIX's verification infrastructure comprises multiple
    intermediate languages and verification approaches, all
    implemented in the Coq proof assistant. In particular, it uses
    verified term rewriting, translation validation, metaprogramming,
    verified compilation, layered monadic interpreters; it also
    supports application-specific uses of (verified) numerical
    analysis as we demonstrate via the running example.
  \end{abstract}

\maketitle

\tableofcontents


\sloppy

\section{Introduction}
\label{sec:intro}

\subsection{Motivation and background}
\label{sec:problem}

The increased dependence of modern society on computer programs
combined with the growing sophistication of software systems and
hardware architectures poses a significant challenge in keeping
computer systems reliable and correct. For example, Boeing's 787
airplane contains about 6.5 million lines of code to operate its
avionics and onboard support systems~\citep{charette2009car}. According
to the same source, the premium class automobile contains more than
100 million lines of code. Parts of this code control
mission-critical systems, which, if they malfunction, could endanger human
lives. There are multiple reported cases or costly recalls when such
bugs were discovered and deemed critical. As a consequence, there are significant
ongoing engineering efforts in developing reliable embedded software
at such a scale~\citep{10.1109/MC.2009.118}.

Beyond safety and reliability, there are other constraints at play in such embedded systems.
Most vehicular embedded software is run on Electronic
Control Units (ECUs)---these are simple processors that control various aspects of car functionality.
Lower-end models utilize between 30-50 ECUs, while higher-end
luxury cars might use many more. These ECUs exhibit
different Instruction Set Architectures (ISAs), memory architectures, clock speeds, and data bus
widths. This makes the problem of developing software for them 
formidable: not only does one have to write and maintain potentially millions of lines of code, but this code will be targeting multiple hardware
architectures. Due to price and energy consumption constraints, these
embedded systems are typically underpowered compared to modern desktop
computers. At the same time, they often control functions that require
high-performance computations. For example, an ECU that controls airbag
deployment must react within 15 to 40 milliseconds
\citep{charette2009car}.

High-performance numerical computing is an essential part of
cyber-physical systems. In the course of their operation, they employ
a number of numerical computations such as PID
controllers~\citep{johnson2005pid}, DFT~\citep{macian2006dft}, Kalman
filter~\citep{kalman1960new}, in real-time, oftentimes on
performance-constrained hardware. Hence, there is a need for
high-performance implementation of such algorithms for diverse
underlying hardware. The traditional approach is to implement such
algorithms in a high-level language like C and use the C compiler to
generate optimized code for target platforms. An alternative approach
is code synthesis, where from a high-level numerical algorithm
description a code is generated that is optimized for the target
hardware. It has been demonstrated
\citep{Franchetti:09,Franchetti:02,Moura:01,franchetti:18} that such
early optimization could yield better performance results than
compiler optimizations such as loop unrolling, algebraic
simplification, vectorization, etc. used in the traditional approach.

Another key requirement for such software is correctness. Instead of
proving the correctness of one imperative implementation of the
algorithm, we are now dealing with multiple versions of the same
algorithm generated for different hardware architectures. During code
synthesis, the target platform capabilities (parallelism,
vectorization, the timing of different operations) may inform how
mathematical computations may be reshaped to best map to target
hardware. The correctness of the required transformations is based on
algebraic reasoning and thus require different proof techniques than
those used for standard compiler optimization correctness. These
techniques need to combine algebraic reasoning with numerical
stability, error bounds, etc. Ultimately, the correctness proofs need
to extend to synthesized code compiled to machine assembly (via a
low-level representation such as LLVM IR).

Therefore problem we addressed lies in this intersection of
high-performance and high-assurance numerical computing. Our goal was
to build a system that enables the generation of high-performance code
which is proven to be correct for a class of numeric algorithms,
useful for practical applications, such as code generation for the ECU
vehicular control domain described above.

To that end, this paper presents HELIX, a framework for generating
high-assurance numerical software, implemented and fully verified in
the Coq interactive theorem prover.

This style of formal verification, in which a program is proved meet a desired
specification, is a promising way to enable the construction of high-assurance
software.  In this approach, a HELIX source program's behaviors are described
using an \textit{interactive theorem prover} (in the case of HELIX, Coq) and
then compiler optimization and code transformation passes are proved (with a
machine-checked proof) to be correct, that is, the resulting low-level code's
behavior is shown to \textit{refine} that of the source program.

One challenge in building such a framework is that the design space is quite
large: there are many choices and tradeoffs to be made about how to represent a
program, how to define programming language semantics, how to structure the
proofs of correctness, etc. The HELIX compiler pipeline uses a novel combination
of \textit{translation validation} (in which the correctness of an optimization
is proved correct for a specific input program of interest) and \textit{verified
  compilation} (in which a compiler transformation itself is proved to work
correctly for \textit{all} inputs) to produce an end-to-end proof of
correctness.

Another challenge in this context is correctly optimizing numerical code. For
this, HELIX uses a sequence of verified program transformations. The order in
which these optimizations are applied is determined by an (untrusted)
oracle---the previously existing SPIRAL framework~\citep{pueschel:05}---guides the
optimizer.  These transformations exploit a novel sparse-vector representation
amenable for algebraic reasoning.  HELIX also facilitates the use of
application-specific analyses that can be used to justify the numerical
precision properties that relate the ``real-valued'' computations of the source
algorithm to the ``floating-point'' computations of the target.

HELIX paves a novel path through this design space , and the key contributions
of this paper are:

\begin{enumerate}
\item We demonstrate how to build and verify in Coq a compilation
  pipeline for SPIRAL-like mathematical expressions down to LLVM IR.
  This pipeline involves a variety of intermediate representations,
  crossing from shallow, functional embedding down to a deeply
  embedded representation of imperative LLVM IR code.  In doing so, we
  combine vastly different approaches to compiler verification:
  oracle-based metaprogramming paired with rewriting-based translation
  validation, traditional semantic-preserving results between big-step
  interpreters, and bisimulation-based results between monadic
  embedding of languages.
\item We structure HELIX to facilitate the compilation to
  high-performance and numerically sound low-level code.  In
  particular, we develop a methodology to algebraically reason about
  partial computations using sparse vectors, which would enable
  downstream parallelization and vectorization.  Furthermore, we
  demonstrate how to apply domain-specific numerical analyses that
  provably characterize the precision of floating-point computation
  relative to its source semantics with respect to real values.
\item The final step of the compilation chain targets LLVM IR, the
  formalization of which is given by the preexisting Vellvm
  project~\citep{vellvmicfp}. This part of our development constitutes
  the largest stress test to date for both Vellvm's metatheory, as
  well as the underlying semantic foundations it relies on, i.e.,
  Interaction Trees.
\end{enumerate}

All our contributions are formalized and proven in Coq. The
development is openly available on
GitHub\footnote{\url{https://github.com/vzaliva/helix}}.

\subsection{HELIX overview}
\label{sec:helixintro}

With the current level of sophistication of hardware architectures,
the problem of high-performance implementation of numerical algorithms
becomes challenging for manual implementation even when using
optimizing compilers and is often solved by specialized code
generation systems, such as SPIRAL~\citep{pueschel:05}. Developed over
the last 20 years, the SPIRAL system has been used to generate,
synthesize, and autotune programs and libraries. It works by
translating rule-encoded high-level specifications of mathematical
algorithms into highly optimized/library-grade implementations. SPIRAL
has been used to formalize a variety of computational kernels from the
signal and image processing domain, including graph algorithms,
robotic vehicle control, software-defined radio (SDR), and numerical
solution of partial differential equations. SPIRAL is capable of
generating code for multiple platforms ranging from mobile devices and
multicore (desktop and server) processors to high-performance and
supercomputing systems~\citep{franchetti:18}.

When SPIRAL is applied to generate high-performance libraries used in
mission-critical software, the question arises as to what kind of
assurances could be made about the correctness of the generated
code. The goal of HELIX, as a part of the High Assurance SPIRAL
project~\citep{Franchetti:17,Low:17}, is to complement
SPIRAL-generated high-performance code with formally verified
correctness guarantees.

\begin{figure}[h]
  \centering
  \includegraphics[width=\columnwidth]{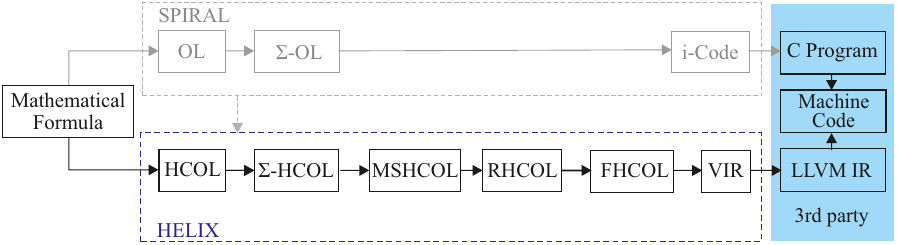}
  \caption{HELIX transformation stages}
  \label{fig:helixstages}
\end{figure}

At its core, \helix constitutes \emph{a verified compilation chain}, formalized in
the \coqlang proof assistant, establishing a theorem of semantic preservation
between high-level specifications of mathematical algorithms in a format similar
to the one taken by \spiral as input and \llvm-compliant code.
In order to achieve this result, \helix introduces a series of intermediate
languages (as shown at Figure~\ref{fig:helixstages}), lowering down progressively
the original expression.
Each translation step corresponds to the concretization of a new level of abstraction:

\begin{enumerate}
\item Mathematical formula
\item Dataflow (\HCOL\footnote{\HCOL stands for Hybrid Control
    Operator Language.})
\item Dataflow with implicit loops (\SHCOL)
\item Dataflow with implicit loops, using memory blocks (\MSHCOL)
\item Imperative program (\RHCOL)
\item Imperative program using machine numeric types (\textit{floats} and \textit{ints}) (\FHCOL)
\item Mainstream low-level assembly language (LLVM IR)
\end{enumerate}

Each language lowers the level of abstraction and introduces an
additional level of detail, narrowing down the space of possible
computational solutions for the given problem toward an exact
algorithm. An expression is not only converted between the languages
but a series of transformations is performed at each language
level. These transformations are optimization steps, performed in
algebraic form. For example, some computations on matrices could be
re-arranged into blocks fitting the target CPU SIMD instruction size.

Presently, HELIX works together with SPIRAL, which it uses as an
Oracle. SPIRAL, using its extensive library of (unverified) algorithms
and heuristics, decides how the original expression will be compiled
into optimal code. HELIX mirrors corresponding SPIRAL transformations,
adding formal verification assurances (translation validation
approach~\citep{pnueli1998translation}). The HELIX input language,
called \HCOL, is designed to be a formalization of SPIRAL's input
language (\OLL), and the input program can be easily syntactically
translated between the two.  The \HCOL language is very close to
mathematical notation and can represent a wide class of relevant
mathematical formulae. As a first step, SPIRAL attempts to deconstruct
the original expression into simpler expressions, which, combined by a
function composition, represent a data-flow graph of the
computation~\citep{Franchetti:2005:FLM:1065010.1065048}. The resulting
expression is then translated into another language, called {\SOL} which adds
the implicit representation of iterative computations. Next, the {\SOL}
expression is rewritten using a series of rewrite rules, driven by the
extensive knowledge base of SPIRAL's optimization algorithms, into a
shape which lends itself to generating the most efficient code for the
target platform. Subsequently, a {\SOL} expression is compiled into an
intermediate imperative language. By doing this, SPIRAL converts the
dataflow graph into a sequence of loops and arithmetic
operations. Finally, an intermediate imperative language
representation, after some additional transformations, yields a C
program which is compiled with an optimizing compiler, producing an
executable high-performance machine code implementation of the
original expression.

\begin{itemize}
\item The projects have different primary goals. SPIRAL's main objective
  is high-performance code generation, while HELIX focuses on high assurance.
\item There are two problems: 1) finding the optimal translation of a given
  formula to machine code for target architecture (search) and 2) verifying this translation. SPIRAL performs the
  former while HELIX is responsible for the latter. Currently, HELIX depends on
  SPIRAL as on a search oracle and just verifies SPIRAL results. In
  future, using HELIX as a foundation, some search functions could be
  transferred to it. This would open a door to formal verification of
  search/optimization algorithms instead of verifying their results.
\item The final steps in the SPIRAL system are to generate a program
  in C language and to compile it using a C compiler. In HELIX, we
  generate LLVM IR code, which verified compilation to machine code
  will be handled by 3rd party projects, like Vellvm. We feel that
  LLVM IR\footnote{The ``IR'' stands for \textit{intermediate
      representation.}} is a better intermediate language for machine
  code generation. The use of LLVM opens a few interesting
  possibilities, such as code generation for a wide list of hardware
  platforms via a growing list of LLVM backends and support for
  additional platform-specific optimization steps which could be
  verified using the same semantics framework. Instead of generating
  raw IR code, HELIX creates an AST in VIR, a Vellvm-supported subset
  of IR. It is used in verification and subsequently to pretty-print
  IR source code, which could be compiled with the LLVM IR compiler.
\item SPIRAL is implemented in a heavily modified~\cite{GAP4}
  computer algebra system, while HELIX is implemented in Coq proof
  assistant.
\item Originally, SPIRAL was built on the foundation of linear and
  multilinear operator theory. In early versions, all SPIRAL operators
  had to be multilinear. In later stages, SPIRAL added some
  support for non-linear operators~\citep{franchetti-SPL}. HELIX, from the
  very beginning, was designed without the linearity assumption and
  supports non-linear operators.
\item Due to the linear algebra lineage of SPIRAL, many of its core
  concepts were expressed using terminology and ideas from this
  field. HELIX, being a programming language (PL) project at its core,
  uses terminology and concepts from PL and type theories, formal
  methods, and functional programming. In other words, a computer
  science PL researcher will feel much at ease reading HELIX papers
  and code while an algebraist will find the SPIRAL literature more
  tractable. With this paper, we aim to bridge this gap.
\end{itemize}

In order to reason about the properties of programs in the languages
used in HELIX, we first need to formally define their syntax and
semantics.  HELIX is implemented in Coq with all its languages
embedded in Gallina. All translation steps are also implemented in Coq
or Template-Coq~\citep{sozeau2020metacoq}.

A quick summary of the languages shown in Figure~\ref{fig:helixstages} is
presented in Table~\ref{tab:languages}. We chose to embed all HELIX
languages in Coq Proof assistant~\citep{Coq}, which allowed us to
reason about them in Coq's Calculus of Inductive Constructions using
its \emph{Ltac} tactics language.

\begin{table}[h!]\centering
  \setlength{\tabcolsep}{4.5pt}
  \begin{tabular}{ l l l l l l }\toprule
    Name & Scalars & Vectors & Embedding & Paradigm & Error Handling  \\
    \midrule
    \HCOL & $\ctype$ & dense vector & shallow & declarative & no \\
    \SHCOL & $\ctype_\theta$ & sparse vector & mixed & functional & no \\
    \MSHCOL & $\ctype$ & memory blocks & mixed & functional & yes \\
    \RHCOL & $\R$ & env. + memory & deep & imperative & yes \\
    \FHCOL & IEEE double & env. + memory & deep & imperative & yes \\
    LLVM IR & IEEE double & env. + memory & deep & imperative & yes \\
    \bottomrule
  \end{tabular}
  \caption{Summary of HELIX languages}
  \label{tab:languages}
\end{table}

\yz{Need to think a bit more about the cohesion of this section: we alternate
  details of the nature of each IR with comparisons with Spiral and details
  about Coq.}
As can be seen from the table, the type representing numerical data
differs among the languages. We start from $\ctype$, which abstracts
$\R$ in \HCOL, and end up with IEEE floating-point numbers in LLVM
IR. Similarly, vector data representation also evolves. We start with
dense finite-size vectors as the main data type of \HCOL which, in the
final translation, are represented as memory blocks referenced by
variables in LLVM IR. Although all languages are embedded in Gallina,
different types of embedding are used, as shown in
Table~\ref{tab:languages}. Also, as we proceed down the translation
chain, we transition from purely functional to imperative
languages. Finally, Coq's dependent type system enforces the type correctness of
\HCOL, specifically in terms of the dimensionality of input and output
vectors. As a result, an \HCOL program that successfully typechecks
cannot trigger any runtime errors. However, once such a program is
translated into an imperative form, error handling is introduced.
The details of all these implementation aspects will be discussed for each
HELIX language in detail in \crefrange{sec:hcoltoplevel}{sec:dhcol}.

Aside from the general principle of lowering the abstraction level of
each step, the choice of languages was driven by several
considerations:
\begin{itemize}
\item The first two languages, \HCOL and \SHCOL, were designed as
  formalizations of SPIRAL \OLL and \SOL languages. They have to match
  these languages modulo syntax to allow us to use SPIRAL results in the
  translation validation approach.
\item The decision to use LLVM IR instead of C affected our choice of
  the previous language in the chain. Both in SPIRAL and HELIX, the
  last intermediate language is a high-level imperative language which
  is capable of expressing the algorithms we synthesize but is
  high-level enough to allow reasoning and proofs about these
  algorithms without going into lower-level details (like memory and
  pointers) of languages like C or IR. SPIRAL uses the {\tt i-Code}
  language for this purpose which is quite close to C. Since in HELIX,
  we no longer use C, we have more flexibility in defining this
  intermediate imperative language. The language we devised, \FHCOL,
  is indeed imperative but at a higher level of abstraction than {\tt
    i-Code}, more specialized, and better suited for formal reasoning
  and translation to IR. For example, it uses de Bruijn indices for
  variables, has logically scoped memory allocation, and employs a
  memory model similar \footnote{Both \FHCOL and
    \VIR use the Compcert-style memory model with integer-indexed
    memory blocks that contain integer-indexed memory cells. The only
    difference is that \FHCOL memory cells are floats, while VIR's are
    bytes} to the one used in \VIR.
\item The split between \RHCOL and \FHCOL is motivated by our desire
  to clearly define the boundary between abstract numeric types
  (\RHCOL) and concrete machine types (\FHCOL). To simplify the
  numeric analysis, we want these languages to be very similar,
  representing the same sequence of computation steps but for
  different types.
\item The \MSHCOL language was added for pure convenience to bridge the
  gap between functional \SHCOL and the imperative \RHCOL. Much of the
  proof for this transition has to do with representing \SHCOL
  vectors in memory. To simplify these proofs, we found it convenient
  to introduce the memory block abstraction first, while keeping the
  language functional.
\end{itemize}

All languages involved are embedded in Coq, but through different
kinds of embedding. We chose a \textit{shallow embedding} for \HCOL as it reflects its algebraic nature most
naturally. \SHCOL uses functional abstraction, which also fits well
with the shallow embedding (as Gallina is a functional language). It
is especially convenient to represent the operator families used in
iterative operators (see Section~\ref{sec:opfam}) as
functions. However, because of sparsity \yz{define?}, \SHCOL operators
need to carry some additional information---in particular, the
\textit{sparsity contract}. Thus, we chose the \textit{mixed
  embedding} where each operator is a record which contains a shallow
embedded operator implementation along with two sets representing the
sparsity contract. This form still allows us to use the \textit{setoid
  rewriting} technique\yz{Explain? Or push to much later.}.  \MSHCOL
representation is similar to that of {\SHCOL}. To express additional
logical properties of operators for all these languages, we used
\textit{typeclasses}. Finally, for \RHCOL, we switched to \textit{deep
  embedding}. This allowed us to clearly separate the language's
abstract syntax from its semantics and, in particular, to assign more
than one semantics to the language \yz{This is a bit mysterious to
  read at this point, should be explained}.

The LLVM IR language formalization which we use (provided by Vellvm
project~\citep{vellvmicfp}) is using the deep embedded representation
of IR syntax. In the following, we refer to the project
itself as Vellvm, and to the Coq formalization of LLVM IR that it provides as
VIR, for Verified IR.

\begin{figure}[H]
  \centering
  \includegraphics[width=\columnwidth]{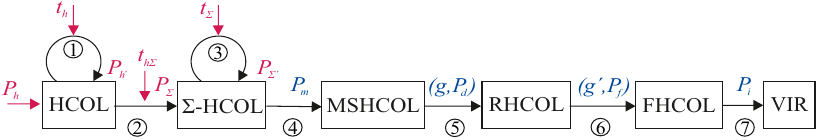}
  \caption{HELIX chain of verification}
  \label{fig:verificationchain}
\end{figure}

The sequence of verification steps in HELIX is shown in
Figure~\ref{fig:verificationchain} and briefly described below and
again in more detail in subsequent sections.

\noindent
Artefacts provided by SPIRAL are shown in red and HELIX-generated
artefacts are shown in blue:
\begin{description}[itemindent=1em,nolistsep]
\item[$P_h$] - source program in \HCOL
\item[$t_h$] - SPIRAL trace containing list of ``breakdown'' steps
\item[$P_{h'}$] - program in \HCOL after ``breakdown'' step (generated by SPIRAL)
\item[$t_{h\Sigma}$] - SPIRAL trace containing list of \OLL to \SOL translation steps
\item[$P_{\Sigma}$] - program in \SHCOL (generated by SPIRAL)
\item[$t_{\Sigma}$] - SPIRAL trace containing list of \SOL rewriting steps
\item[$P_{\Sigma'}$] - program in \SHCOL after rewriting step (generated by SPIRAL)
\item[$P_m$] - program in \MSHCOL
\item[$P_d$] - program in \RHCOL
\item[$g$] - list of global variables upon which $P_d$ depends along with their \RHCOL types
\item[$P_f$] - program in \FHCOL
\item[$g'$] - list of global variables upon which $P_f$ depends along with their \FHCOL types
\item[$P_i$] - AST of the program in LLVM IR
\end{description}

\noindent The numbered arrows in Figure~\ref{fig:verificationchain} depict the translation steps, listed below along with brief descriptions of how they are validated:

\begin{description}

\item[\textcircled{1}] The \HCOL ``breakdown'' step constitutes
  application of a sequence of semantics-preserving rewriting steps
  from $t_h$ to $P_h$ resulting in $P_{h'}$. Each element of $t_h$ corresponds
  to a single SPIRAL \OLL breakdown rule application. For each such
  rule, we formulate and prove a lemma. The semantic equivalence
  of $P_h$ and $P_{h'}$ is then proven by automated sequential
  application of breakdown rule lemmas. See
  Section~\ref{sec:hcolbreakdown} for details.

\item[\textcircled{2}] The \HCOL to \SHCOL translation step constitutes
  application of a sequence of semantics-preserving rewriting steps from
  $t_{h\Sigma}$ to $P_{h'}$ lifted to \SHCOL resulting in a full native (without lifted operators) \SHCOL version $P_{\Sigma}$. Each element
  of $t_{h\Sigma}$ corresponds to a single SPIRAL \SOL rewriting rule
  application. For each such rule, we formulate and prove a lemma. The semantic equivalence of lifted $P_{h'}$ and $P_{\Sigma}$ is
  then proven by automated sequential application of rewriting rule
  lemmas. Additionally, we prove the \textit{structural correctness} of the
  resulting expression. See Section~\ref{sec:hcol2shcol} for details.

\item[\textcircled{3}] After the initial \HCOL to \SHCOL translation, an
  additional translation step is performed by applying another series
  of rewriting rules from $t_{h\Sigma}$ to $P_{\Sigma}$ resulting in
  $P_{\Sigma'}$.  It is proven in the exact same manner as the previous
  step. See Section~\ref{sec:shcol2shcol} for details.

\item[\textcircled{4}] The \MSHCOL program $P_m$ is generated from
  $P_{\Sigma'}$ with the help of a Template-Coq metaprogram. For the resulting program, an {\tt
    MSHOperator\_Facts} instance is proven using proof automation.
  Then, the semantics preservation property is validated by proving
  an instance of an {\tt SH\_MSH\_Operator\_compat} typeclass for a $P_m$
  and $P_{\Sigma'}$ pair. This automated proof relies on {\tt MSHOperator\_Facts} as
  well as a proof of the structural correctness of the $P_{\Sigma'}$ generated
  during the previous step. See
  Section~\ref{sec:shcol2mhcol} for details.

\item[\textcircled{5}] The \RHCOL program $P_d$ is generated from $P_m$
  with the help of a Template-Coq metaprogram, which also produces a list $g$
  of the global variables $P_d$ depends upon. Then, the {\tt DSH\_pure} and {\tt
    MSH\_DSH\_compat} typeclass instances are proven using proof
  automation. See Section~\ref{sec:mhcol2rhcol} for details.

\item[\textcircled{6}] The \RHCOL to \FHCOL translation is implemented in
  Gallina. It translates program $P_d$ and a list of global variables
  $g$ to the corresponding $P_f$ and $g'$. This step is not validated.
  See Section~\ref{sec:rhcol2fhcol} for discussion of our reasoning.

\item[\textcircled{7}] The final step of translation from \FHCOL to
  LLVM IR is performed using a certified compiler that we wrote in
  Gallina. We have proven this compiler to be correct (with
  caveats). See Section~\ref{sec:compiler} for details.

\end{description}

Thus, given an original \HCOL expression, a SPIRAL trace file
containing the transformation steps, and the intermediate SPIRAL code
synthesis results (all shown in red in
Figure~\ref{fig:verificationchain}), HELIX will either fail or
provide the following high-level results: \yz{We'll have to make clear which
  manual, input-specific, work needs to be provided}

\begin{enumerate}
\item LLVM IR version of the \HCOL program, which includes
  SPIRAL-guided code optimizations.
\item Correctness guarantee that the IR
  program on IEEE floating-point inputs that do not contain NaN values
  will generate the same results as the \HCOL program on real numbers
  up to guarantees provided by the numeric analysis step.
\end{enumerate}

If HELIX fails, it could be that SPIRAL-generated steps and
intermediate results are either inconsistent, use unproved rules, or
that the HELIX LLVM compiler is missing the functionality required to
compile the given expression. The first indicates a bug in the SPIRAL
system, which needs to be fixed. The latter two reasons point to HELIX
implementation shortcomings that could be cured by extending HELIX
with additional rules or the compiler capabilities using the existing
framework.

We discuss how these results are achieved in subsequent sections.

\subsection{Motivating example}
\label{sec:motivatingexample}

\begin{figure}[h]
  \centering
  \includegraphics[width=\columnwidth]{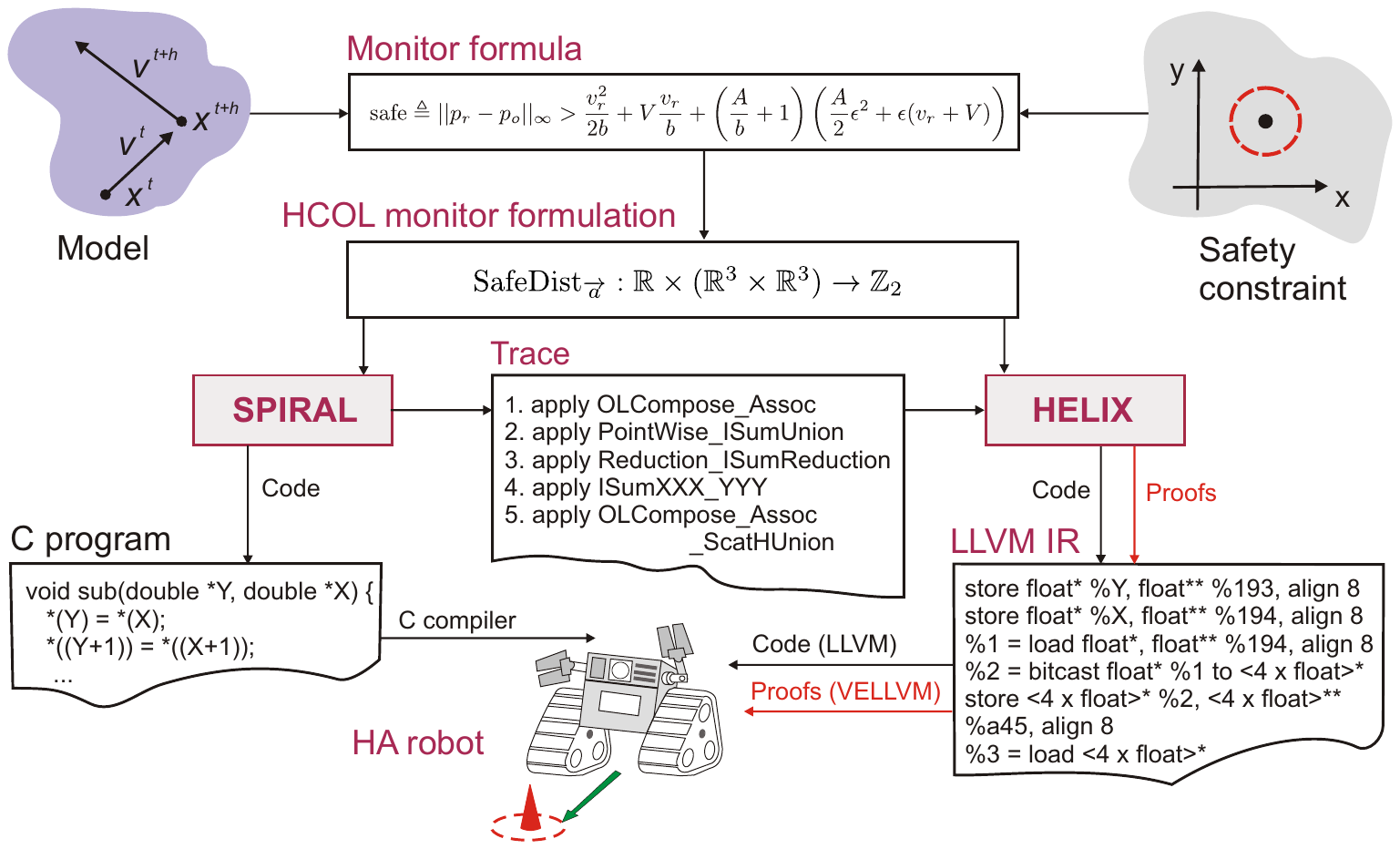}
  \caption{HELIX application to high-assurance vehicle control}
  \label{fig:bigpic}
\end{figure}

Let us consider an application of HELIX to a real-life situation of
high-assurance vehicle control~\citep{Franchetti:17} using a dynamic
window vehicle control approach~\citep{fox1997dynamic}, as shown in
Figure~\ref{fig:bigpic}.

Given a physical model of vehicle dynamics, we generate a code to check a safety
constraint. In this example, we will be checking whether it's safe to proceed
given the vehicle's position, speed, and acceleration, with the location of an
obstacle. The function will be invoked by the vehicle controller, and if it
returns ``false,'' the vehicle will enter ``fail-safe'' mode, which would
trigger emergency braking.

We will consider a ground robot driving on a flat, even surface. The robot is
equipped with a distance measuring sensor (e.g. Lidar) which can detect and
measure distance to obstacles. The obstacle sensor is sampled periodically, for
example every 20ms and based on these measurements, the decisions on control
outputs, such as steering, acceleration, and braking, are made. Once such
decisions are applied to actuators controlling the robot, no further control
changes can be made until the next sampling cycle. The obstacle does not have to
stay static and can move.

This model allows the development of a control system which will help a robot
avoid obstacles by stopping or steering around them. However, in this example,
we will be considering only a \textit{safety monitor} part of the system which
ensures the \textit{passive safety} property. Informally, that means there will
be no collisions while the robot is driving, and collisions may occur only if
the obstacle runs into the robot.

Without discussing detailed dynamic model of the system for which we
refer interested readers to~\cite{Franchetti:17}, we present the final
formula for a dynamic window safety monitor in
Equation~\eqref{eq:dynwin}. The soundness of the monitor formula from the
point of view of cyber-physical control systems was proven in KeYmaera~X~\citep{fulton2015keymaera}.
\begin{equation}
  \mathrm{safe} \triangleq || p_r - p_o||_\infty > \frac{v_r^2}{2b}+V\frac{v_r}{b}+\left( \frac{A}{b} + 1 \right)
  \left(
    \frac{A}{2}\epsilon^2+\epsilon(v_r+V)
  \right)
  \label{eq:dynwin}
\end{equation}

The variables used in the safety monitor formula are:

\begin{description}[leftmargin=1cm,labelindent=2cm,noitemsep]
  \item[$p_r$] - robot position
  \item[$p_o$] - obstacle position
  \item[$v_r$] - robot longitudinal speed
  \item[$V$] - maximum obstacle speed
  \item[$b$] - maximum braking (negative acceleration)
  \item[$A$] - maximum acceleration (positive)
  \item[$\epsilon$] - sampling period
\end{description}

To complete the high-assurance chain for this monitor, we would like to
have its implementation as machine code suitable for execution
onboard, which is proven to be correct with respect to the mathematical
formulation. Additionally, since this monitor will be executed in
real-time and at high frequency, we would like the synthesized
implementation to be efficient with respect to the target hardware.

We will use HELIX to build such an implementation, while applying and certifying
optimization steps produced by SPIRAL. Both HELIX and SPIRAL will start with the
same input formula. Unlike SPIRAL, HELIX does not make decisions about what
transformation steps to perform, but rather it relies on SPIRAL's trace for the
list of steps to perform. However, in HELIX, each of these steps is backed up by
a formally proven lemma which guarantees that it preserves the semantics of the
expression. HELIX will re-create SPIRAL's transformation of the expression, but
this time, each step, and as a result, the whole sequence of transformations,
will be formally proven to be correct. In the unlikely event that SPIRAL would
suggest a non-semantically preserving transformation, HELIX will reject this
step and report an error.

Finally, the optimized expression will be compiled by HELIX into LLVM
IR assembly language (as shown in Listing~\ref{lst:dynwinllvm}). This
compilation will also be proven correct, and the generated code
will be guaranteed to correctly correspond to the computations described
by the input expression.

This final correspondence cannot be exact, due to the bounds of numerical
accuracy arising from the use of floating-point calculations instead of
the real precision of the original formula. Instead, HELIX provides the means for
proving arbitrarily complex, abstract properties of the compilation, without
focusing of numerical precision alone. In the case of \emph{DynWin}, we take
the high-level statement of ``robot safety'', and, relying on
Equation~\ref{eq:dynwin} being a correct mathematical model of it, prove that,
when operating on the algorithm compiled through HELIX, the robot will in fact
remain ``safe'' at all times.

The LLVM IR program could be further compiled by the LLVM compiler toolchain,
into machine code for the target hardware platform. At this time, there are
no formally verified IR compiler backends, but they are included in
the roadmap of the DeepSpec project~\citep{appel2017position}.

As shown by this example, HELIX provides a framework of automated
translation from a mathematical formula to an efficient and formally
verified implementation in machine code.

\section{\HCOL}
\label{sec:hcoltoplevel}

\subsection{\HCOL language}
\label{sec:hcol}
HELIX {\HCOL} language is based on the SPIRAL \OLL language which was
originally designed to represent linear algebra expressions on real or
complex vectors. The primitive \HCOL operators are functions from
vectors to vectors. Higher-order operators, such as function
composition, allow the building of more complex \HCOL expressions.

Since HELIX uses SPIRAL as an oracle, the \HCOL and \OLL languages must
be compatible. \HCOL is a formalization of \OLL, shallow embedded in
Coq. Any well-formed \OLL expression could be trivially mechanically
translated into the corresponding \HCOL expression. Unlike \OLL, \HCOL
expressions are always well-formed as we use Coq's powerful dependent
type system to enforce this. Thus, for example, vector dimensions will
always match when constructing complex \HCOL expressions from
elementary operators. Similarly, constants and arithmetic expressions
representing indices of vector elements will be properly bound.

By varying the dimensions of vectors, \HCOL could represent computations with different levels of granularity. The
structure of such expressions represents the dataflow graph of
computation. By applying a set of rewriting rules, an {\HCOL}
expression could be gradually ``broken down,'' synthesizing the dataflow to match the hardware architecture of the target system.

{\HCOL} is a shallow-embedded language in Coq proof assistant. All
{\HCOL} operators are represented as functions in Coq's host
language, \textit{Gallina}. The limitation of this approach is that
operators must be \textit{total functions} which can be proven to
terminate. This poses no problem in practice, as all \HCOL
operators we defined fit this definition. The following are the
data types used in {\HCOL}:

\begin{description}[style=unboxed,leftmargin=0cm]
  
\item[The \textit{Carrier Type}:] Unlike {\OLL}, the main data type is
  abstract instead of using concrete types, such as $\R$ or $\mathbb{C}$. \HCOL $\ctype$ is
  an abstract representation of such a numeric type, expressed in terms of
  its algebraic properties. See description below in
  Section~\ref{sec:carriera} for details.
\item[Finite-dimensional Vectors:] To represent vectors, we use the
  inductively-defined {\tt Vector} type from Coq's standard
  library. Vector elements have type $\ctype$. We will use Coq
  notation {\tt avector n} or mathematical notation $\ctype^n$
  interchangeably to describe vectors of $\ctype$ of length $n$.
\item[Finite natural numbers:] Some \HCOL operators use finite natural
  numbers as vector offsets. They are upper-bounded to ensure not to exceed the
  expected target vector size. They are encoded using Coq's {\tt sig}
  type to represent a number along with the proof that it has passed the
  range check. In this paper, we sometimes use the shorthand notation
  $\finNat{n}$ to denote $\{x:\mathbb{N} \mid x<n\}$ type.
\end{description}

The dimensions of input and output vectors of an {\HCOL} operator are
encoded as indices of the {\tt vector} type family, and vector type
$\ctype^n$ corresponds to \coqe{(vector $\ctype$ n)} in Coq. When
constructing a complex {\HCOL} expression, Coq's type system ensures
that the dimensions of all components match.

\subsubsection{Carrier type}
\label{sec:carriera}
This is an abstract representation of a numeric type, expressed in
terms of its operations and algebraic properties. Definitions and
proofs formulated for the carrier type can be applied, for example, to
$\R$, $\Q$, or $\Z$, as they satisfy these properties.

We denote the carrier type as $\ctype$. To make it abstract we define
it in a type class {\tt Carrierefs} which also postulates the
existence of the following for this type (via corresponding typeclass
instances):

\begin{itemize}[nolistsep]
\item Equality: {\tt Equiv} (see discussion below in Section~\ref{sec:eq}.).
\item Comparisons: {\tt Lt}, {\tt Le}.
\item Decidability of the equality and comparisons: {\tt Decision
    (x=y)}, {\tt Decision (x<y)}.
\item Constants: {\tt Zero}, {\tt One}.
\item Operators: {\tt Plus}, {\tt Mult}, {\tt Negate}, {\tt Abs}.
\end{itemize}

Additional abstract algebra properties of $\ctype$, expressed using
the corresponding typeclass instances from the \emph{MathClasses}
library~\citep{spitters2011type} are defined in a separate typeclass
{\tt CarrierProperties}:

\begin{itemize}[nolistsep]
\item Setoid Equality: {\tt Setoid} (see discussion below in Section~\ref{sec:eq}.).
\item Abstract algebra structures: {\tt Ring}.
\item Ordering: {\tt TotalOrder}, {\tt StrictSetoidOrder}, {\tt FullPseudoOrder}.
\item Inequality of constants: {\tt Zero $\neq$ One}.
\end{itemize}

The reason for splitting $\ctype$ properties into two typeclasses is
to minimize the burden of typeclass resolution. The operators
definitions require only basic properties from {\tt CarrierDefs},
whereas proofs of \HCOL program transformations add additional
typeclass constraint {\tt CarrierProperties} as a pre-condition.

In other words, we require that $\ctype$, along with the corresponding
operations, forms an algebraic ring, has a total ordering, and has
decidable \textit{equality}. Some operators do not require all these
properties, but to be able to construct homogeneous {\HCOL}
expressions, the carrier type imposes the superposition of all the
constraints required by all {\HCOL} operators. Additionally, it is
assumed that the $\ctype$ type is populated with some special values, like
{\tt zero} and {\tt one}.

\subsubsection{Equality}
\label{sec:eq}

The definition of equality is essential for {\HCOL} operator
rewriting. The Coq default notion of equality ({\tt eq}) is too
restrictive for our purposes. For example, it would not allow us to
work with rational numbers represented by non-reduced integer
fractions. Depending on what concrete type is used in place of the
abstract carrier type, we would like to be able to define a meaningful
equality relation for this type. In general, we would like to work on a
carrier type equipped with an equivalence relation, which is also
called a \emph{setoid}.

Operational typeclass {\tt Equiv} defines an {\tt equiv} relation for a
given type. its subclass, {\tt Setoid}, additionally requires this
relation to be an \textit{equivalence relation} by presenting proofs
that it is transitive, commutative, and reflexive.

Since we have declared our carrier type $\ctype$ to be an instance of a
{\tt Setoid} typeclass, we can define {\tt equiv} for vectors of this
type as a \textit{pointwise relation} which makes them also a setoid:
\coqinline{forall (n:nat), Setoid (vector$\,\ctype$ n)}.

From that follows the natural definition of the {\HCOL} operator
\textit{extensional equality} which states that two operators $F$ and
$G$ are equal if for all possible input vectors $x$, the values of
$(F\,x)$ and $(G\,x)$ are also equal.

As we work mostly with setoid equality instead of Coq's
standard equality, we follow the MathClasses library convention of using the
$(=)$ notation to refer to the {\tt equiv} relation instead of Coq's
standard {\tt eq}. To refer to standard Coq's {\tt eq} equality, we
use the notation $(\equiv)$. These notations are followed throughout this paper.

\subsubsection{{\tt HPointwise} operator}

Here is an example of an \HCOL operator, {\tt HPointwise}. The full list of
\HCOL operators is shown in Appendix~\ref{sec:hcoloperators}.

\paragraph*{\underline{\underline{$\hoperator{HPointwise}\,
  (n \colon \N)\,(f \colon\ \finNat{n} \rightarrow \ctype \rightarrow \ctype)\colon\
  {\ctype^n \rightarrow \ctype^n}$}}\label{sec:HPointwiseExample}}

\begin{figure}[h]
  \centering
  \includegraphics[keepaspectratio=true]{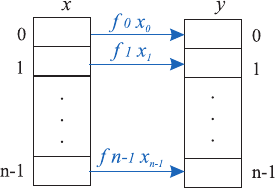}  
  \caption{{\tt HPointwise} operator}
  \label{fig:examplehpointwise}
\end{figure}

This operator applies a function $f$ to each element of the input
vector, as shown in Figure~\ref{fig:examplehpointwise}. The function
$f$ takes two arguments: the element's index and its value. The output
is the vector of the same length as the input vector.

Let us look more closely at the definition of the {\tt HPointwise}
operator in Coq:

\begin{lstlisting}[language=Coq, mathescape=true,basicstyle=\ttfamily\footnotesize,columns=flexible,label={lst:hpointwise},caption={{\tt HPointwise} operator definition}]
Definition HPointwise {n: nat} (f: $\finNat{n}$ -> $\ctype$ -> $\ctype$) (x: vector $\ctype$ n) : vector $\ctype$ n
  := Vbuild (fun j jd => f (mkFinNat jd) (Vnth x jd)).
\end{lstlisting}

The operator is implemented using the {\tt Vbuild} function from the
\emph{CoLoR} librar~\citep{blanqui2011color}. The definition is fairly
straightforward; {\tt HPointwise} generates a vector of length $n$ where an element
with index $j$ is the result of the application of the $j$-th function
from family $f$ to the input vector $x$.

\subsubsection{Running example}

An \HCOL formulation of the dynamic window monitor expression
\eqref{eq:dynwin1} introduced in Section~\ref{sec:motivatingexample}
is shown in Listing~\ref{lst:dwhcol}.

\dynwinlisting[language=Coq,label=lst:dwhcol,
caption={Dynamic Window Monitor in \HCOL},
basicstyle=\fontsize{8}{8}\ttfamily]{dynwin_hcol.v}

We will use it as our running example, showing how it is translated at
each step of the HELIX compilation and verification pipeline.

\subsection{\HCOL breakdown}
\label{sec:hcolbreakdown}
At the first translation step, HELIX performs semantically preserving
modifications of the \HCOL expressions. The goal is to break down more
complex operations into elementary ones, representing a fully terminated
computation dataflow graph optimized for target hardware taking into
account such target architecture parameters as number of words in SIMD
instructions, number of cores, and CPU cache size. The breakdown
steps are determined by SPIRAL and validated by HELIX. Each step is
implemented in SPIRAL as an application of a ``breakdown rule.''
Per the translation validation approach discussed earlier, we
prove a lemma for each rule and apply them following the trace
generated by SPIRAL. Application of the breakdown rules is done using
\textit{setoid rewriting}~\citep{Sozeau2010} together with \HCOL operator
equational theory.

The result of the \HCOL rewriting steps of the expression from
Listing~\ref{lst:dwhcol} is shown in Listing~\ref{lst:dwhcol1}

\dynwinlisting[language=Coq,label=lst:dwhcol1,
caption={Dynamic Window Monitor in \HCOL after rewriting},
basicstyle=\fontsize{8}{8}\ttfamily]{dynwin_hcol1.v}

\begin{figure}[H]
  \centering
  \includegraphics[width=\textwidth]{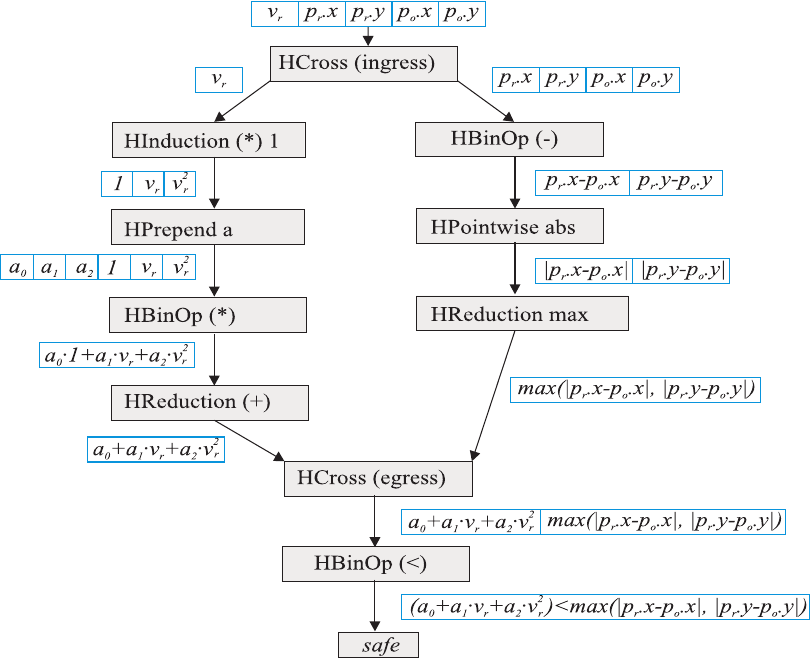}
  \caption{Dynamic Window Monitor dataflow graph in \HCOL}
  \label{fig:dataflow}
\end{figure}

It corresponds to the dataflow graph shown in
Figure~\ref{fig:dataflow}.

\subsubsection{Breakdown rules as lemmas}
\label{sec:breadownrule}

SPIRAL breakdown rules are expressed as lemmas in HELIX. Each lemma
states the equality (using {\tt equiv} relation) between two \HCOL
expressions. Typically, the lemma is used in a left-to-right direction in the {\tt
  setoid\_rewrite} Coq tactic, replacing the left hand side expression
with the right hand side equivalent. Let us look at a couple of rules.

\subsubsection{{\tt HScalarProd} breakdown rule}

Here is an example of a SPIRAL breakdown rule and the corresponding
lemma in HELIX for the {\tt HScalarProd} operator. See Appendix~\ref{sec:hcoloperators}
for the description of all \HCOL operators.

The {\tt HScalarProd} operator
calculates the dot product of two vectors. The input vectors are
concatenated and passed as a single vector of size $n+n$. The result
is returned as a single-element vector. For an input vector
$\vec{x} = [x_0, x_1, \dotsc, x_{n+n-1}]$, it computes
[$x_0 \cdot x_n + x_1 \cdot x_{n+1} + \dotsb + x_{n-1} \cdot
x_{n+n-1}]$.

The SPIRAL breakdown rule for this operator states that it could be
represented as a composition of {\tt HReduction} and {\tt HBinOp}
operators. First, {\tt HBinOp} is applied multiplying corresponding
elements in the first and second halves of the input vector producing, as an
intermediate result, a vector of size $n$ with values
[$x_0 \cdot x_n, x_1 \cdot x_{n+1} , \dotsc , x_{n-1} \cdot
x_{n+n-1}]$. The operator takes as parameter $f$ a function with
type
$\finNat{n} \rightarrow \ctype \rightarrow \ctype \rightarrow \ctype$
which will be applied to an index and a pair of elements from the
first and the second halves of the vector. Since in this case, we only
want to multiply elements without using the index, we wrap the multiplication
function in {\tt IgnoreIndex2}, which discards the first argument, and
use it as the parameter $f$. 

Next, with {\tt HReduction (+) 0}, we compute
$[0 + x_0 \cdot x_n + x_1 \cdot x_{n+1} + \dotsb + x_{n-1}
\cdot x_{n+n-1}]$ as a right fold. The corresponding lemma and the definition of the helper function are shown in
Listing~\ref{lst:scalarbreak}.

\begin{lstlisting}[language=Coq,
  mathescape=true,basicstyle=\ttfamily\footnotesize,columns=flexible,caption={{\tt
      HScalarProd} breakdown rule},label={lst:scalarbreak}]
Definition IgnoreIndex2 {A B:Type} (f:A -> A -> A) := const (B:=B) f.

Fact breakdown_OScalarProd: forall {n:nat},
    HScalarProd (n:=n) = HReduction  (+) 0 $\circ$ HBinOp (IgnoreIndex2 mult).
\end{lstlisting}

\subsubsection{\HCOL semantics preservation verification framework}
\label{sec:rewriting}

We define our semantics preservation property as an equivalence
relation on \HCOL expressions. To prove that \HCOL expression $A$ could
be broken down into \HCOL expression $B$ while preserving its
semantics, we need to prove $A = B$.

In the case of simple operators, we can just prove a lemma
stating the equality of the two exact expressions. For complex
expressions consisting of a composition of multiple operators, such
proof can be performed in a series of automated steps. Each step corresponds to an application of a ``breakdown
rule'' modifying all or a part of an expression. For each rule,
there is a lemma in the form $A=B$. It is applied using the {\tt
  setoid\_rewrite} tactic, which searches the current expression for patterns matching $A$ and replaces their occurrences with $B$. Because $(=)$
relation is transitive, proving each rewriting step will guarantee the equality
between the initial expression and the results of an application of a
sequence of rules. The rewriting rules in the HELIX library must be manually
proven once but after that, these proofs can be reused to
automatically prove the correctness of any sequence of their
applications.

For example, to prove that $A \circ B = D \circ E$, we may first apply
rule $A=D$ to get $D \circ B = D \circ E$ and then apply rule $B=E$
to $D \circ E = D \circ E$, which is true because
our equality is reflexive.

To make this machinery work, we need to impose some additional
requirements on operators. This is done by making them all instances
of the {\tt HOperator} typeclass:

\begin{lstlisting}[language=Coq,
  mathescape=true,basicstyle=\ttfamily\footnotesize,columns=flexible,caption={{\tt
      HOperator} class}] 
Class HOperator {i o:nat} (op: vector $\ctype$ i -> vector $\ctype$ o) :=
    op_proper :> Proper ((=) ==> (=)) op.
\end{lstlisting}

Currently, this typeclass does not contain any additional fields
except the one it inherits from the {\tt Proper} typeclass, which is
required for the {\tt setoid\_rewrite} tactic to work. The theory of
generalized setoid rewriting and related typeclasses is discussed
in~\cite{Sozeau2010}. Informally, it could be said that this {\tt Proper}
typeclass instance guarantees that for any two inputs of an operator
that are related (via {\tt equiv} relation), the results of respective applications of the operator to these inputs will also be in the same relation.

Breakdown rule proofs frequently make use of the algebraic properties of $\ctype$ and
linear algebra identities.

The key techniques of our semantics preservation
verification framework for {\HCOL} rewriting are:

\begin{itemize}
\item Abstract the data type on which {\HCOL} operates as carrier type $\ctype$.
\item Assume an equivalence relation $(=)$ on $\ctype$.
\item Assume some algebraic properties of $\ctype$. 
\item Define $(=)$ on vectors of $\ctype$ as a pointwise relation.
\item Define \HCOL operators as functions from vectors to vectors (of a carrier type) which are instances of the {\tt HOperator} typeclass.
\item Define extensional equality of \HCOL operators.
\item Define breakdown rules as lemmas stating equality between \HCOL expressions.
\end{itemize}

Using this framework, given the original and the final \HCOL
expressions, $h$ and $h'$, and the trace (list) of breakdown rules
applied to get from $h$ to $h'$, the HELIX \HCOL rewriting proof
engine can prove that an applied sequence of breakdown rules is
\textit{semantically preserving} and that $h = h'$.

\begin{figure}[h!]
  \centering
  \includegraphics{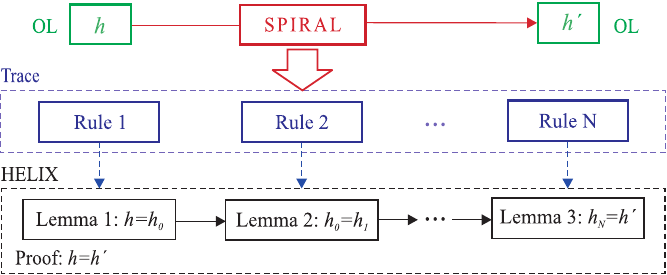}
  \caption{HELIX translation validation of SPIRAL {\HCOL} rewriting}
  \label{fig:translationvalidation}
\end{figure}

This approach, shown in Figure~\ref{fig:translationvalidation}, is an
extension of the \textit{translation validation} technique. A sequence
of rewriting steps is generated outside of HELIX by SPIRAL. Instead of
proving that SPIRAL will always transform an expression correctly,
HELIX formally verifies the correctness of the produced results. Given
that SPIRAL and HELIX use the same library of breakdown rules, the
proof of the goal $h = h'$ is a sequence of applications of setoid
rewrites using already proven per-rule lemmas from the HELIX
library. We can automatically generate such proof from the trace and
if Coq accepts it, the rewriting is proven correct. If, for some
reason, the trace contains a non-semantically preserving rewriting
sequence, Coq will not accept the proof.

\section{\SHCOL}

\subsection{\SHCOL language}
\label{sec:shcol}
Most vector and matrix operations are naturally expressed as iterative
computations on their elements. To generate machine code for such
computations, we transform our expressions into a form where these
iterations become explicit.

The goal of our next language, \SHCOL is to represent algebraically
iterative computations on vectors. Where \HCOL operates on whole
vectors, \SHCOL allows for finer granularity introducing operations
on individual elements.

\begin{figure}[h]
  \centering
  \includegraphics[height=2.5cm]{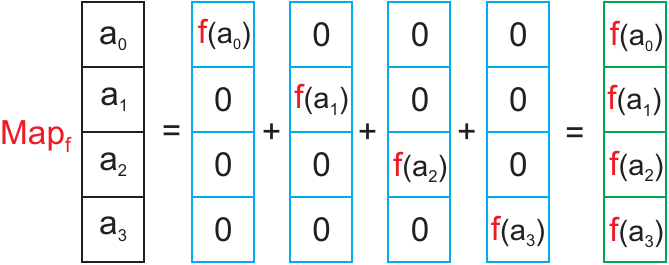}
  \caption{Dense vector as sum of four sparse vectors}
  \label{fig:diag}
\end{figure}

An iterative computation on vectors can be viewed as superposition of
computations performed during each step which processes only a subset
of elements. The vector positions not used during an iteration step
can be left undefined. This can be represented naturally with sparse
vectors. For example, an element-wise function application to elements
of a dense vector can be represented as the sum of columns of a
diagonal sparse matrix, as shown in Figure~\ref{fig:diag}. In this
example, for simplicity, we use $\ctype^n$ type to represent sparse
real-valued vectors of length $n$ and assume that sparse cells hold a
special \textit{structural zero} value, which is treated as regular
zero under addition. Later in this section, we will give a more formal
treatment of how we represent and reason about sparsity in HELIX.

\begin{figure}[H]
  \centering
  \includegraphics[height=2.5cm]{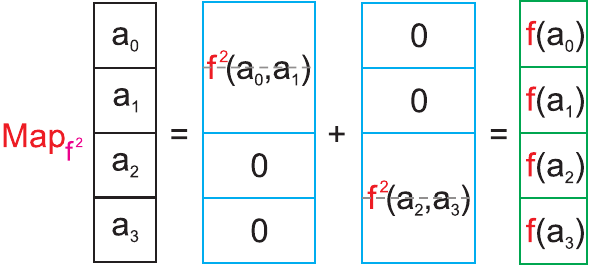}
  \caption{Dense vector as sum of two sparse vectors}
  \label{fig:diag2}
\end{figure}

Assuming that $f$ is implemented in C as \lstinline[language=C,mathescape=true]!void f($\ctype$ *src, $\ctype$ *dst)!,
it roughly corresponds to the following loop:
\begin{lstlisting}[language=C,mathescape=true, basicstyle=\ttfamily\footnotesize]
  for(int i=0;i<4;i++)
  {
    $\color{red}f$(src+i,dst+i);
  }
\end{lstlisting}

\noindent which requires four iterations. If we have a vectorized implementation of
$f^2$ with type $f^2: \R^2 \rightarrow \R^2$ which is implemented in C as \lstinline[language=C,mathescape=true]!void f2($\ctype$ src[2], $\ctype$ dst[2])!, the sum would look like Figure~\ref{fig:diag2}.

It roughly corresponds to the following loop, which now requires only two iterations:
\begin{lstlisting}[language=C,mathescape=true, basicstyle=\ttfamily\footnotesize]
  for(int i=0;i<2;i++)
  {
    $\color{red}f2$(src+2*i,dst+2*i);
  }
\end{lstlisting}

In these examples, $f$ can be viewed as an abstraction for a scalar CPU
instruction, such as x86 \emph{FADD}, and $f2$ can be a SIMD version of
it, similar to \emph{ADDPS} x86 SSE instruction.

In essence, sparsity allows us to represent partial computations. For
instance, in Figure~\ref{fig:diag}, we use an operation
(e.g. addition) and the default value for the sparse elements
(e.g. zero), which form a \textit{monoid}, to represent superposition
of partial computations algebraically. Maintaining algebraic
abstraction allows us to transform and to prove equality of operations
on vectors, representing various computation flows.

\subsubsection{Modelling missing values and collisions}
\label{sec:sparsitytracking}

In our formalization, each sparse vector's element could be either an
actual value or a \textit{structural} value. One can think about a
structural value as an empty cell with a default placeholder value
assigned to it.

To ensure proper factorization of a complex computation into
superposition of elementary ones, we need to make sure that the
calculation of each vector's element is assigned to exactly one
elementary computation. That means that when combining vectors
representing the results of two computations, we should never combine two
non-empty vector elements. When combining vector
elements pairwise, one of the values in each pair must be structural. If both
values are non-structural, we call this a \textit{collision}, which indicates
there are two conflicting paths trying to perform computation of the same
value. Any collision detected should be tracked down the computation tree, and
any operation with a value produced as a result of that collision should be
marked as colliding as well. Normally, a well-formed \SHCOL expression triggers
no collisions (see Section~\ref{sec:structuralproperties} on how we ensure
this).

The first, na\"ive approach to tracking structural and collision flags is
to use a product type which contains the actual value and two flags:
$\ctype\times \B \times \B$. However in many situations, we only care
about actual values and want to avoid dealing with structural flags
implicitly. Our solution is to use the \emph{Writer Monad} to track
structural properties of carrier type $\ctype$, as described below.

\begin{lstlisting}[language=Coq, mathescape=true,basicstyle=\ttfamily\footnotesize]
Record $\ctype_{\mathrm{flags}}$ : Type := mk$\ctype_{\mathrm{flags}}$ {is_struct: bool ; is_collision: bool}.
\end{lstlisting}

In the snippet above, we first define a record data type
$\ctype_{\mathrm{flags}}$ which holds structural and collision
flags. To use these flags in Writer Monad (using
\textit{ExtLib}~\citep{extlibgithub} library), we also need to define a
\emph{Monoid} record:

\begin{lstlisting}[language=Coq, mathescape=true,basicstyle=\ttfamily\footnotesize,
  caption=Exclusive union monoid,label=lst:unionflags]
Definition mzero := mkRthetaFlags $\top$ $\bot$.
Definition mappend (a b: $\ctype_{\mathrm{flags}}$) : $\ctype_{\mathrm{flags}}$ :=
    mkRthetaFlags
      (is_struct a && is_struct b)
      (is_collision a || is_collision b ||
                     (negb (is_struct a || is_struct b))).
Definition Monoid_$\ctype_{\mathrm{flags}}$ : Monoid $\ctype_{\mathrm{flags}}$ := Build_Monoid mappend mzero.
\end{lstlisting}

The initial flags' value called \textit{mempty} has the structural flag
\emph{true} and the collision flag \emph{false}. The \textit{mappend} operation
shown in Listing~\ref{lst:unionflags} is used to combine the two sets of flags.
It works as follows. If one of the operands is non-structural, the result is
also non-structural. The collision flags are "sticky;" once set for either
operand, they are propagated into the result. Finally, attempting to combine two
non-structural elements raises a new collision.

It can be shown that the following Monoid laws are satisfied:
\begin{align}
\label{eq:monoidpz}
 \forall a,&\quad a\oplus 0 = 0 \oplus a = a \\
\label{eq:monoidpa}
\forall a,\forall b,\forall c,&\quad (a\oplus b) \oplus c = a \oplus (b \oplus c)
\end{align}
 
Here, we use $0$ for \textit{mempty} and $\oplus$ for
\textit{mappend}. In Coq, we just declare an instance of the
\emph{MonoidLaws} typeclass for our newly defined Monoid record and
prove properties from the Equations~\eqref{eq:monoidpz} and
\eqref{eq:monoidpa}.

\begin{lstlisting}[language=Coq, mathescape=true,basicstyle=\ttfamily\footnotesize]
Instance MonoidLaws_$\ctype_{\mathrm{flags}}$: MonoidLaws Monoid_$\ctype_{\mathrm{flags}}$.
\end{lstlisting}

To track the flags while performing operations on the values of type
$\ctype$, we use Writer Monad, parametrized by a Monoid which
defines how flags will be handled:

\begin{lstlisting}[language=Coq, mathescape=true,basicstyle=\ttfamily\footnotesize]
Variable fm:Monoid $\ctype_{\mathrm{flags}}$.
Definition Monad_$\ctype_{\mathrm{flags}}$ := writer fm.
Definition $\ctype_{\mathrm{fm}}$ := Monad_$\ctype_{\mathrm{flags}}$ $\ctype$.
\end{lstlisting}

To construct values of the type $\ctype_{\mathrm{fm}}$, we define two
convenience functions:

\begin{lstlisting}[language=Coq, mathescape=true,basicstyle=\ttfamily\footnotesize]
Variable fm:Monoid $\ctype_{\mathrm{flags}}$.
Definition mkStruct (v:$\ctype$) : $\ctype_{\mathrm{fm}}$ fm := ret v.
Definition mkValue (v:$\ctype$) : $\ctype_{\mathrm{fm}}$ fm := tell (mk$\ctype_{\mathrm{flags}}$ $\bot$ $\bot$) ;; ret v.
\end{lstlisting}

For most \SHCOL operators, we are interested in the
$\ctype_{\mathrm{flags}}$ type parametrized by a Monoid with
\textit{exclusive union} as a {\tt mappend} operation, as shown in
Listing~\ref{lst:unionflags}. We will call this type $\rtheta$. A
commonly used constant $\pmzerodot$ of this type holds $\ctype$ ring's
\textit{additive identity} ($0$) as a value and has the structural flag
set and the collision flag unset:

\begin{lstlisting}[language=Coq, mathescape=true,basicstyle=\ttfamily\footnotesize]
Definition $\rtheta$ := $\ctype_{\mathrm{fm}}$ Monoid_$\ctype_{\mathrm{flags}}$.
Definition $\pmzerodot$ : $\rtheta$ := mkStruct 0.

\end{lstlisting}

To illustrate how Writer Monad is used to track the flags, let us examine how they are combined when a binary operation is performed on
underlying values. Recall the \emph{execWriter} will return the flags value of type $\ctype_{\mathrm{flags}}$ for a given monadic
value of type $\rtheta$. To apply the binary operation {\tt op} to the underlying $\ctype$ values, we use \emph{liftM2} to
promote it to a monad. Now, just by unfolding the underlying definitions,
it could be trivially shown that:

\begin{lstlisting}[language=Coq, mathescape=true,basicstyle=\ttfamily\footnotesize]
forall (op: $\ctype$ -> $\ctype$ -> $\ctype$) (a b: $\rtheta$),
  execWriter (liftM2 op a b) =  mappend (execWriter a) (execWriter b).
\end{lstlisting}

In other words, that arguments' flags will be combined using the
\emph{mappend} operation. Similarly, using {\tt evalWriter} to unwrap the
writer monad to extract the underlying value, it could be shown that
lifting a binary operation will result in a value computed by applying
it to the unwrapped arguments:

\begin{lstlisting}[language=Coq, mathescape=true,basicstyle=\ttfamily\footnotesize]
forall (op: $\ctype$ -> $\ctype$ -> $\ctype$) (a b: $\rtheta$),
  evalWriter (liftM2 op a b) = op (evalWriter a) (evalWriter b).
\end{lstlisting}

The $\rtheta$ is not the only parametrisation of
$\ctype_{\mathrm{fm}}$ we use. Other monoids could be used
to provide different strategies for combining flags. For example, another useful monoid uses the same \emph{mzero} value as
in Listing~\ref{lst:unionflags} except with the \emph{mappend'} function
which tracks the
preexisting collisions without generating new ones:

\begin{lstlisting}[language=Coq, mathescape=true,basicstyle=\ttfamily\footnotesize,caption={``Safe'' union monoid}, label=lst:safeunionflags]
Definition mappend' (a b: $\ctype_{\mathrm{flags}}$) : $\ctype_{\mathrm{flags}}$ :=
    mkRthetaFlags
      (is_struct a && is_struct b)
      (is_collision a || is_collision b).
Definition Monoid_$\ctype_{\mathrm{flags}}$' : Monoid $\ctype_{\mathrm{flags}}$ :=
    Build_Monoid mappend' mzero.
\end{lstlisting}

This monoid instance allows us to define a ``safe'' variant of the $\rtheta$
type as $\rstheta \triangleq \ctype_{\mathrm{fm}}\;
\mathrm{Monoid}\_\ctype_{\mathrm{flags}'}$. This type could be used, for
example, in scenarios where iteration does not represent partial computations.
To mix these different types of iterations, a conversion between $\rtheta$ and
$\rstheta$ is defined, which preserves both flags and values (see {\tt SafeCast}
and {\tt UnSafeCast} operators in Appendix~\ref{sec:shcoloperators}).

Some definitions do not depend on the monoid used to specialize
$\ctype_{\mathrm{fm}}$, and in the rest of this section, we use
type $\actype \triangleq \ctype_{\mathrm{fm}}\; \mathrm{fm}$ and
assume that all equations using this type are universally quantified
over $\mathrm{fm} \in \mathrm{Monoid}\; \ctype_{\mathrm{flags}}$.

Frequently, we need to combine vectors element-wise using the provided
binary scalar function $\mathrm{dot}: \ctype \rightarrow \ctype \rightarrow \ctype$. For
each output vector element, the values are computed applying
$\mathrm{dot}$ to the corresponding elements of the two input vectors. The
flags are combined using the {\tt mappend} operation from the provided
monoid. This operation is called {\tt Vec2Union}:

\begin{lstlisting}[language=Coq,
  mathescape=true,basicstyle=\ttfamily\footnotesize,
  caption={{\tt Vec2Union} vector combining operation},label=lst:vec2union]
  Definition Vec2Union {n:nat } (dot: $\ctype$ -> $\ctype$ -> $\ctype$)
  : svector fm n -> svector fm n -> svector fm n.
\end{lstlisting}

When \coqe{fm = Monoid_$\ctype_{\mathrm{flags}}$}, this operation
represents the combination of two partial computations. On the other hand,
when \coqe{fm = Monoid_$\ctype_{\mathrm{flags}'}$}, it is just an
element-wise combination of the two sparse vectors using the provided binary
operation. For example, if $\mathrm{dot} = plus$, it is just vector addition.

\subsubsection{Index mapping functions}

Sometimes, we will use functions to express the relationship between the indices of two vectors. We call such functions  \textit{index mapping functions}.

An index mapping function $f$ has a domain of natural numbers
$\mathbb{N}$ in interval $[0,m)$ (denoted as $\finNat{m}$) and the
range of $\mathbb{N}$ in interval $[0,n)$ (denoted as $\finNat{n}$ and
encoded as $\{x:\mathbb{N} \mid x<n\}$ type in Coq).

\begin{equation*}
\ifunc{f}{m}{n}:{\finNat{m}}\rightarrow{\finNat{n}}
\end{equation*}

Such a function, for example, could be used to establish a relation
between the indices of two vectors with respective sizes $m$ and $n$.

\subsubsection{Families of index mapping functions}
\label{sec:indexfamily}

We can extend our notion of an index mapping function into a
\textit{family of index mapping functions}. We define a
\textit{family} $f$ of $k$ index mapping functions:
\begin{equation}
\label{eq:indexfamily}
\forall j<k,\quad
\ifunc{f_j}{m}{n}:{\finNat{m}}\rightarrow{\finNat{n}}
\end{equation}

Such families could be used, for example, to represent the individual
index maps used per loop iteration. The non-collision property corresponds to the \textit{injectivity} of a family of index mapping functions, and the totality of computation corresponds to \textit{bijectivity}.

The family is called \textit{injective} if it satisfies:
\begin{equation}
\label{eq:finjective}
\forall a,\forall b,\forall i,\forall j,\quad f_{a}(i) = f_{b}(j) \implies (i = j) \land (a = b)
\end{equation}

The family is called \textit{surjective} if it satisfies:
\begin{equation}
\label{eq:surjective}
\forall j, \exists a, \exists i, f_a(i) = j
\end{equation}

The family is called \textit{bijective} if it is both \textit{injective} and \textit{surjective}.

The subscript indices in the mathematical notation used in formulas
\ref{eq:indexfamily}, \ref{eq:finjective}, \ref{eq:surjective} are just
additional arguments, and the actual type of the function is:
\begin{equation}
f:\finNat{k} \rightarrow\finNat{m}\rightarrow{\finNat{n}}
\end{equation}

Or in \textit{uncurried} form:
\begin{equation}
\ifunc{f}{m}{n}:{(\finNat{k} \times\finNat{m})}\rightarrow{\finNat{n}}
\end{equation}

In which case, the standard definitions of surjectivity and injectivity
apply.

\subsubsection{Operator type}

\SHCOL operators are defined using mixed embedding. By that, we mean
that an operator's implementation is a Gallina function from vectors to vectors,
which is wrapped up in a record that holds some additional information. The full
definition is shown in Listing~\ref{lst:shoperator}.

\begin{lstlisting}[language=Coq,basicstyle=\fontsize{8}{12}\selectfont\ttfamily,
   columns=flexible,caption={{\tt SHOperator} type},label=lst:shoperator]
Record SHOperator {i o: nat} {svalue: CarrierA} {fm:Monoid RthetaFlags} : Type :=
mkSHOperator {
    op: svector fm i -> svector fm o ;
    op_proper: Proper ((=) ==> (=)) op;
    in_index_set: FinNatSet i ;
    out_index_set: FinNatSet o;
    svalue_at_sparse: forall v,
        (forall j (jc:j<o), ¬ out_index_set (mkFinNat jc) -> evalWriter (Vnth (op v) jc) = svalue);
}.
\end{lstlisting}

\noindent The operator record type is indexed by:

\begin{description}[style=unboxed,leftmargin=1cm,nolistsep]
\item[i,o] Input and output vector dimensions. Vector sizes are
  static and must match when building complex expressions from
  elementary operators.
\item[svalue] The default value which will be used to initialize new
  sparse cells.
\item[fm] Flags monoid instance. It defines how sparsity flags will be
  handled.
\end{description}
\bigskip
The fields of the {\tt SHOperator} record are:

\begin{description}[style=unboxed,leftmargin=1cm,nolistsep]
\item[op] Functional, shallow-embedded, implementation of the operator.
\item[op\_proper] Proper morphism instance for \emph{op} function.
\item[in\_index\_set, out\_index\_set] Sparsity patterns for input and
  output vectors, encoded as sets of finite natural numbers with
  bounds corresponding to dimensions of input and output vectors,
  respectively.
\item[svalue\_at\_sparse] The guarantee (proof) that the operator's
  output will contain the \emph{svalue} at sparse indices.
\end{description}

It should be noted that this definition of an operator provides no
guarantees that the implementation will respect sparsity patterns. This will be ensured via
\textit{structural properties}, discussed next.

\subsubsection{Structural correctness}
\label{sec:structuralproperties}
Expressions must be in a certain shape which lends itself to efficient code generation. Ensuring such a shape is a 
problem distinct from semantics preservation, and we have defined a separate set of properties to ensure what we call ``structural correctness.'' It
involves reasoning about the underlying operations performed on sparse
vectors using a monad to track sparsity and detect structural
errors.

Each operator definition includes two sets, {\tt in\_index\_set} and
{\tt out\_index\_set}, representing its \textit{sparsity
  contract}. They define the expected sparsity patterns of input
vectors and the guaranteed sparsity patterns of output vectors.

We have also defined the following structural properties which guarantee
that a {\SHCOL} expression is in a form which is suitable for optimal
and correct code generation:

\begin{enumerate}
\item The sparsity contract ({\tt in\_index\_set} and {\tt out\_index\_set} membership) is decidable.
\item Only the values at indices from the {\tt in\_index\_set} of the input vector affect the output.
\item During operator evaluation, a sufficiently filled input vector
  (values at all indices in the {\tt in\_index\_set}) guarantees a
  properly filled output vector (values at all indices in the {\tt
    out\_index\_set}).
\item An operator evaluation will never generate values at indices
  which are not present in the {\tt out\_index\_set}.
\item As long as there are no collisions at indices in the {\tt
    in\_index\_set} in the input vector, none will be produced at
  indices in the {\tt out\_index\_set} in the output vector.
\item An operator evaluation will never generate collisions at indices outside the
  {\tt out\_index\_set} of the output vector.
\end{enumerate}

We have grouped these properties in a {\tt SHOperator\_Facts} type class and have
proven its instances for all \SHCOL operators that we have
defined. The proof of these properties for higher-order operators is
\textit{compositional}; as long as all operators involved are
instances of {\tt SHOperator\_Facts}, it can be shown that all
\SHCOL higher-order operators are also instances of {\tt
  SHOperator\_Facts}. That gives us a structural correctness proof
``by construction'' for any \SHCOL expression.

\subsubsection{Operator families}
\label{sec:opfam}
Similar to families of index mapping functions, we can have families
of operators. We define a \textit{family} $F$ of $k$ operators as:
\begin{equation}
\label{eq:operatorfamily}
\forall \mathrm{fm}, \forall i, \forall o, \forall \mathrm{svalue},\quad F:\finNat{k}\rightarrow\mathrm{SHOperator}\, \mathrm{fm}\, i\, o\, \mathrm{svalue}
\end{equation}

All operators in the family use the same monoid, the
same input and output dimensions, and the same default structural value.

\subsubsection{Equality}
\label{sec:seq}
For \SHCOL, we need to define the notion of equality for scalar
values, vectors, and operators, as we did for \HCOL in
Section~\ref{sec:eq}. Here again, we use {\tt Equiv}
typeclass to define our equality relation for various types as
described below.

For scalar values of type $(\ctype_{\mathrm{fm}}\, \mathrm{fm})$, the
equality relation is defined for any monoid
$\mathrm{fm} \in \mathrm{Monoid}\; \ctype_{\mathrm{flags}}$. It is
defined as an equality of the underlying values of type $\ctype$:

\begin{lstlisting}[language=Coq, mathescape=true,
  basicstyle=\ttfamily\footnotesize, label={lst:seq}]
Instance $\ctype_{\mathrm{fm}}$_equiv: Equiv ($\ctype_{\mathrm{fm}}\; \mathrm{fm})$) :=
  fun am bm => (evalWriter am) = (evalWriter bm).
\end{lstlisting}

For sparse vectors of this type, we use pointwise equality.

Finally for \SHCOL operators, the equality is defined as extentional
equality of the underlying shallow-embedded implementations:

\begin{lstlisting}[language=Coq, mathescape=true,
  basicstyle=\ttfamily\footnotesize]
Instance SHOperator_equiv {i o: nat} {svalue: $\ctype$}:
     Equiv (@SHOperator i o svalue) := fun a b => op a = op b.
\end{lstlisting}

At first glance, the definition is missing the comparison of sparsity patterns.
It can be shown that, as defined, the relation is strong enough to
guarantee that the sparsity patterns will also match, assuming both the
operators are \textit{structurally correct}.

\subsubsection{Sparse embedding}
\label{sec:sparseembedding}
One class of \SHCOL expressions that we are particularly interested in has the
following form:
\begin{equation}
  \hoperator{SparseEmbedding}_{f,g,K} \triangleq (\lambda i.\, \hoperator{Scat}_{f_i} \fCompose K_i \fCompose \hoperator{Gath}_{g_i})
  \label{eq:sparseembd}
\end{equation}

The parameters are:

\begin{itemize}
  \item A family of $k$ index mapping function $\ifunc{g}{m}{t}$ 
  \item A family of $k$ ``kernel'' operators $K$
  \item An \textit{injective} family of $k$ index mapping function $\ifunc{f}{\ell}{n}$ 
\end{itemize}

This form is called a \textit{sparse embedding} of an operator family
$K$ (the \textit{kernel}) and could be used as a step in iterative
processing of a vector's elements. It corresponds to the body of a
loop with $k$ iterations in which \textit{gather} picks the input
vector's elements, which are then processed by $K$, and the results
are dispatched to appropriate positions in the output vector
using \textit{scatter}. The index function family $f$ must be
injective. The {\tt SparseEmbedding} is monoid-agnostic and defined
for vectors of $\actype$.

This is a very flexible and powerful construct. We can process vector
elements one by one or in groups. The order of processing is
controlled by index mapping functions. It allows us to model various
memory access patterns useful for SIMD or CPU cache related
optimizations.

\subsubsection{Map-Reduce}
\label{sec:mr}

The {\tt IUnion} and {\tt IReduction} \SHCOL operators are variants of
the same operation, which we will call \textit{map-reduce}. The
higher-order \textit{map-reduce} operation $\mr_{k,f,z}$ takes an
indexed family of $k$ operators (typically a \textit{sparse embedding}) and
produces a new operator. 
It has the following type:

\begin{equation}
  \label{eq:hmr}
  \mr_{k,f,z}\colon\ (\mathbb{N} \rightarrow (\actype^n \rightarrow \actype^m)) \rightarrow \actype^n \rightarrow \actype^m
\end{equation}

When evaluated, \textit{map-reduce} applies all family members with
indices between $0$ and $k-1$ (inclusive) to an input vector, and the
resulting $k$ vectors are folded element-wise using a binary function
($f: \ctype \rightarrow \ctype \rightarrow \ctype)$. The initial value $(z: \actype)$ is used in the first folding step and treated as a \textit{structural} value.

A simple example applies a function $f$ to all elements of a
vector of size 2:
\begin{equation}
  \iUnion{2}{+}{0}{i}{\hoperator{Scat}_{\lambda x.i} \fCompose \llbracket f \rrbracket \fCompose \hoperator{Gath}_{\lambda x.i}}
  \label{eq:pw2}
\end{equation}

\begin{figure}[h]
  \centering
  \includegraphics{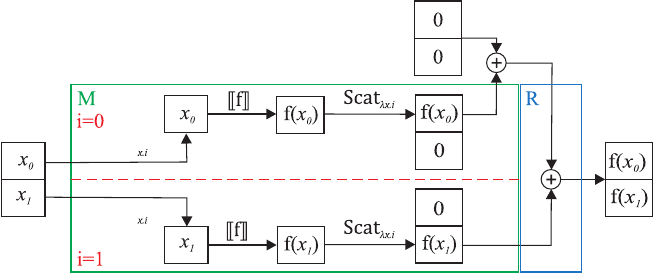}
  \caption{Map-Reduce of a sparse embedding}
  \label{fig:pw2}
\end{figure}

When using \textit{map-reduce} in {\tt IReduction}, the results of family members'
applications must be dense. In the case of {\tt IUnion}, the body of
\textit{map-reduce} should be a family of \textit{sparse
  embeddings}. The dataflow of expression \eqref{eq:pw2} is shown in
Figure~\ref{fig:pw2}.

\subsubsection{Relation between \HCOL and \SHCOL}

Lifting \HCOL operators to be used in \SHCOL 
allows a temporary mixture of abstractions, corresponding to embedding
mathematical formulae in a functional program. A \SHCOL expression
can be gradually transferred to a purely functional form by applying a
series of rewriting rules.

In iterative factorization of operations on vectors, each iteration
represents a partial computation which outputs a sparse vector. In
SPIRAL, the sparsity is represented by default values (typically $0$)
assigned to sparse cells. No tracking is performed which historically
leads to many difficult to find implementation bugs. In \SHCOL, we have
implicit sparsity tracking and a special sparse vector type. Thus,
while in SPIRAL, {\SOL} is a superset of {\OLL}, in HELIX, \SHCOL and
\HCOL are two distinct languages operating on different data types:
sparse vs. dense vectors.

\SHCOL operators represent partial computations and are defined
on sparse vectors, unlike \HCOL operators which represent total
computations and are defined on dense vectors.

\subsubsection{Running example}

A result of the initial translation of our running example of an \HCOL
formulation of the dynamic window monitor from
Listing~\ref{lst:dwhcol} to \SHCOL is shown in
Listing~\ref{lst:dwshcol}. It has been abridged to hide some
non-essential parameters.

The full list of \SHCOL operators can be found in
Appendix~\ref{sec:shcoloperators}.

\dynwinlisting[language=Coq,label=lst:dwshcol,
caption={Dynamic Window Monitor in \SHCOL},
basicstyle=\fontsize{8}{8}\ttfamily]{dynwin_shcol.v}

\subsection{\HCOL to \SHCOL translation}
\label{sec:hcol2shcol}
As mentioned in Section~\ref{sec:shcol}, \HCOL operators can be
embedded in \SHCOL using the {\tt liftM\_HOperator} wrapping operator.
When lifting, we use the {\tt Monoid\_RthetaFlags} monoid to detect
potential collisions. Thus, a trivial translation of an \HCOL expression
{\tt h} to \SHCOL is simply \coqe{(liftM_HOperator Monoid_RthetaFlags h)}.

However, this first na\"ive translation can be further refined by the additional rewriting steps which replace \HCOL operators with similar
\SHCOL equivalents (e.g. lifted {\tt HBinOp} with {\tt SHBinOp}). To
do that for complex \HCOL operators, we need to exploit the facts that {\tt liftM\_HOperator}
distributes over operator composition, and that the operator compostion is associative. Translating a \SHCOL expression to a ``normal
form'' (without occurrences {\tt liftM\_HOperator}) is the goal of the
\HCOL to \SHCOL translation step.

The reasoning about semantics preservation during \SHCOL rewriting is
similar in approach to our reasoning about \HCOL rewriting. The main
difference is that \SHCOL operators work on sparse rather than dense vectors,
and a different equality relation is used (see \ref{lst:seq}).

Finally, we need to make sure that the resulting \SHCOL expression is
\textit{structurally correct}, as discussed in
Section~\ref{sec:structuralproperties}. This is done by proving an
instance of the {\tt SHOperator\_Facts} typeclass for the resulting
expression. As mentioned earlier, structural correctness proofs are
compositional and thus easy to automate. Provided that we have
instances of {\tt SHOperator\_Facts} for all basic operators,
obtaining structural correctness for a composite operator is simply a
matter of applying respective typeclass instances. We do this with simple \emph{Ltac} automation. While automation solves all goals related to
{\tt SHOperator\_Facts} instances, some of these goals introduce
additional obligations which must also be proven. These have not
been fully automated yet, but they could be in future, as discussed in
Section~\ref{sec:setproofs}.

In addition to semantics preservation and structural correctness, there
are some additional properties which we want to verify for
the final \SHCOL expression:

\subparagraph*{Sparsity contract ``subtyping.''} It guarantees that the
resulting expression's {\tt in\_index\_set} is included in the
original expression's {\tt in\_index\_set}, while the {\tt
  out\_index\_set} of each expression is the same. This permits
potential optimization (dead code elimination) during rewriting, when
indices of input vectors which were used by the original expression
are no longer used by the resulting expression. This is proven
compositionally by constructing respective sparsity contracts of input
and output expressions.

\subparagraph*{Totality of the computation.} In general, {\SHCOL} operators
work on sparse vectors. However, the sparsity is used only internally
to represent partial computation. The whole composite computation
should be total by taking the dense input and producing the dense
output. That means that for top-level {\SHCOL}
expressions, we want to prove that both {\tt in\_index\_set} and
{\tt out\_index\_set} are the full sets. This is proven
compositionally as well, by constructing respective sparsity contracts of the
input and output expressions.


\subsection{\SHCOL rewriting}
\label{sec:shcol2shcol}
Once an \HCOL expression is translated to the \SHCOL ``normal form,''
additional rewriting steps are performed on the \SHCOL expression to
optimize it for efficient code-generation for target architecture. The
optimization steps are determined by SPIRAL and validated by
HELIX. Each step is implemented in SPIRAL as an application of a
``rewriting rule.'' Following the translation validation approach
discussed earlier, we prove a lemma for each rule and apply them
following the trace provided by SPIRAL.  The mechanics of \SHCOL
rewriting are similar to \HCOL breakdown, described in
Section~\ref{sec:rewriting}.

\subsubsection{Running example}

The result of the \SHCOL rewriting steps of the expression from
Listing~\ref{lst:dwshcol} is shown in Listing~\ref{lst:dwshcol1}

\dynwinlisting[language=Coq,label=lst:dwshcol1,
caption={Dynamic Window Monitor in \SHCOL after rewriting},
basicstyle=\fontsize{8}{8}\ttfamily]{dynwin_shcol1.v}

\section{\MSHCOL}

\subsection{\MSHCOL language family}
\label{sec:mhcol}
\MSHCOL is an intermediate step in the HELIX transformation chain between purely
functional \SHCOL and imperative \DHCOL languages. Imperative language
semantics, as is shown in Appendix~\ref{sec:dhcolbigstep}, ultimately describes
how program execution steps update the memory state. Sparse vectors in {\SHCOL}
are an algebraic abstraction for \textit{memory blocks}. We make this explicit
in the intermediate mixed-embedded language, \MSHCOL (M stands for
\textit{memory}). While \MSHCOL is still a functional language, we bring it
closer to the next language transformation step by changing data representation
from vectors to memory blocks. Each memory block is represented as a finite map
from memory offsets to values of a carrier type. There is no mappings for keys
corresponding to structural values. Besides physical data representation, the
main change from \SHCOL is that we no longer maintain algebraic abstraction
which we used to transform and optimize \SHCOL expressions. There are no
``default'' values for sparse cells. That means that reading a sparse element is
an error which leads to the introduction of implicit error handling in \MSHCOL.

\begin{figure}[h!]
  \centering
  \includegraphics{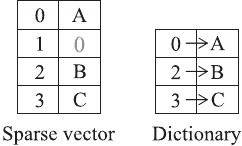}
  \caption{Sparse vectors as dictionaries}
  \label{fig:dict}
\end{figure}

An example of both representations is shown in
Figure~\ref{fig:dict}. It shows a sparse vector with three initialized
cells, A, B, and C, and one sparse cell with default value $0$.
The memory representation of the same vector uses a dictionary with
three elements. There is no mapping for key $1$ corresponding to
vector's sparse cell.

\begin{figure}[h!]
  \centering
  \includegraphics{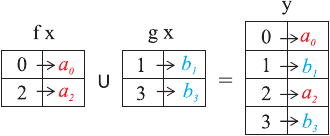}
  \caption{{\tt MApply2Union} in {\MSHCOL}}
  \label{fig:munion}
\end{figure}

Generally speaking, there is a 1-to-1 correspondence between \SHCOL and
\MSHCOL operators with the main difference in the input and output container data
types. \MSHCOL uses memory blocks, where \SHCOL uses sparse vectors.
An example application of the {\tt MApply2Union} operator in {\MSHCOL}
is shown in Figure~\ref{fig:munion}. It is similar to the {\tt
  Apply2Union} \SHCOL operator in Figure~\ref{fig:apply2union} but operates on memory blocks instead
of vectors. Each of the two operators {\tt f} and {\tt g} are applied to
the input memory block {\tt x} producing corresponding dictionaries
with disjoint keys $\{0,2\}$ and $\{1,3\}$, respectively.  They are then
merged into the final resulting dictionary {\tt y}. Unlike \SHCOL,
merging memory blocks does not involve combining cells with matching
keys using a binary operation. The memory blocks are simply merged. If
there is a value associated with a key in both input blocks, it is
considered an error.

With this change of data representation, we move away from the algebraic nature
of {\SHCOL} towards a lower-level representation. In this representation, an
actual value must be associated with a key in a dictionary before it can be
accessed. Trying to access an uninitialized key is an error. It means that
{\MSHCOL} operators could return errors and thus have the type:
$\mathrm{mem\_block} \rightarrow \mathrm{option\ mem\_block}$.
However, we will prove later that our translation of a structurally
correct \SHCOL program produces an \MSHCOL program that does not err when applied
to an input memory block that matches the expected input sparsity patterns.

In the title of this section, we have referred to \MSHCOL as a
``language family''. The reason for this is it is implemented as a
module, parametrized by the type of the data values. It is similar to
$\ctype$ abstraction we used in \HCOL and \SHCOL, but instead of
typeclasses we are using here Coq's modules mechanism. We define {\tt
CType} module type which contains the following:

\begin{itemize}[nolistsep]
  \item Carrier Type: {\tt CType.t}
  \item Constants: {\tt Zero}, {\tt One}.
  \item Equality: {\tt Equiv}
  \item Operations: {\tt Plus}, {\tt Neg}, {\tt Mult}, {\tt Abs}, {\tt Zless}, {\tt Min}, {\tt Max}, {\tt Sub}.
  \item Inequality of constants: $\texttt{Zero} \neq \texttt{One}$.
  \item Operations respect equality: $\forall \texttt{op}, \; \texttt{Proper} \; ((=) \Longrightarrow (=)) \; \texttt{op}$.
\end{itemize}








For {\tt CType}, we define equality and equivalence relations;
constants for \textit{additive identity} ({\tt Zero}) and
\textit{multiplicative identity} ({\tt One}); and basic algebraic
operations like addition and multiplication. Additionally, we require
all these operations to be proper with respect to defined equality.
{\tt CType} is in essence a subset of what we had assumed and could
easily be instantiated for $\ctype$. The set of properties is smaller
since at this stage, we no longer require some algebraic properties
like \textit{ring} or \textit{total order}. The MHCOL family consists
of only one language, with {\tt CType} instantiated for the set of
real numbers $\R$. As we will see later in Section~\ref{sec:fhcol},
{\tt CType} can be also instantiated for IEEE floating point numbers.

Like in {\SHCOL}, we use \textit{mixed embedding}~\citep{chlipala2017frap}
(a combination of \textit{shallow} and \textit{deep embedding}) to represent
{\MSHCOL} operators. We use {\tt MSHOperator} record, which is indexed by
dimensions $i, o \in \N$ of input and output memory blocks and contains the
following:


\begin{itemize}[nolistsep]
  \item $\texttt{mem\_op} \colon \texttt{mem\_block} \rightarrow \texttt{option mem\_block}$
  \item $\texttt{mem\_op\_proper} \colon \texttt{Proper} \; ((\texttt{equiv}) \Longrightarrow (\texttt{equiv})) \; \texttt{mem\_op}$
  \item $\texttt{m\_in\_index\_set} \colon \finNat{i} \rightarrow \mathbb{B}$
  \item $\texttt{m\_out\_index\_set} \colon \finNat{o} \rightarrow \mathbb{B}$
\end{itemize}

The fields include: a function {\tt mem\_op} implementing the
operation on memory blocks which can fail (returning {\tt None}); a
witness for the function being a \textit{proper morphism}~\citep{casteran2012gentle}
with respect to the setoid equality {\tt equiv} (required because the carrier
type is still abstract); and the two sets (represented as characteristic
functions) which define input and output memory access patterns.

Additionally, all {\MSHCOL} operator implementations must satisfy
certain \textit{memory safety} properties. We have formulated these
properties as the typeclass, {\tt MSHOperator\_Facts}, and have proven
instances of it for all operators. This is a similar approach to what
we took with \SHCOL \textit{structural properties}, but the properties
are different:

\begin{enumerate}
\item When applied to a memory block which has mappings present for all keys in {\tt m\_in\_index\_set}, {\tt mem\_op} will not return an error.
\item The {\tt mem\_op} must assign a value to each element with a key
  in {\tt
    m\_out\_index\_set} and must not assign a value to any element
  with a key not in {\tt m\_out\_index\_set}.
\item The output block of {\tt mem\_op} is guaranteed to contain no mappings for keys outside of an operator's declared output size.
\end{enumerate}

\subsubsection{Memory model}
\label{sec:memmodel}

\MSHCOL definitions rely on \textit{memory block} abstraction. 
In \MSHCOL, memory blocks are immutable and
transient and are passed as arguments to the operators and returned as the
results. However, the same type of memory block can be made
persistent and organized into a \textit{memory} which will maintain
the state of a collection of such blocks, whereas each block will
maintain the state of its cells. This hierarchical two-level memory
organization was inspired by the CompCert \citep{Leroy2009,leroy:hal-00703441}
compiler and by the Vellvm project.

As we re-use this memory model for other HELIX languages, it is
generalized to support various value types. This is implemented using
a Coq module system. Parameterized by the module type {\tt CType}, the
module type {\tt MBasic} defines memory model basics. It includes the
abstract type {\tt memory}, which represents a mapping from
``addresses'' (represented as natural numbers) to memory blocks and
provides the essential operations for adding, removing, or looking up
memory blocks. It also provides a constant {\tt memory\_empty}, which
represents the initial empty memory state. An abridged definition of
{\tt memory} interface from the {\tt MBasic} module type is shown in
Listing~\ref{lst:mbasicmemory}.

\begin{lstlisting}[language=Coq,basicstyle=\fontsize{8}{12}\selectfont\ttfamily,
  columns=flexible,label={lst:mbasicmemory},caption={Memory interface}]
Definition memory : Type.
Definition mem_block : Type.
  
Definition memory_empty : memory.
Definition memory_lookup : memory -> nat -> option mem_block.
Definition memory_set : memory -> nat -> mem_block -> memory.
Definition memory_remove : memory -> nat -> memory.
Definition memory_keys_lst : memory -> list nat.
Definition memory_next_key : memory -> nat.

Definition mem_block_exist${}$s : nat -> memory -> Prop.
Parameter decidable_mem_block_e${}$xists :
    forall (k : nat) (m : memory), decidable (mem_block_e${}$xists k m).
\end{lstlisting}

The memory address space is unbounded. In a given state, some addresses
could be already associated with memory blocks while others may be not
initialized yet. Operation {\tt memory\_set} assigns a memory block to a given memory address. It could be also used to change a
memory state by replacing a block at the given address. Decidable
predicate {\tt mem\_block\_exists} checks whether a given memory
address has been initialized. A block can be freed with {\tt
  memory\_remove}. The function {\tt memory\_next\_key} returns the
address of the next unallocated address.

The memory block has abstract type {\tt mem\_block}, which in turn is also
a mapping of ``offsets'' (represented as natural numbers) to values of
type {\tt CType.t}. Memory blocks are also unbounded and have
interface similar to {\tt memory}. A constant {\tt mem\_empty}
represents an empty memory block which contains no values. An abridged
definition of {\tt mem\_block} interface from the {\tt MBasic} module is
shown in Listing~\ref{lst:mbasicblock}.

\begin{lstlisting}[language=Coq,basicstyle=\fontsize{8}{12}\selectfont\ttfamily,
  columns=flexible,label={lst:mbasicblock},caption={Memory block interface}]
Definition mem_empty : mem_block.
Definition mem_lookup : nat -> mem_block -> option CType.t.
Definition mem_add : nat -> CType.t -> mem_block -> mem_block.
Definition mem_delete : nat -> mem_block -> mem_block.

Definition mem_in : nat -> mem_block -> Prop.
Parameter decidable_mem_in : forall (k : nat) (m : mem_block), decidable (mem_in k m).

Definition mem_keys_lst : mem_block -> list nat.
Definition mem_value_lst : mem_block -> list CType.t.
\end{lstlisting}

The {\tt MMemSetoid} module type extends {\tt MBasic} with a
definition of setoid equality for memory blocks and proofs of proper
morphism for memory operations with respect to setoid equality.

Currently, we define only one language from \MSHCOL language family:
\MSHCOL. This language uses type $\R$ to represent numerical values. It
is implemented by using a {\tt CType} instance for $\R$ to specialize
\MSHCOL. Up to this point, we kept the data type abstract enough to
allow instantiating it with types like $\R$ or even $\Complex$. At
this stage, we are intentionally narrowing it down to $\R$ in line
with our intention to generate code working with floating-point
machine numbers.

\subsubsection{Running example}

A result of a translation of the \SHCOL formulation of a dynamic window
monitor from Listing~\ref{lst:dwshcol1} to \MSHCOL is shown in
Listing~\ref{lst:dwmhcol}. It has been abridged to hide the
non-essential operator parameters.

\dynwinlisting[language=Coq,label=lst:dwmhcol,
caption={Dynamic Window Monitor in \MSHCOL},
basicstyle=\fontsize{8}{8}\ttfamily]{dynwin_mhcol.v}

\subsection{\SHCOL to \MSHCOL translation}
\label{sec:shcol2mhcol}
\subsubsection{Implementation}
Translation from \SHCOL to \MSHCOL is implemented using the Coq
meta-programming plugin, Template-Coq~\citep{sozeau2020metacoq}. The
translation is fairly straightforward, as there is a one-to-one 
correspondence between language operators, with the exception of {\SHCOL} {\tt
  SafeCast} and {\tt UnSafeCast} operators which are translated as identities.
In addition to mapping the operators'
names, the translation changes their input types from vectors to memory blocks. The return type of all \MSHCOL operators is a memory block wrapped in an {\tt option} type to facilitate error handling.

\SHCOL operators are instances of the {\tt SHOperator} typeclass. They
can be parameterized by some constants which technically means they
can be enclosed in additional lambdas, introducing corresponding
variables which could be used inside an operator's shallow embedded
definition. For example, our dynamic window monitor \SHCOL definition
from Listing~\ref{lst:dwshcol} is parametrized by parameter $a$ and
has the actual type:

\coqe{forall (a: avector 3), @SHOperator Monoid_RthetaFlags (1+(2+2))$\,$ 1 zero}. 

Such optional parameters will be detected during translation to \MSHCOL
and mapped to corresponding binders in the resulting expression. Thus,
the result of the \MSHCOL translation of the dynamic window monitor \SHCOL
definition from Listing~\ref{lst:dwshcol} will also be parametrized by
$a$ and have the actual type:

\coqe{forall (a: avector 3), @MSHOperator (1 + 4) 1}. 

\subsubsection{Proof of semantics preservation}

The semantic equivalence between an {\SHCOL} and an {\MSHCOL} operator is
defined as the {\tt SH\_MSH\_Operator\_compat} typeclass. It ensures
that they have the same dimensionality and input and output
patterns (index sets) and are both structurally correct (by the presence of respective {\tt
  SHOperator\_Facts} and {\tt MSHOperator\_Facts} instances). In
addition to these properties, it also states the main semantic equivalence property:

\begin{lstlisting}[language=Coq,basicstyle=\fontsize{8}{12}\selectfont\ttfamily,
   columns=flexible,caption={\SHCOL and \MSHCOL main semantic equivalence 
    property}]
mem_vec_preservation:
forall (x:svector i),
  (forall (j: nat) (jc: j < i), in_index_set sop (mkFinNat jc) -> Is_Val (Vnth x jc)) ->
  Some (svector_to_mem_block (op sop x)) = mem_op mop (svector_to_mem_block x)
\end{lstlisting}

\noindent Informally it can be stated as:
\begin{quote}For any vector which complies with the input sparsity
  contract of the \SHCOL operator, an application of the \MSHCOL
  operator to such vector, converted to a memory block, must succeed
  and return a memory block which must be equal to the memory
  block produced by converting the result of the \SHCOL operator.
\end{quote}

For regular operators, {\tt SH\_MSH\_Operator\_compat} instances can
be proven directly. For higher-order operators, the proofs are
predicated on {\tt SH\_MSH\_Operator\_compat} assumptions for all
operators involved. Some operators may have additional
prerequisites. For example, for {\tt Apply2Union}, the output index sets of
{\tt f} and {\tt g} must be disjoint.

For translated programs, we use proof automation to prove that {\tt
  SH\_MSH\_Operator\_compat} holds between the original and the
compiled programs. Assuming one-to-one operator correspondence, the
syntax-driven proof automation applies manually proven per-operator
{\tt SH\_MSH\_Operator\_compat} instances. This should succeed for all
\MSHCOL{} generated from the corresponding \SHCOL{} ones by our
translation Template-Coq metaprogram.

\section{\DHCOL language family}
\label{sec:dhcol}
\DHCOL is a family of imperative languages, deep-embedded in Coq proof
assistant. Languages in the family share their memory model with those
of \MSHCOL. In addition to dynamic memory, \DHCOL has lexically scoped variables, which
can hold {\tt CType.t} values, pointers to memory blocks, and integers
(used for loop bounds and memory offset computations).

All variables are immutable. The language is statically scoped and de
Bruijn indices are used to reference variables from an
\textit{evaluation context}.

It should be noted that \DHCOL does not define general-purpose
programming languages. It serves as an intermediate representation
language for the class of problems that HELIX is designed for. As
such, it primarily focuses on operations on $\ctype$-valued
vectors. Only the features vital to represent corresponding HELIX
abstractions are incorporated into the language. Given these design
objectives, it may seem fairly esoteric compared to general-purpose
programming languages.

The \DHCOL memory model, shared with \MSHCOL, is described in
Section~\ref{sec:memmodel}. New memory blocks could be allocated and
freed by a \DHCOL program, and their elements can be modified
repeatedly. This changing memory state represents an imperative aspect
of the \DHCOL design.

\DHCOL supports typed expressions on integer values and {\tt CType.t}
(discussed in the next section) values, constant memory blocks, and
pointers to memory blocks. {\tt CType.t} and integer value expressions
allow constants and provide a set of arithmetic operations such as
addition, subtraction, and division. Additionally, expressions could
reference variables from the environment, which are also
typed. Evaluation of an expression has no side effects but could fail.

The \crefrange{sec:typemodules}{sec:dhcoleval} provide more technical details on
the formalisation of \DHCOL in Coq. The language syntax, list of operators and
formal semantics are provided in Appendix~\ref{sec:dhcollang}.

\subsection{Type parametrization}
\label{sec:typemodules}

Like the previous language family in the HELIX transformation chain, \DHCOL also
uses {\tt CType} for data stored in memory addresses. Unlike \MSHCOL however,
\DHCOL also defines a new type for integer values, {\tt NType}. The primary
reason for this is that with the languages of the \DHCOL family we also
transition to lower-level datatypes: from $\R$ to binary floats, and from $\N$
to machine integers. In fact, the difference in these types is the only thing
separating languages of the \DHCOL family from each other.

Our {\tt NType} module type definition is similar to {\tt CType} with a few
additions:

\begin{itemize}[nolistsep]
  \item Constant: {\tt zero}.
  \item Operations: {\tt plus}, {\tt minus}, {\tt mult}, {\tt div}, {\tt mod}, {\tt min}, {\tt max}.
  \item Conversion to strings: {\tt to\_string}.
  \item Conversion to natural numbers: {\tt to\_$\N$}.
  \item Conversion from natural numbers: {\tt from\_$\N$}.
\end{itemize}

A method to convert {\tt NType.t} values to strings is provided mostly for
debugging convenience. We additionally require this type to always be
convertible to $\N$. Conversion from $\N$ is also defined, but it could return
an error, because for fixed-size parametrizations, it might not be possible to
fit arbitrary-size natural numbers to a fixed number of bits.

A couple of properties must be proven for instances of this module type:

\begin{itemize}[nolistsep]
  \item $\forall x \; x_i \; z,
         \mathtt{from}\_\N \; x = \mathtt{inr} \; x_i \rightarrow
         y < x \rightarrow
         \exists y_i, \mathtt{from}\_\N \; y = \mathtt{inr} \; y_i$
  \item $\exists z, \mathtt{from}\_\N \; 0 = \mathtt{inr} \; z$
\end{itemize}

The first property states that if conversion from $\N$ succeeds for a given
natural number, it must also succeed for all natural numbers below that one.
This property could be considered a form of monotonicity. The second property
simply states that the natural number $0$ could always be successfully converted
to {\tt NTtype.t}. This makes the type inhabited, with at least one value
corresponding to $0\in\N$. This module type could easily be instantiated for
natural numbers as well as for fixed-length machine integers.

It should be noted that while {\tt NType.t} is used to represent loop
indices in iterative operators, loop bounds are expressed as natural
numbers. This implementation decision is meant to simplify proofs by
induction.

\subsection{Expressions and operators}

The DHCOL language family is defined as a module type parametrized by modules
{\tt CT} for {\tt CType} and {\tt NT} for {\tt NType}. It defines \DHCOL
expressions and operators. There are four expression types, each of which can be
evaluated to the values of the corresponding types:

\begin{description}[style=unboxed,leftmargin=0cm]
\item[NExpr] is a type of integer expressions which evaluates to {\tt
    NType.t} values.
\item[PExpr] type represents pointers to blocks in memory. These
  pointers can be used to modify the objects they point to, changing
  the memory state. It evaluates to a tuple with type
  $\N \times \mathrm{NT.t}$. The first element of the tuple is an
  address of a block in memory and the second is the size of this
  block.
\item[MExpr] also refers to memory blocks but can only be used to
  access, not modify data. In addition to referencing blocks in
  memory, it can also represent standalone constant memory blocks. It
  evaluates to a tuple with type
  $\mathrm{mem\_block}\times\mathrm{NT.t}$.
\item[AExpr] expressions evaluate to scalar values of {\tt CType.t}.
\end{description}

Syntax of {\tt AExpr} expressions is shown in Figure~\ref{fig:dhcolaexprbnf}.
The evaluation semantics for expressions will be discussed later in
Section~\ref{sec:dhcoleval}.

\begin{figure}[h]
  \begin{bnf*}
    \bnfprod{AExpr}{
      \bnfts{AVar} \bnfsp \bnfpn{Nat} \bnfor
      \bnfts{AConst} \bnfsp \bnfpn{CTypeConst} \bnfor
      \bnfts{ANth} \bnfsp \bnfpn{MExpr} \bnfsp \bnfpn{NExpr} \bnfor}\\
    \bnfmore{
      \bnfts{AAbs} \bnfsp \bnfpn{AExpr} \bnfor
      \bnfts{APlus} \bnfsp \bnfpn{AExpr} \bnfsp \bnfpn{AExpr} \bnfor
      \bnfts{AMinus} \bnfsp \bnfpn{AExpr} \bnfsp \bnfpn{AExpr} \bnfor}\\
    \bnfmore{
      \bnfts{AMult} \bnfsp \bnfpn{AExpr} \bnfsp \bnfpn{AExpr} \bnfor
      \bnfts{AZless} \bnfsp \bnfpn{AExpr} \bnfsp \bnfpn{AExpr} \bnfor}\\
    \bnfmore{
      \bnfts{AMin} \bnfsp \bnfpn{AExpr} \bnfsp \bnfpn{AExpr} \bnfor
      \bnfts{AMax} \bnfsp \bnfpn{AExpr} \bnfsp \bnfpn{AExpr}}\\
  \end{bnf*}
\caption{Syntax of {\tt AExpr} expressions}
\label{fig:dhcolaexprbnf}
\end{figure}


Syntax for \DHCOL operators is shown in Figure~\ref{fig:dhcoloperatorsbnf}. In
addition to the expressions shown in Figure~\ref{fig:dhcolaexprbnf}, some \DHCOL
operators take {\tt MemRef} values as arguments. {\tt MemRef} value is a memory
pointer combined with an offset as a natural number: \coqe{(PExpr, NExpr)}.


\begin{figure}[h]
  \begin{bnf*}
    \bnfprod{DSHOperator}{\bnfts{DSHNop} \bnfor}\\
    \bnfmore{\bnfts{DSHAssign} \bnfsp \bnfpn{MemRef} \bnfsp \bnfpn{MemRef} \bnfor}\\
    \bnfmore{
      \bnfts{DSHIMap} \bnfsp
        \bnfpn{Nat} \bnfsp
        \bnfpn{PExpr} \bnfsp
        \bnfpn{PExpr} \bnfsp
        \bnfpn{AExpr} \bnfor}\\
    \bnfmore{
      \bnfts{DSHBinOp} \bnfsp
        \bnfpn{Nat} \bnfsp
        \bnfpn{PExpr} \bnfsp
        \bnfpn{PExpr} \bnfsp
        \bnfpn{AExpr} \bnfor}\\
    \bnfmore{
      \bnfts{DSHMemMap2} \bnfsp
        \bnfpn{Nat} \bnfsp
        \bnfpn{PExpr} \bnfsp
        \bnfpn{PExpr} \bnfsp
        \bnfpn{PExpr} \bnfsp
        \bnfpn{AExpr} \bnfor}\\
    \bnfmore{
      \bnfts{DSHPower} \bnfsp
        \bnfpn{NExpr} \bnfsp
        \bnfpn{MemRef} \bnfsp
        \bnfpn{MemRef} \bnfsp}\\
    \bnfmore{
      \quad
        \bnfpn{AExpr} \bnfsp
        \bnfpn{CTypeConst} \bnfor}\\
    \bnfmore{\bnfts{DSHLoop} \bnfsp \bnfpn{Nat} \bnfsp \bnfpn{DSHOperator} \bnfor}\\
    \bnfmore{\bnfts{DSHAlloc} \bnfsp \bnfpn{NTypeConst} \bnfsp \bnfpn{DSHOperator} \bnfor}\\
    \bnfmore{\bnfts{DSHMemInit} \bnfsp \bnfpn{PExpr} \bnfsp \bnfpn{CTypeConst} \bnfor}\\
    \bnfmore{\bnfts{DSHSeq} \bnfsp \bnfpn{DSHOperator} \bnfsp \bnfpn{DSHOperator}}\\
  \end{bnf*}
\caption{Syntax of \DHCOL operators}
\label{fig:dhcoloperatorsbnf}
\end{figure}

\noindent Example of a \DHCOL operator, DSHIMap. Detailed description
of all \DHCOL operators can be found in Appendix~\ref{sec:dhcoloperators}.

\paragraph*{\underline{\underline
  {\lstinline[language=Coq]|DSHIMap (n: nat) (x y: PExpr) (f: AExpr)|}}}

It iterates the index from $0$ to $n-1$. For each iteration, the values from {\tt x} at the index offset are
added to the evaluation context, and the expression {\tt f} is
evaluated. It could be viewed as calling function $f$ with the two
arguments: an index $i$ of type {\tt NType.t} and a value
{\tt x[i]} of type {\tt CType.t}. The result is written to
{\tt y[i]}.

\subsection{Evaluation}
\label{sec:dhcoleval}

\DHCOL evaluation is performed within an \textit{evaluation context}: 

\begin{lstlisting}[language=Coq,basicstyle=\ttfamily\footnotesize,
  columns=flexible,label=lst:evalcontext]
Inductive DSHVal :=
  | DSH$nat$Val (n:NT.t): DSHVal
  | DSHCTypeVal (a:CT.t): DSHVal
  | DSHPtrVal (a:nat) (size:NT.t): DSHVal.

Definition evalContext := list (DSHVal*bool).
\end{lstlisting}

It is a list where each position corresponds to the \textit{de Bruijn}
 index of a variable and holds its value and \textit{protection flag}. The function of the \textit{protection flag} is to ensure that some variables, such as loop indices, are not accessible within loop bodies where they are not supposed to be exposed.

Recursive definition \emph{evalDSHOperator} takes an operator to
evaluate, an evaluation context $\sigma$, and the initial memory state
{\tt mem}. If successful, it returns the final memory state after the
evaluation:

\begin{lstlisting}[language=Coq,basicstyle=\ttfamily\footnotesize,
  columns=flexible]
Fixpoint evalDSHOperator ($\sigma$: evalContext) (op: DSHOperator)
  (mem: memory) (fuel: nat): option (err memory).
\end{lstlisting}

It is implemented via structural recursion over the structure of the
\DHCOL expression. For error handling, it is wrapped in
\textit{exception monad}, for which we use {\tt err} type defined by
Vellvm. All \DHCOL programs are guaranteed to terminate, but
\textit{fuel} is used to simplify termination proofs. The evaluation
result is double wrapped in an \textit{option monad} (on top of the {\tt err}
monad) to account for errors due to insufficient fuel. There is a
matching {\tt estimateFuel} function which estimates the fuel required
to execute a given \DHCOL operator, and we've proven a lemma showing that the
estimated fuel is always sufficient for a successful {\tt evalDSHOperator}
 completion. In other words, with estimated fuel it will
never return {\tt None} (but may still return an error).

The \emph{evalDSHOperator} function defines big-step operation
semantics, formally presented in Appendix~\ref{sec:dhcolbigstep}.

\subsection{\RHCOL language}
\label{sec:rhcol}

\RHCOL is the first imperative language used in HELIX's compilation
chain. It is the instantiation of the \DHCOL family with real numbers
$\R$ as the carrier type {\tt CType.t} and natural numbers $\N$ as the integer type {\tt NType.t}. The arithmetic operators on \emph{CType}
and \emph{NType} are instantiated following standard mathematical
definitions for $\R$ and $\N$ types. This means that no special
handling of numerical computation is required at this step: iterators
do not overflow, memory and pointer sizes are unlimited, and operations
and comparisons are exact, among other things.

\subsubsection*{Running example}

The result of a translation of the \MSHCOL formulation of a dynamic window
monitor from Listing~\ref{lst:dwmhcol} to \RHCOL is shown in
Listing~\ref{lst:dwrhcol}. For simplicity, we use standard mathematical notation
for arithmetic operations over {\tt CType.t} and {\tt NType.t}, array
subscription notation for access to memory blocks and semicolon notation for
{\tt DSHSeq} operator within Listing~\ref{lst:dwrhcol}.

\dynwinlisting[language=Coq,label=lst:dwrhcol,
caption={Dynamic Window Monitor in \DHCOL},
basicstyle=\fontsize{8}{8}\ttfamily]{dynwin_rhcol_pretty.v}

\subsection{\MSHCOL to \RHCOL translation}
\label{sec:mhcol2rhcol}
\subsubsection{Implementation}

Translation from \MSHCOL to \RHCOL is also implemented in
Template-Coq as a recursive descent on {\MSHCOL}'s
AST. The top-level translation function has the following type:

\begin{lstlisting}[language=Coq,
  mathescape=true,basicstyle=\ttfamily\footnotesize,columns=flexible]
Fixpoint compileMSHCOL2DSHCOL
         (res: var_resolver)
         (vars: varbindings)
         (t: term)
         (x_p y_p: PExpr)
  : TemplateMonad (varbindings*(DSHOperator)).
\end{lstlisting}

Since we know that $t$ is Gallina AST of the \MSHCOL operator which is a
function with type
\coqe{(mem_block -> option mem_block)}, the
compiler is parameterized by two memory pointer expressions specifying where an
imperative \RHCOL program corresponding to this function should read
input from ({\tt x\_p}) and write output to ({\tt y\_p}).

The translation, if it succeeds, returns a \RHCOL program and the list of
``global'' variables it depends on. Recall that \MSHCOL operators are
{\tt MSHOperator} records, which may depend on additional
parameters. All such additional parameters will be detected during
translation to \RHCOL and treated as ``global'' variables which must
be present in the evaluation context prior to evaluating the \RHCOL
program. The {\tt varbindings} part of a tuple returned by the
compiler is a list of these variable names with their respective types.

The translation procedure is fairly straightforward, as each \MSHCOL
operator is compiled to a \RHCOL fragment, and the translation is
compositional. For example, both {\tt MSHEmbed} and {\tt MSHPick} from
\MSHCOL translate to {\tt DSHAssign} in \RHCOL. Some \MSHCOL operators
translate not to a single \RHCOL operator but rather to a \RHCOL program
fragment. Let us look more closely at an example of how {\tt
  MSHCompose} is translated to \RHCOL. The relevant part of the {\tt
  match} statement on the {\MSHCOL} operator's name and arguments from AST
is shown in Listing~\ref{lst:compose2dhcol}:

\begin{lstlisting}[language=Coq,
  mathescape=true,basicstyle=\ttfamily\footnotesize,columns=flexible,label={lst:compose2dhcol},caption={{\tt
      MSHCompose} operator translation to \RHCOL}]
      | Some n_SHCompose, [i1 ; o2 ; o3 ; op1 ; op2] =>
        ni1 <- tmUnquoteTyped nat  i1 ;;
        no2 <- tmUnquoteTyped nat  o2 ;;
        no3 <- tmUnquoteTyped nat  o3 ;;
        (* freshly allocated, inside alloc *)
        let t_i := PVar 0 in
        (* single inc. inside alloc *)
        let x_p' := incrPVar 0 x_p in
        let y_p' := incrPVar 0 y_p in
        let res1 := Fake_var_resolver res 1 in
        '(_, cop2) <- compileMSHCOL2DSHCOL res1 vars op2 x_p' t_i ;;
        '(_, cop1) <- compileMSHCOL2DSHCOL res1 vars op1 t_i y_p' ;;
        tmReturn (vars, DSHAlloc no2 (DSHSeq cop2 cop1))
\end{lstlisting}

Using double square brackets as a shortcut for the {\tt
  compileMSHCOL2DSHCOL} call, compiling the \MSHCOL operator $\llbracket \mathrm{op1}\, \circ\, \mathrm{op2}\rrbracket(x, y)$ will
result in the \RHCOL program {\tt DSHAlloc no2 (DSHSeq
  $\llbracket\mathrm{op2}\rrbracket(x,t)$
  $\llbracket\mathrm{op1}\rrbracket(t,y)$)} which can be summarized with the following pseudo-code:

\begin{lstlisting}[basicstyle=\ttfamily\footnotesize,columns=flexible]
  t := allocate o2
  t := op2 x
  y := op1 t
  free t
\end{lstlisting}

The variables in both \MSHCOL AST and \RHCOL are referenced by de
Bruijn indices. However, since the lexical structure of the two languages is
different, a non-trivial mapping between two indices must be
established. This is implemented with the help of the ``variable resolvers''
mechanism, defined in Listing~\ref{lst:resolvers}.

\begin{lstlisting}[language=Coq,
  mathescape=true,basicstyle=\ttfamily\footnotesize,columns=flexible,label={lst:resolvers},caption={Variable
    resolvers}]
Definition var_id := nat.
Definition var_resolver := nat -> var_id.

Definition ID_var_resolver : var_resolver := id.
Definition Fake_var_resolver (parent: var_resolver) (n:nat) : var_resolver
  := fun r => (parent r) + n.
Definition Lambda_var_resolver (parent: var_resolver) (n:nat) : var_resolver
  := fun r => if lt_dec r n then r else (parent (r-n)) + n.
\end{lstlisting}

The \textit{var\_resolver} is a map from \MSHCOL to \RHCOL
variable indices. The most basic one is {\tt ID\_var\_resolver} which
establishes identity mapping between the two. The next useful resolver
is {\tt Fake\_var\_resolver} which is used when \RHCOL introduces one
or more new variables, which have no counterparts in \MSHCOL. This
resolver is stacked on top of an existing resolver modifying its behavior
to accommodate for such new variables. This is what is happening in
Listing~\ref{lst:compose2dhcol}. We introduce a new {\tt
  Fake\_var\_resolver} corresponding to one new variable used to store the
allocated temporary memory block. Since this variable is lexically
scoped by {\tt DSHAlloc}, the new resolver will be used only while
compiling the operators constituting the {\tt DSHAlloc} body. If these operators
reference the other variables declared earlier, their references will be
computed by adding a unit offset to take into account the new
temporary variable introduced by {\tt DSHAlloc}.

In the rest of Listing~\ref{lst:compose2dhcol}, we ``unquote''\footnote{
Template-Coq term which means decoding from AST representation.} memory
block sizes to natural numbers. Then, we compile both operators using
the {\tt Fake\_var\_resolver} with $n=1$. When compiling two operators,
we need to specify the input and output variable indices
for each operator. The index of the newly allocated temporary variable
will be 0, being the most recently introduced. To reference the original
{\tt x\_p} and {\tt y\_p}, we need to increment their indices by 1
using the {\tt incrPVar} function to take into account the new temporary
variable introduced by {\tt DSHAlloc}. This might look similar to what
we do in the resolver, but these are \RHCOL variable indices, which
have already been resolved earlier and thus, the resolver mechanism
will not be used, and manual adjustment is required. Finally, we
construct the resulting expression, which consists of allocation of a temporary memory block followed by sequential execution of the \RHCOL code
corresponding to {\tt op2} and {\tt op1}.

Another resolver we use elsewhere is {\tt Lambda\_var\_resolver} which
is required when compiling functions with $n$ arguments. It is
typically used when compiling the function argument of operators like {\tt
  MSHPointwise} or {\tt MSHBinOp}. It
will introduce $n$ new variables with de Bruijn indices $0\ldots n-1$
representing the parameters of the function. While compiling a function,
{\tt Lambda\_var\_resolver} will map these indices as unchanged. However,
when \MSHCOL variables with indices outside this range are mapped, $n$ is first subtracted, the
parent resolver is applied, and then $n$ is added back to the result.

\begin{figure}[h]
  \centering
  \includegraphics[keepaspectratio=true]{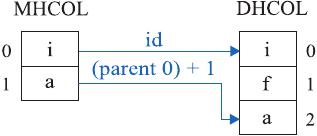}  
  \caption{Variable resolver example}
  \label{fig:resolver}
\end{figure}

To illustrate this, let us
consider an \MSHCOL expression which contains a function with one argument
$\lambda i.i+a$. We assume the current resolver to be {\tt
  parent\_resolver = Fake\_var\_resolver (ID\_Var\_resolver) 1} and
the de Bruijn index of $a$ in \MSHCOL is $1$. When compiling our lambda function, we use
{\tt(Lambda\_var\_resolver parent\_resolver 1)}. When resolving $i$
using this resolver, we get $0$ because of the identity mapping in the lambda arguments. To resolve $a$ with index 1, we  first pass
$1-1$ to the parent resolver which returns $1$ as an index of $a$
before entering lambda. Then $1$ is added to compensate for the
function argument resulting in the final mapping of $2$. Variable
index mapping for this example is shown in Figure~\ref{fig:resolver}

\subsubsection{Proof of semantics preservation}

While the regular \MSHCOL operators translate to a \RHCOL program
fragment, the higher-order operators translate into a sequence of
instructions, with placeholders filled with \RHCOL translations of
their respective parameters. For example, {\MSHCOL}'s {\tt
  (MSHIReduction i o n z f op\_family)} operator is compiled to the
following {\RHCOL} program:

\begin{lstlisting}[language=Coq,basicstyle=\fontsize{8}{12}\selectfont\ttfamily,
  columns=flexible]
DSHSeq
  (DSHMemInit o y_p z)
  (DSHAlloc o (DSHLoop n (DSHSeq dop_family (DSHMemMap2 o y_p$'$ (PVar 1) y_p$'$ df)))))
\end{lstlisting}

The parameters of the {\tt MSHIReduction} operator are the dimensions of the
input and the output vectors ($i$ and $o$ respectively), the size $n$ of the
operator family {\tt op\_family}, and the initialization value $z$. The
parameters  {\tt df} and {\tt dop\_family} in \RHCOL correspond to {\tt f} and
{\tt op\_family}  in \MSHCOL, respectively.

Operators {\tt DSHAlloc} and {\tt DSHLoop} introduce two new
variables: the pointer to a newly allocated memory block and the loop
index. Inside the loop, they can be referenced by their respective
de Bruijn indices as {\tt (PVar 1)} and {\tt (PVar 0)}. To evaluate each iteration, the {\tt
  dop\_family} takes the loop index to access the family operator
member, which is then executed and writes
output to a temporary memory block. The output of {\tt MSHIReduction} is
assumed to be written to a memory block referenced by variable {\tt
  y\_p}, and {\tt y\_p$'$} is the same variable with the de Bruijn
index increased by two to accommodate for the loop index and a new
variable holding a reference to the newly allocated temporary memory
block.

We want to prove that our translation from \MSHCOL to \RHCOL
preserves the semantics. As with other HELIX languages, we use an
automated \textit{translation validation} approach. To allow automatic
proof of translation results, we need to prove correctness lemmas for
each \MSHCOL operator and its \RHCOL translation. Then, these
lemmas can be applied recursively, descending the
structure of the reified \MSHCOL expression hierarchically.

The first step in the process is to formalize the notion of semantic
equivalence between a purely functional language with denotational
semantics (\MSHCOL) and an imperative language with operational
semantics (\RHCOL). Each \MSHCOL operator is a function
$x \mapsto y$ where {\tt x} and {\tt y} are memory blocks\footnote{We
  omit error handling for now.}. These functions are \textit{pure functions}
without side effects, whose output {\tt y} depends on {\tt x} and
other variables in scope. On the other hand, a \RHCOL translation of
this \MSHCOL operator is an imperative program that can read
variables available in the \textit{evaluation context} and can also
 read and modify the memory. One block from this memory will
correspond to {\tt x}, and some other block will correspond to {\tt
  y}. Being a translation of a pure function, the operator can modify
only {\tt y}. The formalization of the class of \RHCOL programs
representing pure functions is expressed as a {\tt DSH\_pure} typeclass:

\begin{lstlisting}[language=Coq,basicstyle=\fontsize{8}{12}\selectfont\ttfamily,
  columns=flexible,literate={exists}{exists}6,
  label=lst:dshpure,caption={{\tt DSH\_pure} typeclass}]
  Class DSH_pure  (d: DSHOperator) (y: PExpr) := {
      mem_stable: forall $\sigma$ m m$'$ fuel,
        evalDSHOperator $\sigma$ d m fuel = Some (inr m$'$) ->
        forall k, mem_block_exists k m <-> mem_block_exists k m$'$;

      mem_write_safe: forall $\sigma$ m m$'$ fuel,
          evalDSHOperator $\sigma$ d m fuel = Some (inr m$'$) ->
          (forall y_i , evalPexp $\sigma$ y = inr y_i ->
                   memory_equiv_except m m$'$ y_i)
    }.
\end{lstlisting}

\noindent It has the following two properties:
 
\begin{enumerate}[label=$\circ$]
\item \textit{memory stability} states that the operator does not free
  or allocate any memory blocks.
\item \textit{memory safety} states that the operator modifies only the memory block
  referenced by the pointer variable {\tt y}, which must be valid in the environment, $\sigma$.
\end{enumerate}

Now, we can proceed to formulate the semantic equivalence between an
\MSHCOL operator and a ``pure'' \RHCOL program. Since the \MSHCOL
part of this relation is a function, we need to universally quantify
on all possible inputs. Since \RHCOL operators read and modify
memory, the input and output of this function must correspond to some
existing memory blocks. In \RHCOL memory, locations can be
accessed via pointer variables only, so we state that there are two
pointer variables in the evaluation context corresponding to the input
and output memory block locations. For convenience, we define semantics
equivalence as a type class parameterized by the respective \MSHCOL
and \RHCOL operators, the evaluation context, and by the name of the
input and output pointer variables in this context. Additionally, the
purity of the \RHCOL operator must be guaranteed by providing a {\tt
  DSH\_pure} instance.

\begin{lstlisting}[language=Coq,basicstyle=\fontsize{8}{12}\selectfont\ttfamily,columns=flexible,
  label=lst:mshdshcompat,caption={{\tt MSH\_DSH\_compat} typeclass}]
Class MSH_DSH_compat
      {i o: nat} ($\sigma$: evalContext) (m: memory)
      (mop: @MSHOperator i o) (dop: DSHOperator)
      (x_p y_p: PExpr) `{DSH_pure dop y_p} :=
      {
      eval_equiv: forall (mx mb: mem_block),
          (lookup_Pexp $\sigma$ m x_p = inr mx) -> (lookup_Pexp $\sigma$ m y_p = inr mb) ->
          (h_opt_opterr_c
             (fun md m' => err_p (fun ma => SHCOL_DSHCOL_mem_block_equiv mb ma md)
                              (lookup_Pexp $\sigma$ m' y_p))
             (mem_op mop mx)
             (evalDSHOperator $\sigma$ dop m (estimateFuel dop)));
      }.
\end{lstlisting}

In the Listing~\ref{lst:mshdshcompat}, {\tt h\_opt\_opterr\_c} deals with error
handling. While {\tt mem\_op} has simple error reporting via option
type, {\tt evalDSHOperator} has two-level error handling,
distinguishing between running out of fuel and other errors. The equality
is defined if both operators err (for whatever reason) or both
succeed, in which case, their results must satisfy a provided
sub-relation. The sub-relation (expressed via lambda) does additional
error handling via {\tt err\_p} to ensure that {\tt y\_p} lookup
succeeds in {\tt m$'$}. Finally, the equality is reduced to the predicate
{\tt SHCOL\_DSHCOL\_mem\_block\_equiv} relating memory blocks {\tt mb}, {\tt ma},
and {\tt md}.

\begin{figure}[H]
  \centering
  \includegraphics{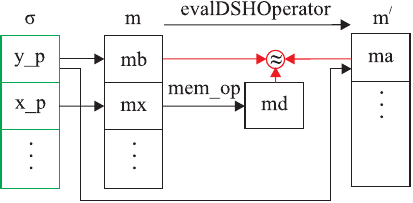}
  \caption{{\RHCOL} and {\MSHCOL} equality relation}
  \label{fig:compat}
\end{figure}

Figure~\ref{fig:compat} shows the origin of these values in a case
where no errors occur. Legend: $\sigma$ is an evaluation context, and {\tt
 m} and {\tt m$'$} are memory states before and after execution of
the {\tt evalDSHOperator}. The {\tt ma} corresponds to a memory block
in {\tt m$'$} referenced by {\tt y\_p}. The {\tt md} is the result of
applying the {\MSHCOL} operator to {\tt mx}.

To understand this relation, we must recall, that in \SHCOL,
sparse vectors represent the results of partial computation. Sparse
elements correspond to as yet uncomputed values, while dense elements
are already computed. Performing a union of the resulting sparse
vectors represents the combining of several partial computations. Replacing
immutable vectors with mutable memory blocks allows us to replace
the operation of combining computation results with a simple memory
update. Following this reasoning, the result of the \MSHCOL operator
application (called {\tt md}, where ``d'' stands for \textit{delta})
is a memory block containing values only at the indices that we need to
update. The values at all other indices must remain unchanged. On
the other hand, in \RHCOL, we know the memory state before the operator
evaluation and the updated state after it has been evaluated. Thus,
{\tt SHCOL\_DSHCOL\_mem\_block\_equiv} represents the relation
between:
\begin{itemize}[label=$\circ$]
\item {\tt mb} - memory state of the output block before {\RHCOL} execution
\item {\tt ma} - memory state of the output block after {\RHCOL} execution
\item {\tt md} - values of changed output block elements after {\MSHCOL} evaluation
\end{itemize}

This relation is implemented via the element-wise relation, {\tt MemOpDelta}, which is 
lifted to memory blocks as {\tt SHCOL\_DSHCOL\_mem\_block\_equiv}:

\begin{lstlisting}[language=Coq,basicstyle=\fontsize{8}{12}\selectfont\ttfamily,
  columns=flexible]
  Definition SHCOL_DSHCOL_mem_block_equiv (mb ma md: mem_block) : Prop
  := forall i, MemOpDelta
         (mem_lookup i mb)
         (mem_lookup i ma)
         (mem_lookup i md).

  Inductive MemOpDelta (b a d: option CarrierA) : Prop :=
  | MemPreserved: is_None d -> b = a -> MemOpDelta b a d 
  | MemExpected: is_Some d -> a = d -> MemOpDelta b a d.
\end{lstlisting}      

\noindent Informally, it could be stated as:
\begin{quote}
  For all memory indices in {\tt md} where a value is present, the
  value at the same index in {\tt ma} should be the same. For indices
  not set in {\tt md}, the value in {\tt ma} should remain as it was in {\tt mb}.
\end{quote}

Once we have proven {\tt SH\_MSH\_Operator\_compat} instances for all
\MSHCOL operators and their corresponding \RHCOL equivalents, we
can automatically generate proof for the result of any \MSHCOL to
\RHCOL translation as an instance of this class for top-level
\MSHCOL and \RHCOL expressions. During this proof automation, we
need to recursively descend on an \MSHCOL expression. The reason for
this is that mapping between the two is not injective, and compiling
two different \MSHCOL operators could result in similar \RHCOL
constructs not easily distinguishable by simple matching on the
structure. Whereas \MSHCOL operators can be uniquely matched.

\subsection{\FHCOL language}
\label{sec:fhcol}
The \FHCOL language is an instantiation of \DHCOL with machine numeric
types. This brings us one abstraction step closer to machine code from
\DHCOL.

We instantiate {\tt NType} module type with {\tt MInt64asNT} module to
represent unsigned 64-bit machine integers.  The underlying machine
integer (type {\tt int}) type definition from the Vellvm library,
which can be traced back to a similar CompCert definition, is defined
as an arbitrary-precision integer (type {\tt Z}) plus the proof that
it is in the range from 0 (inclusive) to {\tt modulus} (exclusive):

\begin{lstlisting}[language=Coq, mathescape=true,basicstyle=\ttfamily\footnotesize,columns=flexible]
Record int: Type := mkint { intval: Z; intrange: -1 < intval < modulus }.
\end{lstlisting}

For example, the concrete type {\tt Int64} that we use, is an
instantiation of {\tt int} with $\mathrm{modulus}=2^{64}$. Definitions
of the remaining fields of the {\tt NType} module and required proofs
are straightforward. To extend HELIX to generate code for other
hardware platforms, {\tt int} could be similarly instantiated for
32-bit or even 16-bit integers.

We instantiate the {\tt CType} module type with the {\tt MFloat64asCT}
module, representing fixed-length 64-bit IEEE floating-point
numbers. To work with IEEE floating-point numbers, we use the
Flocq~\citep{boldo2011flocq} library, which provides a comprehensive
formalisation of IEEE floating-point arithmetic in Coq. Flocq supports
a range of binary float representations. We are using Flocq's {\tt
  binary64} type which corresponds to the {\tt double} type in the C
language. We instantiate the {\tt CType} module type for the {\tt
  binary64} type, which represents 64-bit IEEE floating-point
numbers. As with integers, the definitions of all the required module
fields are straightforward. Whenever the rounding mode needs to be
specified, we always use \emph{round to the nearest, ties to even}.

\subsubsection*{Running example}

A result of the translation of a \RHCOL formulation of the dynamic window
monitor from Listing~\ref{lst:dwrhcol} to \FHCOL is shown in
Listing~\ref{lst:dwfhcol}. For simplicity, we use standard mathematical notation
for arithmetic operations over {\tt CType.t} and {\tt NType.t}, array
subscription notation for access to memory blocks and semicolon notation for
{\tt DSHSeq} operator within Listing~\ref{lst:dwfhcol}.

\dynwinlisting[language=Coq,label=lst:dwfhcol,
caption={Dynamic Window Monitor in \FHCOL},
basicstyle=\fontsize{8}{8}\ttfamily]{dynwin_fhcol_pretty.v}

\subsection{\RHCOL to \FHCOL translation}
\label{sec:rhcol2fhcol}
\subsubsection{Implementation}
\label{sec:dhcoltranslator}

Translation between any two languages of the \DHCOL family is straightforward,
since they are implemented by different parametrizations of the same module and
thus have the same syntax and semantics, up only to the primitive values
and operations.

Translation is defined as a module parametrized by two \DHCOL languages. We
implement the translation function directly in Gallina. It works by
recursively traversing the structure of {\tt DSHOperator}, utilizing one-to-one
correspondence between \DHCOL and \FHCOL language constructs. 

\begin{lstlisting}[language=Coq, mathescape=true,
                   basicstyle=\ttfamily\footnotesize,
                   columns=flexible,
                   label={lst:dshtypetranslator},
                   caption={\DHCOL family translator}]
Module MDHCOLTypeTranslator
       (Import CT: CType)
       (Import CT': CType)
       (Import NT: NType)
       (Import NT': NType)
       (Import L: MDSigmaHCOL(CT)(NT))
       (Import L': MDSigmaHCOL(CT')(NT'))
       (Import LE: MDSigmaHCOLEval(CT)(NT)(L))
       (Import LE': MDSigmaHCOLEval(CT')(NT')(L')).

[...]

Fixpoint translate (d: L.DSHOperator): err L'.DSHOperator := [...]
\end{lstlisting}

The only reason a translation may fail is if a constant of the source language
is not representable as a constant of the target language:

\begin{enumerate}
\item
  Certain {\tt CType.t} values may not be directly translatable. For example,
  during the \RHCOL to \FHCOL compilation pass, we might need to represent $\pi
  : \R$ exactly as a {\tt binary64} number, which is not possible. Currently,
  only known constants can be translated, and just two are presently defined,
  corresponding to {\tt CTypeZero} and {\tt CTypeOne}. Any other constant if
  encountered will cause translation to fail. The list of known constants could
  be extended in the future, if needed.

\item
  Translation of {\tt NType.t} constants faces a similar issue. In the
  \RHCOL to \FHCOL translation case, values of type $\N$ need to be translated
  to {\tt Int64}. The mapping is not total, as some natural numbers can not fit
  64-bit integer representation. The natural numbers which need to be translated
  appear in {\tt NExpr} constants and memory block sizes. If such an
  out-of-range natural number value is encountered, the translation will fail.
\end{enumerate}

%

\subsubsection{Proof of semantics preservation}

Semantic preservation is proven generically for a translation between any two
\DHCOL family languages. It is parametrized by \emph{heterogeneous equivalence
relations} on {\tt CType.t} and {\tt NType.t} of the respective languages
(Listing~\ref{lst:dshheq}). The particular instantiation of these relations will
influence resultant correctness guarantees: the more restricted the relation, the
more restricted the guarantees.

This parametrization enables us to prove different statements for
different translation passes. Most notably, an exact equivalence
cannot be established for the $\R$ to {\tt binary64} translation of
the \RHCOL to \FHCOL step. Exactly equivalent real and floating-point
numbers may not preserve this quality over equivalent operations due
to rounding, preventing us from establishing
such an invariant throughout the execution of respective
programs.

For example, consider two real numbers, 0.1 and 0.2, and their sum
0.3, in real arithmetic. However, in IEEE 754 \textit{binary64} (which
corresponds to the \texttt{double} type in the C language) floating
point arithmetic, $0.1 + 0.2 \neq 0.3$.

Thus for this step, instead of an exact relation, we use a
trivial on which allows us to establish other necessary properties of
the translation (such as the preservation of memory and stack
operations), while offloading numerical analysis to be performed
later. One example of such verified numerical analysis is shown in
Section~\ref{sec:dynwin}.

\begin{lstlisting}[language=Coq, mathescape=true,
                   basicstyle=\ttfamily\footnotesize,
                   columns=flexible,
                   label={lst:dshheq},
                   caption={DHCOL primitive type equivalence classes}]
(* Heterogeneous equivalence on values before and after translation *)
Class NTranslation_heq :=
  {
    heq_NType : NT.t -> NT'.t -> Prop ;
    heq_NType_proper : Proper ((=) ==> (=) ==> iff) heq_NType;
    (* Value mapping should result in "equal" values *)
    translateNTypeConst_heq_NType :
      forall x x', translateNTypeConst x = inr x' -> heq_NType x x';
  }.

Class CTranslation_heq :=
  {
    heq_CType' {env : Env} : CT.t -> CT'.t -> Prop ;
    heq_CType_proper : Proper (($\equiv$) ==> (=) ==> (=) ==> iff) (@heq_CType');
    (* Value mapping should result in "equal" values *)
    translateCTypeConst_heq_CType :
      forall x x', translateCTypeConst x = inr x' ->
            forall env, @heq_CType' env x x';
  }.
\end{lstlisting}

Proof of semantic preservation of the \RHCOL to \FHCOL translation also involves
some numerical analysis, discussed below in \cref{sec:floatproofs,sec:intproofs}.

\subsubsection{Floating-point numerical analysis}
\label{sec:floatproofs}

The fact that the HELIX compilation chain targets efficient and portable (in
LLVM IR) machine code, imposes a restriction on the set of data representations
used by the final compiled program. In the compilation chain leading up to
FHCOL, all representations of data and expressions have been either fully
abstract ($\ctype$ in \HCOL to \MSHCOL) or infinitely precise ($\R$ in
\RHCOL).

Actual hardware usually compromises when representing real values and
operations on them in favour of fixed representation size in memory
and registers, and prioritises performance over ultimate precision. As
a consequence, fixed or floating-point types used by hardware cannot
represent arbitrary real values exactly (in particular, irrational
numbers, such as $\sqrt{2}$). Methods exist for achieving exact
results using finite hardware, such as multiple-precision arithmetic,
capable of perfectly capturing any computations on rational numbers;
or symbolic computation which allows analysing abstract expressions
over arbitrary variables. However, these are all much more
resource-intensive than fixed precision arithmetic built into any
processor; and so in HELIX we choose to use standard finite-precision
IEEE-754 floating point arithmetic for operations in the compiled
program.

This choice poses a problem for another stated goal of HELIX -- safety. Almost
any computation carried out in floating point arithmetic will necessarily
accumulate an error that is only likely to grow, as more atomic operations are
performed by the processor.

For most everyday applications, this imprecision is negligible. However, to
provide safety guarantees on mission-critical systems HELIX is designed for, it
must be quantified and analyzed. An unaccounted-for rounding error can lead to
catastrophic results~\citep{general1992patriot}.

At this point, it is worth noting that the HELIX chain does not include any
optimizations aimed at improving the numerical stability of the compiled
algorithm.

In fact, HELIX also does not provide any general guarantees relating the
numerical stability of the compiled program to that of the source expression
(treated in a floating-point paradigm). \HCOL breakdown
(Section~\ref{sec:hcolbreakdown}) and \SHCOL rewriting
(Section~\ref{sec:shcol2shcol}) steps of the chain, as well as some of the
compilation passes, while validated to be perfectly semantically preserving
under exact arithmetic assumed by {\tt CarrierProperties}, are not analyzed for
rounding errors they might introduce at later steps.

Establishing a numeric boundary on the rounding error introduced by a particular
program can not, by itself, be sufficient to prove that the program is safe for
use in the real world. No constant error bound $\epsilon$ is universally safe for all
applications; this part of the analysis necessarily requires understanding the
physical meaning of the numbers.

For example, numerical analysis can establish that the rounding error in an
algorithm does not exceed $0.001$. But in order to establish that this number is
safe, we need to first understand whether it measures millimeters or kilometers,
as well as what tolerances are involved in our application -- is it acceptable
to miss by a meter?

For mission critical applications where the user may wish to verify the
numerical properties of their algorithm, HELIX provides a foundation for
statically analyzing error bounds on the outputs and proving custom safety
properties of programs.

The mechanism for dealing with floating-point inaccuracy in HELIX is a symbolic
execution engine. The output of any \FHCOL program generated by HELIX can be
provably reduced to a set of arithmetic expressions over that program's
respective inputs. 

Certain characteristics of HELIX-generated programs mean they lend themselves
well to analysis via symbolic execution; some of the biggest challenges commonly
associated with the approach can be overcome:

\begin{itemize}

\item \emph{Path explosion.} While for general-purpose languages symbolic
  execution requires keeping track of many possible branches of execution, often
  resulting in exponential growth of the search space, HELIX-generated \FHCOL
  programs are known to always take the same path over the execution tree. This
  means that a single pass is sufficient to explore the entire search space, or,
  in other words, that any \FHCOL program corresponds to a single pre-defined
  set of arithmetic expressions.

\item \emph{Environment interactions.} Any HELIX-generated FHCOL program is pure,
  meaning no external calls need to be modeled.

\item \emph{Modeling memory, symbolic store updates.} The modular architecture
  of \DHCOL enables symbolic execution of any program to be defined in terms of
  \DHCOL's regular semantics. In particular, it is sufficient to instantiate
  the Carrier Type (see Section~\ref{sec:typemodules}) as the type of arithmetic
  S-expressions, to produce effectively a ``symbolic-valued'' language.  This
  way, the language's memory itself becomes the equivalent of a symbolic store,
  and the regular execution procedure can be seen as a symbolic execution
  pass.

  Any \DHCOL-family language can be converted into such a model using an
  instance of the standard translator (described in
  Section~\ref{sec:dhcoltranslator}), and since all translations within \DHCOL
  are proven to produce semantically equivalent programs, we can infer that the
  output of any \DHCOL program is exactly the output of its symbolic
  counterpart, evaluated with the program's concrete input.

\end{itemize}

Combined with automated numerical analysis tools such as Gappa~\citep{de2008certifying},
this means that error bounds can be inferred automatically for all outputs of
all \FHCOL programs. See Section~\ref{sec:dynwinfloatproofs} for a case study on
how it may be used in practice.

\subsubsection{Integer range analysis}
\label{sec:intproofs}

Several \DHCOL operators, such as \texttt{DSHIMap}, \texttt{DHSBinOp}
and \texttt{DSHLoop} are parameterised by a natural number, which
controls the number of iterations during iterative computations and is
usually related to the input vectors' dimensions. During their evaluation,
the index's current value, upper-bounded by this parameter, will be
assigned to an environment variable of type \texttt{NType.t}, which
could be used in integer expressions.

While \RHCOL maps \texttt{NType.t} to $\N$, in \FHCOL it is mapped to
unsigned fixed-width machine integers. This makes the transition from
the unbounded $\N$ in \RHCOL to bounded machine integers in \FHCOL
problematic as it is susceptible to an integer overflow. Additionally,
evaluation of \texttt{NType.t} expressions in \RHCOL and \FHCOL might
result with an integer underflow, which is handled differently within these
languages. In \RHCOL result of $a - b$ where $b > a$ is defined to be
$0$, while in \FHCOL the result would be computed modulo $2^n$.

Integer overflow and underflow can lead to incompatible behavior
between an \RHCOL and an \FHCOL program, which might result in runtime
errors such as out-of-bounds vector indices or wrong numbers of loop
iterations, etc. The only way to guarantee the correct execution of a
program is to prove that no such incompatible behavior is introduced
during \RHCOL to \FHCOL translation. This property is not necessarily
true of all HELIX-generated programs, but can still be automatically
inferred for a given one, assuming it does hold.

A practical consideration worth mentioning here is that HELIX is
designed to operate on dense vectors in memory. Consequently, for most
of the practical applications we envision, the individual vectors
typically have relatively small sizes. Since generated loop bounds and
vector sizes represented as $\N$ in \RHCOL are usually linked to data
dimensions, that also makes them relatively small, and the overflow
when translating into \FHCOL is unlikely to occur.

The general approach for ensuring integer bounds guarantees for \RHCOL
to \FHCOL translation is as follows: the main observation is that the
structure of computation will be the same in both languages. At each
step of the evaluation, there is a one-to-one relation between their
corresponding evaluation environments. For this analysis, we do not
concern ourselves with \texttt{CType.T} variables and only consider
variables holding \texttt{NType.t} values. We aim to prove that an
integer overflow never occurs and that for all $\N$ variables in
\RHCOL are represented exactly as fixed-width integers in \FHCOL.

However, ensuring the exact representation of integer variables is not
enough. During the evaluation of iterative operators such as
\texttt{DSHIMap}, the index will be used in integer expressions
(\texttt{NExpr} type) which could exhibit an integer overflow during
evaluation. For that reason, we need to ensure that the evaluation of
such expressions, with all index values in the expected range, will
not cause incompatible behavior due to an integer overflow or
underflow.

In order to establish equivalence between \RHCOL and \FHCOL programs,
we define \emph{closure trace} semantics for \DHCOL programs. Firstly,
we consider an evaluation context that contains ranges of values for
integer variables. Pointers and Carrier Type values in the context are
ignored by replacing them with placeholders to preserve the de Bruijn
indices. Listing~\ref{lst:dshindexrange} shows the type of values that
are stored in the context. \coqinline{DSHIndex n} denotes an integer
variable in the range $[0, n]$, while \texttt{DSHOtherVar} is used as
a placeholder for non-integer variables.

\begin{lstlisting}[language=Coq, mathescape=true,
                   basicstyle=\ttfamily\footnotesize,
                   columns=flexible,
                   label={lst:dshindexrange},
                   caption={Reduced {\tt DSHVal}}]
Inductive DSHIndexRange : Type :=
| DSHIndex : NT.t → DSHIndexRange
| DSHOtherVar: DSHIndexRange.
\end{lstlisting}

We use the term \emph{range closure} to mean a pair consisting of an
\texttt{NExpr} and an integer evaluation context with
\texttt{DSHIndexRange} values for all \texttt{NType.t} variables it
may reference.

While computing the closure trace semantics no numeric computations of any
kind (not even integer operations) are performed. Instead, any time an
integer expression would normally be evaluated, it is instead saved
along with the ranges of its inputs as a ranged closure. Evaluating an
operator then results in a list of all such closures, which we refer
to as a \emph{closure trace}. \DHCOL's lack of branching and
statically-known loop bounds and vector sizes allow to compute the
closure trace statically. The same trace will always be encountered
for the same operator, independent of input. For example, during the
evaluation of {\tt DSHIMap} and {\tt DSHBinOp} operators, we require
the value of the variable used to access the index of the current
iteration to be within the input vector size and during the evaluation
of the {\tt DSHLoop} operator we require the value of the variable
that contains the loop index to be within the loop bounds.

Secondly, as shown in the Listing~\ref{lst:nexprclosureeq}, we
consider two range closures (before and after \RHCOL to \FHCOL
translation) to be equivalent if their corresponding integer
expressions $n$ and $n'$ are structurally equivalent
(\texttt{heq\_NExpr}) and evaluate (\texttt{evalNExpr}) to equivalent
(\texttt{heq\_NType}) results under all equivalent
(\texttt{heq\_evalContext}) contexts which are compatible with the
range restrictions (\texttt{evalContext\_in\_range}).

Specifically, for \RHCOL and \FHCOL pairs, the \texttt{heq\_NType}
predicate relates natural numbers to fixed-width machine integers and
defines that the second argument converted to $\N$ should match the
first argument exactly. This ensures that no incompatible behavior
occurs due to integer overflow or underflow during the evaluation of a
\texttt{NExpr} expression.

\begin{lstlisting}[language=Coq, mathescape=true,
        basicstyle=\ttfamily\footnotesize,
        columns=flexible,
        label={lst:nexprclosureeq},
        caption={Equivalence of ranged closures}]
Definition evalNExpr_closure_equiv
  '(($\sigma$n, n) : LE.evalNatClosure)
  '(($\sigma$n', n') : LE'.evalNatClosure)
  : Prop :=
  forall $\sigma$ $\sigma$',
    LE.evalContext_in_range $\sigma$ $\sigma$n ->
    LE'.evalContext_in_range $\sigma$' $\sigma$n' ->
    heq_evalContext $\sigma$ $\sigma$' ->
    heq_NExpr n n'
    ->
    herr_c heq_NType (LE.evalNExpr $\sigma$ n) (LE'.evalNExpr $\sigma$' n').
\end{lstlisting}

Finally, two operators (before and after translation) are equivalent
if all range closures from their closure traces are equivalent. The
full semantical preservation statement is proven by relying on this
assumption. For a given operator, it becomes a proof obligation,
discharged using simple proof automation (reducing the goal to a set
of linear inequalities for all the ranges).

Listing~\ref{lst:intdhcolexample} shows an example of a \DHCOL
program. In this example, {\tt DSHIMap} operator is used inside {\tt
  DSHLoop} operator. During the evaluation of both of these operators,
an additional variable used to access the current iteration index is
added to the context. This program is equivalent to the C program
shown in Listing~\ref{lst:intcexample}.

\begin{figure}[H]
  \begin{minipage}[t]{.45\textwidth}
    \begin{lstlisting}[language=Coq, mathescape=true,
  basicstyle=\ttfamily\footnotesize,
  columns=flexible]
DSHLoop 3
  (DSHIMap 3
     (PVar 1) (PVar 2)
     (APlus (ANth (MPtrDeref (PVar 3)) (NVar 1))
            (ANth (MPtrDeref (PVar 5))
                  (NMult (NConst 3) (NVar 2))))).
    \end{lstlisting}
  \end{minipage}\hfill
  \begin{minipage}[t]{.45\textwidth}
    \begin{lstlisting}[language=C, mathescape=true,
  basicstyle=\ttfamily\footnotesize,
  columns=flexible, captionpos=b]
for (uint64_t i = 0; i < 3; i++)
    for (uint64_t j = 0; j < 3; j++)
        y[i] = x[i] + a[3 * j];
    \end{lstlisting}
  \end{minipage}
  \begin{minipage}[b]{.45\textwidth}
    \captionof{lstlisting}{Example of a DHCOL program}
    \label{lst:intdhcolexample}
  \end{minipage}\hfill
  \begin{minipage}[b]{.45\textwidth}
    \captionof{lstlisting}{Equivalent C program}
    \label{lst:intcexample}
  \end{minipage}
\end{figure}

Listing~\ref{lst:inttraceexample} shows the computed closure trace for
this program. In this example, the {\tt DSHIMap} index variable (\coqinline{NVar 1}, which corresponds to \cinline{j} in the C program) is bounded
by the number of iterations that {\tt DSHIMap} performs. For the {\tt
  DSHLoop} operator, range closures are computed individually for each
iteration: on every iteration, the loop index variable (\coqinline{NVar 2}, which corresponds to \cinline{i} in the C program) is bounded by
the index of the current iteration.

\begin{lstlisting}[language=Coq, mathescape=true,
  basicstyle=\ttfamily\footnotesize,
  columns=flexible,
  label={lst:inttraceexample},
  caption={Closure trace of a DHCOL program}]
[(* loop iteration 3 *)
([DSHOtherVar; /&DSHIndex 3&/; /&DSHIndex 2&/; DSHOtherVar; DSHOtherVar; DSHOtherVar],
 NVar 1); (* i *)
([DSHOtherVar; /&DSHIndex 3&/; /&DSHIndex 2&/; DSHOtherVar; DSHOtherVar; DSHOtherVar],
 NMult (NConst 3) (NVar 2)); (* 3*j *)
 (* loop iteration 2 *)
([DSHOtherVar; /&DSHIndex 3&/; /&DSHIndex 1&/; DSHOtherVar; DSHOtherVar; DSHOtherVar],
 NVar 1); (* i *)
([DSHOtherVar; /&DSHIndex 3&/; /&DSHIndex 1&/; DSHOtherVar; DSHOtherVar; DSHOtherVar],
 NMult (NConst 3) (NVar 2)); (* 3*j *)
 (* loop iteration 1 *)
([DSHOtherVar; /&DSHIndex 3&/; /&DSHIndex 0&/; DSHOtherVar; DSHOtherVar; DSHOtherVar],
 NVar 1); (* i *)
([DSHOtherVar; /&DSHIndex 3&/; /&DSHIndex 0&/; DSHOtherVar; DSHOtherVar; DSHOtherVar],
NMult (NConst 3) (NVar 2)) (* 3*j *)]
\end{lstlisting}

The example program contains two integer expressions that correspond to \cinline{i} and \cinline{(3 * j)}, which are used to access values from the input vector \cinline{x} and the constant parameter \cinline{a} respectively. Thus, the resulting closure trace contains six range closures: range closures for the two integer expressions are computed for each of the three loop iterations.

\section{LLVM IR}
\label{sec:fhcol2llvm}







We are reaching the final step in our chain: \FHCOL programs are
further compiled to LLVM IR. This is a low-level language of the LLVM
toolchain which can be further compiled to machine code for a variety
of supported instruction sets. At the time of this writing, LLVM
18.1.1 supports code generation for IA-32, x86-64, ARM, Qualcomm
Hexagon, LoongArch, M68K, MIPS, NVIDIA Parallel Thread Execution,
PowerPC, AMD TeraScale, SPARC, z/Architecture, and
XCore. Using LLVM toolchain terminology, our \FHCOL to IR
  compiler can be considered an LLVM \textit{front end} for the \FHCOL
  language.\footnote{Conversely, from the HELIX point
    of view, the compiler can be considered a \textit{back end}.}

This compilation step is notable as the LLVM IR is a general-purpose language not specifically designed for this project. 
In particular, it is the first Turing-complete language in HELIX compilation chain. As such, the gap to bridge is arguably
larger than in the previous step, especially in terms of semantic preservation.

At a high level, LLVM IR programs are sets of mutually recursive
functions, where functions are control flow graphs with blocks of code
that define local variables following the static single assignment
(SSA) invariant. We refer the interested reader to the reference
manual\footnote{\url{https://llvm.org/docs/LangRef.html}} for an
extensive description of the language.

To build our verified compilation chain, we naturally need a formal
version of LLVM IR: HELIX relies to this end on the Vellvm project~
\citep{znmz12,znmz13,vellvmicfp}. Vellvm exposes a formal language,
dubbed \VIR, deeply embedded in Coq, that models a large fragment of
LLVM IR. Its infrastructure offers (1) a parser for valid LLVM IR
syntax into its internal representation of (VIR) programs, (2) a
formal semantics equipped with a battery of lemmas to reason about
refinement of programs, (3) a pretty printer into valid LLVM IR syntax
for LLVM toolchain interoperability, and (4) an executable interpreter
proven correct with respect to \VIR semantics, for testing purposes.

Its semantics is relatively unconventional: programs are modelled as
monadic computations. Intuitively and to a first approximation, one
can think of it as writing a monadic interpreter in a language like
Haskell. The major caveat, however, is that unlike Haskell, our host
language, Gallina, does not allow for divergence. This issue, which
may seem as a showstopper, is resolved by representing programs as
coinductive objects, elements to a generalization of Capretta's delay
monad~\citep{capretta}: no divergence is built internally to Coq, and a
lazy interpreter is recovered by extraction to OCaml.  This semantics
is built using a modern library providing a wealth of semantic tools
for building and reasoning about such monadic interpreters, the
\emph{Interaction Trees} (ITrees)
library~\citep{xia2019interaction,yzz22}.

Verifying the translation of \FHCOL to \VIR constitutes one of the largest stress
test to Vellvm's meta-theory, as well as more generally to the itree-based
approach to verified compilation. Through this section, we aim to describe the
technique, but also document the lessons for both of these projects we believe
have emerged from the experience.

We first focus on the description of the compilation pass itself, before
discussing the formal theorems we establish, and major elements of their proof.

\subsection{Compiling from \FHCOL to \VIR}
\label{sec:compiler}
\subsubsection{Type mapping}
\label{sec:irtypes}

Both \FHCOL and LLVM use IEEE-754 64-bit binary floats, and during compilation, {\FHCOL}'s {\tt CType.t} type is mapped to {\tt TYPE\_Double} in {\VIR}'s AST. Both languages use Flocq's {\tt binary64} type for floating-point values.

{\FHCOL}'s {\tt NType.t} is mapped to {\tt TYPE\_I64} in {\VIR}'s AST. Similarly, {\FHCOL}'s {\tt Int64.int} values are mapped to {\VIR}'s own {\tt int} type, which implements 64-bit unsigned integers.
The LLVM type system
makes no distinction between signed and unsigned integers, unlike C
language. IR integers can be interpreted either as signed or unsigned
depending on the operations used. For instance, the {\tt udiv}
instruction will perform unsigned integer division, while the {\tt
  sdiv} instruction performs signed division. Since \FHCOL integers
are always non-negative, we will always emit unsigned instructions
when generating IR code.

\FHCOL Memory blocks are mapped to IR memory regions and accessed
 using pointers to {\tt [n x double]} where $n$ is the block size. The {\tt getelementptr} instruction is used to address individual values of floating point arrays. All new memory blocks allocated by the compiled \FHCOL program are allocated on the stack frame using {\tt alloca} instruction.

\subsubsection{Translation units}
\label{sec:cu}
A top level translation unit is an \textit{\FHCOL program} which
corresponds to an \FHCOL operator which may depend on one or more
global variables. Programs are described by the record shown in
Listing~\ref{lst:fhcol-prog}.

\begin{lstlisting}[language=Coq,basicstyle=\ttfamily\footnotesize,
  columns=flexible,caption={\FHCOL Program definition},label={lst:fhcol-prog}]
Record FSHCOLProgram := mkFSHCOLProgram {
    i o: Int64.int;
    name: string;
    globals: list (string * DSHType);
    op: DSHOperator;
}.
\end{lstlisting}

The record contains the name of the function to generate, a list of
global variable names with their corresponding types, and the {\tt
 DSHOperator}, which represents the program code.

To recall from \ref{sec:dhcoleval}, \FHCOL evaluation context is a De-Bruijn-indexed list of typed variables. In a well-formed \FHCOL program, the {\tt op} assumes that the evaluation context contains global variables starting from index 0 and matching the order and types from {\tt globals}, followed by two variables which are pointers to allocated memory blocks of sizes $i$ and $o$, holding the input and output data vectors, respectively.

All global \FHCOL variables are mapped to corresponding \VIR global variables, which, depending on the compiler invocation flags, are declared with either the \textit{external} or \textit{internal} linkage type. Specifically, when compiler flags indicate code generation for testing, global variables are initialized with random data, as discussed in Section~\ref{sec:cctests}. Local \FHCOL variables are similarly mapped to \VIR local variables.

\subsubsection{Compiler organisation}
\label{sec:comp-orga}
The \FHCOL to \VIR compiler is written in Gallina. It translates
\FHCOL programs into a corresponding \VIR AST. The main compilation
logic is encoded in the {\tt genIR} function, which translates {\tt
  DSHOperator} into a non-empty sequence of LLVM \textit{blocks}. It
takes as a parameter a successor block id, where control should be
passed at the end of the compiled operator execution (also known as
``destination-passing style'').

\begin{lstlisting}[language=Coq,basicstyle=\ttfamily\footnotesize,
columns=flexible]
(* List of blocks with entry point *)
Definition segment:Type := block_id * list (block typ).

Fixpoint genIR (fshcol: DSHOperator) (nextblock: block_id) : cerr segment.
\end{lstlisting}

The top-level compiler entry point used for testing and verification
is {\tt compile\_w\_main}, which performs some additional steps compared
to standalone compilation before invoking {\tt genIR}, such as
declaring and initializing global variables, generating function
declaration for the compiled operator, and generating the \textit{main}
function.

There are a few situations where \FHCOL program compilation may
fail. The most obvious example is an invalid \FHCOL program that attempts to access an undeclared variable or one of the wrong type. Another reason for errors is the integer size restriction: the loop
bounds in \FHCOL operators are expressed as natural numbers which may
not fit into the target platform integer size (e.g. {\tt int64}). For
\FHCOL programs generated by HELIX, most of these errors are guaranteed
not to occur, but instead of requiring formal guarantees as a
pre-condition to invoking the compiler, we add error handling and later prove that the compiler never fails on
HELIX-generated programs. The error handling is implemented via the
\textit{exception monad}. The compiler returns either a compiled program or
an error message as a \textit{string}.

During compilation, the compiler maintains the compiler state. The state consists
of integer counters to generate unique LLVM names and the \textit{typing
  context} $\Gamma$. The state data type is shown in Listing~\ref{lst:irstate}

\begin{lstlisting}[language=Coq,basicstyle=\ttfamily\footnotesize,
  float=htpb,
  caption={Compiler state},
  columns=flexible,label={lst:irstate}]
Record IRState := mkIRState {
    block_count: nat;
    local_count: nat;
    void_count : nat;
    $\Gamma$: list (ident * typ)
}.
\end{lstlisting}

The types of all \FHCOL variables are recorded in
$\Gamma$ since \FHCOL does not distinguish between global and local
variables as they both live the in evaluation context. When compiling the top-level \FHCOL operator,
all pre-defined variables are assumed to be \textit{global} in \VIR while all
new variables created when opening scopes (e.g. memory pointers in
{\tt DSHAlloc} or loop indices in {\tt DSHLoop}) are considered
\textit{local} in \VIR.

The counters in {\tt IRState} are used to generate unique identifiers
for blocks, local identifiers, and void identifiers.\footnote{\textit{void} identifiers are used in Vellvm's SSA form to assign results of \VIR operators, such as {\tt store}, which do not return any values.}
The new identifiers are generated by appending the corresponding
counter number to an arbitrary string prefix.  The counters are
initialized with zeroes and increased after each new variable
generation. The combination of counter type and counter value is
guaranteed to be unique for each generated identifier. The string
prefix is arbitrary and sometimes used to give the identifier a
meaningful human-readable prefix to make the generated IR code more
legible.

In contrast, global identifiers already have pre-assigned names in
{\tt FSHCOLProgram.globals}, and these names are used in the generated
 \VIR code. It is assumed that they do not use the same prefix as locally generated variables, and the compiler performs additional checks to ensure they are unique and do not collide with the names of built-in LLVM intrinsics.
\VadimTODO{I can see checks for uniqueness and non-collision with intrinsics names in the compiler, but I do not see checks that they do not collide with generated locals. Perhaps we are handling this in proofs?}

Since \FHCOL is lexically scoped, as the compiler proceeds with structural recursion over the \FHCOL structure, $\Gamma$ at each \FHCOL variable's de Bruijn index holds the LLVM global variable names or compiler-assigned register names and their \VIR types.

The compiler state is maintained in a \textit{state monad}. To combine
it with error handling a new monad, combining features of the \textit{state
  monad} and the \textit{exception monad} is defined with instances for
both ExtLib's {\tt MonadExc} and {\tt MonadState} classes.

To give an example of internal compiler workings, we will now discuss how loops with integer indices are compiled. Such loops are used to compile \FHCOL operators such as {\tt DSHIMap}, {\tt DSHBinOp}, {\tt DSHMemMap2}, and others. The generated \VIR code should follow the shape described by the following pseudo-code:

\begin{lstlisting}[language=C,basicstyle=\ttfamily\footnotesize,
  columns=flexible]
  int i, from, to;

  init();
  i=from;

  while(i<to)
  {
    body();
    i++;
  }
\end{lstlisting}

{\noindent}which corresponds to the following IR code:

\begin{lstlisting}[language=llvm,basicstyle=\ttfamily\footnotesize,
  columns=flexible]
.entry:
    (init)
    %c0 = icmp ult i32 %start, %n
    br i1 %c0, label %.loop, label %.nextblock
.loop:
    %i = phi i32 [ %next_i, .loopcontblock], [ %start, .entry ]
   (body)
.loopcontblock:
    %next_i = add nsw i32 %i, 1
    %c = icmp ult i32 %next_i, %n
    br i1 %c, label %.loop, label nextblock
nextblock:
\end{lstlisting}

In particular, the IR code generated for the following \FHCOL
operators makes use of this construct: {\tt DSHMemInit},
{\tt DSHPower}, {\tt DSHIMap}, {\tt DSHBinOp}, {\tt DSHMemMap2},
and {\tt DSHLoop}.

To avoid duplication, the code generation for such loops was
abstracted into the {\tt genWhileLoop} function:

\begin{lstlisting}[language=Coq,basicstyle=\ttfamily\footnotesize,
  columns=flexible]
Definition genWhileLoop
           (prefix: string)
           (from to: exp typ)
           (loopvar: raw_id)
           (loopcontblock: block_id)
           (body_entry: block_id) (body_blocks: list (block typ))
           (init_code: (code typ))
           (nextblock: block_id)
  : cerr (block_id * list (block typ))
\end{lstlisting}

The loop initialization code is passed as \coqe{init_code}. The loop
body is passed as a pre-compiled list of blocks as
\coqe{body_blocks}. Its entry point is \coqe{body_entry}. After
execution, it is expected to pass control to {\tt loopcontblock}. To
be able to access the loop index variable within the body, its name
must be known in advance and thus, it is passed as {\tt loopvar}. Upon
completion, the generated loop code will pass control to
{\tt nextblock}. The loop bounds are passed as Vellvm expressions
{\tt from} and {\tt to}. Finally, the string {\tt prefix} is used
to generate new identifiers (using counters from {\tt IRState}). Upon
success, {\tt genWhileLoop} returns a list of blocks for the loop and
block id of the loop entry point. This loop generalization, in addition
to simplifying compiler implementation, is also useful for verification as it allows formulating and proving lemmas about loops
in general.

\subsection{Verification: background}

\subsubsection{A primer on Interaction Trees}


\newcommand{\itreen}{ITree\xspace}
\newcommand{\itreesn}{ITrees\xspace}
\newcommand{\ret}[1]{\ensuremath{\tm{ret}~#1}}
\newcommand{\bind}{~\tm{;\!;}~}
\newcommand{\itreef}[1]{\ensuremath{\tm{itree}~#1}}
\newcommand{\itree}[2]{\ensuremath{\itreef{#1}~#2}}
\newcommand{\translate}{\tm{tr}}
\newcommand{\trigger}[1]{\ensuremath{\tm{trigger}~(#1)}}
\newcommand{\triggernoparens}[1]{\ensuremath{\tm{trigger}~#1}}
\newcommand{\polytrigger}[1]{\ensuremath{\tm{polytrigger}~#1}}
\newcommand{\mrec}{\tm{mrec}}
\newcommand{\tloop}{\tm{loop}}
\newcommand{\titer}{\tm{iter}}
\newcommand{\tmrec}{\tm{mrec}\xspace}
\newcommand{\cloop}{\tm{loop}\xspace}
\newcommand{\citer}{\tm{iter}\xspace}
\newcommand{\interp}{\tm{interp}}
\newcommand{\caseitree}{\tm{case\_}}

\emph{Interaction Trees}~\citep{XZHH+20} (\itreesn) have emerged in the
Coq ecosystem as a rich toolbox for building compositional and modular
monadic interpreters for first order languages.  \itreesn are first
and foremost a data structure for representing computations
interacting with an external environment through \textit{visible
  events}. Concretely, \itreesn are defined as:

\begin{lstlisting}[language=Coq,float=htpb,basicstyle=\footnotesize\ttfamily]
CoInductive itree (E: Type -> Type) (R: Type) : Type :=
  | Ret (r: R)           (* pure computation *)
  | Tau (t: itree E R)   (* "silent" tau transition *)
  | Vis {A: Type} (e : E A) (k : A -> itree E R). (* event e yielding an answer in A *)
\end{lstlisting}

The datatype takes two parameters: a signature \coqe{E} that specifies
the set of interactions the computation may have with the environment,
and the type \coqe{R} of values that it may return. \itreen
computations can be thought of as trees built out of three
constructors.  Leaves, via the \coqe{Ret} constructor, model pure
computations, carrying \coqe{R} values.  \coqe{Vis} nodes model an
effect \coqe{e} being performed, before yielding to the continuation
\coqe{k} with the value resulting from \coqe{e}. The \coqe{Tau}
constructor represents a non-observable internal step that
occurs. Notably, \itreesn are defined coinductively, allowing them to
model diverging computations as non-well-founded trees.

\itreesn form a convenient model of computation. They are a monad, supporting
the usual \coqe{(ret : A -> itree E A)} operation to embed pure
computations into the monad, and the sequencing of computations via
\coqe{(bind (c : itree E A) (k : A -> itree E B) : itree E B)}.
But they furthermore allow for modeling loops and recursion. In particular, the
\coqe{(iter (body : I -> itree E (I + A)) (i : i) : itree E A)} combinator builds
the operation computing first \coqe{(body i)}, and then pursuing based on the
value returned: if it is a new accumulator (a value \coqe{(inl j)}), it reenters
the body, if it is a final result (a value \coqe{(inr a)}), it terminates on this value.

These operations come with the expected laws: \coqe{bind} is
associative, \coqe{ret} acts as a unit on both side for the former,
and \coqe{iter} satisfies the slightly more involved but unsurprising
laws capturing the behavior of a loop.\footnote{I.e., the associated
  Kleisli category is traced.}  The precise meaning of these equations
depends on the notion of equality --- generally \itreesn are
considered equivalent up to weak bisimulation. Trees co-terminate,
i.e. both diverge or both terminate on the same value, while
additionally having the same set of traces of external events.

Weak bisimilarity captures an adequate notion of equality over
computations.  For many purposes, however, logical equality on return
values is too restrictive. For instance, computations which build
memory states may need to be compared with a more extensional
equivalence relation. More generally when reasoning about
optimization, arbitrary relations are quickly necessary in order to
emulate something akin to Benton's relational program
logic~\citep{10.1145/982962.964003}.  The interaction tree library
provides such facilities, specifically, \coqe{eutt R t u} expresses
that the computations \coqe{t} and \coqe{u} coterminate, with the same
traces of external events, and the returned value in each leaf is
related by \coqe{R}. Crucially, this extension makes it possible to compare
heterogeneous computations, in particular, to relate distinct memory 
models: \FHCOL's and \VIR's.

\itreesn are \emph{free} 
in the sense that external events can be given a
concrete model into any monad,\footnote{More precisely, any iterative monad: the
monad must be able to represent diverging computations.} i.e., can be implemented
after the fact.
This property is embodied in the \coqe{interp} function: given any \emph{handler}
\coqe{(h: forall X, E X -> M X)} implementing the events \coqe{E} in the monad
\coqe{M}, \coqe{interp h : forall X, itree E X -> M X} lifts this implementation
to computations.

Based on this notion of interpretation, the \itreen{} library is designed to structure the semantics of languages by first representing their syntax as trees whose events abstract all effects, and implement these effects into monads progressively. 
The approach has proven to scale to realistic language, as we will see for instance in the case of Vellvm in Section~\ref{sec:vellvm}. We illustrate here the idea on a minimal example.

\paragraph*{Example.}
As a mean to illustrate, let us imagine
we model a simple imperative language, with commands including assignments, sequencing and loops:
\[c ::= skip~\mid~c;c~\mid~x:=e~\mid~\mathtt{while}~b~\mathtt{do}~c\]
A first perspective on
modelling commands  is to represent them as elements of \coqe{itree MemE unit}
where \coqe{MemE} describes read and write operations. The
representation is built \emph{by recursion on the structure of
  commands}: the difficult case, the \texttt{while} loop, is
resolved thanks to the \coqe{iter} combinator.  However, at this
stage, equality of computations, defined as bisimilarity of the trees representing the commands, is too
tight: the memory model is left unspecified. Expected equations, such
as two consecutive writes to the same variables being equivalent to
only performing the second one, are not satisfied.  Indeed, the
bisimilarity observes the trace of all reads and writes that occur during the computation---reading twice from the same variable is different from reading only once.  We
hence refine the model by providing an implementation of reads and
writes in a simple state monad: by \emph{interpreting} the previous
representation, we obtain a model of commands in
\coqe{StateT mem (itree void) unit}, that is stateful computations that may diverge
silently, or eventually return a final memory.  On this model, pointwise
bisimilarity provides a suitable notion of equivalence of programs.

\subsubsection{A primer on Vellvm}
\label{sec:vellvm}

Interaction Trees have been put to work in a variety of contexts: the
verification of a web server~\citep{cpp19-ns,DBLP:conf/itp/ZhangHK0LXBMPZ21}, of
transactional objects~\citep{DBLP:journals/pacmpl/LesaniXKBCPZ22}, as semantic
foundations for contextual refinement~\citep{DBLP:journals/pacmpl/SongCLHSD23},
and even RoboChart~\citep{DBLP:journals/corr/abs-2303-09106}. They perhaps find
their largest application to date as part of the Vellvm~\citep{vellvmicfp}
project: a realistic sequential fragment of LLVM IR is given a denotational
semantics based on Interaction Trees.

\newcommand{\tm}[1]{\ensuremath{\mathtt{#1}}} 
\newcommand{\undefv}{\tm{undef}}
\newcommand{\dvpoison}{\tm{poison}\xspace}

The version of \VIR against which we interface in this work is essentially
based on the one presented in~\cite{vellvmicfp}, largely extended with new
contributions to its meta-theory.
It covers most features of
the core sequential fragment of LLVM IR 11.0.0  as per its informal
specification, including:
the basic operations on 1-, 8-, 32-, and 64-bit integers, \tm{Double}s,
\tm{Float}s, structs, arrays, pointers, and casts; \undefv{} and \dvpoison;
SSA-structured control-flow-graphs, global data, mutually-recursive functions,
and support for intrinsics.
The main features that are currently unsupported
are: some block terminators (\tm{switch}, \tm{resume}, indirect branching,
\tm{invoke}), the \tm{landing\_pad} and \tm{va\_arg} instructions,
architecture-specific floats and opaque types. The list of supported intrinsics
is small, but user-extensible.

At a bird's eye view, the denotation process is exactly the same as in the case of the toy 
imperative language from the previous section: we represent,
by structural recursion on the syntax, programs into trees filled with
interactions, and then plug a monadic interpretation for these interactions into
the model.
At a more technical level, things get significantly more complex: we highlight two aspects of this complexity.

First, the representation of \VIR programs must operate at two distinct levels: 
from the perspective of a single function, and from the perspective of a complete program.
In order to represent a \VIR function, that is a control flow graph,
one needs simply to define the semantics of jumps: following a denotational
approach, they are resolved using the \coqe{iter} combinator---intuitively,
one can think of a jump at the end of a block as a tail recursive call.
However, a complete LLVM IR program is a set of statically known mutually recursive
functions. The function-level representation therefore exposes function calls
as external events, and the top-level representation resolves the mutual recursion.
The interplay between these two perspectives is source of a significant overhead in reasoning.

\begin{figure}
  \centering{\includegraphics[width=4.5in]{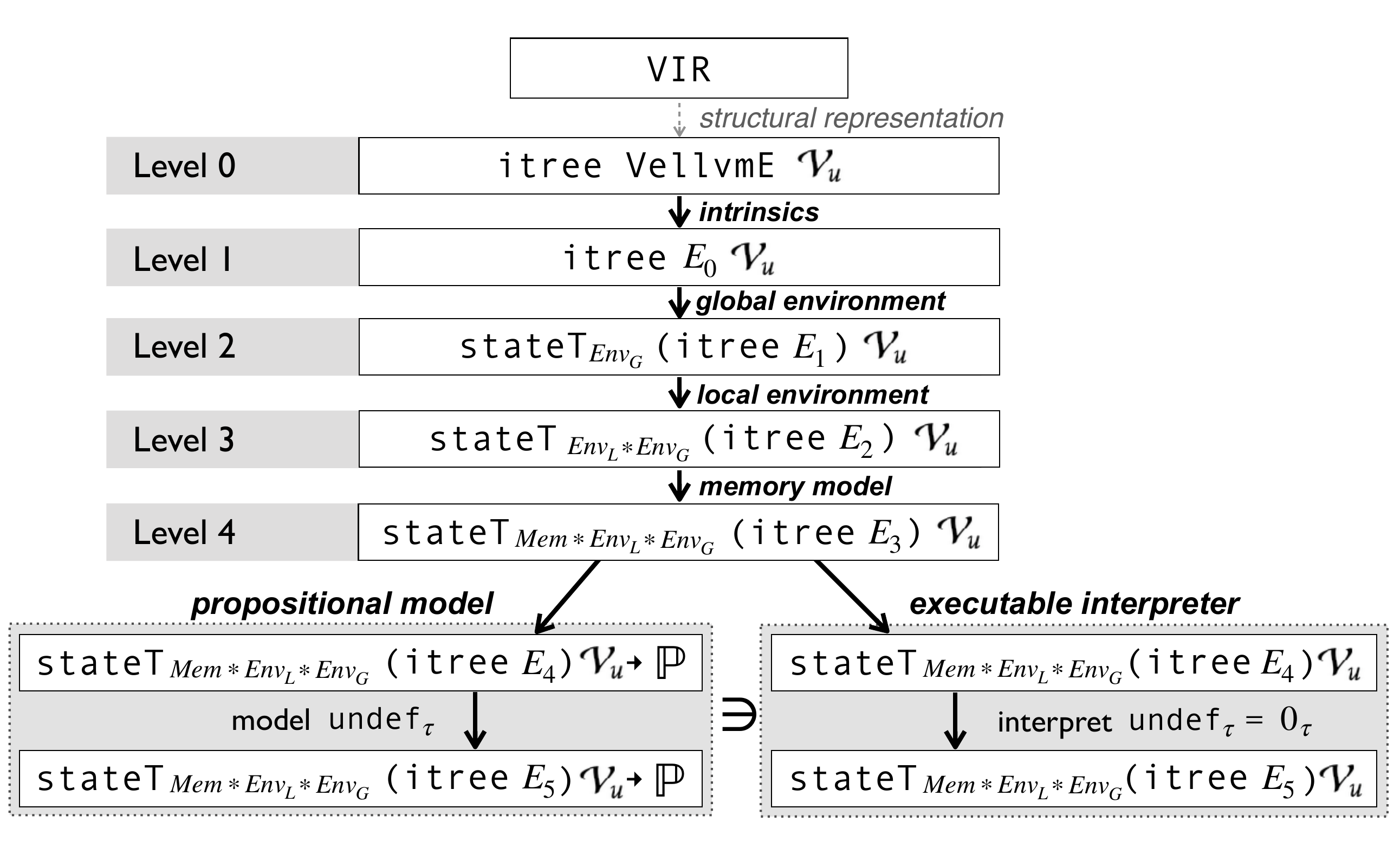}}
  \caption{Vellvm: Levels of interpretation}
  \label{fig:interp}
\end{figure}

Second, the list of effects a \VIR program can exhibit goes far beyond simple
read and writes. Rather, one needs to consider read/writes to global
variables, read/writes to registers, interactions with the memory (including
allocation and casts between pointers and integers), manipulation of the call
stack, calls to intrinsics, non-deterministic access to under-defined values,
and undefined behaviors.
While we refer the interested reader to~\cite{vellvmicfp} for details, the
situation is summed up graphically on Figure~\ref{fig:interp}.
Rather than a simple monolithic implementation of the effects as in the toy
example described above, a stack of interpreters is incrementally layered atop
of the initial representation of programs, as described in more details by~\cite{yzz22}.
The fork in the stack happens where non-determinism comes into play: the
semantics itself (the left path) captures sets of valid computations, while
a deterministic, executable, alternative (the right path) is provided for
testing purposes, and proved against the semantics to be a valid execution.

Crucially, each level of the layer induces its own notion of refinement of
programs, and the refinements get increasingly coarser. We shall exploit this
property in the present development: since the compiler from \FHCOL to \VIR
does not introduce any non-deterministic effect, we are able to ``pretend'' that
the semantics is deterministic by reasoning w.r.t. programs interpreted only
until before the fork.

\subsubsection{Setting the scene: an itree-based semantics for \FHCOL}

Interaction Trees offer a very flexible framework for establishing refinements
of programs, including between heterogeneous computations operating over
distinct memory models. However, they do assume both semantics of interest to be
modeled using \itreesn: before getting started, we hence need a bit of glue.

We provide an itree-based model for \FHCOL. For simplicity,
we stay as close as possible to the original fuel-based big step
interpreter, but in a monadic style in the \coqe{itree Event} monad
where \coqe{Event}.  The interface \coqe{Event} is defined as the sum
\coqe{MemEvent + StaticFailE + DynamicFailE}.  We hence distinguish
two types of failures: \textit{static} and \textit{dynamic}. The
former refers to errors indicating an ill-formed program which could
be detected at the compilation stage, such as accessing an undefined
variable, which causes \DHCOL compilation to fail. In contrast,
\textit{dynamic errors} refer to errors that cannot cause compilation
failure but do cause the evaluation to fail. The
denotation of a \DHCOL program can fail with either static or dynamic
failures, although this distinction is not currently used, because our
compiler correctness proof is stated only for programs which evaluate
successfully. However, this mechanism will allow us to extend the
correctness formulation in future to reason about failing programs as
well. For example, we can try to prove that evaluation and compilation
would fail with the same errors, which could be a useful security
property for a compiler.  Both failures are implemented as exception
monads and, combined with HELIX memory events, represent a complete
model of the \DHCOL interaction with its environment, including error
handling:

\begin{lstlisting}[language=Coq,basicstyle=\ttfamily\footnotesize,
  columns=flexible]
  Definition StaticFailE := exceptE string.
  Definition DynamicFailE := exceptE string.
  Definition Event := MemEvent +$'$ StaticFailE +$'$ DynamicFailE.
\end{lstlisting}

The \coqe{MemEvent} interface captures \FHCOL-level memory event (read, writes, allocations and free):

\begin{lstlisting}[language=Coq,basicstyle=\ttfamily\footnotesize,columns=flexible,caption={DHCOL events},label={lst:dhcol-events}]
Variant MemEvent: Type -> Type :=
  | MemLU  (msg: string) (id: mem_block_id): MemEvent mem_block
  | MemSet (id: mem_block_id) (bk: mem_block): MemEvent unit
  | MemAlloc (size: NT.t): MemEvent mem_block_id
  | MemFree (id: mem_block_id): MemEvent unit.
\end{lstlisting}

We define an event handler which maps memory events to the state monad on HELIX
memory extended with dynamic and static failures:

\begin{lstlisting}[language=Coq,basicstyle=\ttfamily\footnotesize,
  columns=flexible]
  Definition Mem_handler:
  MemEvent ~> Monads.stateT memory (itree (StaticFailE +$'$ DynamicFailE))
  := fun T e mem =>
  match e with
  | MemLU msg id  =>
  lift_Derr (Functor.fmap (fun x => (mem,x)) (memory_lookup_err msg mem id))
  | MemSet id blk => ret (memory_set mem id blk, tt)
  | MemAlloc size => ret (mem, memory_next_key mem)
  | MemFree id    => ret (memory_remove mem id, tt)
  end.
\end{lstlisting}

The {\tt lift\_Derr} function converts failed memory lookups into
dynamic errors. The memory interpretation can only produce dynamic
errors.
The events describe the
operations of allocation and de-allocation of memory blocks as well as
reading and writing them.

To remain close to the interpreter, the evaluation context is passed around as an explicit
argument to the representation function
\coqe{denoteDSHOperator : evalContext -> DSHOperator -> itree Event unit}.

Naturally, to connect this new model to the upper layers of the compiler, we
prove it sound against the big step semantics. I.e.:

\begin{lstlisting}[language=Coq,basicstyle=\ttfamily\footnotesize,
  columns=flexible,caption={The monadic model for \FHCOL is sound against the
    interpreter},label={lst:dhcol-itree}]
Theorem Denote_Eval_Equiv:
  forall (s: evalContext) (op: DSHOperator) (mem: memory) (fuel: nat) (mem': memory),
  evalDSHOperator s op mem fuel = Some (inr mem') ->
  eutt equiv (interp_Mem (denoteDSHOperator s op) mem) (Ret (mem', tt)).
\end{lstlisting}

We consider any \FHCOL{} operator \coqe{op} and any pair of
evaluation context \coqe{s} and memory \coqe{mem} such that the big step
semantics succeeds with final memory \coqe{mem'}. Then the monadic model of \coqe{op}
initialized with \coqe{s} and \coqe{mem} is weakly bisimilar to a pure
computation returning \coqe{mem'}.
To be perfectly precise, we establish that the final memory is only
extensionally equivalent to \coqe{mem'} (using the \coqe{equiv} argument given to
\coqe{eutt}) so that we do not have to rely on functional extensionality.
This technicality can be soundly ignored.
Finally, we write \coqe{interp_helix} for the composition of \coqe{interp_Mem} with an interpreter
sending dynamic failure into the failure monad.

We can now abstract away from the big step semantics, and dedicate our attention to
the problem of establishing the soundness of the compilation of \FHCOL{} operators by
relating their respective monadic models.

\newcommand{\euttn}{\tm{eutt}\xspace}
\newcommand{\eutt}{\approx}
\newcommand{\euttR}[1]{\approx_{#1}}
\newcommand{\euttRel}[4]{\{#1\}(#2,#3)\{#4\}}
\newcommand{\eqit}{\cong}


\newcommand{\uvalue}{\ensuremath{\mathcal V_u}}

\newcommand{\memV}{\ensuremath{\mathtt{M_V}}}
\newcommand{\localV}{\ensuremath{\mathtt{L_V}}}
\newcommand{\globalV}{\ensuremath{\mathtt{G_V}}}
\newcommand{\glm}{~g~l~m}
\newcommand{\mlg}[1]{(m,(l,g))}
\newcommand{\mlgv}[1]{(m,(l,(g,#1)))}
\newcommand{\sfofnat}[1]{\ensuremath{\tm{nat2i64}(#1)}}
\newcommand{\gac}[4]{\ensuremath{\tm{get\_cell}~#1~#2~#3~#4}}
\newcommand{\readm}[4]{\ensuremath{\tm{read}~#1~#2~#3=\iinr{#4}}}
\newcommand{\writem}{\ensuremath{\tm{write}}}
\newcommand{\allocatem}[1]{\mathtt{allocate}~#1}

\newcommand{\sem}[1]{\ensuremath{\llbracket #1 \rrbracket}}
\newcommand{\seme}[1]{\ensuremath{\llbracket #1 \rrbracket_{e}}}
\newcommand{\semet}[2]{\ensuremath{\llbracket (#1,#2) \rrbracket_{e}}}
\newcommand{\semi}[1]{\ensuremath{\llbracket #1 \rrbracket_i}}
\newcommand{\semt}[1]{\ensuremath{\llbracket #1 \rrbracket_t}}
\newcommand{\semb}[2]{\ensuremath{\llbracket #1 \rrbracket^{#2}_{bk}}}
\newcommand{\semc}[1]{\ensuremath{\llbracket #1 \rrbracket_c}}
\newcommand{\sembs}[1]{\ensuremath{\llbracket  \mathit{#1}  \rrbracket_{\mathsf{bks}}}}
\newcommand{\semphi}[2]{\ensuremath{\llbracket  \mathit{#2}  \rrbracket_{\Phi}^{#1}}}
\newcommand{\semphis}[2]{\ensuremath{\llbracket \mathit{#2}  \rrbracket_{{\Phi_s}}^{#1}}}
\newcommand{\semcfg}[1]{\ensuremath{\llbracket \mathit{#1} \rrbracket_{\mathsf{cfg}}}}
\newcommand{\semfun}[1]{\ensuremath{\llbracket \mathit{#1} \rrbracket_{\mathsf{fun}}}}
\newcommand{\semmcfg}[1]{\ensuremath{\llbracket \mathit{#1} \rrbracket_{\mathsf{mcfg}}}}
\newcommand{\semphiaux}[3]{\tm{resolve\_phi(#1,#2,#3)}}
\newcommand{\semtop}[1]{\ensuremath{\llbracket #1 \rrbracket}_{\mathsf{\VIR}}}
\newcommand{\semvir}[1]{\ensuremath{\llbracket \mathit{#1} \rrbracket_\mathsf{VIR}}}
\newcommand{\concretizeorpickn}{\tm{concretize\_or\_pick}}
\newcommand{\concretizeorpick}[1]{\concretizeorpickn{}(#1)}
\newcommand{\EV}{\ensuremath{\mathtt{E_V}}} 
\newcommand{\voidE}{\ensuremath{\mathtt{\emptyset}}} 
\newcommand{\stateVC}{\ensuremath{\memV \rightarrow \localV \rightarrow \globalV}} 
\newcommand{\stateVU}{\ensuremath{(\memV \times \localV \times \globalV)}} 
\newcommand{\nt}[1]{\ensuremath{\mathit{#1}}} 

\newcommand{\calls}[2]{\ensuremath{\mathtt{call}~(#1,~#2)}}
\newcommand{\allocas}[1]{\ensuremath{\mathtt{alloca}~(#1)}}
\newcommand{\stores}[2]{\ensuremath{\mathtt{store}~(#1,~#2)}}
\newcommand{\loads}[2]{\ensuremath{\mathtt{load}~(#1,~#2)}}
\newcommand{\phis}[1]{\ensuremath{\mathtt{phi}~(#1)}}
\newcommand{\geps}[3]{\ensuremath{\mathtt{GEP}}~(#1,~#2,~#3)}
\newcommand{\rets}[1]{\ensuremath{\mathtt{return}}~(#1)}
\newcommand{\branchs}[3]{\ensuremath{\mathtt{branch}}~(#1,~#2,~#3)}
\newcommand{\nat}{\ensuremath{\mathcal N}}
\newcommand{\iphi}{\nt{phi}}
\newcommand{\expr}{\nt{exp}}
\newcommand{\texpr}{\nt{exp}}
\newcommand{\instr}{\nt{instr}}
\newcommand{\term}{\nt{term}}
\newcommand{\sphi}{\nt{phi}}
\newcommand{\block}{\nt{block}}
\newcommand{\id}{\nt{id}}
\newcommand{\bid}{\nt{bid}}
\newcommand{\ocfg}{\nt{ocfg}}
\newcommand{\cfg}{\nt{cfg}}
\newcommand{\cfgs}{\nt{cfg}s}
\newcommand{\mcfg}{\nt{mcfg}}
\newcommand{\mcfgs}{\nt{mcfg}s}
\newcommand{\isf}{\nt{i64}}
\newcommand{\ibool}{\nt{i1}}
\newcommand{\lid}[1]{\ensuremath{\mathtt{\%}#1}}
\newcommand{\gid}[1]{\ensuremath{\mathtt{@}#1}}
\newcommand{\arrayt}[1]{\ensuremath{\mathtt{[}#1\mathtt{]}}}
\newcommand{\ptrt}[1]{\ensuremath{#1\mathtt{*}}}
\newcommand{\inputs}[1]{\texttt{inputs}(#1)}
\newcommand{\outputs}[1]{\texttt{outputs}(#1)}
\newcommand{\noreentrance}[2]{\outputs{#2}\cap \inputs{#1} = \emptyset}
\newcommand{\independent}[2]{\texttt{independent}~#1~#2}

\newcommand{\vP}{\ensuremath{\mathtt{mcfg}}} 
\newcommand{\dop}{\ensuremath{\mathtt{op}}} 

\newcommand{\memH}{\ensuremath{\mathtt{M_H}}}
\newcommand{\semH}[2]{\ensuremath{\llbracket \mathit{#1} \rrbracket_\mathsf{H}^{#2}}}
\newcommand{\nofail}[1]{\ensuremath{\mathtt{nofail}(#1)}}
\newcommand{\stinv}[4]{\ensuremath{\mathtt{state\_invariant}~#1~#2~#3~#4}}
\newcommand{\gsafe}[3]{\ensuremath{\mathtt{scoped}~#1~#2~#3}}
\newcommand{\lsm}[4]{\ensuremath{\mathtt{local\_modif}~#3~#4\subseteq [#1,#2]}}
\newcommand{\option}[1]{\ensuremath{\tm{option(#1)}}}
\renewcommand{\ret}[1]{\ensuremath{\tm{ret}~#1}}

\newcommand{\coqcomment}[1]{(*~\mathtt{#1}~*)}
\newcommand{\genir}{\ensuremath{\mathtt{genIR}}}
\newcommand{\iinr}[1]{\ensuremath{\tm{inr}~#1}}
\newcommand{\iinl}[1]{\ensuremath{\tm{inl}~#1}}
\newcommand{\none}{\ensuremath{\tm{None}}}
\newcommand{\some}[1]{\ensuremath{\tm{Some}~#1}}

\newcommand{\tryret}[2]{\tm{tryret}\ #1\ \tm{in}\ #2}
\newcommand{\match}[1]{\tm{match}\ #1\ \tm{with}}
\newcommand{\matchbranch}[2]{\mid #1\Rightarrow #2}

\newcommand{\rname}[1]{\mbox{\textsc{#1}}}
\newcommand{\IMP}{\ensuremath{\Rightarrow}}
\newcommand{\Rule}[3]{\infer[\mbox{\rname{#1}}]{#3}{#2}}
\newcommand{\Ret}[1]{\ensuremath{\tm{Ret}~#1}}
\newcommand{\gallina}{Gallina\xspace}

\subsection{Verification: proof technique}

The verification of a compiler for a general purpose language seeks to establish
that the set of observable behaviors for the compiled program is included in the
set of behaviors for the source program. In order to be termination-sensitive---to prove that
diverging programs are compiled to diverging code---behaviors must include
infinite traces, whether silent or reactive.

A standard proof method to establish such inclusion of behavior 
is the one popularized notably by CompCert.
Semantics are assumed to be formalised as labelled transition systems,
and elementary simulation diagrams are established for each program transformation considered. 
An isolated, generic,
coinductive argument stitches these diagrams together to derive the
inclusion of behavior.

This inclusion is established in a different manner for denotational
semantics built upon \itreesn{}, though the result is semantically
equivalent. Users directly establish weak bisimilarities defined as a
coinductive predicate---the \euttn{} relation mentioned above.  Of
course this relation only relate uninterpreted computations,
i.e. trees: one lifts this relation to the monadic domain at hand by,
typically, quantifying over states.

Through this subsection, we highlight the important elements of proof methodology
underlying the approach.

\subsubsection{A relational program logic for termination and observation sensitive refinements of programs}

Xia et al. introduced \euttn as a way to define a
notion of equivalence of computations modeled as \itreesn.
Here, we emphasize that, especially once the trees get interpreted into a state monad
transformer, this relation benefits from being read as the fundamental
building block for a relational program logic for termination-and-trace
sensitive refinements.

Consider for instance the type of the denotation for open \VIR{} control flow graphs, before
introduction of the non-deterministic effects of the language:
$\semcfg{cfg}:~
\stateVC \rightarrow
\itree{\EV}
{(\stateVU \times
  ((\bid\times\bid) + \uvalue)} $. Similarly, consider the type of the
denotation of \FHCOL programs: $\semH{\dop}{\sigma}:~\memH\rightarrow
\itree{\voidE}{\option{\memH \times {\tt unit}}}$, where $\voidE$ denotes the empty signature.
Building a refinement between \FHCOL operators and \VIR cfgs amounts to
assuming that initial memories $m_H$ and $m_V$ are appropriately related by the precondition $P$ (which also has access to \VIR local and global state), proving the computations are bisimilar,
and expressing as a postcondition $Q$ the relation between the resulting states, as well as
the dynamic value computed on the \VIR side.

The relational program logic nature is further emphasized by the following notation:
$$\euttRel P {\mathit{op}} {\mathit{cfg}} Q\triangleq
\forall~m_H~m_V~l~g,~P~(m_H,(m_V,(l,g)))\rightarrow \semH{\mathit{op}}~m_H
\euttR{Q}\semcfg{cfg}~m_V~l~g.$$

While this notation provides the right intuition, we keep things more shallow in
our formal development as our language of assertions itself is shallow, and as
we need to work at various levels of denotation whose types differ slightly.

Figure~\ref{fig:euttR} describes the main relational rules we need at the
level of interaction trees to conduct the proof.
Pure computations are  related if they return values related by the postcondition.
The \textsc{EBind} rule acts as a
cut, or a traditional sequence rule: one can introduce any intermediate relation
acting as a postcondition for the first part of the computations, and
precondition for the following part. Note that in our case of interest, we
apply this rule after having provided our stateful computation with its initial
state, the return type therefore encompasses both the pure values computed, as
well as the intermediate states.
Separately established postconditions can be combined via the usual intersection
of relations in \textsc{EAnd}\footnote{It is worth noting that we lose this rule when we move
  into the non-deterministic monad.}. The logic supports the expected
weakening rule \textsc{EMon}: it is crucial in order to allow compositional reasoning.
Finally, the $\eutt$ relation is a congruence on either side of a refinement proof: we
exploit this fact to compute symbolically during the refinement proofs.

\begin{figure}[t]
  {\small
    $
      \Rule{ERet}{
        R(r_1, r_2)
      }{
        \ret{r_1} \eutt_{R} \ret{r_2}
      }
      \hfil
      \Rule{EBind}{
        t_1 \eutt_{U} t_2
        \qquad
        \forall \; u_1, u_2, U(u_1,u_2) \IMP (k_1\; u_1)$ $\eutt_{R} (k_2\; u_2)
      }{
        (x \gets t_1 ;; (k_1\; x)) \eutt_{R} (x \gets t_2 ;; (k_2 \; x))
      }
      $
      \vspace{0.2cm}

      $
      \Rule{EMon}{
        t_1 \eutt_{R_1} t_2 \quad R_1 \subseteq R_2
      }{
        t_1 \eutt_{R_2} t_2
      }
     \hfil
     \Rule{EAnd}{
        t_1 \eutt_{R_1} t_2 \quad t_1 \eutt_{R_2} t_2
     }{
       t_1 \eutt_{R_1\sqcap R_2} t_2
     }
      \hfil
      \Rule{EuttCong}{
           t_1 \eutt t_1' \quad t_1' \eutt_R t_2' \quad t_2' \eutt t_2
      }{
         t_1 \eutt_R t_2
      }
      $
  }
  \caption{Relational reasoning principles}
  \label{fig:euttR}
\end{figure}

\newcommand{\kw}[1]{\ensuremath{\mathsf{#1}}}
\newcommand\append{\mathbin{+\mkern-5mu+}}

\subsubsection{A symbolic interpreter for \VIR}
\label{sec:symbolic}

Zakowski et al. introduced
with \VIR~\citep{vellvmicfp} how the equational theory supported by the \itreen
combinators---particularly the iterator used to resolve jumps--could be used to
derive rich reasoning principles about open control flow graphs.
We further commit to this approach by characterizing equationally the behavior of all syntactic
constructs of the language and sealing off the denotation functions.

This proof theory of \VIR can be organized into three main groups. For readability, we write these proof rules 
as inference rules: under the conditions above the line, the (possibly existentially quantified) equation below the line holds.

Structural rules, described in Figure~\ref{fig:eutt-vir-struct}, are
mostly administrative, yet crucial: they spell out the compositionality of the
semantics.
The semantics of each syntactic construct can be unfolded in terms of the semantics
of its sub-components, allowing for inductive hypotheses and refinements of smaller syntactic
components to be invoked during the proof.

Figure~\ref{fig:eutt-vir-cf} describes the rules governing the semantics of open
control flow graphs in terms of its sub-graphs. Most of these rules are
conditioned by static preconditions expressing notions of separation between the
sets of input and output labels of the sub-graphs at play. The \texttt{independent}
predicate asserts that all labels present in each graphs are disjoint, while
\texttt{wf\_cfg} ensures that all bound labels are unique. These lemmas are
sufficiently expressive to reason about all cfgs built by the compiler,
including the complex parametric looping construct described in Section~\ref{sec:fhcol2llvm}
which plugs a body represented as an arbitrary open control flow graph inside of
another one performing the iteration.

Finally, we provide atomic semantic rules for \VIR expressions, instructions, and
terminators -- a minimal excerpt is provided on Figure~\ref{fig:eutt-vir-inst}.
These equations relate the denotation of atomic constructions to pure
computations, i.e. of the shape $\Ret{\_}$. Because
$\Ret{}$is a left unit for \texttt{bind}, these rules symbolically
reduce. For instance if $i$ is an instruction, $\semi{i}\glm\bind k$, binds $i$'s
result in the continuation $k$. In contrast to the
semantics itself, which is implemented as a computable total function in \gallina,
these rules can be partial and rely on propositional preconditions.

We exploit this partiality in particular to provide a
layer of abstraction when reasoning about \VIR's memory over a sane
subset of the language. The memory model in all its generality is
complex: it has to allow for accesses to arbitrary structured data, as
well as pointer-to-int and int-to-pointer casts. These low-level
memory features are implemented via a quasi-concrete memory model,
strongly inspired by~\cite{KHM+15}, but most of the complexities of
this memory model (such as precisely how dynamic values are serialized
into bytes) are irrelevant to the semantics of the reasonably
well-behaved programs that \FHCOL compiles down to. We're therefore
able to provide higher level memory accesses (\texttt{read},
\texttt{write}, \texttt{allocated},..) whose relative behavior is
specified through lemmas. In particular when it comes to the kind of
\kw{getelementptr} instruction we generate, we provide the illusion of
accessing array-like structures.

\begin{figure}[t]\small

  $
  \Rule{CNil}{
  }{
    \semc{[]} \eutt \ret{tt}
  }
  \hfil
  \Rule{CCons}{
  }{
    \semc{i :: c} \eutt  \semi{i}\bind \semc{c}
  }
  \hfil
  \Rule{CodeApp}{
  }{
    \semc{c1 \append c2} \eutt  \semc{c1}\bind \semc{c2}
  }
  $
  \vspace{0.2cm}

  $
  \Rule{BkUnfold}{
  }{
    \semb{\{b,\Phi_s, c,t\}}{b_f} \eutt \semphis{b_f}{\Phi_s}\bind\semc{c}\bind\semt{t}
  }
  \hfil
  \Rule{Out}{
    bks[b_s] = \none
  }{
    \sembs{bks}{(b_f,b_s)} \eutt \Ret{\iinl{(b_f,b_s)}}
  }
  $
  \vspace{0.2cm}

  $
    \hfil
    \Rule{In}{
      bks[b_s] = \some{bk}
    }{
      \sembs{bks}{(b_f,b_s)} \eutt fov\gets\semb{bk}{b_f}\bind \tryret{fov}{\sembs{bks}}
    }
    \hfil
    $

 \vspace{0.2cm}

 $
 \tryret{fov}{f} \triangleq \match{fov} \matchbranch{\iinl{x}}{f(x)} \matchbranch{\iinr{v}}{\Ret{v}}
 $
    \caption{Structural \VIR equations (excerpt)}
    \label{fig:eutt-vir-struct}
    \vspace{-3ex}
\end{figure}

\vspace{2\baselineskip} 

\begin{figure}[t]\small

\[
\begin{array}{c}
\Rule{OcfgSeq1}{
\noreentrance{bks_1}{bks_2}\quad b_s \not\in \texttt{inputs}(bks_1)
}{
\sembs{bks_1 \append bks_2}~(b_f,b_s) \eutt \sembs{bks_2}~(b_f,b_s)
} \\ \\
\Rule{OcfgSeq2}{
\noreentrance{bks_1}{bks_2}
}{
\sembs{bks_1 \append bks_2}~(b_f,b_s) \eutt fov\gets \sembs{bks_1}~(b_f,b_s)\bind \tryret{fov}{\sembs{bks_2}}
} \\ \\
\Rule{OcfgBrL}{
\independent{bks_1}{bks_2}\quad b_s\in\inputs{bks_1}
}{
\sembs{bks_1 \append bks_2}{(b_f,b_s)}\eutt \sembs{bks_1}{(b_f,b_s)}
} \\ \\
\Rule{OcfgBrR}{
\independent{bks_1}{bks_2}\quad b_s\in\inputs{bks_2}
}{
\sembs{bks_1 \append bks_2}{(b_f,b_s)}\eutt \sembs{bks_2}{(b_f,b_s)}
} \\ \\
\Rule{OcfgSub}{
bks = (prefix \append bks' \append postfix) \quad \texttt{wf\_cfg}(bks)
}{
\sembs{bks}~(b_f,b_s) \eutt
fov\gets \sembs{bks'}~(b_f,b_s)\bind \tryret{fov}{\sembs{bks}}
  }
  \end{array}
\]
  \caption{\VIR proof rules for reasoning about open programs}
  \label{fig:eutt-vir-cf}
\end{figure}

\begin{figure}[t]\small
  \[
\begin{array}{c}    
  \Rule{Br}{
\seme{e}~g~l~m \eutt \Ret{(m,(l,(g,i)))}\quad i\in\{0,1\}
}{
\semt{\kw{br}~b_0~b_1}\glm \eutt \Ret{(m,(l,(g,\iinl{b_i})))}
} \\ \\
\Rule{Br1}{
}{
\semt{\kw{br1}~b}\glm \eutt \Ret{(m,(l,(g,\iinl{b})))}
} \\ \\
\Rule{Op}{
\seme{op}\glm \eutt \Ret (m, (l, (g, uv)))
}{
  \semi{id = op}\glm \eutt \Ret (m, (l[id\gets uv], (g, tt)))
} \\ \\
\Rule{Al}{
\mathtt{size}(\tau) > 0
}{
\begin{array}{rl}
  \exists m'~a,&\hspace{-0.2cm}\allocatem{m}\tau = \iinr{(m',a)}~\land\\
  &\hspace{-0.2cm}\semi{id= \mathtt{alloca}~\tau}\glm \eutt\Ret{(m',(l[id\gets a],(g,tt)))}
\end{array}
} \\ \\
\Rule{gepA}{
\begin{array}{l}
\seme{ptr}\glm\eutt\Ret{\mlgv{a}} \\
  \seme{e_{ix}}\glm\eutt\Ret{\mlgv{\sfofnat{ix}}}\land\gac{m}{a}{ix}{\tau} = \iinr{val}
\end{array}
}{
\begin{array}{rl}
  \exists ptr,&\hspace{-0.2cm}~\readm{m}{ptr}{\tau}{val}~\land\\
  &\hspace{-0.2cm}\semi{id= gep~[\tau]~ptr~[0_{i64};e_{ix}]}\glm\eutt\Ret{(m,(l[id\gets ptr],(g,tt)))}
\end{array}
}
\end{array}
\]
  \caption{Specifications for \VIR (excerpt)}
\label{fig:eutt-vir-inst}
\vspace{-3ex}
\end{figure}

All put together, this proof theory achieves several goals. First, it
allows us to keep the representation functions and interpreters opaque, hiding their internal representation as itrees,
recovering reasonably performant and convenient setoid-based rewrites when reasoning about \VIR's
syntax.
Second, it lifts the reasoning to the level of syntactic \VIR{} program, i.e., yielding notably a more convenient way to provide abstract interfaces with the memory model.
Finally, we define a symbolic interpreter for \VIR that can be run \emph{inside}
of refinement proofs between \VIR programs and arbitrary other computations.

All equations are indeed established with respect to $\eutt$: they can be rewritten under
$\euttR{R}$ for any arbitrary $R$. We hence provide a \emph{reduction} tactic matching on a
refinement goal to rewrite the adequate equations and clean up the goal: the
denotation of a \VIR syntactic component gets reduced into the \texttt{bind} of the
denotation of an instruction, or a terminator, followed by its continuation.
The atomic instruction at the head can then be symbolically
\emph{stepped} --- we provide a partial tactic to perform this as well, but leave
it in the hands of users.
Naturally, this symbolic reduction can only be performed over concrete code, but
it allows us to reduce code such as that generated by the compiler as far as
possible: typical compiled components start with concrete code, followed by some
parametric code resulting from the call to other compilation sub-routines.

\subsection{Verification: compilation of operators}

We now turn our focus to the \coqe{genIR} segment of the compilation pass.
Recall from Section~\ref{sec:comp-orga} that this function
compiles an \FHCOL{} operator to a \VIR \emph{open} control flow graph.
More specifically, \coqe{genIR} outputs the \VIR function implementing the compiled operator,
save for a missing final return block to the main function that only gets inserted at the
top-level, in the \coqe{LLVMGen} unit.

Monadic semantics built using the \itreen framework can seamlessly express the semantics
of open programs, such as the ones generated by \coqe{genIR}. We hence can focus on the correctness of the \coqe{genIR} compilation unit in isolation, without having to capture their syntactic context.

The statement we establish is, formally:

\begin{lstlisting}[language=Coq,basicstyle=\ttfamily\footnotesize,
  columns=flexible,caption={Correctness statement for {\tt genIR}},label={lst:genir-correct}]
Theorem compile_FSHCOL_correct :
  forall (** Compiler bits *) (s1 s2: IRState)
    (** Helix bits    *) (op: DSHOperator) ($\sigma$ : evalContext) (memH : memoryH)
    (** Vellvm bits   *) (nextblock bid_in bid_from : block_id) (bks : list block)
                         (g : global_env) ($\rho$ : local_env) (memV : memoryV),
  (* (1) Evaluation succeeds *)
  no_failure (interp_helix (denoteDSHOperator $\sigma$ op) memH) ->
  (* (2) Compilation succeeds *)
  genIR op nextblock s1 = inr (s2,(bid_in,bks)) ->
  (* (3) Preconditions on IRStates *)
  bid_bound s1 nextblock ->
  Gamma_safe $\sigma$ s1 s2 ->
  (* (4) Memory invariant *)
  state_invariant $\sigma$ s1 memH (memV, ($\rho$, g)) ->
  (* (5) Equivalence of computations, and postconditions *)
  eutt (succ_cfg (genIR_post $\sigma$ s1 s2 nextblock $\rho$ g))
       (interp_helix (denoteDSHOperator $\sigma$ op) memH)
       (interp_cfg   (denote_ocfg bks (bid_from,bid_in)) g $\rho$ memV).
\end{lstlisting}

This statement is a mouthful: we first unpack it informally,
before delving formally into its components and proofs in the remainder of this subsection.
It reads in English: if (1) for any \FHCOL{} operator \coqe{op} and initial state from which evaluation succeeds,
(2) assuming that the compilation of this operator succeeds, (3) if the \coqe{IRStates} satisfy the adequate preconditions,
and (4)  the initial memories are adequately related,
then (5) the monadic denotations of the operator and compiled code are bisimilar, and produce adequately related results.

We now detail some technicalities hidden behind this theorem and its proof.

\subsubsection{On freshness of names, and well-formedness of graphs}

We focus first our attention on (3), the assumptions made on \coqe{IRStates}.
Recall from Section~\ref{sec:comp-orga} that an \coqe{IRState} $s$ is the compiler state recording essentially two pieces of information:
a collection of monotonic counters used to generate fresh names and an environment $s.\Gamma$
recording the type of the previously generated \VIR identifiers. In the following,
we pretend there is a single high watermark counter $s.n$ for simplicity.

From a programmatic side, this provides a compiler engineer with a convenient, simple abstraction:
essentially a freshness monad with an additional log $\Gamma$ to retrieve typing information.
When it comes to reasoning about the compiler, invariants must be made explicit.
Posing $s_1$ and $s_2$ as the respective \coqe{IRState} of the input and output of the compilation unit, these invariants can be abstracted as:
\begin{enumerate}
  \item the \VIR-level typing information tracked in $s_1.\Gamma$ must soundly reflect the
\FHCOL-level evaluation context \coqe{$\sigma$};
  \item variables recorded in $s_1.\Gamma$ should have been generated by counters below $s_1.n$;
  \item the \coqe{nextblock} passed to the compiler is generated by a counter below $s_1.n$;
  \item local variables introduced in the compilation of the operator must have been generated by counters comprised in the interval $[s_1.n, s_2.n]$---this fact, among others, is captured in the post-condition (5).
\end{enumerate}
This collection of facts ensures that our freshness monad actually yields fresh names, and that configurations are always well-formed.
Finally, an additional memory invariant, elaborated upon in Section~\ref{sec:mem-inv}, 
relates the data stored in memory at the \FHCOL-level to the data stored in registers and memory at the \VIR-level. 

A considerable amount of boilerplate and reasoning is necessary to
maintain these invariants. 
Setting aside the memory invariant, 
the set of invariants necessary to formulate to capture the behavior of the freshness monad 
and the consequent well-formedness of the generated graphs is surprisingly intricate. Naturally, the effort
to preserve them as part of the correctness
proof for a compilation unit is on par. We had to develop over
600 lines of specification and 1,000 lines of proofs in isolation just
to handle the simple arithmetic involved, ensuring that the compiler
generates variables from counters within the range of its input/output
freshness counters.

This experience leads us to draw Lesson~\ref{lesson:graph-combinators}: the need for a library of graph combinators.

\begin{lesson}{Graph combinators for \VIR}

The effort of checking that the freshness monad and its use in writing the compiler are correct should be internalized inside a library. We conjecture that this could be achieved by a well-designed graph combinators library exposing a suitable DSL for writing frontends targeting \VIR{}. The library should have the following characteristics:
\begin{itemize}
\item each combinator is characterized by a semantic equation;
\item any graph built out of the surface DSL should be well-formed (in particular, in SSA form with no collusion of names) by construction;
\item the set of combinators should be expressive enough to support writing compilers such as the one presented here.
\end{itemize}
We believe the design of this library to be a non-trivial but worthwhile endeavor.
\label{lesson:graph-combinators}
\end{lesson}

\subsubsection{On characterizing successful executions}

Let us now consider a seemingly innocuous aspect of the theorem in Listing~\ref{lst:genir-correct}: Premise (1) states that the \FHCOL{} operator considered should have a valid semantics. Hypotheses of this flavor are routine in the context of verified compilation, as notably embodied by the famous absence of undefined behavior assumption made by the CompCert compiler~\citep{compcert}.

The adoption of an unconventional semantic model compels us to reevaluate commonplace approaches. When working with a small-step semantics, the assumption is typically expressed as a safety predicate: all reachable states, w.r.t. the transitive closure of our reduction, are safe, i.e., cannot trigger undefined behavior, or failure.
In this setting, when proceeding by induction on the syntax of the source program, deriving this safety assumption for the intermediate states at play is straightforward:
these states are reached by reduction, and hence are safe as well by definition.

We are here tasked with expressing the same semantic intuition, but in the context of a monadic computation. We have three fairly natural choices to draw from:
\begin{enumerate}
\item explicit use the small step transition system denoting the monadic computation, and piggyback on the usual approach;
\item leave dynamic failure in the computation as an external event, and craft the coinductive predicate stating the absence of these events from the tree;
\item interpret failure into the fail monad and express safety as a post-condition of the computation.
\end{enumerate}

Making explicit the underlying transition system has been leveraged in similar context when reasoning on non-deterministic computations, notably by~\cite{ctrees} on CTrees and~\cite{dtrees} on DTrees. It is however a heavyweight device, especially for our sequential context.
The two other options are in essence quite similar. Having experienced with both, we have chosen to favor (3) in order to reuse as much as possible of the existing infrastructure surrounding the \coqe{eutt} relation.
Let us be more precise.

As observed in~\cite{yzz22}, one can derive a unary logic for \itreen{} computations by simply considering the diagonal of the binary relation \coqe{eutt}. I.e., given a computation \coqe{c: itree E X} and a postcondition \coqe{Q : X -> Prop}, \coqe{Q} is a postcondition for \coqe{c}, written \coqe{c [Q]}, if \coqe{eutt (fun x _ => Q x) c c}.
With this definition, we can simply define the \coqe{no_failure t} predicate over computations in the failure monad transformer by ensuring in their postcondition they did not fail:
\coqe{no_failure t := t [fun x => x <> None]}.

While the definition is trivial, its use has ramifications.
Of course, the generic program logic defined over arbitrary \itreesn{} is minimalist: one typically lifts it after stateful interpretation,
and derive richer rules for the source language under consideration. However, it does inherit a generic cut rule from \coqe{eutt}:
\coqe{c [S] -> (forall x, S x -> k x [Q]) -> c >>= k [Q]}.

This is however a \emph{forward} proof rule. In order to manipulate \coqe{no_failure} during a proof, say to derive that the semantic of two subexpressions are safe given that the semantics of their addition is safe, one needs a backward rule, inverting a fact of the shape \coqe{c >>= k [Q]}. However, one can easily observe that the proof rule above is not complete. Consider \coqe{c = ret true} and \coqe{k = fun b => if b then ret 0 else ret 1}: the computation \coqe{c >>= k} satisfies the postcondition \coqe{fun x => x = 0}, but certainly \coqe{k} does not at all points.

This observation led~\cite{yzz22} to introduce the notion of the image of a monadic computation in order to precisely capture the points in the continuation that are reachable, restricting in the previous example the safety in \coqe{k} to the \coqe{true} branch. We refer the interested reader to~\cite{yzz22} for further details.

\begin{lesson}{The image of a monadic computation is crucial to inverting equations between binds, and hence predicates in particular}
  \label{lesson:image}
While the \coqe{eutt} family of relation provides the adequate ingredient to derive a unary logic suitable to express invariants such as \coqe{no_failure}, great care must be taken to invert invariants over monadic sequences. The invariant can only be derived in the continuation at reachable points, as captures by the notion of image of a monadic computation in~\cite{yzz22}.
\end{lesson}

\subsubsection{On the overall proof structure}

The use of denotational techniques brings in two standard conveniences: (1)
expressing the equivalence of open programs is straightforward and (2),
we can reason by induction on the syntax despite the presence of general
recursion.

These two principles guide the overall structure of the proof: for each
translation unit of the compiler, we express its semantic correctness,
and prove it by induction on the syntax of the component handled by this unit.
These correctness lemmas culminate in \coqe{compile_FSHCOL_correct}, which is itself
proven by induction on the operator \coqe{op}.

Each of these lemmas roughly follow the same pattern: in each case of the
induction, terms on both side are executed symbolically---using the symbolic
interpreter described in Section~\ref{sec:symbolic} on the \VIR side, and the
much simpler interpreter on the \FHCOL side.
When hitting a parameter, the cut rule combined with either an inductive proof, or
a previously establish lemma for the translation unit, is used to make progress.
Finally, the invariants are reestablished at the leaves.

\begin{lesson}{Languaged-level reasoning principles are mandatory}
  \label{lesson:abstraction}
  Smaller developments based on \itreesn{} have had the tendency to simply
  expose the model and reason over it directly. Once reaching the scale of
  proofs such as the one at hand here, this approach simply becomes intractable.
  Sealing off all denotation functions, as pointed out in
  Section~\ref{sec:symbolic}, and building the combination of an equational
  theory and a relational program logic at the level of the source languages quickly becomes mandatory.

  This experiment suggests the perspective of interfacing \VIR with a modern
  logic such as Iris.
\end{lesson}

\subsubsection{On the memory invariant}
\label{sec:mem-inv}

Compositional reasoning techniques absolve us from maintaining complex
invariants about the execution context. A well designed library of combinators, as suggested in Lesson~\ref{lesson:graph-combinators} would go one step further in this direction.
The heart of the job however remains: we need to reason about the invariants
relating the states of the source and target languages.
We give here an intuitive overview of the state
invariant, referring to the formal development for its full complexity.
We maintain the following facts:
(1) the DHCOL and VIR memory state agree
(2) the typing context is well-formed and the identifiers in the context are bound,
(3) no pointers or identifiers are aliased,
(4) all referenced identifiers are appropriately allocated, and
(5) the static context is well-scoped.

The trickiest invariant is (1), the \textit{memory invariant}, which
describes how the dynamic values stored on the \FHCOL side in a state $\sigma$ are mapped into memory on the \VIR side.


\begin{figure}[H]
\includegraphics[width=0.5\columnwidth]{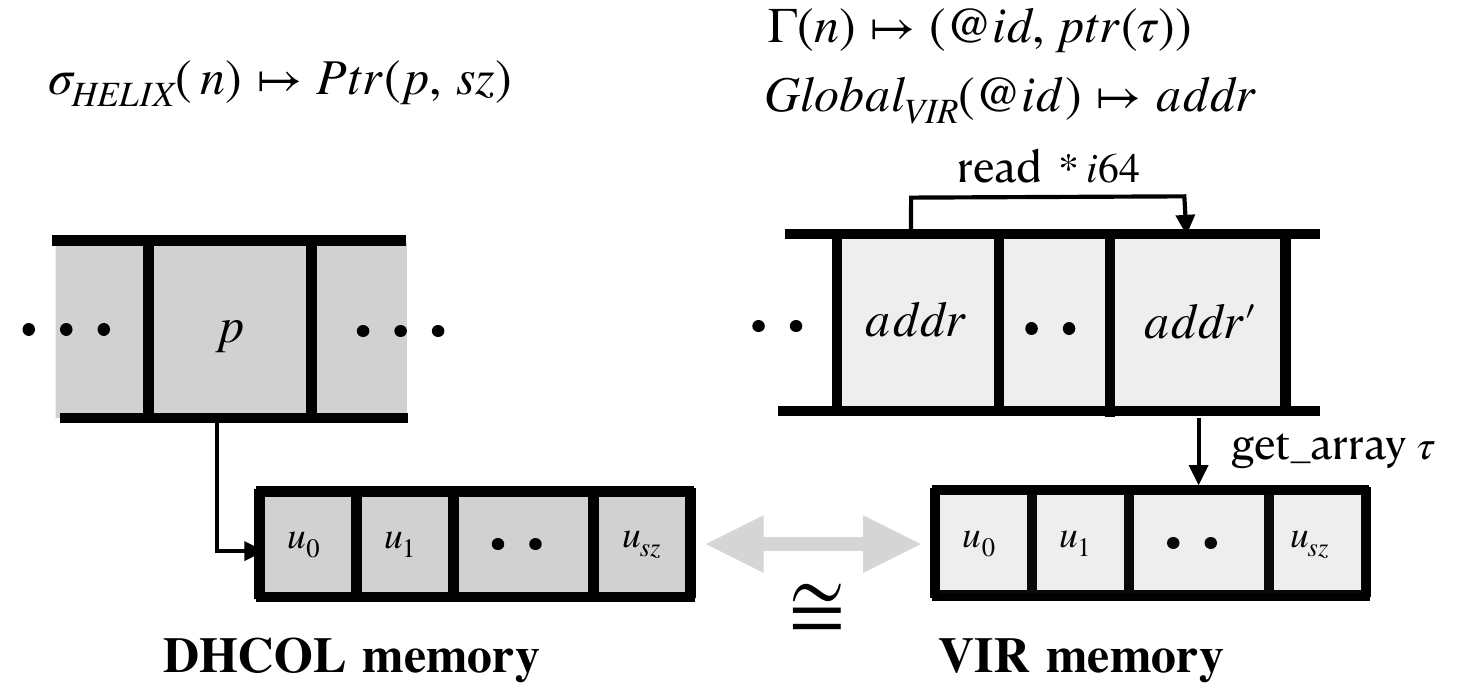}
  \caption{Global pointer memory invariant}
\label{fig:memory-inv}
\end{figure}

\noindent For each variable stored in $\sigma$, the invariant expresses how the static context $\Gamma$
associates a \VIR identifier storing in memory the same data---the details of this mapping depends on
the type of data concerned.
Figure~\ref{fig:memory-inv} illustrates the situation for the case of a \FHCOL
variable containing a pointer to a vector stored in memory.
The \VIR identifier associated in $\Gamma$ is global, storing (after
a level of indirection) an address to memory such that reading at this address
via the \FHCOL-specific \kw{getelementptr} abstraction mentioned in Section~\ref{sec:symbolic} leads to decoding
a list of dynamic values isomorphic to the source vector.

\section{Dynamic Window Monitor}
\label{sec:dynwin}
In Section~\ref{sec:motivatingexample}, we have given a general overview of the
\emph{Dynamic Window Safety Monitor} approach and how its mathematical expression
would be compiled using HELIX's verified chain. In the following section, we
discuss in more detail the full extent of {\tt DynWin}'s compilation chain and
the end-to-end correctness guarantees that have been proven for it. It is
intended as a mostly self-contained glossary of the techniques introduced
earlier in the paper, all applied, in sequence, to a single example operator.

\subsection{Overview}
\label{sec:dwoverview}

Let us begin by more closely examining the original arithmetic form of the
{\tt DynWin} expression and its inputs. Following \cite{Franchetti:17},
Equation~\eqref{eq:dynwin} is first rewritten as the following system of
equations, a form more suitable (for reasons discussed later in this section)
for \HCOL representation.

\begin{align}
  \label{eq:dynwin1}
  \mathrm{safe}_{V,A,b,\epsilon} &: \R \times \R^2 \times \R^2 \rightarrow \Z_2 ; \\
  &(v_r,p_r,p_o) \mapsto (p(v_r) < d_\infty(p_r,p_o)) \nonumber \\
  \label{eq:dynwin_cheb}
  d_\infty(\vec{x}, \vec{y}) &\triangleq || \vec{x} - \vec{y} ||_\infty \\
  \label{eq:dynwin_poly}
  p(x) &\triangleq a_2 x^2 + a_1 x + a_0 \\
  \label{eq:dynwin_a0}
  a_0 &= \left(\frac{A}{b}+1\right)\left(\frac{A}{2}\epsilon^2+\epsilon V\right) \\
  \label{eq:dynwin_a1}
  a_1 &= \frac{V}{b}+\epsilon\left(\frac{A}{b}+1\right) \\
  \label{eq:dynwin_a2}
  a_2 &= \frac{1}{2b} \\
\end{align}

Table~\ref{tab:dwinputs} lists all the input variables of the
expression along with range constraints on them. The constraints have
been selected to include most vehicle architectures on which the
algorithm might be deployed (LandShark robot by Black-I Robotics and
an American built car), while eliminating absurd situations such as
faster-than-light travel. From an analytical point of view, this has
two benefits: knowing that inputs are small allows us to automatically
eliminate many concerns regarding floating-point overflow, while
having tighter bounds on the numbers in turn greatly tightens the
error bounds of output expressions at the numerical analysis step.

\begin{table}[h!]\centering
  \begin{tabular}{ l l l }
    \toprule
    Maximum obstacle velocity & $V \in [0, 20] \unit{\metre\per\second}$ & up to 72 \unit{\kilo\metre\per\hour} \\
    Maximum robot acceleration & $A \in [0, 5] \unit{\metre\per\second^2}$ & 0-60 \unit{\si{mph}} in 8\unit{\second}\footnotemark \\
    Maximum robot breaking & $b \in [1, 6] \unit{\metre\per\second^2}$ & 0.1-0.61g \\
    Sampling period & $\epsilon \in [0.01, 0.1] \unit{\second}$ & 100-10 times per second \\
    \midrule
    Robot longitudinal speed & $v_r \in [0, 20] \unit{\metre\per\second}$ & up to 72 \unit{\kilo\metre\per\hour} \\
    Robot $x$ coordinate & $x_r \in [-5000, 5000] \unit{\metre}$ & 100 square km window \\
    Robot $y$ coordinate & $y_r \in [-5000, 5000] \unit{\metre}$ & 100 square km window \\
    Obstacle $x$ coordinate & $x_o \in [-5000, 5000] \unit{\metre}$ & 100 square km window \\
    Obstacle $y$ coordinate & $y_o \in [-5000, 5000] \unit{\metre}$ & 100 square km window \\
    \bottomrule
  \end{tabular}
  \caption{{\tt DynWin} parameters and their ranges}
  \label{tab:dwinputs}
\end{table}
\footnotetext{Same as 2011 BMW 520d.}

Informally, the inputs of the algorithm can be divided into two parts: the
static (general problem-specific constants, the properties of the robot and its
environment) and the dynamic (the information obtained by the robot during
operation). We consider $V$, $b$, $A$ and $\epsilon$ to be problem-specific
constants, unchanging in any particular deployment of the algorithm. Notice also
that in the breakdown (Equations \eqref{eq:dynwin_a0} - \eqref{eq:dynwin_a2})
the values of $a_i$ only depend on these constants and can therefore be
precomputed. The monitor function will take $v_r$, $p_r = (x_r, y_r)$, and $p_o
= (x_o, y_o)$ as inputs and return a Boolean value corresponding to the safety
property ({\tt True} standing for ``safe to proceed'', {\tt False} - ``unsafe,
emergency stop required)''.

\subsection{Unverified compilation with SPIRAL}
\label{dwspiral}
Running an \HCOL expression from Listing~\ref{lst:dwhcol} representing
Equation~\eqref{eq:dynwin1} through SPIRAL will produce an unverified
C-language implementation. During the processing, SPIRAL will invoke a
number of sophisticated heuristics to find an optimal implementation
for the target platform. The heuristics result in the application of a
sequence of semantically preserving transformation steps to the
original expression to reshape it into its optimized form. The list of
transformation steps is recorded in a ``trace'' file produced by
SPIRAL. The produced C code will depend on parameters specific to the
target platform. For the particular case where the target platform is
set to generic x86 without SIMD instructions, the resulting C
implementation of the dynamic monitor is shown in
Listing~\ref{lst:cdynwin}. Its parameters, arrays {\tt D} and {\tt X}
are expected to contain the vector $a$, and operating values of speed
and position of the robot, respectively. Its return values are encoded
as $1$ for {\tt True}, $0$ for {\tt False}.

For comparison, in Listing~\ref{lst:cdynwin1}, we show the C code which is
generated for an x86 platform with Intel Streaming SIMD Extensions (SSE)
enabled. This version includes a runtime floating point error analysis using
the \textit{online uncertainty propagation} approach, as described in
Section~\ref{sec:floatproofs}. In addition to $1$ for {\tt True} or $0$ for {\tt
False}, this version could also return $-1$ to represent that the safety
property could not be reliably computed for the given input values.

\subsection{Verified compilation with HELIX}
\label{dwhelix}

With HELIX, we begin by using the \HCOL expression to encode the
function from Equation~\eqref{eq:dynwin1}, utilising shallow embedded
syntax ({\tt dynwin\_orig} in Listing~\ref{lst:dwhcol}). The
expression is parameterised by two vectors: one encodes the
problem-specific constants $a_i$, and the other encodes the data
received from the sensors. This expression is then processed through a
series of verified transformation steps, as shown in
Figure~\ref{fig:verificationchain} in Section~\ref{sec:helixintro} and
discussed below.

The first transformation applied to the original expression is
\emph{\HCOL breakdown}. Based on the trace of optimizations obtained
by running SPIRAL on a program equivalent to our \HCOL input, the
expression is transformed via a number of semantically preserving
modifications. Currently, the sequence of steps is transferred from
SPIRAL's trace manually, although this process could easily be
automated in the future. On the Coq side, each elementary
transformation takes the form of a setoid rewrite, applied with a
tactic, validated by HELIX.

This process results in an optimized \HCOL representation of the program, {\tt
dynwin\_HCOL}, displayed in Listing~\ref{lst:dwhcol1}. At this stage, the
soundness of the transformation means that the two Gallina functions have the
same types and are extensionally equivalent.  The only difference is that the
new HCOL term is expected (by merit of SPIRAL's heuristics) to compile to more
efficient machine code for the target architecture.

The next intermediate language of the chain of compilation is
\SHCOL. Translating {\tt dynwin\_HCOL} to \SHCOL is a matter of ``lifting''
the \HCOL expression, since any \HCOL operator has a trivial corresponding
\SHCOL version. This is done by wrapping the \HCOL operator with
{\tt liftM\_HOPerator}: the input, a sparse vector, is first ``densified'' - all
structural flags are dropped and corresponding indices are filled with a default
\emph{svalue}, the \HCOL operator is evaluated, and the result is ``sparsified''
by simply treating a dense vector as a sparse one with no structural elements.

However, such a translation is not useful on its own: the embedded
\HCOL expression would continue to operate on dense vectors, failing
to make use of \SHCOL's sparsity contracts and barring future
optimizations making use of iterative computations. Once the operator
is lifted (using {\tt liftM\_HOperator}), the lift function is further
recursively distributed over the structure of {\tt dynwin\_HCOL} and
individual constituent \HCOL operations are replaced with their \SHCOL
counterparts. Similar to the application of SPRIAL optimizations, this
step is currently hardcoded, in the form of applying a series of
rewriting rules to the operator. The result of this process, called
{\tt dynwin\_SHCOL} is shown in Listing~\ref{lst:dwshcol}. Since
lifting is proven to preserve outputs exactly, the resultant operator
is known to be equivalent to its predecessor. Additionally, {\tt
  dynwin\_SHCOL} is proven to be structurally well-formed for
parallelization by proving {\tt SHOperator\_Facts} class instance for
it.

Once the initial \SHCOL operator is built, we can perform further optimizations
inside it, this time making use of vector sparsity. Similarly to the \emph{\HCOL
breakdown} step, we use SPIRAL's optimization trace as an oracle, only
validating the results with HELIX. The final optimized \SHCOL expression,
{\tt dynwin\_SHCOL1}, is shown in Listing~\ref{lst:dwshcol1}.

The next step for {\tt DynWin} is the \MSHCOL language and the
introduction of memory. The translation is fairly straightforward:
using Template-Coq, the \SHCOL language operators are mapped (mostly)
one-to-one into \MSHCOL counterparts. The result ({\tt dynwin\_MHCOL},
shown in Listing~\ref{lst:dwmhcol}) is structurally identical to the
original \SHCOL operator, though, of course, with updated,
memory-block-based semantics. Additionally, in this step, we
specialize $\ctype$ as $\R$ by proving that $\R$ is indeed a
legitimate specialization of $\ctype$ and poses all required
properties.

With the introduction of memory blocks in \MSHCOL, we shift toward an implicit
model of environment and memory and an imperative paradigm in the languages of
the \DHCOL family. The chain goes through two of them: \RHCOL and \FHCOL.

First, with \RHCOL, we transition to imperative code, while leaving the types
unchanged. This is performed, as in the previous step, using Template-Coq. Unlike
the \SHCOL to \MSHCOL step, however, you may notice (see
Listing~\ref{lst:dwrhcol}) that the correspondence is far from one-to-one: a
single \MSHCOL function may compile into a construct of multiple nested \RHCOL
operators. In addition to returning the converted \RHCOL operator, our MetaCoq
compiler also generates a list of ``global variables''; in the case of {\tt
DynWin}, this corresponds to the vector $a$. The semantic preservation proof for
this step is based on the assumption that precomputed values from $a$ are already
in memory before {\tt DynWin} is executed.

The penultimate step, transition to \FHCOL, is simple computationally: since
both languages are instances of \DHCOL, compilation between them is a trivial
cast. The compiled \FHCOL version of {\tt DynWin} is syntactically exactly the
same as the \RHCOL one, the only difference is the types of constants.

Proving this transition correct however, poses a unique problem. As we have
shown, in the compilation chain leading up to \FHCOL, the outputs of all
intermediate representations of the {\tt DynWin} algorithm are exactly
equivalent. At the \FHCOL step, floating-point imprecision takes hold, and an
exact computation is no longer achievable.

\subsubsection{Numerical analysis}
\label{sec:dynwinfloatproofs}

As discussed in Section~\ref{sec:floatproofs}, HELIX provides a mechanism for
performing numerical analysis on \FHCOL algorithms. Acceptable boundaries of
precision vary by application, and thus the operator must make a choice on
whether a particular compiled program is suitable for use, on a case-by-case
basis.

In the case of {\tt DynWin}, the output is always a single value,
corresponding to one of two outcomes: ``the robot is safe to proceed'' or ``the
robot is not safe to proceed''. We can use this, in conjunction with a degree of
understanding of the underlying formula's physical meaning, to postulate a
safety property which is sufficient for us to consider the \FHCOL version of
{\tt DynWin} suitable for its purpose.

This property can be stated simply as follows: \emph{if the infinitely precise
version of DynWin deems the robot unsafe to proceed, the FHCOL version
must also return ``unsafe''}.

Let us elaborate. The mathematical model of robot safety, described in
Equation~\ref{eq:dynwin}, was proven correct in KeYmaera
X~\citep{fulton2015keymaera}. This means that, interpreted with infinitely precise
real-valued semantics, it can be treated as ground truth for evaluating the
robot's situation in the physical world. We will call it the oracle. The
algorithm we produce by compiling it with HELIX is different and could, in
principle, produce an output that differes from the oracle for the same state of
the robot. In table~\ref{tab:dwoutputs}, we list all possible combinations of
outputs from the oracle and the compiled algorithm at a given point in time.

\begin{table}[h!]\centering
  \begin{tabular}{ l l l l }
    \toprule
      No & Real world & Computation & Acceptable? \\
      & (oracle) & (compiled algorithm) & \\
    \midrule
    1 & SAFE & SAFE & Yes: perfect sync \\
    2 & SAFE & UNSAFE & Yes: algorithm overly cautious \\
    3 & UNSAFE & SAFE & No: algorithm not cautious enough \\
    4 & UNSAFE & UNSAFE & Yes: perfect sync \\
    \bottomrule
  \end{tabular}
  \caption{Possible situations arising at any point during the operation of the robot}
  \label{tab:dwoutputs}
\end{table}

Ideally, we would like only situations (1) and (4) to be possible - an algorithm
in perfect sync with the oracle. As discussed however, this is not achievable in
practice due to floating-point imprecision. We need to formulate a more
permissive property that would still be sufficient to ensure the behavior of the
robot guided by our program remains safe. One such property can be defined by
including situation (2) from the Table~\ref{tab:dwoutputs} as acceptable: if the
algorithm classifies some safe situations as unsafe, the robot may sometimes
stop unnecessarily\footnote{How common such occurrence will be with our approach
is discussed in more detail at the end of this section.}, but so long as this is
the only disagreement between the algorithm and the oracle (i.e. (3) does not
occur), the robot will still keeps a safe course at all times.


In order to establish the discussed property, let us begin by calculating the
error bounds on all floating-point computations performed by the {\tt
DynWin} \FHCOL program.

The first step is to extract the arithmetic form of the computation from the
full operator (see Listing~\ref{lst:dwrhcol}). We can do so (as discussed in
Section~\ref{sec:floatproofs}) by means of symbolic execution: translating the
original operator to the ``symbolic-valued'' version of the language (named
``\emph{SFHCOL}''), executing it, and looking up the S-expression corresponding
to the relevant part of the output in resultant memory.
Listing~\ref{lst:dwsymbolic} shows such a chain (the code is simplified,
omitting some error handling for readability) and
Listing~\ref{lst:dwsymbolicres} shows the computed S-expression.  {\tt (SVar i)}
corresponds to a symbolic variable numbered $i$, while {\tt
\verb|dynwin_S_|$\sigma$} and {\tt \verb|dynwin_S_memory|} designate an input
environment filled with such arbitrarily (but uniquely) numbered variables.

\begin{lstlisting}[language=Coq, mathescape=true,
                    basicstyle=\ttfamily\footnotesize,
                   columns=flexible,
                   label={lst:dwsymbolic},
                   caption={DynWin symbolic execution}]
Definition DynWin_S_out : option SFHCOL.memory :=
  (* translate into symbolic language *)
  DynWin_SFHCOL <- FHCOLtoSFHCOL.translate DynWin_FHCOL ;;
  (* evaluate with symbolic inputs *)
  s_m <- evalDSHOperator dynwin_S_$\sigma$ DynWin_SFHCOL dynwin_S_memory ;;
  (* look up relevant block in memory *)
  s_ymb <- memory_lookup s_m dynwin_y_addr ;;
  (* look up relevant cell in block *)
  sexpr <- mem_lookup dynwin_y_offset s_ymb ;;
  ret sexpr.
\end{lstlisting}

\begin{lstlisting}[language=Coq, mathescape=true,
  basicstyle=\ttfamily\footnotesize,
 columns=flexible,
 label={lst:dwsymbolicres},
 caption={DynWin symbolic execution result}]
Some
  (SZLess
    (SPlus
      (SPlus (SPlus SConstZero (SMult SConstOne (SVar 0)))
             (SMult (SMult SConstOne (SVar 3)) (SVar 1)))
      (SMult (SMult (SMult SConstOne (SVar 3)) (SVar 3)) (SVar 2)))
    (SMax (SMax SConstZero (SAbs (SSub (SVar 4) (SVar 6))))
          (SAbs (SSub (SVar 5) (SVar 7)))))
\end{lstlisting}

Each of the variables in the input memory corresponds to some physical value (or
an expression over such values) and has size constraints. In
Listing~\ref{lst:dwsymbolicres}, variables {\tt (SVar 0)} through {\tt (SVar 2)}
correspond to elements of the vector $a$. For the scope of this problem, we
assume values of $a_i$ are precomputed using floating-point arithmetic.  This is
important, as the imprecision of this computation, while falling outside the
focus area of our analysis (i.e. being performed outside the languages in the
HELIX compilation chain), still impacts the precision of outputs obtained from
compiled \FHCOL programs.

To get around this issue, we require that $a$ be computed using strictly
IEEE-754 compliant arithmetic and exactly by the formulas \ref{eq:dynwin_a0},
\ref{eq:dynwin_a1}, \ref{eq:dynwin_a2} shown in Section~\ref{sec:dwoverview}.
Relying on that, we are able to infer the errors accumulated in $a$ before it is
passed to HELIX and use those in further analysis.

The first step of our analysis is to observe that the arithmetic form of {\tt
DynWin} obtained by symbolic execution is a single comparison: on the left-hand
side is an expression corresponding to the polynomial (\ref{eq:dynwin_poly}), on
the right-hand side - the Chebyshev distance (\ref{eq:dynwin_cheb}) between the
robot and the obstacle. With all inputs of these two expressions constrained, we
can formulate a Gappa error analysis problem (see Listing~\ref{lst:dwgappa}
\footnote{Supplying Gappa with direct input is not necessary, all Gappa
invocations can be performed from Ltac. The listing is provided purely for
demonstration purposes.}\textsuperscript{,}\footnote{You may notice that
Chebyshev distance is represented simply as subtraction in Gappa. This is
because the other components of that computation, $max$ and $abs$, are perfectly
precise on floats and are not fully supported by Gappa.}).

\begin{lstlisting}[language=Gappa, mathescape=true,
                    basicstyle=\ttfamily\footnotesize,
                   columns=flexible,
                   label={lst:dwgappa},
                   caption={DynWin polynomial error bounds}]
@rnd64 = float<ieee_64, ne>;

poly = ((0.0 + (1.0*a0)) + ((1.0*v)*a1)) + (((1.0*v)*v)*a2);         # error-free lhs
poly64 rnd64 = ((0.0 + (1.0*a0)) + ((1.0*v)*a1)) + (((1.0*v)*v)*a2); # rounded lhs

cheb = x - y;                      # error-free rhs
cheb64 rnd64 = x - y;              # rounded rhs

{
  # input bounds
    a0 in [0, 12.15]
  /\ a1 in [0.01, 20.6]
  /\ a2 in [0.0833333, 0.5]
  /\ v in [0, 20]
  /\ x1 in [-5000, 5000]
  /\ x2 in [-5000, 5000]
  ->
    |poly64 - poly| in ?           # error in calculating lhs
  /\ |cheb64 - cheb| in ?          # error in calculating rhs
}
\end{lstlisting}

The resulting error bounds are $2 \times 10^{-13}$ on the left-hand side and
$9.1 \times 10^{-13}$ on the right-hand side. We can interpret these as sizes of
the intervals on the real line around infinitely precise results, in which
floating-point results may fall.

\begin{figure}[h]
  \centering
  \includegraphics[width=0.6\columnwidth]{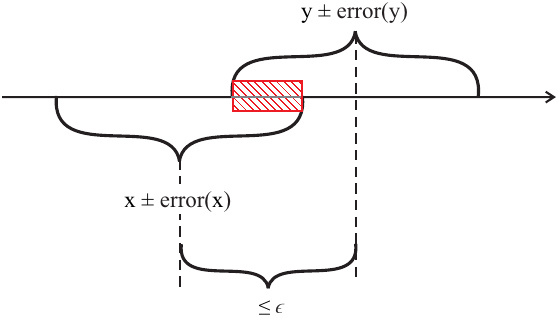}
  \caption{Comparing two imprecise values can be unsafe if their error bounds
  intersect.}
  \label{fig:errorintervals1}
\end{figure}

\begin{figure}[h]
  \centering
  \includegraphics[width=0.6\columnwidth]{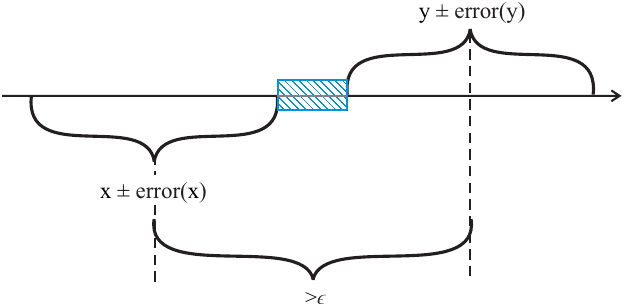}
  \caption{It is safe to compare two imprecise values of sufficiently different
  magnitudes, as the error cannot have an influence on the comparison.}
  \label{fig:errorintervals2}
\end{figure}

Next, we notice that when trying to compare two numbers by comparing their
floating-point representations, we can only get the wrong result if their
respective error intervals intersect (Figure~\ref{fig:errorintervals1}). We
can make use of this: by restricting the comparison operator to only pairs of
numbers separated by a sufficiently large interval (the ``safety margin'',
denoted $\epsilon$; see Figure~\ref{fig:errorintervals2}), we can ensure that no
two numbers are compared if their error intervals intersect. This behavior is
equivalent to stating that any two numbers within $\epsilon$ of each other are
``equal''.

This approach - comparing floating-point numbers using some absolute or relative
error margin, instead of relying on a direct comparison - is common practice in
floating-point computations~\citep{dawson2012comparing}.

In the case of absolute errors, a sufficient safety margin is the sum of error
intervals of the two numbers being compared; for {\tt DynWin} we have: $\epsilon
= 2 \times 10^{-13} + 9.1 \times 10^{-13} = 1.11 \times 10^{-12}$. The comparison
operator is then implemented as shown in Listing~\ref{lst:safezless}.

\begin{lstlisting}[language=Coq, mathescape=true,
                    basicstyle=\ttfamily\footnotesize,
                   columns=flexible,
                   label={lst:safezless},
                   caption={Comparison operator with a safety margin}]
(* [a <? b = 1.0 <-> b - a > epsilon] *)
Definition CTypeZLess (a b: binary64) : binary64 :=
  match Bcompare _ _ epsilon (CTypeSub b a)  with
  | Some Datatypes.Lt => CTypeOne
  | _ => CTypeZero
  end.
\end{lstlisting}

Expressed differently, such a comparison operator has an important property not
possessed by the standard IEEE-754 comparison: if the two floats compare as {\tt
LT}, then any real-valued prototypes which they are approximations of must also
compare as {\tt LT}.

\begin{figure}[h]
  \centering
  \includegraphics[width=0.6\columnwidth]{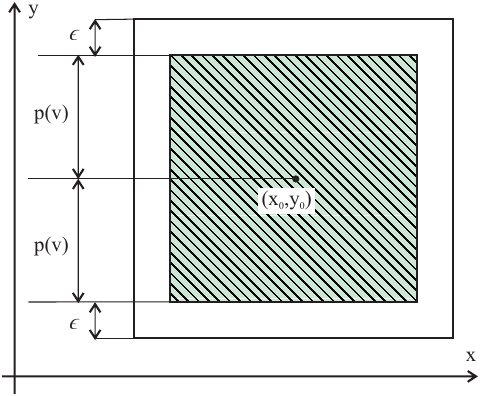}
  \caption{DynWin safety margin on the robot's plane}
  \label{fig:dwborder}
\end{figure}

The introduction of a safety margin raises a question of false positives. After
all, our definition theoretically allows excessively large choices of
$\epsilon$, which could, in practice, result in an unmoving, constantly
trivially safe robot. We address this issue in Figure~\ref{fig:dwborder}. Any
invocation of the {\tt DynWin} algorithm ultimately compares the Chebyshev
distance between the robot and the obstacle to some number dependent on the
robot's velocity $p(v)$; if the distance is smaller than $p(v)$, the robot is
considered unsafe to proceed. On the robot's coordinate plane this corresponds
to an ``unsafe area'' in the shape of a square with side $2p(v)$ centered around
the obstacle. Introducing an absolute safety margin as described, effectively
means increasing $p(v)$ by $\epsilon$. Thus, the unsafe area as computed by our
floating-point algorithm, will be a square with side $2p(v) + 2\epsilon$
centered on the obstacle. In other words, the safety margin we introduced is a
uniform border of thickness $\epsilon$ around the ``actually unsafe''
area. Considering the previously calculated value of $\epsilon = 1.11 \times
10^{-12}$, the safety margin we added is \SI{0.00111}{\nm} thick,
which falls well within any practical limits for deployment of {\tt
DynWin}. Furthermore, it is worth nothing that even this margin could be
tightened significantly, by basing the comparison operator on relative errors
instead of absolute and by making more stringent requirements about the
precision of precomputed values in the constant vector $a$.

\subsubsection{Final step: end-to-end verification for DynWin}

After numerical analysis, we have therefore expressed and proved sound the 
compilation of the DynWin source operator down
to the lowest representation of the \helix{} chain, namely \FHCOL.
Remains to exit \helix{} per se and compile down to \VIR.

Theorem in Listing~\ref{lst:genir-correct} establishes the generic correction of the compilation of
arbitrary \FHCOL{} operator down to \VIR, i.e., of the \coqe{gen_IR} compilation
function.
Our work is not quite done: the code produced is a return-less open control flow
graph, and we assume from the start memories satisfying our invariant.
To fully close the gap, we need to consider the top-level compilation function
that additionally:
\begin{itemize}
  \item allocate globals, and notably two arrays of memories to read the input
    and output;
  \item closes the graph generated by \coqe{gen_IR} with a final block returning;
  \item generates a \emph{main} control-flow-graph calling the main one.
\end{itemize}

This may seem like a very short distance to close, and in some regards it is indeed.
Yet, it hides a considerable length of additional necessary developments.

\paragraph*{Theory at the level of mutually recursive graphs.}
First, we had to develop major generic missing infrastructures on the Vellvm
side. Indeed, all previously established results had been developed intra-thread.
Although our inter-thread behavior is completely trivial---entering the main,
performing a function call, and returning twice---it completely changes the
semantic context. Indeed, the semantic model considered in
Theorem in Listing~\ref{lst:genir-correct}, built from the composition of \coqe{denote_ocfg}
and \coqe{interp_cfg}, treats function calls as abstract, uninterpreted external
events. It does not see the ambient call stack, nor the context required to
resolve mutual recursion.
In contrast, the semantics of top-level programs ties this recursive knot.
This extra infrastructure has been incorporated to Vellvm during the development of this project.

\paragraph*{Initialization.} Theorem in Listing~\ref{lst:genir-correct} assumes initial
configurations satisfying the memory invariant.
On the \FHCOL{} side, initialization is rather straight forwards: it is
implemented as a shallow Coq function computing the initial memory as a function 
of the \FHCOL{} globals, input and output vectors, and input data.
On the Vellvm side however, initialization is itself a quite generic process implemented 
itself as an \itreen{} computation. It is responsible notably for allocating and initializing 
globals and function pointers.
Here again, we have contributed to Vellvm the necessary infrastructure to reason about this process
of initialization, and establish that the invariant is indeed satisfied after initialization.

\paragraph*{End-to-end correctness.}
We are finally ready to put all the pieces together: establishing the correctness of the compilation 
of the DynWin program down to \VIR. 

\begin{lstlisting}[language=Coq,basicstyle=\ttfamily\footnotesize,
  columns=flexible, caption={Top-level correctness theorem}, label={th:top}, float]
Theorem DynWin_correct :
  forall (a : Vector.t CarrierA 3)           (* parameter *)
    (x : Vector.t CarrierA dynwin_i)    (* input *)
    (y : Vector.t CarrierA dynwin_o), (* y - output *)
        (* Evaluation of the source program. *)
        dynwin_orig a x = y $\rightarrow$
        (* if the input data is within bounds for numerical stability, *)
        input_inConstr a x $\rightarrow$
        (* and if compilation of the FHCOL program generated
           by the framework compiles down to VIR, *)
        forall s pll, compile_w_main (dynwin_fhcolp DynWin_FHCOL_hard)
                                 (a++x) newState $\equiv$ inr (s,pll) $\rightarrow$
        (* then the resulting code is observationally pure,
           returning a value within bound with the expected output *)
        exists g l m r, semantics_llvm pll $\approx$ Ret (m,(l,(g, r))) $\lor$ final_rel_val y r.
\end{lstlisting}
 
This final statement is slightly complicated by two factors:
\begin{itemize}
  \item compilation to \FHCOL{} is not performed via a shallow Coq function, but via meta-programming, we hence need to explicitly refer to the generated code (\coqe{DynWin_FHCOL_hard}),
  \item and we need to account for numerical approximation, proving that inputs with bounds lead to results within bound rather than a simple equality.
\end{itemize}
But accounting for these two remarks, it captures the desired result.

The proof of this theorem is not completely trivial.
It essentially reduces to chaining the following schematic steps:
\begin{itemize}
  \item Rewriting the source evaluator by the one of \FHCOL{} by correctness of the \helix{} toolchain;
  \item Substituting the FHCOL{}'s evaluator by its \itreesn{} denotation;
  \item Leveraging the proof of relative soundness of initialization to symbolically process \VIR's initialization phase, establishing the memory invariant;
  \item Symbolically processing on the \VIR{} side the initial call to main, and its internal call to the operator's function;
  \item Leveraging Theorem in Listing~\ref{lst:genir-correct} to fully process the \FHCOL{} side, and the subgraph generated by the compilation of the operator on the \VIR{} side;
  \item Symbolically processing the jump to the final block, and both returns, to close the \VIR{} computation. This final step requires yet another bit of meta-theory for Vellvm: well-parenthesizing of function calls, capturing the invariants induced by its calling convention.
\end{itemize}

We stress that the final theorem is specialized to the program of interest, DynWin in this case.
In this sense, we could qualify the approach we offer through this project of framework.
Almost all of its pieces are entirely generic and reusable. However, the presence of meta-programming forces some manual interaction
to wire the produced code. Furthermore, the application-specific numerical, the initialization phase, and the process of stitching together
all the results for establishing this final theorem of correctness are intertwined, and must therefore be patched for each application.                     

Overall, although proving correct a compilation from a source DSL to a general-purpose language such as LLVM IR should be expected to be a challenge,
the process has turned out more complex than we had initially anticipated.

\section{Results and discussion}
\label{sec:results}
\subsection{Evaluation}
\label{sec:cctests}

To test our \FHCOL to \VIR compiler during development before
formally verifying it, we developed a testing framework and defined
several compiler tests. Each test was defined in Gallina
as an {\tt FSHCOLProgram} record which contains an \FHCOL program; its
input and output vector dimensions; and a list of the global variable names
and types it depends upon. The \FHCOL operator in the test is intended to
be produced by HELIX and must follow all of its conventions on variable
naming, sparsity, etc., even though this is not enforced by the testing
framework.

A single test execution for a given {\tt FSHCOLProgram} performs the following steps:

\begin{enumerate}
\item Initialize a data buffer with some random floating-point values.
  This buffer will be used as a data pool to initialize all global
  variables and the input memory array during the test. The required
  data pool size is computed based on the types of global variables
  and the input vector size from {\tt FSHCOLProgram}. As an additional
  precaution, the buffer will be treated as cyclic, so if the size is
  insufficient, some of the random values will be re-used.
\item Compile \FHCOL program to \VIR, represented as an in-memory
  AST. The program contains a self-enclosed IR \textit{module} with global
  variables and an input array initialized with data from the provided data
  pool, as discussed in Section~\ref{sec:cu}.
\item Run Vellvm interpreter on the generated IR program, executing
  the {\tt main} function and recording the value it returns.
\item Run the \FHCOL big-step evaluator giving it sufficient
  fuel (as estimated by {\tt estimateFuel}) and an initial memory state with $\sigma$ initialized with global variables, an input vector
  holding the data taken from the data pool, and the placeholder for output data. Upon successful evaluation, note the contents of the
  memory block designated to hold the result of the computation.
\item Compare results of successful \FHCOL evaluation and IR
  interpretation steps.
\end{enumerate}

\begin{figure}[H]
  \centering
  \framebox[9.5cm]{\includegraphics[width=9cm]{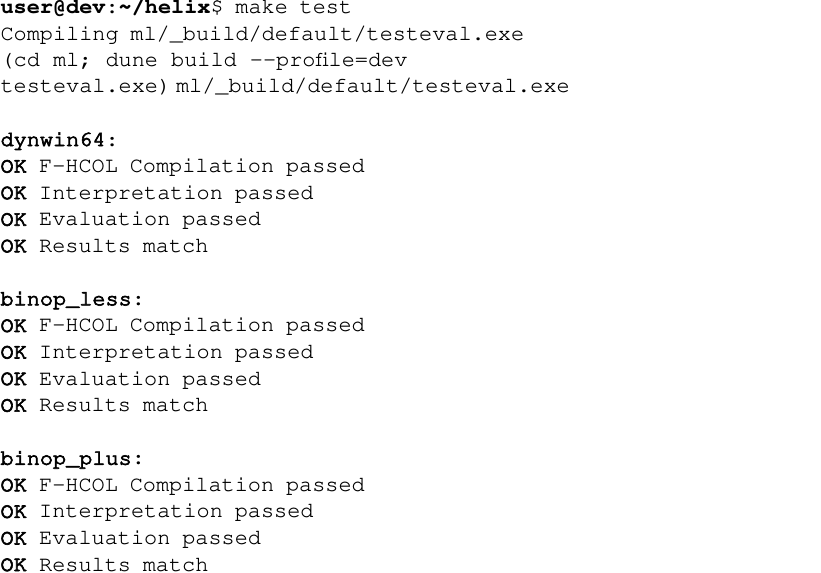}}
  \caption{A screenshot of the compiler test suite execution}
  \label{fig:testscreen}
\end{figure}

Most of the testing logic, like running the evaluator and interpreter,
is implemented in Gallina and extracted to OCaml. Native OCaml code is
used to generate random numbers, call extracted test drivers, and to
compare the results. For a test to pass, all steps above must
succeed. For the test suite to pass, all tests must succeed. A partial
screenshot of a successful run of the test suite is shown in
Figure~\ref{fig:testscreen}.

We've defined twelve tests with different levels of complexity, that cover all
\FHCOL operators. Some test a single \FHCOL operator, while more complex ones
compute the \textit{Chebyshev distance} or calculate the dynamic window monitor
expression. The latter, being the most complex test in the suite, produces 368
lines of IR code shown in Appendix~\ref{sec:dynwinllvm}.

\subsection{Implementation}
\label{sec:metrics}

The current size of HELIX development is 80 KLOC of Coq. The
proof-to-specifications ratio is approximately 3:1. There is an
insignificant amount of OCaml code (less than 500 lines) implementing
the test harness.


\begin{table}[h!]\centering
  \begin{tabular}{ r  l  l }\toprule
    Module & loc & Percent   \\
    \midrule
    \HCOL & 1,160 & 1.45\% \\
    \SHCOL & 10,902 & 13.62\% \\
    \MSHCOL & 9,023 & 11.28\% \\
    \DHCOL & 7,218 & 9.02\% \\
    \FHCOL & 593 & 0.74\% \\
    \RHCOL & 7,977 & 9.97\% \\
    LLVM & 33,985 & 42.47\% \\
    Tactics & 93 & 0.12\% \\
    Util & 5,676 & 7.09\% \\
    Dynamic Window Monitor & 3,390 & 4.24\% \\
    \midrule
    TOTAL & 80,017 & 100.00\% \\
    \bottomrule
  \end{tabular}
  \caption{Implementation code size}
  \label{tab:loc}
\end{table}

The distribution of code in the lines of code (loc) between modules is shown in
Table~\ref{tab:loc}. Modules corresponding to HELIX languages (e.g. \HCOL,
\SHCOL) include both formalization of the given language as well as translation
with the proof from the previous language in the chain. For example, the \SHCOL
module includes: 1) \SHCOL language formalization; 2) \HCOL to \SHCOL compiler
implementation and proofs; and 3) \SHCOL rewriting engine implementation and
proofs. The \emph{Dynamic Window Monitor} module includes a complete dynamic
window monitor example, first defined in HCOL and then translated through all
intermediate steps to LLVM IR with end-to-end proofs. The
Table~\ref{tab:dynwincoverage} shows the percentage of operators used within the
DynWin example for each of the HELIX's intermediate languages.


\begin{table}[h!]\centering
  \begin{tabular}{ l  l  l  l }\toprule
    Language & Operators in the DynWin example & Operators in the language & Percent \\
    \midrule
    \HCOL & 9 & 18 & 50\% \\
    \SHCOL & 11 & 14 & 79\% \\
    \MSHCOL & 9 & 9 & 100\% \\
    \RHCOL & 9 & 10 & 90\% \\
    \FHCOL & 9 & 10 & 90\% \\
    \bottomrule
  \end{tabular}
  \caption{Percentage of operators covered by the DynWin example}
  \label{tab:dynwincoverage}
\end{table}

HELIX depends on several 3rd-party tools and libraries such as
\emph{ExtLib}, \emph{MathClasses}, \emph{CoLoR}, \emph{Vellvm},
\emph{MetaCoq}, and \emph{Gappa}. It uses \emph{dune} as a build
system. As of this writing, a full compilation of HELIX on a modern
laptop (13th Generation Intel Core i7-1365U vPro CPU, quad-core, 32GB
RAM, SSD) takes about 1 hour. We anticipate that this could be
optimised, and compilation time could be significantly shortened, but
this is left for future work.

\subsection{Future work}
\label{sec:future}

HELIX is a prototype system and as such has some limitations. It could
be extended in the future and developed further to become a
production-level counterpart of SPIRAL. A fair amount of engineering
work is required to make it such. However, it is an entirely feasible
engineering project and HELIX can eventually be developed into a
code syntehsis system for general practical use. A few limitations and
shortcomings of the current implementation are worth mentioning:

\paragraph*{SPIRAL integration.} HELIX uses intermediate \OLL and \SOL
expressions as well as a sequence of rule applications computed by
SPIRAL. Our original plan was to parse the SPIRAL log file to extract
the sequence of rules, map them to corresponding HELIX lemmas, and use
them to automatically generate a proof. Due to time constraints, we
have not implemented SPIRAL log file parsing. For our running example
and tests, we have manually extracted and applied these
rules. Similarly, SPIRAL-generated intermediate \OLL and \SOL
expressions were manually translated to \HCOL and \SHCOL,
respectively. This translation could be trivially automated, as it is
purely syntactic.
  
\paragraph*{Numerical analysis.} \DHCOL to \FHCOL translation has been
proven for \textit{dynamic window monitor} program. The numerical
analysis it involves is described in Section \ref{sec:floatproofs}.
This is a generic analysis suitable for many problems, but some
problems may benefit from custom numerical analysis implementations.
To support that, numerical analysis is HELIX is implemented as a
customizable module.

\vspace{1em}
\noindent Additionally, we identified several interesting future research
directions, discussed below.

\paragraph*{LLVM vector instructions.} To generate high-performance
code, SPIRAL uses C compiler intrinsics to generate SIMD
instructions. In LLVM IR language, they correspond to using vector
arguments in arithmetic operators. For example, a {\tt fadd} operator
could take two arguments of type {\tt <4 x double>} to perform
pointwise addition of two vectors of four double-precision floating
point values. When IR is compiled to machine code for a platform
supporting SIMD instructions (e.g. Intel SSE, AMD 3DNow!, Motorola
AltiVec, or IBM SPU), {\tt fadd} could be compiled to a single CPU
instruction which performs such addition, while for platforms without
SIMD support, it will be translated to a sequence of four scalar
additions.  HELIX does not currently use IR vector arithmetic as it is
not yet fully supported by the formal IR semantics of the Vellvm
project. Once support for these instructions is added, HELIX can be
modified to generate even more efficient code.

\paragraph*{Finite-sets proofs.}
\label{sec:setproofs}
In automated proof scenarios involving the application of per-rule
lemmas to \SHCOL, we encounter the generation of sub-goals for
additional pre-conditions, predominantly arithmetic goals related to
index bounds and set-theoretic goals concerning the (non)intersection,
equality, or inclusion of sparsity patterns, represented as finite
sets. While arithmetic goals are typically and easily resolved
automatically using tactics like {\tt lia}, the set-theoretic goals,
which deal with more complex set operations on sparsity patterns, are
presently solved manually. It is pertinent to note that these sparsity
patterns are represented using the type \coqe{Ensemble (FinNat n)},
signifying a set of finite numbers bounded by $n\in\N$. Given the
decidable nature of set membership and the generally small size of
$n$, correlating to vector size, the proof obligations arising from
set membership are relatively straightforward. Thus, automating the
resolution of these sub-goals is not only feasible but
practical. While classic Coq automation techniques using \emph{Ltac}
could be utilized for this purpose, an alternative and more innovative
approach might involve the use of \textit{computation reflection}, as
suggested by~\cite{malechareflection}, offering a potentially more
efficient and elegant solution.

\subsection{Related work}
\label{sec:related}

\paragraph*{Term rewriting.} Encoding program transformations as
rewriting rules using the algebraic properties of the language is a
well-established approach in the field (see, for example,
\cite{borovansky1998overview, 10.1145/2858949.2784754,
  10.1145/3408974}). These systems use functional program rewriting
without our sparse vector parallelism abstraction, and none of them
are formally verified.

\paragraph*{Formal reasoning about SPIRAL.} Another attempt to formally
verify SPIRAL is described in~\cite{brinich2020}. In this work, the
author limits himself to a subset of SPIRAL \OLL language where all
operators are linear and thus could be evaluated to matrices. This
informs his linear algebra approach to operator equivalence and rule
validation. He also stops one step short of actual machine code
generation (directly or via an intermediate language like CompCert
Clight) with the output language to which SPIRAL expressions are
compiled being a simple imperative language with arrays called
IMP+V. The formalization and proofs related to \SOL are listed as
future work. Parts of the work in progress in HELIX design and
implementation were reported in~\cite{helix-vstte20, helixfhpc18,
  zaliva2015formal, zaliva:19, zaliva2020helix}.

\paragraph*{Certified compilation of imperative languages.} Probably the
most influential project in the field of certified compilation
research is CompCert~\citep{compcert}. It has been used alone or in
conjunction with the Verified Software Toolchain (VST)~\citep{vst} to
verify various applications~\citep{vst-hmac, cpp19-ns, vst-mailboxes,
  zhk+21}, and has spawned many related projects that extend it in
various directions~\citep{beringer2014sharedmem, compcertx,
  cascompcert, compcerttso, compcertm, bourke20}.

\paragraph*{Verified compilation of functional languages.} There are
several projects for certified compilation from functional to
imperative languages. \emph{CertiCoq} translates Gallina programs to
CompCert's \textit{Clight}. The goal is much more ambitious than ours,
as the aim is to translate not a domain-specific language like
{\SHCOL} but a dependently typed general-purpose
language. Interestingly, all three guiding principles cited
in~\cite{anand2017certicoq} also apply to our approach. Some of their
transformation steps could be related to ours. For example, going from
dependently typed \SHCOL to \MSHCOL, we perform nominal \textit{type
  erasure}. However, there are some differences at later stages. For
example, unlike HELIX, CertiCoq uses a continuation-passing style
representation and proves compiler correctness, while HELIX relies on
automated translation validation. Another related project is
\emph{CakeML}~\citep{kumar2014cakeml}, which also targets a subset of a
general-purpose language (Standard ML). Unlike HELIX, \emph{CakeML}
uses higher-order logic (HOL) to specify \textit{functional big-step
  semantics}~\citep{owens2016functional}. There are similarities with
our approach, as we also use a \textit{definitional
interpreter}~\citep{10.1145/800194.805852} written in Gallina, to define the
semantics of the higher-order language, {\FHCOL}. The main differences
are: small-step versus big-step semantics and translation validation
versus certified compilation. Additionally, our programs can not
diverge, and while we technically use \textit{fuel} to simplify
termination checking, this differs from the \textit{clock} usage in
CakeML.

\paragraph*{Numerical analysis.}
In our example we use Gappa~\citep{de2008certifying} for numerical
analysis. Similar tools, such as FPTaylor~\citep{solovyev2018rigorous}
or PRECiSA~\citep{moscato2017automatic}, exist, which are also capable
of automated error analysis and produce machine-checked proofs of
their results. We chose to use Gappa in HELIX for its tight
integration into the Coq ecosystem~\citep{boldo2009combining}.

\paragraph*{LLVM IR.}
We rely on Vellvm~\citep{vellvmicfp} LLVM IR semantics to implement and prove
our IR code generation step. This is not the only project aiming to provide
formal IR semantics. Other notable projects include Alive~\citep{alive2},
Crellvm~\citep{kksj+18}, and K-LLVM~\citep{lg20}. Vellum is under active
development and some recent publications are~\citep{twophase}
and~\citep{yoon2023modular}.

\paragraph*{Interaction Trees.} ITrees~\citep{itrees} is a comprehensive
Coq library that facilitates formal modeling and logical reasoning of
interactive computations, which are possibly non-terminating, by
representing them as trees of interactions, thus providing a robust
tool for the formal verification of complex, concurrent, and reactive
systems.  Besides Vellvm, there are several other existing
applications for it~\citep{fhw21, lesanixkbcpz22, dimsum, itrees-spec,
sh+23, sclh+23, ye2022formally}, and it is an area of active
research~\citep{itrees-spec, sz21, chhzz23, yzz22, zhhz20}.

\paragraph*{Formal verification of cyber-physical control systems.}
There is some existing work on formal end-to-end verification of
cyber-physical systems. The work closest to ours is
VeriPhy~\citep{DBLP:conf/pldi/BohrerTMMP18}. It uses a
combination of KeYmaera~\citep{fulton2015keymaera} for
\textit{differential dynamic logic} model safety verification,
Isabelle/HOL for KeYmaera proof-term checking and arithmetic soundness
proofs, and CakeML~\citep{kumar2014cakeml} for verified compilation.
From an application point of view, the main difference between HELIX
and VeriPhy is that the former is oriented towards high-performance
code generation, with much of its pipeline designed to verify code
optimisations in intermediate languages. Another difference is that
HELIX-generated code uses IEEE floating-point arithmetic, whereas
VeriPhy uses fixed-precision integers. Other notable related work
includes~\citep{DBLP:conf/rv/DesaiDS17} (\textit{model checking} and
\textit{signal temporal logic}), ROSCoq~\citep{DBLP:conf/itp/AnandK15}
(\textit{logic of events}, \textit{constructive real analysis}, Coq),
and VeriDrone~\citep{DBLP:conf/cpsweek/MalechaRAL16}
(\textit{discrete-time linear temporal logic}, SMT, Coq).

\subsection{Contributions}
\label{sec:contrib}
In this section, we delve into the significant contributions and the
valuable lessons learned from the HELIX project. This project has made
substantial strides in several domains, notably:

\paragraph*{Formal methods.} Our work has been pivotal in formalizing a
class of Operator Languages. We have advanced the field by developing
methods to algebraically reason about partial computations using
sparse vectors and have introduced a dual semantics approach to build
a certified compiler. Additionally, we have explored several
methodologies for handling the inherent imprecision of floating point
representation, enhancing the robustness and versatility of our
computational models.

\paragraph*{System development.} A notable achievement is the
development of an end-to-end system prototype. This prototype is
notable for its integration of multiple deep and shallow embedded
languages, used in translation validation and certified
compilation. Another key accomplishment is the creation of an
LLVM-based verified code generation backend for SPIRAL, which is more
suitable for more nuanced code generation and formal verification than
the existing C-language backend. Furthermore, have demonstrated a
formal verification of a mature and complex existing system as opposed
to designing a system for verification from the ground up.

\paragraph*{Coq developments.} In the realm of Coq, our work in monadic
sparsity has contributed to the development of more efficient
computational models. We have successfully implemented a mechanism to
switch between deep and shallow embedding, which offers flexibility in
programming and verification. Additionally, we have automated the
process of translation validation, simplifying the verification
process. Lastly, the use of typeclasses to formalize the properties of
operators demonstrates a useful formalization and verification
technique.

\paragraph*{Vellvm.} HELIX is the first front end for Vellvm. The proof
 of correctness, described in Section 6 has both contributed heavily to the development of Vellvm’s meta-theory and opened avenues for future work. As hinted in Lesson~\ref{lesson:abstraction}, the sheer scale of the generated proofs has led us to make all denotation functions in Vellvm opaque and introduce a symbolic execution mechanism based on rewriting. Reasoning about the lack of failure in the source program led to the notion of the image of a monadic computation, as described in Lesson~\ref{lesson:image}. Finally, the overhead induced by the need for reasoning about the freshness monad used in the compiler highlights the need for a rich DSL of graph combinators for Vellvm, a perspective we touched in Lesson~\ref{lesson:graph-combinators}. Naturally, at a smaller scale, this project has contributed to enriching the meta-theory of Vellvm as a whole.

Throughout the development of HELIX, we encountered and overcame many
challenges. A key insight that we garnered is that formal
verification, despite the availability of core tools and techniques,
remains a complex and labour-intensive process. This observation
underscores the need for further advancements in this field to
streamline and simplify the verification process.

\subsection{Conclusion}

The diversity and complexity of computer hardware is likely to
continue to grow, making manual implementation of high-performance
numeric algorithms more and more challenging. The optimization space
required to generate high-performance code will grow exponentially
with the number of hardware capabilities. Even now, a good
implementation already needs to take into account the heterogeneous
use of TPUs, GPUs, and CPUs, instruction timing, instruction cache
behavior, SIMD instruction use, the number of CPU threads and cores,
the number of registers, memory access speed, etc.

We believe that the future direction of numerical computing is heading
towards high-level, possibly declarative, languages used to describe
numeric algorithms. These languages could be equipped with strong type
systems and formal semantics and compiled into efficient code for
various target hardware platforms. Compilation could involve going
through several intermediate languages, each gradually decreasing the
level of abstraction towards the target hardware architecture. During
such compilation (or rather optimal code synthesis), the
performance-related code transformations could be performed at every
step but as early as possible. Each level of these transformations
will be backed up by sound theory (algebraic, functional, or
computational) which will allow the automatic generation of proofs of
correctness of the generated code. This is the approach HELIX
demonstrates.

\pagebreak
\appendix
\renewcommand{\thesection}{\Alph{section}}
\renewcommand{\thesubsection}{\thesection.\arabic{subsection}}
\section*{APPENDIX}

\section{\HCOL language}
\label{sec:hcollang}
The \HCOL language comprises operators, each of which is essentially a
function that maps from $\ctype^n$ to $\ctype^m$. Certain \HCOL
operators necessitate additional parameters, which are required at
compile-time. These operators can accept values of type $\ctype$, as
well as functions on $\ctype$ and $\N$ as compile-time
parameters. Currently, HELIX supports two $\ctype$ constants: {\tt
  Zero} and {\tt One}, but support for more $\ctype$ values may be
added in the future. HELIX also provides a set of primary binary and
unary functions on $\ctype$, and it offers the ability to construct
new functions through $\lambda$-abstraction and function
application. Furthermore, some \HCOL operators are higher-order,
meaning they accept other operators as parameters, enabling the
creation of more complex operators. In some \HCOL operators, the input
has a type in the form $\ctype^{n+m}$, which can be viewed as
functions operating on two distinct arguments, $\ctype^n$ and
$\ctype^m$. These are passed as one concatenated vector.

\subsection{\HCOL syntax}
\label{sec:hcolbnf}

\HCOL syntax is a subset of Coq's syntax and is made up of the
following syntactic categories. \bnfd{Var} represents names for
variables that can be introduced by the
$\lambda$-abstraction. \bnfd{NatConst} and \bnfd{CTypeConst} represent
literals for values of types $\N$ and $\ctype$ respectively, while
\bnfd{NTypeExpr} and \bnfd{CTypeExpr} represent syntax for expressions
which evaluate to those types. \bnfd{CTypeUnFun} and
\bnfd{CTypeBinFun} represent unary and binary functions on
$\ctype$. \bnfd{CTypeIUnFun} and \bnfd{CTypeIBinFun} are similar to
those but also take additional argument of type $\N$ which represents
an index in the input vector. \bnfd{NatBinFun} represents binary
functions on $\N$. \bnfd{VectorConst} represents $\ctype^n$
constants. Finally, \bnfd{HOperator} represents \HCOL operators.

\begin{bnf*}
  \bnfprod{Var}{\bnftd{variable names}}\\
  \bnfprod{NatConst}{\bnftd{literals for natural numbers}}\\
  \bnfprod{CTypeConst}{\bnfts{Zero} \bnfor \bnfts{One}}\\
  \bnfprod{VectorConstBase}{\bnfpn{CTypeConst} \bnfor \bnfpn{CTypeConst} \bnfts{,} \bnfsp \bnfpn{VectorConstBase}}\\
  \bnfprod{VectorConst}{\bnfts{[} \bnfpn{VectorConstBase} \bnfts{]}}\\
  \bnfprod{CTypeExpr}{
    \bnfpn{CTypeConst} \bnfor
    \bnfpn{Var} \bnfor}\\
  \bnfmore{
    \bnfpn{CTypeUnFun} \bnfsp \bnfpn{CTypeExpr} \bnfor}\\
  \bnfmore{
    \bnfpn{CTypeBinFun} \bnfsp \bnfpn{CTypeExpr} \bnfsp \bnfpn{CTypeExpr} \bnfor}\\
  \bnfmore{
    \bnfpn{CTypeIUnFun} \bnfsp \bnfpn{NatExpr} \bnfsp \bnfpn{CTypeExpr} \bnfor}\\
  \bnfmore{
    \bnfpn{CTypeIBinFun} \bnfsp \bnfpn{NatExpr} \bnfsp \bnfpn{CTypeExpr} \bnfsp \bnfpn{CTypeExpr} \bnfor}\\
  \bnfmore{
    \bnfts{Vnth} \bnfsp \bnfpn{VectorConst} \bnfsp \bnfpn{NatExpr}}\\
  \bnfprod{CTypeUnFun}{
    \bnfts{abs} \bnfor
    \bnfts{fun} \bnfsp \bnfpn{Var} \bnfsp \bnfts{=>} \bnfsp \bnfpn{CTypeExpr}}\\
  \bnfprod{CTypeBinFun}{
    \bnfts{plus} \bnfor
    \bnfts{sub} \bnfor
    \bnfts{mul} \bnfor
    \bnfts{min} \bnfor
    \bnfts{max} \bnfor
    \bnfts{lt} \bnfor}\\
  \bnfmore{
    \bnfts{fun} \bnfsp \bnfpn{Var} \bnfsp \bnfpn{Var} \bnfsp \bnfts{=>} \bnfsp \bnfpn{CTypeExpr}}\\
  \bnfprod{CTypeIUnFun}{
    \bnfts{fun} \bnfsp
    \bnfpn{Var} \bnfsp
    \bnfpn{Var} \bnfsp
    \bnfts{=>} \bnfsp 
    \bnfpn{CTypeExpr}}\\
  \bnfprod{CTypeIBinFun}{
    \bnfts{fun} \bnfsp
    \bnfpn{Var} \bnfsp
    \bnfpn{Var} \bnfsp
    \bnfpn{Var} \bnfsp
    \bnfts{=>} \bnfsp 
    \bnfpn{CTypeExpr}}\\
  \bnfprod{NatExpr}{
    \bnfpn{NatConst} \bnfor
    \bnfpn{Var} \bnfor
    \bnfpn{NatBinFun} \bnfsp \bnfpn{NatExpr} \bnfsp \bnfpn{NatExpr}}\\
  \bnfprod{NatBinFun}{
    \bnfts{add} \bnfor
    \bnfts{sub} \bnfor
    \bnfts{mul} \bnfor
    \bnfts{fun} \bnfsp \bnfpn{Var} \bnfsp \bnfpn{Var} \bnfsp \bnfts{=>} \bnfsp \bnfpn{NatExpr}}\\
  \bnfprod{HOperator}{
    \bnfts{HPointwise} \bnfsp \bnfpn{CTypeIUnFun} \bnfor
    \bnfts{HAtomic} \bnfsp \bnfpn{CTypeUnFun} \bnfor}\\
  \bnfmore{
    \bnfts{HScalarProd} \bnfor
    \bnfts{HBinOp} \bnfsp \bnfpn{CTypeIBinFun} \bnfor}\\
  \bnfmore{
    \bnfts{HReduction} \bnfsp \bnfpn{CTypeBinFun} \bnfsp \bnfpn{CTypeExpr} \bnfor}\\
  \bnfmore{
    \bnfts{HEvalPolynomial} \bnfsp \bnfpn{VectorConst} \bnfor}\\
  \bnfmore{
    \bnfts{HAppend} \bnfsp \bnfpn{VectorConst} \bnfor
    \bnfts{HPrepend} \bnfsp \bnfpn{VectorConst} \bnfor}\\
  \bnfmore{
    \bnfts{HMonomialEnumerator} \bnfsp \bnfpn{NatExpr} \bnfor}\\
  \bnfmore{
    \bnfts{HInductor} \bnfsp \bnfpn{NatExpr} \bnfsp \bnfpn{CTypeBinFun} \bnfsp \bnfpn{CTypeExpr} \bnfor}\\
  \bnfmore{
    \bnfts{HInduction} \bnfsp \bnfpn{NatExpr} \bnfsp \bnfpn{CTypeBinFun} \bnfsp \bnfpn{CTypeExpr} \bnfor}\\
  \bnfmore{
    \bnfts{HInfinityNorm} \bnfor
    \bnfts{HChebyshevDistance} \bnfsp \bnfpn{NatExpr} \bnfor}\\
  \bnfmore{
    \bnfts{HVMinus} \bnfor
    \bnfts{HTLess} \bnfsp \bnfpn{HOperator} \bnfsp \bnfpn{HOperator} \bnfor}\\
  \bnfmore{
    \bnfts{HCross} \bnfsp \bnfpn{HOperator} \bnfsp \bnfpn{HOperator} \bnfor}\\
  \bnfmore{
    \bnfts{HStack} \bnfsp \bnfpn{HOperator} \bnfsp \bnfpn{HOperator} \bnfor}\\
  \bnfmore{
    \bnfts{HCompose} \bnfsp \bnfpn{HOperator} \bnfsp \bnfpn{HOperator} \bnfor}\\
\end{bnf*}

\subsection{\HCOL operators}
\label{sec:hcoloperators}

Our current formalization of {\HCOL} includes the following operators,
shallow-embedded in Coq. In the signatures below, all arguments before
semicolon are compile-time parameters.

\newcommand{\hopsection}[3]{\paragraph*{\underline{\underline{$\hoperator{#1}\,#2 \colon\ #3$}}\label{sec:#1}}}

\hopsection{HPointwise}
  {(n \colon \N)\,(f \colon\ \finNat{n} \rightarrow \ctype \rightarrow \ctype)}
  {\ctype^n \rightarrow \ctype^n}

\begin{figure}[h]
  \centering
  \includegraphics[keepaspectratio=true]{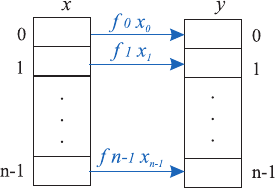}  
  \caption{{\tt HPointwise} operator}
  \label{fig:hpointwise}
\end{figure}

This operator applies a function $f$ to each element of the input
vector, as shown in Figure~\ref{fig:hpointwise}. The function $f$
takes two arguments: the element's index and its value. The output is
the vector of the same length as the input vector.

\hopsection{HAtomic}
  {(f \colon \ctype \rightarrow \ctype)}
  {\ctype^1 \rightarrow \ctype^1}

This operator ``lifts'' a scalar valued function $f$ to an {\HCOL}
operator on single element vectors.

\hopsection{HScalarProd}
  {(n \colon \N)}
  {\ctype^{n+n} \rightarrow \ctype^1}

Calculates the dot product of two vectors. The input vectors must be
concatenated and passed as a single vector of size $n+n$. The result
is returned as a single-element vector. For an input vector $\vec{x} =
[x_0, x_1, \dotsc, x_{n+n-1}]$ it computes [$x_0 \cdot x_n + x_1 \cdot
x_{n+1} + \dotsb + x_{n-1} \cdot x_{n+n-1}]$.

\hopsection{HBinOp}
  {(n \colon \N)\,(f \colon \finNat{n} \rightarrow \ctype \rightarrow \ctype \rightarrow \ctype)}
  {\ctype^{n+n} \rightarrow \ctype^n}

\begin{figure}[h]
  \centering
  \includegraphics[keepaspectratio=true]{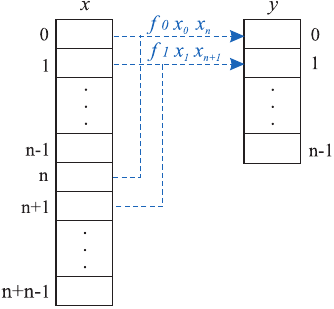}  
  \caption{{\tt HBinOp} operator}
  \label{fig:hbinop}
\end{figure}

This operator applies a function $f$ to pairs of elements from two
halves of an input vector, as shown in Figure~\ref{fig:hbinop}. The
function $f$ additionally takes an index in the range of $0$ to $n-1$
as an argument.

\hopsection{HReduction}
  {(n \colon \N)\,(f \colon \ctype \rightarrow \ctype \rightarrow \ctype)\,(z \colon \ctype)}
  {\ctype^n \rightarrow \ctype^1}

This operator performs a \textit{right fold} of a vector. The two
parameters are a binary function $f$ and the initial value $z$. The
result is returned as a vector size $1$.

\hopsection{HEvalPolynomial}
  {(n \colon \N)\,(\vec{a} \colon \ctype^n)}
  {\ctype^1 \rightarrow \ctype^1}

This operator computes a univariate polynomial of an $n$-th degree. It
is parameterized by a vector $\vec{a}$ of constant coefficients. The
input and output scalar values are represented as vectors of size $1$.
For input $\vec{x} = [x_0]$ and parameter $\vec{a} = [a_0, a_1, \dotsc,
a_n]$, it computes $[a_0 + a_1\cdot x_0 + a_2\cdot x_0^2 +
\dotsb + a_n \cdot x_0^n]$.

\hopsection{HPrepend}
  {(m \, n \colon \N)\,(\vec{a} \colon \ctype^m)}
  {\ctype^n \rightarrow \ctype^{m+n}}

This operator concatenates the input vector of size $n$ with the
constant vector ``prefix'' $\vec{a}$ of size $m$.

\hopsection{HAppend}
  {(m \, n \colon \N)\,(\vec{a} \colon \ctype^m)}
  {\ctype^n \rightarrow \ctype^{n+m}}

This operator concatenates the input vector of size $n$ with the
constant vector ``postfix'' $\vec{a}$ of size $m$.

\hopsection{HMonomialEnumerator}
  {(n \colon \N)}
  {\ctype^1 \rightarrow \ctype^{n+1}}

This operator computes a list of positive integer powers ($0$ to $n$)
of a single variable. Assuming the input is a single-element vector
$\vec{x}=[x_0]$, it returns $[1,x_0,x_0^2, \dotsc, x_0^{n-1}]$.

\hopsection{HInductor}
  {(n \colon \N)\,(f \colon \ctype \rightarrow \ctype \rightarrow \ctype)\,(z \colon \ctype)}
  {\ctype^1 \rightarrow \ctype^1}

This operator is a \textit{recursor}~\citep{moschovakis1989} applied up to the
depth $n$. The parameters are a binary function $f$ and the initial
value $z$. For example, if the input is a single-element vector $\vec{x}=[x_0]$,
for $n=5$, it computes \coqe{[$f$ ($f$ ($f$ ($f$ ($f$ $z$ $x_0$) $x_0$) $x_0$) $x_0$) $x_0$]}.
It is analogous to {\tt Fold} operator in \emph{Wolfram Mathematica}~\citep{wolfram1999mathematica}.

\hopsection{HInduction}
  {(n \colon \N)\,(f \colon \ctype \rightarrow \ctype \rightarrow \ctype)\,(z \colon \ctype)}
  {\ctype^1 \rightarrow \ctype^n}

This operator is similar to {\tt HInductor} but returns all intermediate values for
depths up to $n$. For example, if the input is a single-element vector $\vec{x}=[x_0]$,
for $n=3$, it computes \coqe{[$z$, $f$ $z$ $x_0$, $f$ ($f$ $z$ $x_0$) $x_0$]}.
It is analogous to {\tt FoldList} operator in \emph{Wolfram Mathematica}.

\hopsection{HInfinityNorm}
  {(n \colon \N)}
  {\ctype^n \rightarrow \ctype^1}

This operator computes an \textit{infinity norm} $||\vec{x}||_\infty$ of a
given input vector and returns it as a single-element vector.

\hopsection{HChebyshevDistance}
  {(n \colon \N)}
  {\ctype^{n+n} \rightarrow \ctype^1}

This operator computes the \textit{Chebyshev distance} between two
vectors of size $n$. The vectors are concatenated and passed as a
single input vector of size $n+n$. The resulting scalar value is
returned as a single-element vector.

\hopsection{HVMinus}
  {(n \colon \N)}
  {\ctype^{n+n} \rightarrow \ctype^n}

This operator represents subtraction of two vectors of size $n$. The vectors are
concatenated and passed as an input vector of size $n+n$.

\hopsection{HCross}
  {(n_1\,n_2\,m_1\,m_2\colon\N)\,(f \colon \ctype^{m_1} \rightarrow \ctype^{n_1})(g \colon \ctype^{m_2} \rightarrow \ctype^{n_2})}
  {\ctype^{m_1+m_2} \rightarrow \ctype^{n_1+n_2}}

This is a higher-order operator implementing the \textit{cartesian
product} of two operators, sometimes called a \textit{tupling
combinator}. When applied to operators $f$ and $g$, it produces a new
operator which takes as input a pair of vectors $(\vec{x_0},\vec{x_1})$
(concatenated) and returns $(f\, \vec{x_0},\, g\, \vec{x_1})$ (also
concatenated).

\hopsection{HStack}
  {(n_1\,n_2\,m\colon\N)\,(f \colon \ctype^{m} \rightarrow \ctype^{n_1})\,(g \colon \ctype^{m} \rightarrow \ctype^{n_2})}
  {\ctype^{m} \rightarrow \ctype^{n_1+n_2}}

This is a higher-order operator implementing parallel application of
two operators. When applied to operators $f$ and $g$ it produces a new
operator which for input vector $\vec{x}$ and returns $(f\, \vec{x},\,
g\, \vec{x})$ (concatenated).

\hopsection{HCompose}
  {(m\,n\,t\colon\N)\,(f \colon \ctype^{t} \rightarrow \ctype^{n})\,(g \colon \ctype^{m} \rightarrow \ctype^{t})}
  {\ctype^{m} \rightarrow \ctype^{n}}

This is a higher-order operator for operator composition. When applied to
operators $f$ and $g$ it produces a new operator $(f \circ g)$.

\hopsection{HTLess}
  {(n\,m_1\,m_2\colon\N)\,(f \colon \ctype^{m_1} \rightarrow \ctype^{n})\,(g \colon \ctype^{m_2} \rightarrow \ctype^{n})}
  {\ctype^{m_1+m_2} \rightarrow \ctype^{n}}

This is a higher-order operator implementing element-wise comparison
of the results of the cartesian product of two operators. When applied
to the operators $f$ and $g$, it produces a new operator which takes
as input a pair of vectors $(\vec{x_0},\vec{x_1})$ (concatenated) and
returns a vector produced by element-wise ``less than'' comparison of
$f\, \vec{x_0}$ and $g\, \vec{x_1}$ using a decidable {\tt lt}
predicate, which must be defined for $\ctype$. If the predicate is
{\tt True}, the corresponding output vector's element will be {\tt
zero} or {\tt one} otherwise. The {\tt zero} and {\tt one} are special
values in $\ctype$.

\section{\SHCOL language}
\label{sec:shcollang}
Similarly to \HCOL, \SHCOL language comprises operators, each of which is
essentially a function that maps from $\ctype^n$ to $\ctype^m$. However, unlike
\HCOL operators, \SHCOL operators operate on sparse vectors rather then dense
vectors. \SHCOL is mixed-embedded in Coq: \SHCOL operators are represented with
{\tt SHOperator} record, which contains a function alongside with its sparsity
contract. See Section~\ref{sec:shcol} for more details.

\subsection{\SHCOL syntax}
\label{sec:shcolbnf}

\SHCOL syntax is a subset of Coq's syntax and is made up of the
following syntactic categories. \bnfd{Var} represents names for
variables that can be introduced by the
$\lambda$-abstraction. \bnfd{NatConst} and \bnfd{CTypeConst} represent
literals for values of types $\N$ and $\ctype$ respectively, while
\bnfd{NTypeExpr} and \bnfd{CTypeExpr} represent syntax for expressions
which evaluate to those types. \bnfd{CTypeUnFun} and
\bnfd{CTypeBinFun} represent unary and binary functions on
$\ctype$. \bnfd{CTypeIUnFun} and \bnfd{CTypeIBinFun} are similar to
those but also take an additional argument of type $\N$ which represents
an index in the input vector. \bnfd{NatBinFun} represents binary
functions on $\N$. \bnfd{VectorConst} represents $\ctype^n$
constants. \bnfd{CoqProof} represents Coq proof terms. \bnfd{IndexMap}
represents index-mapping functions. \bnfd{HOperator} and
\bnfd{SHOperator} represent \HCOL and \SHCOL operators. Finally,
\bnfd{SHOperatorFamily} represents families of \SHCOL operators.

\begin{bnf*}
  \bnfprod{Var}{\bnftd{variable names}}\\
  \bnfprod{NatConst}{\bnftd{literals for natural numbers}}\\
  \bnfprod{CTypeConst}{\bnfts{Zero} \bnfor \bnfts{One}}\\
  \bnfprod{VectorConstBase}{\bnfpn{CTypeConst} \bnfor \bnfpn{CTypeConst} \bnfts{,} \bnfsp \bnfpn{VectorConstBase}}\\
  \bnfprod{VectorConst}{\bnfts{[} \bnfpn{VectorConstBase} \bnfts{]}}\\
  \bnfprod{CoqProof}{\bnftd{Coq proof terms}}\\
  \bnfprod{HOperator}{\bnftd{HCOL operators, see Appendix~\ref{sec:hcolbnf}}}\\
  \bnfprod{CTypeExpr}{
    \bnfpn{CTypeConst} \bnfor
    \bnfpn{Var} \bnfor}\\
  \bnfmore{
    \bnfpn{CTypeUnFun} \bnfsp \bnfpn{CTypeExpr} \bnfor}\\
  \bnfmore{
    \bnfpn{CTypeBinFun} \bnfsp \bnfpn{CTypeExpr} \bnfsp \bnfpn{CTypeExpr} \bnfor}\\
  \bnfmore{
    \bnfpn{CTypeIUnFun} \bnfsp \bnfpn{NatExpr} \bnfsp \bnfpn{CTypeExpr} \bnfor}\\
  \bnfmore{
    \bnfpn{CTypeIBinFun} \bnfsp \bnfpn{NatExpr} \bnfsp \bnfpn{CTypeExpr} \bnfsp \bnfpn{CTypeExpr} \bnfor}\\
  \bnfmore{
    \bnfts{Vnth} \bnfsp \bnfpn{VectorConst} \bnfsp \bnfpn{NatExpr}}\\
  \bnfprod{CTypeUnFun}{
    \bnfts{abs} \bnfor
    \bnfts{fun} \bnfsp \bnfpn{Var} \bnfsp \bnfts{=>} \bnfsp \bnfpn{CTypeExpr}}\\
  \bnfprod{CTypeBinFun}{
    \bnfts{plus} \bnfor
    \bnfts{sub} \bnfor
    \bnfts{mul} \bnfor
    \bnfts{min} \bnfor
    \bnfts{max} \bnfor
    \bnfts{lt} \bnfor}\\
  \bnfmore{
    \bnfts{fun} \bnfsp \bnfpn{Var} \bnfsp \bnfpn{Var} \bnfsp \bnfts{=>} \bnfsp \bnfpn{CTypeExpr}}\\
  \bnfprod{CTypeIUnFun}{
    \bnfts{fun} \bnfsp
    \bnfpn{Var} \bnfsp
    \bnfpn{Var} \bnfsp
    \bnfts{=>} \bnfsp 
    \bnfpn{CTypeExpr}}\\
  \bnfprod{CTypeIBinFun}{
    \bnfts{fun} \bnfsp
    \bnfpn{Var} \bnfsp
    \bnfpn{Var} \bnfsp
    \bnfpn{Var} \bnfsp
    \bnfts{=>} \bnfsp 
    \bnfpn{CTypeExpr}}\\
  \bnfprod{NatExpr}{
    \bnfpn{NatConst} \bnfor
    \bnfpn{Var} \bnfor
    \bnfpn{NatBinFun} \bnfsp \bnfpn{NatExpr} \bnfsp \bnfpn{NatExpr}}\\
  \bnfprod{NatBinFun}{
    \bnfts{add} \bnfor
    \bnfts{sub} \bnfor
    \bnfts{mul} \bnfor
    \bnfts{fun} \bnfsp \bnfpn{Var} \bnfsp \bnfpn{Var} \bnfsp \bnfts{=>} \bnfsp \bnfpn{NatExpr}}\\
  \bnfprod{IndexMap}{\bnfts{IndexMap} \bnfsp \bnfpn{NatBinFun}} \bnfsp \bnfpn{CoqProof}\\
  \bnfprod{SHOperator}{
    \bnfts{Embed} \bnfsp \bnfpn{CoqProof} \bnfor
    \bnfts{Pick} \bnfsp \bnfpn{CoqProof} \bnfor}\\
  \bnfmore{
    \bnfts{Scatter} \bnfsp \bnfpn{IndexMap} \bnfor
    \bnfts{Gather} \bnfsp \bnfpn{IndexMap} \bnfor}\\
  \bnfmore{
    \bnfts{liftM\_HOperator} \bnfsp \bnfpn{HOperator} \bnfor}\\
  \bnfmore{
    \bnfts{SHPointwise} \bnfsp \bnfpn{CTypeIUnFun} \bnfor
    \bnfts{SHBinOp} \bnfsp \bnfpn{CTypeIUnFun} \bnfor}\\
  \bnfmore{
    \bnfts{SHInductor} \bnfsp \bnfpn{NatExpr} \bnfsp \bnfpn{CTypeBinFun} \bnfsp \bnfpn{CTypeExpr} \bnfor}\\
  \bnfmore{
    \bnfts{Apply2Union} \bnfsp \bnfpn{CTypeBinFun} \bnfsp \bnfpn{SHOperator} \bnfsp \bnfpn{SHOperator} \bnfor}\\
  \bnfmore{
    \bnfts{SafeCast} \bnfsp \bnfpn{SHOperator} \bnfor
    \bnfts{UnSafeCast} \bnfsp \bnfpn{SHOperator} \bnfor}\\
  \bnfmore{
    \bnfts{SHCompose} \bnfsp \bnfpn{SHOperator} \bnfsp \bnfpn{SHOperator} \bnfor}\\
  \bnfmore{
    \bnfts{IReduction} \bnfsp \bnfpn{CTypeBinFun} \bnfsp \bnfpn{SHOperatorFamily} \bnfor}\\
  \bnfmore{
    \bnfts{IUnion} \bnfsp \bnfpn{CTypeBinFun} \bnfsp \bnfpn{SHOperatorFamily}}\\
  \bnfprod{SHOperatorFamily}{\bnfts{fun} \bnfsp \bnfpn{Var} \bnfsp \bnfts{=>} \bnfsp \bnfpn{SHOperator}}\\
\end{bnf*}

\subsection{\SHCOL operators}
\label{sec:shcoloperators}

There are fourteen \SHCOL operators described below. Unless
specified otherwise, they operate on sparse vectors with elements of
type $(\ctype_{\mathrm{fm}}\; \mathrm{fm})$ for any
$\mathrm{fm} \in \mathrm{Monoid}\, \ctype_{\mathrm{flags}}$.

\newcommand{\shopsection}[3]{\paragraph*{\underline{\underline{$\hoperator{#1}\;#2 \colon #3$}}}}
\newcommand{\shopsectionlong}[3]{
  \paragraph*{\underline{\underline{$\hoperator{#1}\;#2 \colon$}}
  \noindent \underline{\underline{$#3$}}}}

\shopsection{Embed}
  {(s \colon \ctype) \; (n \; b \colon \N) \; (bc \colon b \, < \, n)}
  {\mathtt{SHOperator} \; fm \; 1 \; n \; s}

Takes an element from a single-element input vector and puts it at a
specific index $b$ in a sparse vector of given length.

\begin{figure}[H]
\centering{\includegraphics{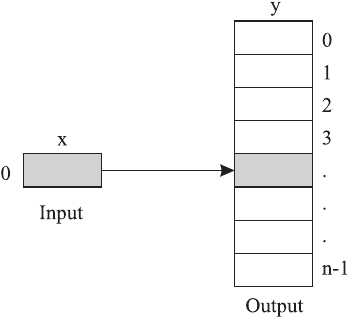}}
\caption{$\hoperator{Embed}$ dataflow}\label{fig:embed}
\end{figure}

\shopsection{Pick}
  {(s \colon \ctype) \; (n \; b \colon \N) \; (bc \colon b < n)}
  {\mathtt{SHOperator} \; fm \; n \; 1 \; s}

Selects an element from the input vector at the given index $b$ and
returns it as a single element vector.

\begin{figure}[H]
\centering{\includegraphics{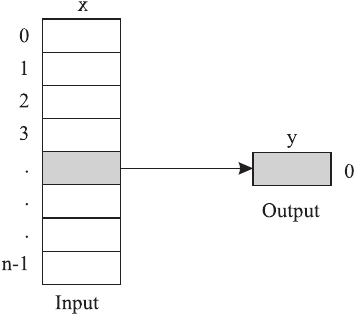}}
\caption{$\hoperator{Pick}$ dataflow}\label{fig:pick}
\end{figure}

\shopsection{Scatter}
  {(s \colon \ctype) \; (n \; m \colon \N) \; (f \colon \mathtt{index{\_}map} \; n \; m)}
  {\mathtt{SHOperator} \; fm \; n \; m \; s}

Embedding can be generalized where more than one element can be
embedded at once. The destination selection is controlled by a
user-provided \textit{index mapping function}.

The operator maps elements of the input vector to the elements of the output
according to an index mapping function $f$. The mapping needs to be
\textit{injective} but not necessarily \textit{surjective}. That means
the output vector could be sparse.

\begin{figure}[H]
\centering{\includegraphics{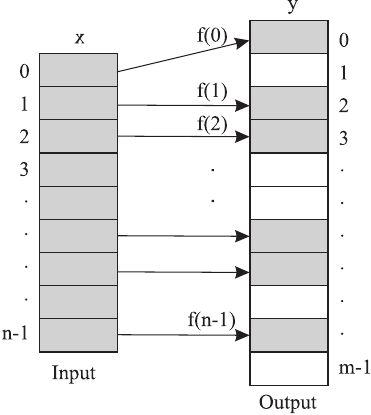}}
\caption{$\hoperator{Scatter}$ dataflow}\label{fig:scatter}
\end{figure}

\shopsection{Gather}
  {(s \colon \ctype) \; (n \; m \colon \N) \; (f \colon \mathtt{index{\_}map} \; m \; n)}
  {\mathtt{SHOperator} \; fm \; n \; m \; s}

Picking can be generalized where more than one element can be picked
at once. The element selection is controlled by a user-provided
\textit{index mapping function} $f$.

\begin{figure}[H]
\centering{\includegraphics{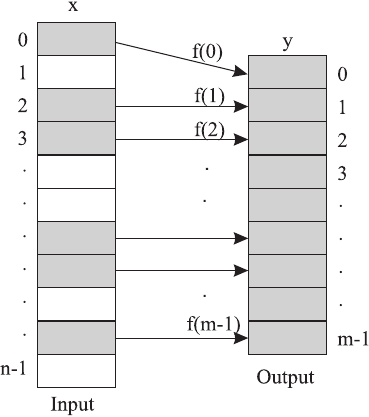}}
\caption{$\hoperator{Gather}$ dataflow}\label{fig:gather}
\end{figure}

\shopsection{liftM{\_}HOperator}
  {(i \; o \colon \N) \; (s \colon \ctype) \; (op \colon \mathtt{avecotr} \; i \rightarrow \mathtt{avector} \; o)}
  {\mathtt{SHOperator} \; fm \; i \; o \; s}

This operator allows ``lifting'' \HCOL operators, so they can be used
in \SHCOL expressions. Since \HCOL operates on dense vectors, the
input vector is first ``densified'' by dropping structural
flags and replacing sparse values with the $s$. During this operation, structural values become indistinguishable from non-structural values. After applying the \HCOL operator, the result is
``sparsified'' (converted to sparse vector format) by marking all
values as non-structural.

\shopsection{SHPointwise}
  {(s \colon \ctype) \; (n \colon \N) \; (f \colon \finNat{n} \rightarrow \ctype \rightarrow \ctype)}
  {\mathtt{SHOperator} \; fm \; n \; n \; s}

This is a \SHCOL version of the {\tt HPointwise} operator from \HCOL (see
Section~\ref{sec:HPointwise}). We could not just lift {\tt HPointwise}
via {\tt liftM\_Hoperator} because we want to preserve structural
flags. However, it can be shown that the two implementations are
equal with respect to {\tt SHOperator\_equiv} which only compares
values and ignores flags.

\shopsection{SHBinOp}
  {(s \colon \ctype) \; (n \colon \N) \; (f \colon \finNat{n} \rightarrow \ctype \rightarrow \ctype \rightarrow \ctype)}
  {\mathtt{SHOperator} \; fm \; (n+n) \; n \; s}

This is a \SHCOL version of the {\tt HBinOp} operator from \HCOL (see
Section~\ref{sec:HBinOp}). Just like with {\tt SHPointwise}, can not
just lift the {\tt HBinOp} via the {\tt liftM\_Hoperator} because we
want to preserve structural flags. However, it can be shown that the
two implementations are equal with respect to {\tt SHOperator\_equiv}
which only compares values and ignores flags.

\shopsection{SHInductor}
  {(s \colon \ctype) \; (n \colon \N) \; (f \colon \ctype \rightarrow \ctype \rightarrow \ctype) \; (z \colon \ctype)}
  {\mathtt{SHOperator} \; fm \; 1 \; 1 \; s}

This is a \SHCOL version of the {\tt HInductor} operator from \HCOL (see
Section~\ref{sec:HInductor}). The initial value $z$ is treated as
\textit{non-structural}. 

\shopsectionlong{Apply2Union}
  {(i \; o \colon \N) \; (s \colon \ctype) \; 
   (dot \colon \ctype \rightarrow \ctype \rightarrow \ctype) \; 
   (f \; g \colon \mathtt{SHOperator} \; fm \; i \; o \; s)}
  {\mathtt{SHOperator} \; fm \; i \; o \; s}

This is a higher-order operator applying two operators to the same input and
combining their results using {\tt Vec2Union} (See
Listing~\ref{lst:vec2union}).

\begin{figure}[H]
\centering{\includegraphics{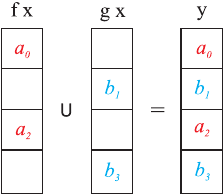}}
\caption{$\hoperator{Apply2Union}$ dataflow}\label{fig:apply2union}
\end{figure}

The additional constraint is that the default value for sparse cells
($s$) is a fixpoint of the provided binary function $dot$,
defined as $dot(s, s) = s$. This is to assure that
the value of sparse cells is preserved when combining them.

\shopsectionlong{SafeCast}
  {(s \colon \ctype) \; (i \; o \colon \N) \;
   (f \colon \mathtt{SHOperator} \; \mathtt{Monoid}\_\ctype_{\mathrm{flags}}' i \; o \; s)}
  {\mathtt{SHOperator} \; \mathtt{Monoid}\_\ctype_{\mathrm{flags}} \; i \; o \; s}

This is a higher-order operator wrapping another \SHCOL operator. While it
does not change the values computed by the wrapped operator, it swaps
the monadic wrapper used to track sparsity properties from
$\mathtt{Monoid}\_\ctype_{\mathrm{flags}}'$ to
$\mathtt{Monoid}\_\ctype_{\mathrm{flags}}$. As we recall from
Section~\ref{sec:sparsitytracking}, the former does not track
collisions, while the latter does.

\shopsectionlong{UnSafeCast}
  {(s \colon \ctype) \; (i \; o \colon \N) \;
   (f \colon \mathtt{SHOperator} \; \mathtt{Monoid}\_\ctype_{\mathrm{flags}} i \; o \; s)}
  {\mathtt{SHOperator} \; \mathtt{Monoid}\_\ctype_{\mathrm{flags}}' \; i \; o \; s}

Similar to {\tt SafeCast} but with wrappers switched in the opposite
direction, from $\mathtt{Monoid}\_\ctype_{\mathrm{flags}}$ to
$\mathtt{Monoid}\_\ctype_{\mathrm{flags}}'$.

\shopsectionlong{SHCompose}
  {(s \colon \ctype) \; (i1 \; o2 \; o3 \colon \N) \;
   (g \colon \mathtt{SHOperator} \; o2 \; o3 \; s) \;
   (f \colon \mathtt{SHOperator} \; i2 \; o2 \; s)}
  {\mathtt{SHOperator} \; i1 \; o3 \; s}

This performs a functional composition of operators. The $f$ is applied to the
input first. The result is then used as an input of $g$. Sometimes,
we use shortcut notation $(g \circledcirc f)$
alluding to the function composition operator $\circ$.

\shopsectionlong{IReduction}
  {(s \colon \ctype) \; (i \; o \; n \colon \N) \;
   (dot \colon \ctype \rightarrow \ctype \rightarrow \ctype) \;
   (F \colon \mathtt{SHOperatorFamily} \; fm \; i \; o \; n \; s)}
  {\mathtt{SHOperator} \; i \; o \; s}

This iteratively applies an indexed family $F$ of $n$ operators to the input and
combines their outputs element-wise using the provided binary function $dot$.
See additional discussion on implementation of this operator in
Section~\ref{sec:mr}.

The additional constraint is that the default value for sparse cells
($s$) is a fixpoint of provided binary function $dot$,
defined as $dot(s, s) = s$. This is to assure that
the value of sparse cells is preserved when combining two of them.

This operator is defined for vectors of $\rstheta$ as its intended use
is to computationally combine the results of a family of operators, and it
must not treat combining non-sparse values as errors (collisions).

\shopsectionlong{IUnion}
  {(s \colon \ctype) \; (i \; o \; n \colon \N) \;
   (dot \colon \ctype \rightarrow \ctype \rightarrow \ctype) \;
   (F \colon \mathtt{SHOperatorFamily} \; fm \; i \; o \; n \; s)}
  {\mathtt{SHOperator} \; i \; o \; s}

Iteratively applies an indexed family $F$ of $n$ operators to the input and
combines their outputs element-wise using the provided binary function $dot$.
See additional discussion on implementation of this operator in
Section~\ref{sec:mr}.

The additional constraint is that the default value for sparse cells
($s$) is a fixpoint of the provided binary function $dot$,
defined as $dot(s, s) = s$. This is to assure that
the value of sparse cells is preserved when combining them.

This operator is defined for vectors of $\rtheta$ and represents an
abstraction for a parallel loop, combining the results of partial
computations. To allow \textit{loop-level parallelism} in a
well-formed \SHCOL expression, partial computations should never
overlap. Such overlapping can not occur as long as the sparsity
patterns of $F$ members are disjoint. The collision
tracking built-in in $\rtheta$ allows us to prove this.


\section{\MSHCOL language}
\label{sec:mhcollang}
\MSHCOL language is similar to \SHCOL. However, unlike \SHCOL operators,
\MSHCOL operators operate on memory blocks rather then sparse vectors. \MSHCOL
is also mixed-embedded in Coq: \MSHCOL operators are represented with {\tt
MSHOperator} record, which contains a function on memory blocks alongside with
two sets that define input and output memory access patterns. See
Section~\ref{sec:mhcol} for more details.

\subsection{\MSHCOL syntax}
\label{sec:mshcolbnf}

\MSHCOL syntax is a subset of Coq's syntax and is made up of the
following syntactic categories. \bnfd{Var} represents names for
variables that can be introduced by the
$\lambda$-abstraction. \bnfd{NatConst} and \bnfd{CTypeConst} represent
literals for values of types $\N$ and $\ctype$ respectively, while
\bnfd{NTypeExpr} and \bnfd{CTypeExpr} represent syntax for expressions
which evaluate to those types. \bnfd{CTypeUnFun} and
\bnfd{CTypeBinFun} represent unary and binary functions on
$\ctype$. \bnfd{CTypeIUnFun} and \bnfd{CTypeIBinFun} are similar to
those but also take additional argument of type $\N$ which represents
an index in the input vector. \bnfd{NatBinFun} represents binary
functions on $\N$. \bnfd{VectorConst} represents $\ctype^n$
constants. \bnfd{CoqProof} represents Coq proof
terms. \bnfd{MSHOperator} and \bnfd{MSHOperatorFamily} represent
\MSHCOL operators and operator families.

\begin{bnf*}
  \bnfprod{Var}{\bnftd{variable names}}\\
  \bnfprod{NatConst}{\bnftd{literals for natural numbers}}\\
  \bnfprod{CTypeConst}{\bnfts{Zero} \bnfor \bnfts{One}}\\
  \bnfprod{VectorConstBase}{\bnfpn{CTypeConst} \bnfor \bnfpn{CTypeConst} \bnfts{,} \bnfsp \bnfpn{VectorConstBase}}\\
  \bnfprod{VectorConst}{\bnfts{[} \bnfpn{VectorConstBase} \bnfts{]}}\\
  \bnfprod{CoqProof}{\bnftd{Coq proof terms}}\\
  \bnfprod{CTypeExpr}{
    \bnfpn{CTypeConst} \bnfor
    \bnfpn{Var} \bnfor}\\
  \bnfmore{
    \bnfpn{CTypeUnFun} \bnfsp \bnfpn{CTypeExpr} \bnfor}\\
  \bnfmore{
    \bnfpn{CTypeBinFun} \bnfsp \bnfpn{CTypeExpr} \bnfsp \bnfpn{CTypeExpr} \bnfor}\\
  \bnfmore{
    \bnfpn{CTypeIUnFun} \bnfsp \bnfpn{NatExpr} \bnfsp \bnfpn{CTypeExpr} \bnfor}\\
  \bnfmore{
    \bnfpn{CTypeIBinFun} \bnfsp \bnfpn{NatExpr} \bnfsp \bnfpn{CTypeExpr} \bnfsp \bnfpn{CTypeExpr} \bnfor}\\
  \bnfmore{
    \bnfts{Vnth} \bnfsp \bnfpn{VectorConst} \bnfsp \bnfpn{NatExpr}}\\
  \bnfprod{CTypeUnFun}{
    \bnfts{abs} \bnfor
    \bnfts{fun} \bnfsp \bnfpn{Var} \bnfsp \bnfts{=>} \bnfsp \bnfpn{CTypeExpr}}\\
  \bnfprod{CTypeBinFun}{
    \bnfts{plus} \bnfor
    \bnfts{sub} \bnfor
    \bnfts{mul} \bnfor
    \bnfts{min} \bnfor
    \bnfts{max} \bnfor
    \bnfts{lt} \bnfor}\\
  \bnfmore{
    \bnfts{fun} \bnfsp \bnfpn{Var} \bnfsp \bnfpn{Var} \bnfsp \bnfts{=>} \bnfsp \bnfpn{CTypeExpr}}\\
  \bnfprod{CTypeIUnFun}{
    \bnfts{fun} \bnfsp
    \bnfpn{Var} \bnfsp
    \bnfpn{Var} \bnfsp
    \bnfts{=>} \bnfsp 
    \bnfpn{CTypeExpr}}\\
  \bnfprod{CTypeIBinFun}{
    \bnfts{fun} \bnfsp
    \bnfpn{Var} \bnfsp
    \bnfpn{Var} \bnfsp
    \bnfpn{Var} \bnfsp
    \bnfts{=>} \bnfsp 
    \bnfpn{CTypeExpr}}\\
  \bnfprod{NatExpr}{
    \bnfpn{NatConst} \bnfor
    \bnfpn{Var} \bnfor
    \bnfpn{NatBinFun} \bnfsp \bnfpn{NatExpr} \bnfsp \bnfpn{NatExpr}}\\
  \bnfprod{NatBinFun}{
    \bnfts{add} \bnfor
    \bnfts{sub} \bnfor
    \bnfts{mul} \bnfor
    \bnfts{fun} \bnfsp \bnfpn{Var} \bnfsp \bnfpn{Var} \bnfsp \bnfts{=>} \bnfsp \bnfpn{NatExpr}}\\
  \bnfprod{MSHOperator}{
    \bnfts{MSHEmbed} \bnfsp \bnfpn{CoqProof} \bnfor
    \bnfts{MSHPick} \bnfsp \bnfpn{CoqProof} \bnfor}\\
  \bnfmore{
    \bnfts{MSHPointwise} \bnfsp \bnfpn{CTypeIUnFun} \bnfor}\\
  \bnfmore{
    \bnfts{MSHBinOp} \bnfsp \bnfpn{CTypeIUnFun} \bnfor}\\
  \bnfmore{
    \bnfts{MSHInductor} \bnfsp \bnfpn{NatExpr} \bnfsp \bnfpn{CTypeBinFun} \bnfsp \bnfpn{CTypeExpr} \bnfor}\\
  \bnfmore{
    \bnfts{MApply2Union} \bnfsp \bnfpn{CTypeBinFun}}\\
  \bnfmore{
    \quad \bnfpn{MSHOperator} \bnfsp \bnfpn{MSHOperator}}\\
  \bnfmore{
    \bnfts{MSHCompose} \bnfsp \bnfpn{MSHOperator} \bnfsp \bnfpn{MSHOperator} \bnfor}\\
  \bnfmore{
    \bnfts{MSHIReduction} \bnfsp \bnfpn{CTypeExpr} \bnfsp \bnfpn{CTypeBinFun}}\\
  \bnfmore{
    \quad \bnfpn{MSHOperatorFamily} \bnfor}\\
  \bnfmore{
    \bnfts{MSHIUnion} \bnfsp \bnfpn{MSHOperatorFamily}}\\
  \bnfprod{MSHOperatorFamily}{\bnfts{fun} \bnfsp \bnfpn{Var} \bnfsp \bnfts{=>} \bnfsp \bnfpn{MSHOperator}}\\
\end{bnf*}

\subsection{\MSHCOL operators}
\label{sec:mshcoloperators}

Recall from Section~\ref{sec:mhcol} that there is a 1-to-1 correspondence
between \MSHCOL operators and operators from the final subset of \SHCOL with the
main difference in the input and output container data types: \MSHCOL uses
memory blocks, where \SHCOL uses sparse vectors.

\section{\DHCOL language family}
\label{sec:dhcollang}
\DHCOL is a language family, parametrised by {\tt CType.t} and {\tt NType.t}
types which are used to represent data values and integer values respectively.
Different instantiations of this parametrisation result in different languages
of \DHCOL family. In HELIX we define two such languages: \RHCOL and \FHCOL.

Unlike all the previous languages in HELIX compilation chain, \DHCOL languages
are imperative. \DHCOL operators can not only compute values, but also modify
the memory state. \DHCOL languages are deep-embedded in Coq: \DHCOL operators
are represented as constructors of the {\tt DSHOperator} inductive type, wich is
used to represent an abstract syntax tree of a \DHCOL program. See
Section~\ref{sec:dhcol} for more details.

\subsection{\DHCOL syntax}
\label{sec:dhcolbnf}

In this section, we present \DHCOL syntax using BNF notation. As \DHCOL language
family is parametrized by {\tt CType.t} and {\tt NType.t} types, \DHCOL syntax
depends on the syntax for literals used to represent values of those types.

\DHCOL syntax is made up of the following syntactic categories. \bnfd{NatConst},
\bnfd{NTypeConst} and \bnfd{CTypeConst} represent literals for values of types
$\N$, {\tt NType.t} and {\tt CType.t} respectively. \bnfd{MemBlockConst}
represents constant memory blocks. \bnfd{CTypeUnFun} and \bnfd{CTypeBinFun}
represent unary and binary functions on {\tt CType.t}. \bnfd{NTypeBinFun}
represents binary functions on {\tt NType.t}. \bnfd{NExpr} and \bnfd{AExpr}
represent syntax for expressions which evaluate to {\tt NType.t} and {\tt
CType.t}. \bnfd{PExpr} represents pointers to memory blocks that can be used to
write to memory. \bnfd{MExpr} represents references to memory blocks that can be
used to read from memory. \bnfd{MemRef} represents pairs of memory references and
block sizes. Finally, \bnfd{DSHOperator} represents \DHCOL operators.

\begin{bnf*}
  \bnfprod{NatConst}{\bnftd{literasls for natural numbers}}\\
  \bnfprod{NTypeConst}{\bnftd{literals for NType.t values}}\\
  \bnfprod{CTypeConst}{\bnftd{literals for CType.t values}}\\
  \bnfprod{MemBlockConstBase}{
    \bnfpn{NatConst} \bnfts{:} \bnfsp \bnfpn{CTypeConst} \bnfor}\\
  \bnfmore{
    \bnfpn{NatConst} \bnfts{:} \bnfsp \bnfpn{CTypeConst} \bnfts{,} \bnfsp \bnfpn{MemBlockBase}}\\
  \bnfprod{MemBlockConst}{\bnfts{\{\}} \bnfor \bnfts{\{} \bnfpn{MemBlockConstBase} \bnfts{\}}}\\
  \bnfprod{NTypeBinFun}{
    \bnfts{NDiv} \bnfor
    \bnfts{NMod} \bnfor
    \bnfts{NPlus} \bnfor
    \bnfts{NMinus} \bnfor
    \bnfts{NMult} \bnfor
    \bnfts{NMin} \bnfor
    \bnfts{NMax}}\\
  \bnfprod{CTypeUnFun}{\bnfts{AAbs}}\\
  \bnfprod{CTypeBinFun}{
    \bnfts{APlus} \bnfor
    \bnfts{AMinus} \bnfor
    \bnfts{AMult} \bnfor
    \bnfts{AMin} \bnfor
    \bnfts{AMax} \bnfor
    \bnfts{AZless}}\\
  \bnfprod{NExpr}{
    \bnfts{NVar} \bnfsp \bnfpn{Nat} \bnfor
    \bnfts{NConst} \bnfsp \bnfpn{NConst} \bnfor}\\
  \bnfmore{
    \bnfpn{NBinFun} \bnfsp \bnfpn{NExpr} \bnfsp \bnfpn{NExpr}}\\
  \bnfprod{PExpr}{\bnfts{PVar} \bnfsp \bnfpn{NatConst}}\\
  \bnfprod{MExpr}{
    \bnfts{MPtrDeref} \bnfsp \bnfpn{PExpr} \bnfor}\\
  \bnfmore{
    \bnfts{MConst} \bnfsp \bnfpn{MemBlockConst} \bnfsp \bnfpn{NTypeConst}}\\
  \bnfprod{AExpr}{
    \bnfts{AVar} \bnfsp \bnfpn{NatConst} \bnfor
    \bnfts{AConst} \bnfsp \bnfpn{CTypeConst} \bnfor}\\
  \bnfmore{
    \bnfts{ANth} \bnfsp \bnfpn{MExpr} \bnfsp \bnfpn{NExpr} \bnfor}\\
  \bnfmore{
    \bnfpn{CTypeUnFun} \bnfsp \bnfpn{AExpr} \bnfor}\\
  \bnfmore{
    \bnfpn{CTypeBinFun} \bnfsp \bnfpn{AExpr} \bnfsp \bnfpn{AExpr}}\\
  \bnfprod{MemRef}{\bnfts{(} \bnfpn{PExpr} \bnfts{,} \bnfsp \bnfpn{NExpr} \bnfts{)}}\\
  \bnfprod{DSHOperator}{\bnfts{DSHNop} \bnfor}\\
  \bnfmore{\bnfts{DSHAssign} \bnfsp \bnfpn{MemRef} \bnfsp \bnfpn{MemRef} \bnfor}\\
  \bnfmore{
    \bnfts{DSHIMap} \bnfsp
      \bnfpn{NatConst} \bnfsp
      \bnfpn{PExpr} \bnfsp
      \bnfpn{PExpr} \bnfsp
      \bnfpn{AExpr} \bnfor}\\
  \bnfmore{
    \bnfts{DSHBinOp} \bnfsp
      \bnfpn{NatConst} \bnfsp
      \bnfpn{PExpr} \bnfsp
      \bnfpn{PExpr} \bnfsp
      \bnfpn{AExpr} \bnfor}\\
  \bnfmore{
    \bnfts{DSHMemMap2} \bnfsp
      \bnfpn{NatConst} \bnfsp}\\
  \bnfmore{
    \quad
      \bnfpn{PExpr} \bnfsp
      \bnfpn{PExpr} \bnfsp
      \bnfpn{PExpr} \bnfsp
      \bnfpn{AExpr} \bnfor}\\
  \bnfmore{
    \bnfts{DSHPower} \bnfsp
      \bnfpn{NExpr} \bnfsp
      \bnfpn{MemRef} \bnfsp
      \bnfpn{MemRef} \bnfsp}\\
  \bnfmore{
    \quad
      \bnfpn{AExpr} \bnfsp
      \bnfpn{CTypeConst} \bnfor}\\
  \bnfmore{\bnfts{DSHLoop} \bnfsp \bnfpn{NatConst} \bnfsp \bnfpn{DSHOperator} \bnfor}\\
  \bnfmore{\bnfts{DSHAlloc} \bnfsp \bnfpn{NTypeConst} \bnfsp \bnfpn{DSHOperator} \bnfor}\\
  \bnfmore{\bnfts{DSHMemInit} \bnfsp \bnfpn{PExpr} \bnfsp \bnfpn{CTypeConst} \bnfor}\\
  \bnfmore{\bnfts{DSHSeq} \bnfsp \bnfpn{DSHOperator} \bnfsp \bnfpn{DSHOperator}}\\
\end{bnf*}

\subsection{\DHCOL operators}
\label{sec:dhcoloperators}

In this section, we present detailed description of all \DHCOL
operators. \DHCOL operators are deeply embedded in Coq: they are
represented as constructors for the `DSHOperator` inductive type.

\newcommand{\dshopsection}[1]{\paragraph*{\underline{\underline{\lstinline[language=Coq]|#1|}}}}

\dshopsection{DSHNop}
This is a ``no-op'' operator. It does nothing.

\dshopsection{DSHAssign (src dst: MemRef)}

An assignment operator copies a single {\tt CType.t} value from memory
location \textit{src} to memory location \textit{dst}.

\dshopsection{DSHIMap (n: nat) (x y: PExpr) (f: AExpr)}

It iterates the index from $0$ to $n-1$. For each iteration, the
values from {\tt x} at the index offset are added to the evaluation
context, and the expression {\tt f} is evaluated. It could be viewed
as calling function $f$ with the two arguments: an index $i$ of type
{\tt NType.t} and a value {\tt x[i]} of type {\tt CType.t}. The result
is written to {\tt y[i]}.

\dshopsection{DSHBinOp (n: nat) (x y: PExpr) (f: AExpr)}

This operator is similar to \emph{DSHIMap}. It also iterates from $0$
to $n-1$. For each iteration, it reads two values: at the index offset
and at $n$ plus the index offset from {\tt x} and adds them to the
evaluation context and then evaluates the expression {\tt f}. It could
be viewed as calling the function $f$ with the two arguments: {\tt x[i]}
and {\tt x[n+i]}. The result is written to {\tt y[i]}.

\dshopsection{DSHMemMap2 (n: nat) (x0 x1 y: PExpr) (f: AExpr)}

This is another iterative map. It also iterates from $0$ to $n-1$. For
each iteration, it reads {\tt x0[i]} and {\tt x1[i]} and adds them to the
evaluation context and evaluates the expression {\tt f}. It could be
viewed as calling the function $f$ with the two arguments: {\tt x0[i]} and
{\tt x1[i]}. The result is written to {\tt y[i]}.

\dshopsection{DSHPower (n:NExpr) (src dst: MemRef) (f: AExpr) (initial: CT.t)}

First, it initializes a memory location pointed by {\tt dst} with
{\tt initial} value and proceeds to iterate $n$ times. On each
iteration, values are loaded from {\tt src} and {\tt dst}, the
{\tt f} is evaluated, and the result of the evaluation is stored
back in {\tt dst}. In C-like pseudo code, it could be expressed as:
\verb|*dst=initial; while(n--){*dst=f(*src,*dst)}|.

\dshopsection{DSHLoop (n:nat) (body: DSHOperator)}

This is a simple loop. It evaluates a body $n$ times. For each
iteration, the value of the loop index (starting from $0$) is put on
the top of the evaluation context before evaluating the \textit{body}.

\dshopsection{DSHAlloc (size:NT.t) (body: DSHOperator)}

Lexically scoped memory allocation. It allocates a new memory block of
size {\tt size}. The newly allocated block is uninitialized. The new
block's offset in memory is put on the top of the evaluation context
and then the {\tt body} is evaluated. The block is only available
while the body is being evaluated.

\dshopsection{DSHMemInit (y: PExpr) (value: CT.t)} Initialize all
elements of block {\tt y} with given {\tt value}.

\dshopsection{DSHSeq (f g: DSHOperator)}

Evaluate $f$ and then, evaluate $g$. Memory modifications performed by
the evaluation of the first operator are visible during the evaluation
of the second.

\subsection{\DHCOL semantics}
\label{sec:dhcolbigstep}

In this section, we present Big-Step Operational Semantics for the
\DHCOL language, informally described in Section~\ref{sec:dhcol}. This
semantics is implemented as a fixpoint evaluator, described in
Section~\ref{sec:dhcoleval}. In presenting \DHCOL semantics, we will
use the following conventions:



\smallskip
\begin{itemize}[nolistsep]
  \item lower-case names with subscript $_s$ will range over {\tt string}
  \item lower-case names with subscript $_\N$ will range over natural
    numbers in $\N$
  \item lower-case names with subscript $_i$ will range over unsigned
    fixed-length machine integers in {\tt NType.t}
  \item lower-case names with subscript $_c$ will range over abstract
    \textit{carrier type} values in {\tt CTtype.t}
  \item lower-case names with subscript $_b$ will range over memory
    blocks in {\tt mem\_block}
  \item lower-case names with subscript $_n$ will range over {\tt NExpr}
  \item lower-case names with subscript $_p$ will range over {\tt PExpr}
  \item lower-case names with subscript $_m$ will range over {\tt MExpr}
  \item lower-case names with subscript $_a$ will range over {\tt AExpr}
  \item upper-case names without subscript will range over {\tt DSHOperator}
\end{itemize}

\subsubsection{The set of states}

The set of states has type
$\mathrm{evalContext} \times \mathrm{memory}$, usually denoted as
$(\sigma,m)$. The {\tt evalContext} is a list of variable types and
bindings of type {\tt DSHVal}:

\begin{lstlisting}[language=Coq,basicstyle=\ttfamily\footnotesize,
  columns=flexible,escapechar=@]
DSHVal := DSHn@@atVal $\mathrm{value}_i$ | DSHCTypeVal $\mathrm{value}_c$ | DSHPtrVal $\mathrm{value}_\N$ $\mathrm{size}_i$
\end{lstlisting}

We use square bracket notation $\sigma[\mathrm{index}_\N]$ to reference a variable with the de Bruijn index $\mathrm{index}_\N$ in the evaluation context $\sigma$. The type of $\sigma[\mathrm{index}_\N]$ is {\tt err DSHVal}.
It will be evaluated to {\tt inr value} in case of successful lookup or {\tt inl \_} if there is no variable with such an index in $\sigma$.

The {\tt memory} represents the state of the memory, per the memory model
described in Section~\ref{sec:memmodel}.
We also use square bracket notation $m[\mathrm{addr}_\N]$ to reference
a memory block at address $\mathrm{addr}_\N$. Thus, the type of $m[\mathrm{addr}_\N]$ is
{\tt err mem\_block}. It will be evaluated to \coqe{inr $\mathrm{value}_b$} in case of successful lookup
or {\tt inl \_} if the address is out of range.

Finally, we write $block_b[\mathrm{offset}_\N]$ to access a value
at offset $\mathrm{offset}_\N$ in memory block $block_b$. The type of
such expression is {\tt err CType.t}. It will be evaluated to \coqe{inr $\mathrm{value}_c$} in case of successful lookup or {\tt inl \_} if the offset is out of range.

\subsubsection{Expressions}

Expressions with types {\tt PExpr}, {\tt MExpr}, {\tt NExpr}, and
{\tt AExpr} are evaluated in a given state $(\sigma,m)$ to their
respective values of type \coqe{(nat$\times$NT.t)}, 
\coqe{(mem_block$\times$NT.t)}, {\tt NT.t}, and {\tt CT.t}. The evaluation can fail, so the {\tt err} monad is used to
wrap the evaluation results. The relation between an expression together with the state and the evaluation result is denoted as $\Downarrow$.

\subsubsection*{{\tt PExpr} evaluation semantics}

\begin{align*}
  \infer[PVarOK]
  {\langle\mathrm{PVar}\; x_\N,\sigma,m\rangle \Downarrow \mathrm{inr}\; (v_\N,\mathrm{size}_i)}
  {\sigma[x_\N]=\mathrm{inr}\; (\mathrm{DSHPtrVal}\; v_\N\; \mathrm{size}_i) }
\\[10pt]
  \infer[PVarErr]
  {\langle\mathrm{PVar}\; x_\N,\sigma,m\rangle \Downarrow \mathrm{inl}\; \textit{"error looking up PVar"}}
  {\sigma[x_\N]=\mathrm{inl}\; \mathrm{msg}_s }
\end{align*}

\subsection*{{\tt MExpr} evaluation semantics}

\begin{align*}
  \infer[MPtrDerefOK]
  {\langle\mathrm{MPtrDeref}\; x_p,\sigma,m\rangle \Downarrow \mathrm{inr}\; (v_b,\mathrm{size}_i)}
  {\langle x_p, \sigma,m\rangle \Downarrow \mathrm{inr}\; (x_\N, \mathrm{size}_i) && m[x_\N] = \mathrm{inr}\; v_b }
\\[10pt]
  \infer[MPtrDerefErr1]
  {\langle\mathrm{MPtrDeref}\; x_p,\sigma,m\rangle \Downarrow \mathrm{inl}\;  \mathrm{msg}_s}
  {\langle x_p, \sigma,m\rangle \Downarrow \mathrm{inl}\; \mathrm{msg}_s}
\\[10pt]
  \infer[MPtrDerefErr2]
  {\langle\mathrm{MPtrDeref}\; x_p,\sigma,m\rangle \Downarrow \mathrm{inl}\; \textit{"MPtrDeref lookup failed"}}
  {\langle x_p, \sigma,m\rangle \Downarrow \mathrm{inr}\; (x_\N, \mathrm{size}_i) && m[x_\N] = \mathrm{inl}\; \mathrm{msg}_s }
\\[10pt]
  \infer[MConst]
  {\langle\mathrm{MConst}\; v_b\;\mathrm{size}_i,\sigma,m\rangle \Downarrow \mathrm{inr}\; (v_b,\mathrm{size}_i)}
  {}
\end{align*}

\subsection*{{\tt NExpr} evaluation semantics}
\begin{align*}
  \infer[NVarOK]
  {\langle\mathrm{NVar}\; x_\N,\sigma,m\rangle \Downarrow \mathrm{inr}\; v_i}
  {\sigma[x_\N]=\mathrm{inr}\; (\mathrm{DSHnatVal}\; v_i)}
\\[10pt]
  \infer[NVarErr1]
  {\langle\mathrm{NVar}\; x_\N,\sigma,m\rangle \Downarrow \mathrm{inl}\; \mathrm{msg}_s}
  {\sigma[x_\N]=\mathrm{inl}\; \mathrm{msg}_s}
\\[10pt]
  \infer[NVarErr2]
  {\langle\mathrm{NVar}\; x_\N,\sigma,m\rangle \Downarrow \mathrm{inl}\; \textit{"invalid NVar type"}}
  {\sigma[x_\N]= \mathrm{inr}\; (\mathrm{DSHCTypeVal}\; v_c)}
\\[10pt]
  \infer[NVarErr3]
  {\langle\mathrm{NVar}\; x_\N,\sigma,m\rangle \Downarrow \mathrm{inl}\; \textit{"invalid NVar type"}}
  {\sigma[x_\N] = \mathrm{inr}\; (\mathrm{DSHPtrVal}\; v_\N\; \mathrm{size}_i)}
\\[10pt]
  \infer[NConst]
  {\langle\mathrm{NConst}\; v_i,\sigma,m\rangle \Downarrow \mathrm{inr}\; v_i}
  {}
\\[10pt]
  \infer[NDivOK]
  {\langle\mathrm{NDiv}\; a_n\; b_n,\sigma,m\rangle \Downarrow \mathrm{inr}\; \frac{a_i}{b_i}}
  {\langle a_n, \sigma,m\rangle \Downarrow \mathrm{inr}\; a_i
  && \langle b_n, \sigma,m\rangle \Downarrow \mathrm{inr}\; b_i
  && b_i \not= 0 }
\\[10pt]
  \infer[NDivErr1]
  {\langle\mathrm{NDiv}\; a_n\; b_n,\sigma,m\rangle \Downarrow \mathrm{inl}\; \mathrm{msg}_s}
  {\langle b_n, \sigma,m\rangle \Downarrow \mathrm{inl}\; \mathrm{msg}_s}
\\[10pt]
  \infer[NDivErr2]
  {\langle\mathrm{NDiv}\; a_n\; b_n,\sigma,m\rangle \Downarrow \mathrm{inl}\; \textit{"Division by 0"}}
  {\langle b_n, \sigma,m\rangle \Downarrow \mathrm{inr}\; b_i
  && b_i = 0 }
\\[10pt]
  \infer[NDivErr3]
  {\langle\mathrm{NDiv}\; a_n\; b_n,\sigma,m\rangle \Downarrow \mathrm{inl}\; \mathrm{msg}_s}
  {\langle b_n, \sigma,m\rangle \Downarrow \mathrm{inr}\; b_i
  && b_i \not= 0
  && \langle a_n, \sigma,m\rangle \Downarrow \mathrm{inl}\; \mathrm{msg}_s  }     
\\[10pt]
  \infer[NModOK]
  {\langle\mathrm{NMod}\; a_n\; b_n,\sigma,m\rangle \Downarrow \mathrm{inr}\; (a_i\bmod b_i)}
  {\langle a_n, \sigma,m\rangle \Downarrow \mathrm{inr}\; a_i
  && \langle b_n, \sigma,m\rangle \Downarrow \mathrm{inr}\; b_i
  && b_i \not= 0 }
\\[10pt]
  \infer[NModErr1]
  {\langle\mathrm{NMod}\; a_n\; b_n,\sigma,m\rangle \Downarrow \mathrm{inl}\; \mathrm{msg}_s}
  {\langle b_n, \sigma,m\rangle \Downarrow \mathrm{inl}\; \mathrm{msg}_s}
\\[10pt]
  \infer[NModErr2]
  {\langle\mathrm{NMod}\; a_n\; b_n,\sigma,m\rangle \Downarrow \mathrm{inl}\; \textit{"Mod by 0"}}
  {\langle b_n, \sigma,m\rangle \Downarrow \mathrm{inr}\; b_i
  && b_i = 0 }
\\[10pt]
  \infer[NModErr3]
  {\langle\mathrm{NMod}\; a_n\; b_n,\sigma,m\rangle \Downarrow \mathrm{inl}\; \mathrm{msg}_s}
  {\langle b_n, \sigma,m\rangle \Downarrow \mathrm{inr}\; b_i
  && b_i \not= 0
  && \langle a_n, \sigma,m\rangle \Downarrow \mathrm{inl}\; \mathrm{msg}_s  }     
\\[10pt]
  \infer[NPlusOK]
  {\langle\mathrm{NPlus}\; a_n\; b_n,\sigma,m\rangle \Downarrow \mathrm{inr}\; (a_i+ b_i)}
  {\langle a_n, \sigma,m\rangle \Downarrow \mathrm{inr}\; a_i
  && \langle b_n, \sigma,m\rangle \Downarrow \mathrm{inr}\; b_i}
\\[10pt]
  \infer[NPlusErr1]
  {\langle\mathrm{NPlus}\; a_n\; b_n,\sigma,m\rangle \Downarrow \mathrm{inl}\; \mathrm{msg}_s}
  {\langle a_n, \sigma,m\rangle \Downarrow \mathrm{inl}\; \mathrm{msg}_s}
\\[10pt]
  \infer[NPlusErr2]
  {\langle\mathrm{NPlus}\; a_n\; b_n,\sigma,m\rangle \Downarrow \mathrm{inl}\; \mathrm{msg}_s}
  {\langle a_n, \sigma,m\rangle \Downarrow \mathrm{inr}\; a_i
  && \langle b_n, \sigma,m\rangle \Downarrow \mathrm{inl}\; \mathrm{msg}_s}
\\[10pt]
  \infer[NMinusOK]
  {\langle\mathrm{NMinus}\; a_n\; b_n,\sigma,m\rangle \Downarrow \mathrm{inr}\; (a_i- b_i)}
  {\langle a_n, \sigma,m\rangle \Downarrow \mathrm{inr}\; a_i
  && \langle b_n, \sigma,m\rangle \Downarrow \mathrm{inr}\; b_i}
\\[10pt]
  \infer[NMinusErr1]
  {\langle\mathrm{NMinus}\; a_n\; b_n,\sigma,m\rangle \Downarrow \mathrm{inl}\; \mathrm{msg}_s}
  {\langle a_n, \sigma,m\rangle \Downarrow \mathrm{inl}\; \mathrm{msg}_s}
\\[10pt]
  \infer[NMinusErr2]
  {\langle\mathrm{NMinus}\; a_n\; b_n,\sigma,m\rangle \Downarrow \mathrm{inl}\; \mathrm{msg}_s}
  {\langle a_n, \sigma,m\rangle \Downarrow \mathrm{inr}\; a_i
  && \langle b_n, \sigma,m\rangle \Downarrow \mathrm{inl}\; \mathrm{msg}_s}
\\[10pt]
  \infer[NMultOK]
  {\langle\mathrm{NMult}\; a_n\; b_n,\sigma,m\rangle \Downarrow \mathrm{inr}\; (a_i\cdot b_i)}
  {\langle a_n, \sigma,m\rangle \Downarrow \mathrm{inr}\; a_i
  && \langle b_n, \sigma,m\rangle \Downarrow \mathrm{inr}\; b_i}
\\[10pt]
  \infer[NMultErr1]
  {\langle\mathrm{NMult}\; a_n\; b_n,\sigma,m\rangle \Downarrow \mathrm{inl}\; \mathrm{msg}_s}
  {\langle a_n, \sigma,m\rangle \Downarrow \mathrm{inl}\; \mathrm{msg}_s}
\\[10pt]
  \infer[NMultErr2]
  {\langle\mathrm{NMult}\; a_n\; b_n,\sigma,m\rangle \Downarrow \mathrm{inl}\; \mathrm{msg}_s}
  {\langle a_n, \sigma,m\rangle \Downarrow \mathrm{inr}\; a_i
  && \langle b_n, \sigma,m\rangle \Downarrow \mathrm{inl}\; \mathrm{msg}_s}
\\[10pt]
  \infer[NMinOK]
  {\langle\mathrm{NMin}\; a_n\; b_n,\sigma,m\rangle \Downarrow \mathrm{inr}\; \min(a_i, b_i)}
  {\langle a_n, \sigma,m\rangle \Downarrow \mathrm{inr}\; a_i
  && \langle b_n, \sigma,m\rangle \Downarrow \mathrm{inr}\; b_i}
\\[10pt]
  \infer[NMinErr1]
  {\langle\mathrm{NMin}\; a_n\; b_n,\sigma,m\rangle \Downarrow \mathrm{inl}\; \mathrm{msg}_s}
  {\langle a_n, \sigma,m\rangle \Downarrow \mathrm{inl}\; \mathrm{msg}_s}
\\[10pt]
  \infer[NMinErr2]
  {\langle\mathrm{NMin}\; a_n\; b_n,\sigma,m\rangle \Downarrow \mathrm{inl}\; \mathrm{msg}_s}
  {\langle a_n, \sigma,m\rangle \Downarrow \mathrm{inr}\; a_i
  && \langle b_n, \sigma,m\rangle \Downarrow \mathrm{inl}\; \mathrm{msg}_s}
\\[10pt]
  \infer[NMaxOK]
  {\langle\mathrm{NMax}\; a_n\; b_n,\sigma,m\rangle \Downarrow \mathrm{inr}\; \max(a_i, b_i)}
  {\langle a_n, \sigma,m\rangle \Downarrow \mathrm{inr}\; a_i
  && \langle b_n, \sigma,m\rangle \Downarrow \mathrm{inr}\; b_i}
\\[10pt]
  \infer[NMaxErr1]
  {\langle\mathrm{NMax}\; a_n\; b_n,\sigma,m\rangle \Downarrow \mathrm{inl}\; \mathrm{msg}_s}
  {\langle a_n, \sigma,m\rangle \Downarrow \mathrm{inl}\; \mathrm{msg}_s}
\\[10pt]
  \infer[NMaxErr2]
  {\langle\mathrm{NMax}\; a_n\; b_n,\sigma,m\rangle \Downarrow \mathrm{inl}\; \mathrm{msg}_s}
  {\langle a_n, \sigma,m\rangle \Downarrow \mathrm{inr}\; a_i
  && \langle b_n, \sigma,m\rangle \Downarrow \mathrm{inl}\; \mathrm{msg}_s}
\end{align*}

\subsection*{{\tt AExpr} evaluation semantics}
\begin{align*}
  \infer[AVarOK]
  {\langle\mathrm{AVar}\; x_\N,\sigma,m\rangle \Downarrow \mathrm{inr}\; v_c}
  {\sigma[x_\N]=\mathrm{inr}\; (\mathrm{DSHCTypeVal}\; v_c)}
\\[10pt]
  \infer[AVarErr1]
  {\langle\mathrm{AVar}\; x_\N,\sigma,m\rangle \Downarrow \mathrm{inl}\; \mathrm{msg}_s}
  {\sigma[x_\N]=\mathrm{inl}\; \mathrm{msg}_s}
\\[10pt]
  \infer[AVarErr2]
  {\langle\mathrm{AVar}\; x_\N,\sigma,m\rangle \Downarrow \mathrm{inl}\; \textit{"invalid AVar type"}}
  {\sigma[x_\N]= \mathrm{inr}\; (\mathrm{DSHnatVal}\; v_i)}
\\[10pt]
  \infer[AVarErr3]
  {\langle\mathrm{AVar}\; x_\N,\sigma,m\rangle \Downarrow \mathrm{inl}\; \textit{"invalid AVar type"}}
  {\sigma[x_\N] = \mathrm{inr}\; (\mathrm{DSHPtrVal}\; v_\N\; \mathrm{size}_i)}
\\[10pt]
  \infer[AConst]
  {\langle\mathrm{AConst}\; v_c,\sigma,m\rangle \Downarrow \mathrm{inr}\; v_c}
  {}
\\[10pt]
  \infer[AAbsOK]
  {\langle\mathrm{AAbs}\; a_a,\sigma,m\rangle \Downarrow \mathrm{inr}\; |v_c|}
  {\langle a_a, \sigma,m\rangle \Downarrow \mathrm{inr}\; v_c}
\\[10pt]
  \infer[AAbsErr]
  {\langle\mathrm{AAbs}\; a_a,\sigma,m\rangle \Downarrow \mathrm{inl}\; \mathrm{msg}_s}
  {\langle a_a, \sigma,m\rangle \Downarrow \mathrm{inl}\; \mathrm{msg}_s}
\\[10pt]
  \infer[APlusOK]
  {\langle\mathrm{APlus}\; a_a\; b_a,\sigma,m\rangle \Downarrow \mathrm{inr}\; (a_c+ b_c)}
  {\langle a_a, \sigma,m\rangle \Downarrow \mathrm{inr}\; a_c
  && \langle b_a, \sigma,m\rangle \Downarrow \mathrm{inr}\; b_c}
\\[10pt]
  \infer[APlusErr1]
  {\langle\mathrm{APlus}\; a_a\; b_a,\sigma,m\rangle \Downarrow \mathrm{inl}\; \mathrm{msg}_s}
  {\langle a_a, \sigma,m\rangle \Downarrow \mathrm{inl}\; \mathrm{msg}_s}
\\[10pt]
  \infer[APlusErr2]
  {\langle\mathrm{APlus}\; a_a\; b_a,\sigma,m\rangle \Downarrow \mathrm{inl}\; \mathrm{msg}_s}
  {\langle a_a, \sigma,m\rangle \Downarrow \mathrm{inr}\; a_c
  && \langle b_a, \sigma,m\rangle \Downarrow \mathrm{inl}\; \mathrm{msg}_s}
\\[10pt]
  \infer[AMinusOK]
  {\langle\mathrm{AMinus}\; a_a\; b_a,\sigma,m\rangle \Downarrow \mathrm{inr}\; (a_c- b_c)}
  {\langle a_a, \sigma,m\rangle \Downarrow \mathrm{inr}\; a_c
  && \langle b_a, \sigma,m\rangle \Downarrow \mathrm{inr}\; b_c}
\\[10pt]
  \infer[AMinusErr1]
  {\langle\mathrm{AMinus}\; a_a\; b_a,\sigma,m\rangle \Downarrow \mathrm{inl}\; \mathrm{msg}_s}
  {\langle a_a, \sigma,m\rangle \Downarrow \mathrm{inl}\; \mathrm{msg}_s}
\\[10pt]
  \infer[AMinusErr2]
  {\langle\mathrm{AMinus}\; a_a\; b_a,\sigma,m\rangle \Downarrow \mathrm{inl}\; \mathrm{msg}_s}
  {\langle a_a, \sigma,m\rangle \Downarrow \mathrm{inr}\; a_c
  && \langle b_a, \sigma,m\rangle \Downarrow \mathrm{inl}\; \mathrm{msg}_s}
\\[10pt]
  \infer[AMultOK]
  {\langle\mathrm{AMult}\; a_a\; b_a,\sigma,m\rangle \Downarrow \mathrm{inr}\; (a_c\cdot b_c)}
  {\langle a_a, \sigma,m\rangle \Downarrow \mathrm{inr}\; a_c
  && \langle b_a, \sigma,m\rangle \Downarrow \mathrm{inr}\; b_c}
\\[10pt]
  \infer[AMultErr1]
  {\langle\mathrm{AMult}\; a_a\; b_a,\sigma,m\rangle \Downarrow \mathrm{inl}\; \mathrm{msg}_s}
  {\langle a_a, \sigma,m\rangle \Downarrow \mathrm{inl}\; \mathrm{msg}_s}
\\[10pt]
  \infer[AMultErr2]
  {\langle\mathrm{AMult}\; a_a\; b_a,\sigma,m\rangle \Downarrow \mathrm{inl}\; \mathrm{msg}_s}
  {\langle a_a, \sigma,m\rangle \Downarrow \mathrm{inr}\; a_c
  && \langle b_a, \sigma,m\rangle \Downarrow \mathrm{inl}\; \mathrm{msg}_s}
\\[10pt]
  \infer[AMinOK]
  {\langle\mathrm{AMin}\; a_a\; b_a,\sigma,m\rangle \Downarrow \mathrm{inr}\; \min(a_c, b_c)}
  {\langle a_a, \sigma,m\rangle \Downarrow \mathrm{inr}\; a_c
  && \langle b_a, \sigma,m\rangle \Downarrow \mathrm{inr}\; b_c}
\\[10pt]
  \infer[AMinErr1]
  {\langle\mathrm{AMin}\; a_a\; b_a,\sigma,m\rangle \Downarrow \mathrm{inl}\; \mathrm{msg}_s}
  {\langle a_a, \sigma,m\rangle \Downarrow \mathrm{inl}\; \mathrm{msg}_s}
\\[10pt]
  \infer[AMinErr2]
  {\langle\mathrm{AMin}\; a_a\; b_a,\sigma,m\rangle \Downarrow \mathrm{inl}\; \mathrm{msg}_s}
  {\langle a_a, \sigma,m\rangle \Downarrow \mathrm{inr}\; a_c
  && \langle b_a, \sigma,m\rangle \Downarrow \mathrm{inl}\; \mathrm{msg}_s}
\\[10pt]
  \infer[AMaxOK]
  {\langle\mathrm{AMax}\; a_a\; b_a,\sigma,m\rangle \Downarrow \mathrm{inr}\; \max(a_c, b_c)}
  {\langle a_a, \sigma,m\rangle \Downarrow \mathrm{inr}\; a_c
  && \langle b_a, \sigma,m\rangle \Downarrow \mathrm{inr}\; b_c}
\\[10pt]
  \infer[AMaxErr1]
  {\langle\mathrm{AMax}\; a_a\; b_a,\sigma,m\rangle \Downarrow \mathrm{inl}\; \mathrm{msg}_s}
  {\langle a_a, \sigma,m\rangle \Downarrow \mathrm{inl}\; \mathrm{msg}_s}
\\[10pt]
  \infer[AMaxErr2]
  {\langle\mathrm{AMax}\; a_a\; b_a,\sigma,m\rangle \Downarrow \mathrm{inl}\; \mathrm{msg}_s}
  {\langle a_a, \sigma,m\rangle \Downarrow \mathrm{inr}\; a_c
  && \langle b_a, \sigma,m\rangle \Downarrow \mathrm{inl}\; \mathrm{msg}_s}
\\[10pt]
  \infer[AZlessOKL]
  {\langle\mathrm{AZless}\; a_a\; b_a,\sigma,m\rangle \Downarrow \mathrm{inr}\; \mathrm{one}}
  {\langle a_a, \sigma,m\rangle \Downarrow \mathrm{inr}\; a_c
  && \langle b_a, \sigma,m\rangle \Downarrow \mathrm{inr}\; b_c
  && a_c < b_c}
\\[10pt]
  \infer[AZlessOKR]
  {\langle\mathrm{AZless}\; a_a\; b_a,\sigma,m\rangle \Downarrow \mathrm{inr}\; \mathrm{zero}}
  {\langle a_a, \sigma,m\rangle \Downarrow \mathrm{inr}\; a_c
  && \langle b_a, \sigma,m\rangle \Downarrow \mathrm{inr}\; b_c
  && a_c \ge b_c}
\\[10pt]
  \infer[AZlessErr1]
  {\langle\mathrm{AZless}\; a_a\; b_a,\sigma,m\rangle \Downarrow \mathrm{inl}\; \mathrm{msg}_s}
  {\langle a_a, \sigma,m\rangle \Downarrow \mathrm{inl}\; \mathrm{msg}_s}
\\[10pt]
  \infer[AZlessErr2]
  {\langle\mathrm{AZless}\; a_a\; b_a,\sigma,m\rangle \Downarrow \mathrm{inl}\; \mathrm{msg}_s}
  {\langle a_a, \sigma,m\rangle \Downarrow \mathrm{inr}\; a_c
  && \langle b_a, \sigma,m\rangle \Downarrow \mathrm{inl}\; \mathrm{msg}_s}
\\[10pt]
  \infer[ANthOK]
  {\langle\mathrm{ANth}\; x_m\; i_n,\sigma,m\rangle \Downarrow \mathrm{inr}\; v_c}
  {{\langle i_n, \sigma,m\rangle \Downarrow \mathrm{inr}\; i_i}
  && {i_i < \mathrm{size}_i} \\[-2pt]
  {\langle x_m, \sigma,m\rangle \Downarrow \mathrm{inr}\; (x_b,\mathrm{size}_i)}
  && {x_b[i_i] = \mathrm{inr}\; v_c \hspace{2em}}}
\\[10pt]
  \infer[ANthErr1]
  {\langle\mathrm{ANth}\; x_m\; i_n,\sigma,m\rangle \Downarrow \mathrm{inl}\; \mathrm{msg}_s}
  {\langle i_n, \sigma,m\rangle \Downarrow \mathrm{inl}\; \mathrm{msg}_s}
\\[10pt]
  \infer[ANthErr2]
  {\langle\mathrm{ANth}\; x_m\; i_n,\sigma,m\rangle \Downarrow \mathrm{inl}\; \mathrm{msg}_s}
  {\langle i_n, \sigma,m\rangle \Downarrow \mathrm{inr}\; i_i
  && \langle x_m, \sigma,m\rangle \Downarrow \mathrm{inl}\; \mathrm{msg}_s}
\\[10pt]
  \infer[ANthErr3]
  {\langle\mathrm{ANth}\; x_m\; i_n,\sigma,m\rangle \Downarrow \mathrm{inl}\; \textit{"ANth index out of bounds"}}
  {\langle i_n, \sigma,m\rangle \Downarrow \mathrm{inr}\; i_i
  && \langle x_m, \sigma,m\rangle \Downarrow \mathrm{inr}\; (x_b,\mathrm{size}_i)
  && i_i \ge \mathrm{size}_i}
\\[10pt]
  \infer[ANthErr4]
  {\langle\mathrm{ANth}\; x_m\; i_n,\sigma,m\rangle \Downarrow \mathrm{inl}\; \textit{"ANth not in memory"}}
  {{\langle i_n, \sigma,m\rangle \Downarrow \mathrm{inr}\; i_i}
  && {i_i < \mathrm{size}_i} \\[-2pt]
  {\langle x_m, \sigma,m\rangle \Downarrow \mathrm{inr}\; (x_b,\mathrm{size}_i)}
  && {x_b[i_i] = \mathrm{inl}\; \mathrm{msg}_s \hspace{2em}}}
\end{align*}

\subsubsection{Operators}

Operators evaluate in a given state $(\sigma,m)$, and the result of the
evaluation is either an error signaled via an {\tt err} monad by
returning $(\mathrm{inl}\; error\_message)$ or the updated memory $m'$
returned as $(\mathrm{inr}\; m')$.

We write $m[\mathrm{addr}_\N/\mathrm{value}_b]$ for the memory
state obtained by replacing a memory block at address $\mathrm{addr}_\N$
with the new memory block $\mathrm{value}_b$. To free a memory at
address $\mathrm{addr}_\N$ and designate this address as currently unassigned
(which would lead to an error trying to access it) we write $m[\mathrm{addr}_\N/]$.

Similarly, $\mathrm{block}_b[\mathrm{offset}_\N/\mathrm{value}_c]$
will result in a memory block with value at $\mathrm{offset}_\N$
replaced with $\mathrm{value}_c$. Notation
$\mathrm{block}_b[0\ldots\mathrm{offset}_\N/\mathrm{value}_c]$ means
to initialize memory block elements with offsets in the range
$[0\ldots \mathrm{offset}_n)$ with value $\mathrm{value}_c$.

We use
$\mathrm{from\_nat}: \N \rightarrow \mathrm{err\; NT.t}$
function to convert natural numbers to {\tt NT.t}. Such a conversion
could return an error because the set of natural numbers is infinite,
while {\tt NT.t} is a finite set. The conversion in the other direction:
from {\tt NT.t} to $\N$ is implicit, and we can use just {\tt NT.t}
where $\N$ is expected.

We use the \textit{cons} operator $::$ to prepend $\sigma$ with a new element which will have index $0$. For example:
$(\mathrm{DSHnatVal}\;n_i)::\sigma$.

Below, we present big-step inference rules for \DHCOL operators. For
brevity, we include only rules describing successful evaluation and
omit rules describing error handling. Since error handling is
always done via the {\tt err} monad, whenever the premise of a rule contains a
clause in the form $x = \mathrm{inr}\; y$, it should be assumed that
there is a corresponding rule for the error case $x = \mathrm{inl}\;
\mathrm{msg}_s$ with $\mathrm{inl}\;\textit{error\_message}$ on the
right side of the $\Downarrow$ in the conclusion.

\begin{align*}
  \infer[Nop]
  {{\langle\mathrm{DSHNop},\sigma,m\rangle \Downarrow \mathrm{inr}\; m}}
  {}
\\[20pt]
  \infer[Assign]
  {\langle\mathrm{DSHAssign}\;(x_p,\mathrm{src}_n)\;(y_p,\mathrm{dst}_n),\sigma,m\rangle \Downarrow \mathrm{inr}\; m[y_\N/(y_b[\mathrm{dst}_i/v_c])]}
  {{\langle x_p,\sigma,m\rangle \Downarrow \mathrm{inr}\;(x_\N,\mathrm{xsize}_i)}
  && {\langle y_p,\sigma,m\rangle \Downarrow \mathrm{inr}\;(y_\N,\mathrm{ysize}_i)} \\[-2pt]
  {m[x_\N] = \mathrm{inr}\; x_b}
  && {m[y_\N] = \mathrm{inr}\; y_b} \\[-2pt]
  {\langle \mathrm{src}_n,\sigma,m\rangle \Downarrow \mathrm{inr}\;\mathrm{src}_i}
  && {\langle \mathrm{dst}_n,\sigma,m\rangle \Downarrow \mathrm{inr}\;\mathrm{dst}_i} \\[-2pt]
  {\mathrm{dst}_i < ysize_i}
  && {x_b[\mathrm{src}_i] = \mathrm{inl}\; v_c}
  }
\\[20pt]
  \infer[IMap]
  {\langle\mathrm{DSHIMap}\;n_\N\;x_p\;y_p\;f_a,\sigma,m\rangle \Downarrow \mathrm{inr}\; m[y_\N/y'_b]}
  {{\langle x_p,\sigma,m\rangle \Downarrow \mathrm{inr}\;(x_\N,\mathrm{xsize}_i)}
  && {\langle y_p,\sigma,m\rangle \Downarrow \mathrm{inr}\;(y_\N,\mathrm{ysize}_i)} \\[-2pt]
  {m[x_\N] = \mathrm{inr}\; x_b}
  && {m[y_\N] = \mathrm{inr}\; y_b} \\[-2pt] 
  {n_i \le \mathrm{xsize_i}}
  && {n_i = \mathrm{from\_nat}\; n_\N}
  \\[-2pt] 
  {\langle\mathrm{evalIMap}\;n_i\;x_b\;y_b\;f_a,\sigma,m\rangle \Downarrow \mathrm{inr}\; y'_b}  
  }
\\[20pt]
  \infer[evalIMap0]
  {\langle\mathrm{evalIMap}\;0\;x_b\;y_b\;f_a,\sigma,m\rangle \Downarrow \mathrm{inr}\; y_b}
  {}
\\[20pt]
  \infer[evalIMapSn]
  {\langle\mathrm{evalIMap}\;(n_i+1)\;x_b\;y_b\;f_a,\sigma,m\rangle \Downarrow \mathrm{inr}\; y'_b}
  {
  {a_c = \mathrm{inr}\;x_b[n_\N]} \\[-2pt]
  {\sigma' = (\mathrm{DSHCTypeVal}\;a_c)::(\mathrm{DSHnatVal}\;n_i)::\sigma} \\[-2pt]
  {\langle f_a,\sigma',m\rangle \Downarrow inr\; v_c} \\[-2pt]
  {\langle \mathrm{evalIMap}\;n_i\;x_b\;y_b\;f_a,\sigma,y_b[n_i/v_c]\rangle \Downarrow inr\; y'_b}
  }
\\[20pt]
  \infer[Map2]
  {\langle\mathrm{DSHMemMap2}\;n_\N\;x0_p\;x1_p\;y_p\;f_a,\sigma,m\rangle \Downarrow \mathrm{inr}\; m[y_\N/y'_b]}
  {{\langle x0_p,\sigma,m\rangle \Downarrow \mathrm{inr}\;(x0_\N,\mathrm{x0size}_i)}
  && {m[x0_\N] = \mathrm{inr}\; x0_b} \\[-2pt]
  {\langle x1_p,\sigma,m\rangle \Downarrow \mathrm{inr}\;(x1_\N,\mathrm{x1size}_i)}
  && {m[x1_\N] = \mathrm{inr}\; x1_b} \\[-2pt]
  {\langle y_p,\sigma,m\rangle \Downarrow \mathrm{inr}\;(y_\N,\mathrm{ysize}_i)} 
  && {m[y_\N] = \mathrm{inr}\; y_b} \\[-2pt]
  {n_i = \mathrm{from\_nat}\; n_\N}
  && {n_i \le \mathrm{ysize_i}} \\[-2pt]
  {\langle \mathrm{evalMap2}\;n_i\;x0_b\;x1_b\;y_b\;f_a,\sigma,m \rangle \Downarrow inr\; y'_b}  }
\\[20pt]
  \infer[evalMap2O]
  {\langle\mathrm{evalMap2}\;0\;x0_b\;x1_b\;y_b\;f_a,\sigma,m\rangle \Downarrow \mathrm{inr}\; y_b}
  {}
\\[20pt]
  \infer[evalMap2Sn]
  {\langle\mathrm{evalMap2}\;(n_i+1)\;x0_b\;x1_b\;y_b\;f_a,\sigma,m\rangle \Downarrow \mathrm{inr}\; y'_b}
  {
  {a_c = \mathrm{inr}\;x0_b[n_\N]} \\[-2pt]  
  {b_c = \mathrm{inr}\;x1_b[n_\N]} \\[-2pt]  
  {\sigma' = (\mathrm{DSHCTypeVal}\;b_c)::(\mathrm{DSHCTypeVal}\;a_c)::\sigma} \\[-2pt]  
  {\langle f_a,\sigma',m\rangle \Downarrow inr\; v_c} \\[-2pt]
  {\langle \mathrm{evalMap2}\;n_i\;x0_b\;x1_b\;y_b[n_\N/v_c]\;f_a,\sigma,m\rangle \Downarrow inr\; y'_b}
  }
\\[20pt]
  \infer[BinOp]
  {\langle\mathrm{DSHBinOp}\;n_\N\;x_p\;y_p\;f_a,\sigma,m\rangle \Downarrow \mathrm{inr}\; m[y_\N/y'_b]}
  {{\langle x_p,\sigma,m\rangle \Downarrow \mathrm{inr}\;(x_\N,\mathrm{xsize}_i)}
  && {\langle y_p,\sigma,m\rangle \Downarrow \mathrm{inr}\;(y_\N,\mathrm{ysize}_i)} \\[-2pt]
  {m[x_\N] = \mathrm{inr}\; x_b}
  && {m[y_\N] = \mathrm{inr}\; y_b} \\[-2pt]
  {n_i = \mathrm{from\_nat}\; n_\N}
  && {n_i \le \mathrm{ysize_i}} \\[-2pt]
  {\langle \mathrm{evalBinOp}\;n_i\;n_i\;x_b\;y_b\;f_a,\sigma,m\rangle \Downarrow inr\; y'_b}
  }
\\[20pt]
  \infer[evalBinOpO]
  {\langle\mathrm{evalBinOp}\;0\;\mathrm{off}_i\;x_b\;y_b\;f_a,\sigma,m\rangle \Downarrow \mathrm{inr}\; y_b}
  {}
\\[20pt]
  \infer[evalBinOpSn]
  {\langle\mathrm{evalBinOp}\;(n_i+1)\;\mathrm{off}_i\;x_b\;y_b\;f_a,\sigma,m\rangle \Downarrow \mathrm{inr}\; y'_b}
  {
  {a_c = \mathrm{inr}\;x_b[n_i]} \\[-2pt]
  {b_c = \mathrm{inr}\;x_b[n_i+\mathrm{off}_i]} \\[-2pt]    
  {\sigma' = (\mathrm{DSHCTypeVal}\;b_c)::(\mathrm{DSHCTypeVal}\;a_c)::(\mathrm{DSHnatVal}\;n_i)::\sigma}  \\[-2pt]
  {\langle f_a,\sigma',m\rangle \Downarrow inr\; v_c} \\[-2pt]
  {\langle\mathrm{evalBinOp}\;n_i\;\mathrm{off}_i\;x_b\;y_b[n_i/v_c]\;f_a,\sigma,m\rangle \Downarrow \mathrm{inr}\; y'_b}  
  }
\\[20pt]
  \infer[Power]
  {\langle\mathrm{DSHPower}\;n_n\;(x_p,\mathrm{src}_n)\;(y_p,\mathrm{dst}_n)\;f_a\;\mathrm{initial}_c,\sigma,m\rangle \Downarrow \mathrm{inr}\; m[y_\N/y'_b]}
  {{\langle x_p,\sigma,m\rangle \Downarrow \mathrm{inr}\;(x_\N,\mathrm{xsize}_i)}
  &&\hspace{-5em} {\langle y_p,\sigma,m\rangle \Downarrow \mathrm{inr}\;(y_\N,\mathrm{ysize}_i)} \\[-2pt]
  {m[x_\N] = \mathrm{inr}\; x_b}
  &&\hspace{-5em} {m[y_\N] = \mathrm{inr}\; y_b} \\[-2pt]
  {\langle \mathrm{src}_n,\sigma,m\rangle \Downarrow \mathrm{inr}\;\mathrm{src}_i}
  &&\hspace{-5em} {\langle \mathrm{dst}_n,\sigma,m\rangle \Downarrow \mathrm{inr}\;\mathrm{dst}_i} \\[-2pt]
  {\mathrm{dst}_i < ysize_i} \\[-2pt]
  {\langle\mathrm{evalPower}\;n_i\;x_b\;y_b\;\mathrm{src}_i\;\mathrm{dst}_i\;f_a\,\sigma,m[\mathrm{dst}_i/\mathrm{initial}_c]\rangle \Downarrow \mathrm{inr}\; y'_b}}
\\[20pt]
  \infer[evalPowerO]
  {\langle\mathrm{evalPower}\;0\;x_b\;y_b\;\mathrm{src}_i\;\mathrm{dst}_i\;f_a\,\sigma,m\rangle \Downarrow \mathrm{inr}\; y_b}
  {}
\\[20pt]
  \infer[evalPowerSn]
  {\langle\mathrm{evalPower}\;(n_i+1)\;x_b\;y_b\;\mathrm{src}_i\;\mathrm{dst}_i\;f_a\,\sigma,m\rangle \Downarrow \mathrm{inr}\; y'_b}
  {
  {a_c = x_b[\mathrm{src}_i]} \\[-2pt]
  {b_c = y_b[\mathrm{dst}_i]} \\[-2pt]
  {\sigma' = (\mathrm{DSHCTypeVal}\;b_c)::(\mathrm{DSHCTypeVal}\;a_c)::\sigma} \\[-2pt]
  {\langle f_a,\sigma',m\rangle \Downarrow inr\; v_c} \\[-2pt]
  {\langle\mathrm{evalPower}\;n_i\;x_b\;y_b[\mathrm{dst}_i/v_c]\;\mathrm{src}_i\;\mathrm{dst}_i\;f_a\,\sigma,m\rangle \Downarrow \mathrm{inr}\; y'_b}
  }
\\[20pt]
  \infer[LoopO]
  {\langle\mathrm{DSHLoop}\;0\;\mathrm{body},\sigma,m\rangle \Downarrow \mathrm{inr}\; m}
  {}
\\[20pt]
  \infer[LoopSn]
  {\langle\mathrm{DSHLoop}\;(n_\N+1)\;\mathrm{body},\sigma,m\rangle \Downarrow \mathrm{inr}\; m''}
  {
  {n_i = \mathrm{from\_nat}\; n_\N} \\[-2pt]
  {\langle\mathrm{DSHLoop}\;n_\N\;\mathrm{body},\sigma,m\rangle \Downarrow \mathrm{inr}\; m'} \\[-2pt]
  {\langle\mathrm{body},((\mathrm{DSHnatVal}\; n_i)::\sigma),m'\rangle \Downarrow \mathrm{inr}\; m''}
  }
\\[20pt]
  \infer[Seq]
  {\langle\mathrm{DSHSeq}\;f\;g,\sigma,m\rangle \Downarrow \mathrm{inr}\; m''}
  {
  {\langle f,\sigma,m\rangle \Downarrow \mathrm{inr}\; m'}
  && {\langle g,\sigma,m'\rangle \Downarrow \mathrm{inr}\; m''}
  }
\\[20pt]
  \infer[MemInit]
  {\langle\mathrm{DSHMemInit}\;y_p\;v_c,\sigma,m\rangle \Downarrow \mathrm{inr}\; m[y_\N/(y_b[0\ldots\mathrm{ysize}_i/v_c])]}
  {
  {\langle y_p,\sigma,m\rangle \Downarrow \mathrm{inr}\;(y_\N,\mathrm{ysize}_i)}
  && {y_b = \mathrm{inr}\;m[y_\N]}
  }
\\[20pt]
  \infer[Alloc]
  {\langle\mathrm{DSHAlloc}\;\mathrm{size}_i\;\mathrm{body},\sigma,m\rangle \Downarrow \mathrm{inr}\; m'}
  {{\exists t_i, m[t_i] = \mathrm{inl}\;\mathrm{msg}_s} \\[-2pt]
  {\langle\mathrm{body},((\mathrm{DSHPtrVal}\;t_i\;\mathrm{size}_i)::\sigma),m[t_i/\mathrm{mem\_empty}]\rangle \Downarrow \mathrm{inr}\; m'[t_i/]}}
\end{align*}

\section{Examples of generated code}
\label{sec:generatedex}
\subsection{Dynamic Window Monitor in LLVM IR}

\label{sec:dynwinllvm}

HELIX-generated LLVM IR code for the dynamic window monitor is shown
in Listing~\ref{lst:dynwinllvm}. For legibility, we wrapped some long
lines using a \textit{backslash} symbol ``\textbackslash'' to indicate
where lines have been split. This is not a part of official IR syntax.

\dynwinlisting[language=llvm,label=lst:dynwinllvm,
basicstyle=\ttfamily\footnotesize,
caption={Dynamic Window Monitor in LLVM IR},captionpos=b]{dynwin.ll}

\subsection{Dynamic Window Monitor in C}

SPIRAL-generated C code for the dynamic window monitor is shown in listings
below. Listing~\ref{lst:cdynwin} shows the code that is generated for generic
x86 target platform without SIMD instructions, while Listing~\ref{lst:cdynwin1}
shows the code for an x86 platform with SIMD instructions extension enabled. See
Section~\ref{dwspiral} for more details about compilation with SPIRAL.

\dynwinlisting[language=C,label=lst:cdynwin,
caption={Dynamic Window Monitor in C (not optimized)},
basicstyle=\fontsize{8}{8}\ttfamily]{dynwin.c}

\dynwinlisting[language=C,label=lst:cdynwin1,
caption={Dynamic Window Monitor in C (optimized)},
basicstyle=\fontsize{8}{8}\ttfamily]{dynwin1.c}

\bibliographystyle{ACM-Reference-Format}
\bibliography{paper}

@article{10.1109/mc.2009.118,
  author     = {Ebert, Christof and Jones, Capers},
  title      = {Embedded Software: Facts, Figures, and Future},
  year       = {2009},
  issue_date = {April 2009},
  publisher  = {IEEE Computer Society Press},
  address    = {Washington, DC, USA},
  volume     = {42},
  number     = {4},
  issn       = {0018-9162},
  url        = {https://doi.org/10.1109/MC.2009.118},
  doi        = {10.1109/MC.2009.118},
  journal    = {Computer},
  month      = apr,
  pages      = {42–52},
  numpages   = {11}
}

@article{dtrees,
  author       = {Minki Cho and
                  Youngju Song and
                  Dongjae Lee and
                  Lennard G{\"{a}}her and
                  Derek Dreyer},
  title        = {Stuttering for Free},
  journal      = {Proc. {ACM} Program. Lang.},
  volume       = {7},
  number       = {{OOPSLA2}},
  pages        = {1677--1704},
  year         = {2023},
  url          = {https://doi.org/10.1145/3622857},
  doi          = {10.1145/3622857},
  bibsource    = {dblp computer science bibliography, https://dblp.org}
}

@article{ctrees,
  author       = {Nicolas Chappe and
                  Paul He and
                  Ludovic Henrio and
                  Yannick Zakowski and
                  Steve Zdancewic},
  title        = {Choice Trees: Representing Nondeterministic, Recursive, and Impure
                  Programs in Coq},
  journal      = {Proc. {ACM} Program. Lang.},
  volume       = {7},
  number       = {{POPL}},
  pages        = {1770--1800},
  year         = {2023},
  url          = {https://doi.org/10.1145/3571254},
  doi          = {10.1145/3571254},
  bibsource    = {dblp computer science bibliography, https://dblp.org}
}

@article{10.1145/2858949.2784754,
  author     = {Steuwer, Michel and Fensch, Christian and Lindley, Sam and
                Dubach, Christophe},
  title      = {Generating Performance Portable Code Using Rewrite Rules:
                From High-Level Functional Expressions to High-Performance
                OpenCL Code},
  year       = {2015},
  issue_date = {September 2015},
  publisher  = {Association for Computing Machinery},
  address    = {New York, NY, USA},
  volume     = {50},
  number     = {9},
  issn       = {0362-1340},
  url        = {https://doi.org/10.1145/2858949.2784754},
  doi        = {10.1145/2858949.2784754},
  journal    = {SIGPLAN Not.},
  month      = aug,
  pages      = {205–217},
  numpages   = {13}
}

@article{10.1145/3408974,
  author     = {Hagedorn, Bastian and Lenfers, Johannes and
                Kundefinedhler, Thomas and Qin, Xueying and Gorlatch,
                Sergei and Steuwer, Michel},
  title      = {Achieving High-Performance the Functional Way: A
                Functional Pearl on Expressing High-Performance
                Optimizations as Rewrite Strategies},
  year       = {2020},
  issue_date = {August 2020},
  publisher  = {Association for Computing Machinery},
  address    = {New York, NY, USA},
  volume     = {4},
  number     = {ICFP},
  url        = {https://doi.org/10.1145/3408974},
  doi        = {10.1145/3408974},
  journal    = {Proc. ACM Program. Lang.},
  month      = aug,
  articleno  = {92},
  numpages   = {29}
}

@inproceedings{10.1145/800194.805852,
  author    = {Reynolds, John C.},
  title     = {Definitional Interpreters for Higher-Order Programming
               Languages},
  year      = {1972},
  isbn      = {9781450374927},
  publisher = {Association for Computing Machinery},
  address   = {New York, NY, USA},
  url       = {https://doi.org/10.1145/800194.805852},
  doi       = {10.1145/800194.805852},
  booktitle = {Proceedings of the ACM Annual Conference - Volume 2},
  pages     = {717–740},
  numpages  = {24},
  location  = {Boston, Massachusetts, USA},
  series    = {ACM ’72}
}

@article{10.1145/982962.964003,
  author     = {Benton, Nick},
  title      = {Simple Relational Correctness Proofs for Static Analyses
                and Program Transformations},
  year       = {2004},
  issue_date = {January 2004},
  publisher  = {Association for Computing Machinery},
  address    = {New York, NY, USA},
  volume     = {39},
  number     = {1},
  issn       = {0362-1340},
  url        = {https://doi.org/10.1145/982962.964003},
  doi        = {10.1145/982962.964003},
  journal    = {SIGPLAN Not.},
  month      = jan,
  pages      = {14–25},
  numpages   = {12}
}

@inproceedings{alive2,
  author    = {Lopes, Nuno P. and Lee, Juneyoung and Hur, Chung-Kil and
               Liu, Zhengyang and Regehr, John},
  title     = {Alive2: Bounded Translation Validation for LLVM},
  year      = {2021},
  isbn      = {9781450383912},
  publisher = {Association for Computing Machinery},
  address   = {New York, NY, USA},
  url       = {https://doi.org/10.1145/3453483.3454030},
  doi       = {10.1145/3453483.3454030},
  booktitle = {Proceedings of the 42nd ACM SIGPLAN International
               Conference on Programming Language Design and
               Implementation},
  pages     = {65–79},
  numpages  = {15},
  location  = {Virtual, Canada},
  series    = {PLDI 2021}
}

@inproceedings{anand2017certicoq,
  title     = {{CertiCoq}: A verified compiler for {Coq}},
  author    = {Anand, Abhishek and Appel, Andrew and Morrisett, Greg and
               Paraskevopoulou, Zoe and Pollack, Randy and Belanger,
               Olivier Savary and Sozeau, Matthieu and Weaver, Matthew},
  booktitle = {The Third International Workshop on Coq for Programming
               Languages (CoqPL)},
  year      = {2017}
}

@article{appel2017position,
  title     = {Position paper: the science of deep specification},
  author    = {Appel, Andrew W and Beringer, Lennart and Chlipala, Adam
               and Pierce, Benjamin C and Shao, Zhong and Weirich,
               Stephanie and Zdancewic, Steve},
  journal   = {Philosophical Transactions of the Royal Society A:
               Mathematical, Physical and Engineering Sciences},
  volume    = {375},
  number    = {2104},
  pages     = {20160331},
  year      = {2017},
  publisher = {The Royal Society Publishing}
}

@inproceedings{beringer2014sharedmem,
  author    = {Lennart Beringer and Gordon Stewart and Robert Dockins and
               Andrew W. Appel},
  title     = {Verified Compilation for Shared-Memory {C}},
  booktitle = {{ESOP}},
  series    = {LNCS},
  volume    = {8410},
  pages     = {107--127},
  year      = {2014},
  doi       = {10.1007/978-3-642-54833-8\_7}
}

@article{blanqui2011color,
  title     = {{CoLoR}: a {Coq} library on well-founded rewrite relations
               and its application to the automated verification of
               termination certificates},
  author    = {Blanqui, Fr{\'e}d{\'e}ric and Koprowski, Adam},
  journal   = {Mathematical Structures in Computer Science},
  volume    = {21},
  number    = {4},
  pages     = {827--859},
  year      = {2011},
  publisher = {Cambridge University Press}
}

@online{dawson2012comparing,
  title   = {Comparing Floating Point Numbers},
  url     = {https://randomascii.wordpress.com/2012/02/25/comparing-floating-point-numbers-2012-edition/},
  author  = {Bruce Dawson},
  year    = {2012},
  urldate = {2023-10-31},
  date    = {2012-02-26}
}

@inproceedings{boldo2009combining,
  title        = {Combining Coq and Gappa for certifying floating-point
                  programs},
  author       = {Boldo, Sylvie and Filli{\^a}tre, Jean-Christophe and
                  Melquiond, Guillaume},
  booktitle    = {International Conference on Intelligent Computer
                  Mathematics},
  pages        = {59--74},
  year         = {2009},
  organization = {Springer}
}

@inproceedings{boldo2011flocq,
  title        = {Flocq: A unified library for proving floating-point
                  algorithms in {Coq}},
  author       = {Boldo, Sylvie and Melquiond, Guillaume},
  booktitle    = {Computer Arithmetic (ARITH), 2011 20th IEEE Symposium on},
  pages        = {243--252},
  year         = {2011},
  organization = {IEEE}
}

@article{borovansky1998overview,
  title     = {An overview of ELAN},
  author    = {Borovansk{\`y}, Peter and Kirchner, Claude and Kirchner,
               H{\'e}lene and Moreau, Pierre-Etienne and Ringeissen,
               Christophe},
  journal   = {Electronic Notes in Theoretical Computer Science},
  volume    = {15},
  pages     = {55--70},
  year      = {1998},
  publisher = {Elsevier}
}

@mastersthesis{brinich2020,
  author  = {Brinich,Patrick J.},
  title   = {Formal Verification of Spiral Generated Code},
  school  = {Drexel University},
  year    = {2020},
  journal = {ProQuest Dissertations and Theses},
  pages   = {145},
  isbn    = {9798684691386}
}

@inproceedings{cakeml,
  author    = {Kumar, Ramana and Myreen, Magnus O. and Norrish, Michael
               and Owens, Scott},
  title     = {CakeML: A Verified Implementation of ML},
  year      = {2014},
  isbn      = {9781450325448},
  publisher = {Association for Computing Machinery},
  address   = {New York, NY, USA},
  url       = {https://doi.org/10.1145/2535838.2535841},
  doi       = {10.1145/2535838.2535841},
  booktitle = {Proceedings of the 41st ACM SIGPLAN-SIGACT Symposium on
               Principles of Programming Languages},
  pages     = {179–191},
  numpages  = {13},
  location  = {San Diego, California, USA},
  series    = {POPL ’14}
}

@inproceedings{compcertx,
  author    = {Gu, Ronghui and Koenig, J{\'e}r{\'e}mie and Ramananandro,
               Tahina and Shao, Zhong and Wu, Xiongnan (Newman) and Weng,
               Shu-Chun and Zhang, Haozhong and Guo, Yu},
  title     = {Deep Specifications and Certified Abstraction Layers},
  booktitle = {Proceedings of the 42nd Annual ACM SIGPLAN-SIGACT
               Symposium on Principles of Programming Languages},
  series    = {POPL '15},
  year      = {2015},
  isbn      = {978-1-4503-3300-9},
  location  = {Mumbai, India},
  pages     = {595--608},
  numpages  = {14},
  url       = {http://doi.acm.org/10.1145/2676726.2676975},
  doi       = {10.1145/2676726.2676975},
  acmid     = {2676975},
  publisher = {ACM},
  address   = {New York, NY, USA}
}

@inproceedings{cascompcert,
  author    = {Hanru Jiang and Hongjin Liang and Siyang Xiao and Junpeng
               Zha and Xinyu Feng},
  title     = {Towards certified separate compilation for concurrent
               programs},
  booktitle = {{PLDI}},
  pages     = {111--125},
  year      = {2019},
  doi       = {10.1145/3314221.3314595}
}

@techreport{casteran2012gentle,
  title       = {A gentle introduction to type classes and relations in
                 {Coq}},
  author      = {Cast{\'e}ran, Pierre and Sozeau, Matthieu},
  year        = {2012},
  institution = {Technical Report hal-00702455, version 1}
}

@article{charette2009car,
  title     = {This car runs on code},
  author    = {Charette, Robert N},
  journal   = {IEEE spectrum},
  volume    = {46},
  number    = {3},
  pages     = {3},
  year      = {2009},
  publisher = {IEEE}
}

@article{chhzz23,
  author   = {Nicolas Chappe and Paul He and Ludovic Henrio and Yannick
              Zakowski and Steve Zdancewic},
  title    = {Choice Trees: Representing Nondeterministic, Recursive,
              and Impure Programs in Coq},
  journal  = {Proceedings of the ACM on Programming Languages},
  year     = 2023,
  volume   = 7,
  number   = {POPL},
  hjournal = {yes},
  plclub   = {yes}
}

@book{chlipala2017frap,
  title  = {Formal reasoning about programs},
  author = {Chlipala, Adam},
  url    = {http://adam.chlipala.net/frap},
  year   = {2017}
}

@article{compcert,
  author    = {Xavier Leroy},
  title     = {Formal verification of a realistic compiler},
  journal   = {Commun. {ACM}},
  volume    = {52},
  number    = {7},
  pages     = {107--115},
  year      = {2009},
  url       = {http://doi.acm.org/10.1145/1538788.1538814},
  doi       = {10.1145/1538788.1538814},
  bibsource = {dblp computer science bibliography, https://dblp.org}
}

@article{compcertm,
  author     = {Song, Youngju and Cho, Minki and Kim, Dongjoo and Kim,
                Yonghyun and Kang, Jeehoon and Hur, Chung-Kil},
  title      = {CompCertM: CompCert with C-Assembly Linking and
                Lightweight Modular Verification},
  year       = {2019},
  issue_date = {January 2020},
  publisher  = {Association for Computing Machinery},
  address    = {New York, NY, USA},
  volume     = {4},
  number     = {POPL},
  url        = {https://doi.org/10.1145/3371091},
  doi        = {10.1145/3371091},
  journal    = {Proc. ACM Program. Lang.},
  month      = dec,
  articleno  = {23},
  numpages   = {31}
}

@article{compcerttso,
  author  = {Jaroslav {\v{S}ev\v{c}\'ik} and Viktor Vafeiadis and
             Francesco Zappa Nardelli and Suresh Jagannathan and Peter
             Sewell},
  title   = {CompCertTSO: A Verified Compiler for Relaxed-Memory
             Concurrency},
  journal = {J. ACM},
  doi     = {10.1145/2487241.2487248},
  volume  = 60,
  number  = 3,
  year    = 2013,
  pages   = 22
}

@manual{coq,
  title        = {The {Coq} proof assistant reference manual},
  author       = {The {Coq} development team},
  organization = {LogiCal Project},
  note         = {Version 8.0},
  year         = {2004},
  url          = {http://coq.inria.fr}
}

@inproceedings{cpp19-ns,
  author    = {Koh, Nicolas and Li, Yao and Li, Yishuai and Xia, Li-yao
               and Beringer, Lennart and Honor{\'e}, Wolf and Mansky,
               William and Pierce, Benjamin C. and Zdancewic, Steve},
  title     = {From C to Interaction Trees: Specifying, Verifying, and
               Testing a Networked Server},
  booktitle = {Proceedings of the 8th ACM SIGPLAN International
               Conference on Certified Programs and Proofs},
  series    = {CPP 2019},
  year      = {2019},
  isbn      = {978-1-4503-6222-1},
  location  = {Cascais, Portugal},
  pages     = {234--248},
  numpages  = {15},
  url       = {http://doi.acm.org/10.1145/3293880.3294106},
  doi       = {10.1145/3293880.3294106},
  acmid     = {3294106},
  publisher = {ACM},
  address   = {New York, NY, USA}
}

@article{de2008certifying,
  title   = {Certifying floating-point implementations using Gappa},
  author  = {De Dinechin, Florent and Lauter, Christoph Quirin and
             Melquiond, Guillaume},
  journal = {arXiv preprint arXiv:0801.0523},
  year    = {2008}
}

@article{dimsum,
  author     = {Sammler, Michael and Spies, Simon and Song, Youngju and
                D'Osualdo, Emanuele and Krebbers, Robbert and Garg, Deepak
                and Dreyer, Derek},
  title      = {DimSum: A Decentralized Approach to Multi-Language
                Semantics and Verification},
  year       = {2023},
  issue_date = {January 2023},
  publisher  = {Association for Computing Machinery},
  address    = {New York, NY, USA},
  volume     = {7},
  number     = {POPL},
  url        = {https://doi.org/10.1145/3571220},
  doi        = {10.1145/3571220},
  journal    = {Proc. ACM Program. Lang.},
  month      = {jan},
  articleno  = {27},
  numpages   = {31}
}

@misc{extlibgithub,
  title        = {{ExtLib} {Coq} library},
  author       = {Malecha, Gregory and others},
  howpublished = {\url{https://github.com/coq-ext-lib/coq-ext-lib}},
  year         = {2012},
  note         = {Accessed: 2020-08-18}
}

@misc{fhw21,
  title         = {Formally Verified Simulations of State-Rich Processes
                   using Interaction Trees in Isabelle/HOL},
  author        = {Simon Foster and Chung-Kil Hur and Jim Woodcock},
  year          = {2021},
  eprint        = {2105.05133},
  archiveprefix = {arXiv},
  primaryclass  = {cs.LO}
}

@article{fox1997dynamic,
  title     = {The dynamic window approach to collision avoidance},
  author    = {Fox, Dieter and Burgard, Wolfram and Thrun, Sebastian},
  journal   = {IEEE Robotics \& Automation Magazine},
  volume    = {4},
  number    = {1},
  pages     = {23--33},
  year      = {1997},
  publisher = {IEEE}
}

@inproceedings{franchetti-SPL,
  author    = {Franz Franchetti and Fr{\'e}d{\'e}ric {de Mesmay} and
               Daniel McFarlin and Markus P{\"u}schel},
  title     = {Operator Language: A Program Generation Framework for Fast
               Kernels},
  booktitle = {IFIP Working Conference on Domain Specific Languages (DSL
               WC)},
  publisher = {Springer},
  series    = {Lecture Notes in Computer Science},
  volume    = {5658},
  pages     = {385--410},
  year      = {2009}
}

@article{franchetti:17,
  author   = {F. Franchetti and T. M. Low and S. Mitsch and J. P.
              Mendoza and L. Gui and A. Phaosawasdi and D. Padua and S.
              Kar and J. M. F. Moura and M. Franusich and J. Johnson and
              A. Platzer and M. M. Veloso},
  journal  = {IEEE Control Systems},
  title    = {High-Assurance {SPIRAL}: End-to-End Guarantees for Robot
              and Car Control},
  year     = {2017},
  volume   = {37},
  number   = {2},
  pages    = {82-103},
  doi      = {10.1109/MCS.2016.2643244},
  issn     = {1066-033X},
  month    = {April}
}

@article{franchetti:18,
  author  = {Franz Franchetti and Tze-Meng Low and Thom Popovici and
             Richard Veras and Daniele G. Spampinato and Jeremy Johnson
             and Markus P{\"u}schel and James C. Hoe and Jos{\'e} M. F.
             Moura},
  title   = {{SPIRAL}: Extreme Performance Portability},
  journal = {Proceedings of the IEEE, special issue on ``From High
             Level Specification to High Performance Code''},
  volume  = {106},
  number  = {11},
  year    = {2018}
}

@inproceedings{franchetti:2005:flm:1065010.1065048,
  author    = {Franchetti, Franz and Voronenko, Yevgen and P\"{u}schel,
               Markus},
  title     = {Formal Loop Merging for Signal Transforms},
  booktitle = {Proceedings of the 2005 ACM SIGPLAN Conference on
               Programming Language Design and Implementation},
  series    = {PLDI '05},
  year      = {2005},
  isbn      = {1-59593-056-6},
  location  = {Chicago, IL, USA},
  pages     = {315--326},
  numpages  = {12},
  doi       = {10.1145/1065010.1065048},
  acmid     = {1065048},
  publisher = {ACM},
  address   = {New York, NY, USA}
}

@inproceedings{fulton2015keymaera,
  title        = {KeYmaera X: An axiomatic tactical theorem prover for
                  hybrid systems},
  author       = {Fulton, Nathan and Mitsch, Stefan and Quesel, Jan-David
                  and V{\"o}lp, Marcus and Platzer, Andr{\'e}},
  booktitle    = {International Conference on Automated Deduction},
  pages        = {527--538},
  year         = {2015},
  organization = {Springer}
}

@manual{gap4,
  key          = {GAP},
  organization = {The GAP~Group},
  title        = {{GAP -- Groups, Algorithms, and Programming, Version
                  4.11.0}},
  year         = 2020,
  url          = {https://www.gap-system.org}
}

@misc{general1992patriot,
  title  = {Patriot Missile Defense: Software Problem Led to System Failure at {Dhahran}, {Saudi} {Arabia}},
  author = {Blair, Michael and Obenski, Sally and Bridickas, Paula},
  year   = {1992}
}

@inproceedings{helix-vstte20,
  author    = {Vadim Zaliva and Ilia Zaichuk and Franz Franchetti},
  title     = {Verified Translation Between Purely Functional and
               Imperative Domain Specific Languages in {HELIX}},
  booktitle = {Proceedings of the 12th Working Conference on Verified
               Software: Theories, Tools, and Experiments (VSTTE)},
  year      = 2020
}

@inproceedings{helixfhpc18,
  author    = {Zaliva, Vadim and Franchetti, Franz},
  title     = {{HELIX}: A Case Study of a Formal Verification of High
               Performance Program Generation},
  booktitle = {Proceedings of the 7th ACM SIGPLAN International Workshop
               on Functional High-Performance Computing},
  series    = {FHPC 2018},
  year      = {2018},
  isbn      = {978-1-4503-5813-2},
  location  = {St. Louis, MO, USA},
  pages     = {1--9},
  numpages  = {9},
  url       = {http://doi.acm.org/10.1145/3264738.3264739},
  doi       = {10.1145/3264738.3264739},
  acmid     = {3264739},
  publisher = {ACM},
  address   = {New York, NY, USA}
}

@article{itrees,
  author  = {{Li-yao} Xia and Yannick Zakowski and Paul He and
             {Chung-Kil} Hur and Gregory Malecha and Benjamin C. Pierce
             and Steve Zdancewic},
  title   = {Interaction Trees},
  journal = {Proceedings of the ACM on Programming Languages},
  year    = 2020,
  volume  = 4,
  number  = {POPL},
  month   = jan
}

@inproceedings{itrees-spec,
  author    = {Silver, Lucas and Westbrook, Eddy and Yacavone, Matthew
               and Scott, Ryan},
  title     = {{Interaction Tree Specifications: A Framework for
               Specifying Recursive, Effectful Computations That Supports
               Auto-Active Verification}},
  booktitle = {37th European Conference on Object-Oriented Programming
               (ECOOP 2023)},
  pages     = {30:1--30:26},
  series    = {Leibniz International Proceedings in Informatics
               (LIPIcs)},
  isbn      = {978-3-95977-281-5},
  issn      = {1868-8969},
  year      = {2023a},
  volume    = {263},
  editor    = {Ali, Karim and Salvaneschi, Guido},
  publisher = {Schloss Dagstuhl -- Leibniz-Zentrum f{\"u}r Informatik},
  address   = {Dagstuhl, Germany},
  url       = {https://drops.dagstuhl.de/entities/document/10.4230/LIPIcs.ECOOP.2023.30},
  urn       = {urn:nbn:de:0030-drops-182239},
  doi       = {10.4230/LIPIcs.ECOOP.2023.30}
}

@inproceedings{khm+15,
  author    = {Kang, Jeehoon and Hur, Chung-Kil and Mansky, William and
               Garbuzov, Dmitri and Zdancewic, Steve and Vafeiadis,
               Viktor},
  title     = {A Formal C Memory Model Supporting Integer-Pointer Casts},
  year      = {2015},
  isbn      = {9781450334686},
  publisher = {Association for Computing Machinery},
  address   = {New York, NY, USA},
  url       = {https://doi.org/10.1145/2737924.2738005},
  doi       = {10.1145/2737924.2738005},
  booktitle = {Proceedings of the 36th ACM SIGPLAN Conference on
               Programming Language Design and Implementation},
  pages     = {326–335},
  numpages  = {10},
  location  = {Portland, OR, USA},
  series    = {PLDI ’15}
}

@inproceedings{kksj+18,
  author    = {Kang, Jeehoon and Kim, Yoonseung and Song, Youngju and
               Lee, Juneyoung and Park, Sanghoon and Shin, Mark Dongyeon
               and Kim, Yonghyun and Cho, Sungkeun and Choi, Joonwon and
               Hur, Chung-Kil and Yi, Kwangkeun},
  title     = {Crellvm: Verified Credible Compilation for LLVM},
  year      = {2018},
  isbn      = {9781450356985},
  publisher = {Association for Computing Machinery},
  address   = {New York, NY, USA},
  url       = {https://doi.org/10.1145/3192366.3192377},
  doi       = {10.1145/3192366.3192377},
  booktitle = {Proceedings of the 39th ACM SIGPLAN Conference on
               Programming Language Design and Implementation},
  pages     = {631–645},
  numpages  = {15},
  location  = {Philadelphia, PA, USA},
  series    = {PLDI 2018}
}

@article{kumar2014cakeml,
  title     = {CakeML: a verified implementation of ML},
  author    = {Kumar, Ramana and Myreen, Magnus O and Norrish, Michael
               and Owens, Scott},
  journal   = {ACM SIGPLAN Notices},
  volume    = {49},
  number    = {1},
  pages     = {179--191},
  year      = {2014},
  publisher = {ACM New York, NY, USA}
}

@article{leroy2009,
  author     = {Leroy, Xavier},
  title      = {Formal Verification of a Realistic Compiler},
  journal    = {Commun. ACM},
  issue_date = {July 2009},
  volume     = {52},
  number     = {7},
  month      = jul,
  year       = {2009},
  issn       = {0001-0782},
  pages      = {107--115},
  numpages   = {9},
  url        = {http://doi.acm.org/10.1145/1538788.1538814},
  doi        = {10.1145/1538788.1538814},
  acmid      = {1538814},
  publisher  = {ACM},
  address    = {New York, NY, USA}
}

@techreport{leroy:hal-00703441,
  title       = {{The CompCert Memory Model, Version 2}},
  author      = {Leroy, Xavier and Appel, Andrew W. and Blazy, Sandrine and
                 Stewart, Gordon},
  url         = {https://hal.inria.fr/hal-00703441},
  type        = {Research Report},
  number      = {RR-7987},
  pages       = {26},
  institution = {{INRIA}},
  year        = {2012},
  month       = jun,
  hal_id      = {hal-00703441},
  hal_version = {v1}
}

@article{lesanixkbcpz22,
  author    = {Mohsen Lesani and Li{-}yao Xia and Anders Kaseorg and
               Christian J. Bell and Adam Chlipala and Benjamin C. Pierce
               and Steve Zdancewic},
  title     = {{C4:} verified transactional objects},
  journal   = {Proc. {ACM} Program. Lang.},
  volume    = {6},
  number    = {{OOPSLA}},
  pages     = {1--31},
  year      = {2022},
  url       = {https://doi.org/10.1145/3527324},
  doi       = {10.1145/3527324},
  bibsource = {dblp computer science bibliography, https://dblp.org}
}

@inproceedings{lg20,
  author    = {Liyi Li and Elsa Gunter},
  title     = {K-LLVM: A Relatively Complete Semantics of LLVM IR},
  doi       = {10.4230/LIPIcs.ECOOP.2020.7},
  booktitle = {34rd European Conference on Object-Oriented Programming,
               {ECOOP} 2020, Berlin, Germany},
  year      = {2020}
}

@inproceedings{low:17,
  author    = {Tze-Meng Low and Franz Franchetti},
  title     = {High Assurance Code Generation for Cyber-Physical
               Systems},
  booktitle = {IEEE International Symposium on High Assurance Systems
               Engineering (HASE)},
  year      = {2017}
}

@misc{malechareflection,
  title        = {Speeding Up Proofs with Computational Reflection},
  author       = {Malecha, Gregory},
  howpublished = {\url{https://gmalecha.github.io/reflections/2017/speeding-up-proofs-with-computational-reflection}},
  year         = {2017},
  note         = {Accessed: 2020-09-01}
}

@inproceedings{moscato2017automatic,
  title        = {Automatic estimation of verified floating-point round-off
                  errors via static analysis},
  author       = {Moscato, Mariano and Titolo, Laura and Dutle, Aaron and
                  Munoz, C{\'e}sar A},
  booktitle    = {Computer Safety, Reliability, and Security: 36th
                  International Conference, SAFECOMP 2017, Trento, Italy,
                  September 13-15, 2017, Proceedings 36},
  pages        = {213--229},
  year         = {2017},
  organization = {Springer}
}

@inproceedings{moschovakis1989,
  title        = {A mathematical modeling of pure, recursive algorithms},
  author       = {Moschovakis, Yiannis N},
  booktitle    = {International Symposium on Logical Foundations of Computer
                  Science},
  pages        = {208--229},
  year         = {1989},
  organization = {Springer}
}

@inproceedings{owens2016functional,
  title        = {Functional big-step semantics},
  author       = {Owens, Scott and Myreen, Magnus O and Kumar, Ramana and
                  Tan, Yong Kiam},
  booktitle    = {European Symposium on Programming},
  pages        = {589--615},
  year         = {2016},
  organization = {Springer}
}

@inproceedings{pnueli1998translation,
  title        = {Translation validation},
  author       = {Pnueli, Amir and Siegel, Michael and Singerman, Eli},
  booktitle    = {Tools and Algorithms for the Construction and Analysis of
                  Systems: 4th International Conference, TACAS'98 Held as
                  Part of the Joint European Conferences on Theory and
                  Practice of Software, ETAPS'98 Lisbon, Portugal, March
                  28--April 4, 1998 Proceedings 4},
  pages        = {151--166},
  year         = {1998},
  organization = {Springer}
}

@article{pueschel:05,
  author  = {M. P\"{u}schel and J. M. F. Moura and J. R. Johnson and D.
             Padua and M. M. Veloso and B. W. Singer and Jianxin Xiong
             and F. Franchetti and A. Gacic and Y. Voronenko and K. Chen
             and R. W. Johnson and N. Rizzolo},
  journal = {Proceedings of the IEEE},
  title   = {{SPIRAL}: Code Generation for {DSP} Transforms},
  year    = {2005},
  volume  = {93},
  number  = {2},
  pages   = {232-275},
  doi     = {10.1109/JPROC.2004.840306},
  issn    = {0018-9219},
  month   = {Feb}
}

@article{sclh+23,
  author     = {Song, Youngju and Cho, Minki and Lee, Dongjae and Hur,
                Chung-Kil and Sammler, Michael and Dreyer, Derek},
  title      = {Conditional Contextual Refinement},
  year       = {2023},
  issue_date = {January 2023},
  publisher  = {Association for Computing Machinery},
  address    = {New York, NY, USA},
  volume     = {7},
  number     = {POPL},
  url        = {https://doi.org/10.1145/3571232},
  doi        = {10.1145/3571232},
  journal    = {Proc. ACM Program. Lang.},
  month      = {jan},
  articleno  = {39},
  numpages   = {31}
}

@inproceedings{sh+23,
  author    = {Lucas Silver and Paul He and Ethan Cecchetti and Andrew K.
               Hirsch and Steve Zdancewic},
  title     = {Semantics for Noninterference with Interaction Trees},
  booktitle = {Proceedings of the 37th Annual European Conference on
               Object-Oriented Programming (ECOOP 2023)},
  year      = {2023b},
  plclub    = {yes},
  hsconf    = {yes}
}

@article{solovyev2018rigorous,
  title     = {Rigorous estimation of floating-point round-off errors
               with symbolic taylor expansions},
  author    = {Solovyev, Alexey and Baranowski, Marek S and Briggs, Ian
               and Jacobsen, Charles and Rakamari{\'c}, Zvonimir and
               Gopalakrishnan, Ganesh},
  journal   = {ACM Transactions on Programming Languages and Systems
               (TOPLAS)},
  volume    = {41},
  number    = {1},
  pages     = {1--39},
  year      = {2018},
  publisher = {ACM New York, NY, USA}
}

@article{sozeau2010,
  author  = {Sozeau, Matthieu},
  journal = {Journal of Formalized Reasoning},
  pages   = {1--12},
  title   = {{A new look at generalized rewriting in type theory}},
  year    = {2010}
}

@article{sozeau2020metacoq,
  title     = {The MetaCoq Project},
  author    = {Sozeau, Matthieu and Anand, Abhishek and Boulier, Simon
               and Cohen, Cyril and Forster, Yannick and Kunze, Fabian and
               Malecha, Gregory and Tabareau, Nicolas and Winterhalter,
               Th{\'e}o},
  journal   = {Journal of Automated Reasoning},
  pages     = {1--53},
  year      = {2020},
  publisher = {Springer}
}

@article{spitters2011type,
  title     = {Type classes for mathematics in type theory},
  author    = {Spitters, Bas and Van der Weegen, Eelis},
  journal   = {Mathematical Structures in Computer Science},
  volume    = {21},
  number    = {4},
  pages     = {795--825},
  year      = {2011},
  publisher = {Cambridge University Press}
}

@article{sz21,
  author     = {Silver, Lucas and Zdancewic, Steve},
  title      = {Dijkstra Monads Forever: Termination-Sensitive
                Specifications for Interaction Trees},
  year       = {2021},
  issue_date = {January 2021},
  publisher  = {Association for Computing Machinery},
  address    = {New York, NY, USA},
  volume     = {5},
  number     = {POPL},
  url        = {https://doi.org/10.1145/3434307},
  doi        = {10.1145/3434307},
  journal    = {Proc. ACM Program. Lang.},
  month      = jan,
  articleno  = {26},
  numpages   = {28}
}

@inproceedings{DBLP:conf/itp/ZhangHK0LXBMPZ21,
  author       = {Hengchu Zhang and
                  Wolf Honor{\'{e}} and
                  Nicolas Koh and
                  Yao Li and
                  Yishuai Li and
                  Li{-}yao Xia and
                  Lennart Beringer and
                  William Mansky and
                  Benjamin C. Pierce and
                  Steve Zdancewic},
  editor       = {Liron Cohen and
                  Cezary Kaliszyk},
  title        = {Verifying an {HTTP} Key-Value Server with Interaction Trees and {VST}},
  booktitle    = {12th International Conference on Interactive Theorem Proving, {ITP}
                  2021, June 29 to July 1, 2021, Rome, Italy (Virtual Conference)},
  series       = {LIPIcs},
  volume       = {193},
  pages        = {32:1--32:19},
  publisher    = {Schloss Dagstuhl - Leibniz-Zentrum f{\"{u}}r Informatik},
  year         = {2021},
  url          = {https://doi.org/10.4230/LIPIcs.ITP.2021.32},
  doi          = {10.4230/LIPICS.ITP.2021.32},
  bibsource    = {dblp computer science bibliography, https://dblp.org}
}

@article{DBLP:journals/pacmpl/LesaniXKBCPZ22,
  author       = {Mohsen Lesani and
                  Li{-}yao Xia and
                  Anders Kaseorg and
                  Christian J. Bell and
                  Adam Chlipala and
                  Benjamin C. Pierce and
                  Steve Zdancewic},
  title        = {{C4:} verified transactional objects},
  journal      = {Proc. {ACM} Program. Lang.},
  volume       = {6},
  number       = {{OOPSLA1}},
  pages        = {1--31},
  year         = {2022},
  url          = {https://doi.org/10.1145/3527324},
  doi          = {10.1145/3527324},
  bibsource    = {dblp computer science bibliography, https://dblp.org}
}

@article{DBLP:journals/pacmpl/SongCLHSD23,
  author       = {Youngju Song and
                  Minki Cho and
                  Dongjae Lee and
                  Chung{-}Kil Hur and
                  Michael Sammler and
                  Derek Dreyer},
  title        = {Conditional Contextual Refinement},
  journal      = {Proc. {ACM} Program. Lang.},
  volume       = {7},
  number       = {{POPL}},
  pages        = {1121--1151},
  year         = {2023},
  url          = {https://doi.org/10.1145/3571232},
  doi          = {10.1145/3571232},
  bibsource    = {dblp computer science bibliography, https://dblp.org}
}

@article{DBLP:journals/corr/abs-2303-09106,
  author       = {Kangfeng Ye and
                  Simon Foster and
                  Jim Woodcock},
  title        = {Formally Verified Animation for RoboChart using Interaction Trees},
  journal      = {CoRR},
  volume       = {abs/2303.09106},
  year         = {2023},
  url          = {https://doi.org/10.48550/arXiv.2303.09106},
  doi          = {10.48550/ARXIV.2303.09106},
  eprinttype    = {arXiv},
  eprint       = {2303.09106},
  bibsource    = {dblp computer science bibliography, https://dblp.org}
}

@article{capretta,
  author    = {Venanzio Capretta},
  title     = {General recursion via coinductive types},
  journal   = {Log. Methods Comput. Sci.},
  volume    = {1},
  number    = {2},
  year      = {2005},
  doi       = {10.2168/LMCS-1(2:1)2005},
  bibsource = {dblp computer science bibliography, https://dblp.org}
}

@article{vellvmicfp,
  author     = {Zakowski, Yannick and Beck, Calvin and Yoon, Irene and
                Zaichuk, Ilia and Zaliva, Vadim and Zdancewic, Steve},
  title      = {Modular, Compositional, and Executable Formal Semantics
                for LLVM IR},
  year       = {2021},
  issue_date = {August 2021},
  publisher  = {Association for Computing Machinery},
  address    = {New York, NY, USA},
  volume     = {5},
  number     = {ICFP},
  url        = {https://doi.org/10.1145/3473572},
  doi        = {10.1145/3473572},
  journal    = {Proc. ACM Program. Lang.},
  month      = {aug},
  articleno  = {67},
  numpages   = {30}
}

@inproceedings{vst,
  author    = {Appel, Andrew W.},
  editor    = {Barthe, Gilles},
  title     = {Verified Software Toolchain},
  booktitle = {Programming Languages and Systems},
  year      = {2011},
  publisher = {Springer Berlin Heidelberg},
  address   = {Berlin, Heidelberg},
  doi       = {10.1007/978-3-642-19718-5_1},
  pages     = {1--17},
  isbn      = {978-3-642-19718-5}
}

@inproceedings{vst-hmac,
  author    = {Beringer, Lennart and Petcher, Adam and Ye, Katherine Q.
               and Appel, Andrew W.},
  title     = {Verified Correctness and Security of OpenSSL HMAC},
  year      = {2015},
  isbn      = {9781931971232},
  publisher = {USENIX Association},
  address   = {USA},
  booktitle = {Proceedings of the 24th USENIX Conference on Security
               Symposium},
  pages     = {207–221},
  numpages  = {15},
  location  = {Washington, D.C.},
  series    = {SEC'15}
}

@article{vst-mailboxes,
  author     = {Mansky, William and Appel, Andrew W. and Nogin, Aleksey},
  title      = {A Verified Messaging System},
  year       = {2017},
  issue_date = {October 2017},
  publisher  = {Association for Computing Machinery},
  address    = {New York, NY, USA},
  volume     = {1},
  number     = {OOPSLA},
  url        = {https://doi.org/10.1145/3133911},
  doi        = {10.1145/3133911},
  journal    = {Proc. ACM Program. Lang.},
  month      = {oct},
  articleno  = {87},
  numpages   = {28}
}

@book{wolfram1999mathematica,
  title     = {The MATHEMATICA{\textregistered} book, version 4},
  author    = {Wolfram, Stephen and others},
  year      = {1999},
  publisher = {Cambridge university press}
}

@article{xia2019interaction,
  title     = {Interaction trees: representing recursive and impure
               programs in {Coq}},
  author    = {Xia, Li-yao and Zakowski, Yannick and He, Paul and Hur,
               Chung-Kil and Malecha, Gregory and Pierce, Benjamin C and
               Zdancewic, Steve},
  journal   = {Proceedings of the ACM on Programming Languages},
  volume    = {4},
  number    = {POPL},
  pages     = {1--32},
  year      = {2019},
  publisher = {ACM New York, NY, USA}
}

@article{xzhh+20,
  author  = {{Li-yao} Xia and Yannick Zakowski and Paul He and
             {Chung-Kil} Hur and Gregory Malecha and Benjamin C. Pierce
             and Steve Zdancewic},
  title   = {Interaction Trees},
  journal = {Proceedings of the ACM on Programming Languages},
  doi     = {10.1145/3371119},
  year    = 2020,
  volume  = 4,
  number  = {POPL}
}

@article{yzz22,
  author   = {Irene Yoon and Yannick Zakowski and Steve Zdancewic},
  title    = {Formal Reasoning About Layered Monadic Interpreters},
  journal  = {Proceedings of the ACM on Programming Languages},
  year     = 2022,
  volume   = 6,
  number   = {ICFP},
  hjournal = {yes},
  plclub   = {yes}
}

@article{zaliva2015formal,
  title     = {Formal Verification of {HCOL} Rewriting},
  author    = {Zaliva, Vadim and Franchetti, Franz},
  booktitle = {FMCAD 2015},
  year      = {2015}
}

@inproceedings{zaliva:19,
  author    = {Vadim Zaliva and Matthieu Sozeau},
  title     = {Reification of Shallow-Embedded {DSLs} in {Coq} with
               Automated Verification},
  booktitle = {International Workshop on Coq for Programming Languages
               (CoqPL)},
  year      = {2019}
}

@inproceedings{zhhz20,
  author    = {Yannick Zakowski and Paul He and Chung-Kil Hur and Steve
               Zdancewic},
  title     = {An Equational Theory for Weak Bisimulation via Generalized
               Parameterized Coinduction},
  doi       = {10.1145/3372885.3373813},
  booktitle = {Proceedings of the 9th ACM SIGPLAN International
               Conference on Certified Programs and Proofs (CPP)},
  year      = 2020
}

@inproceedings{zhk+21,
  author    = {Zhang, Hengchu and Honor\'{e}, Wolf and Koh, Nicolas and
               Li, Yao and Li, Yishuai and Xia, Li-Yao and Beringer,
               Lennart and Mansky, William and Pierce, Benjamin and
               Zdancewic, Steve},
  title     = {{Verifying an HTTP Key-Value Server with Interaction Trees
               and VST}},
  booktitle = {12th International Conference on Interactive Theorem
               Proving (ITP 2021)},
  pages     = {32:1--32:19},
  series    = {Leibniz International Proceedings in Informatics
               (LIPIcs)},
  isbn      = {978-3-95977-188-7},
  issn      = {1868-8969},
  year      = {2021},
  volume    = {193},
  editor    = {Cohen, Liron and Kaliszyk, Cezary},
  publisher = {Schloss Dagstuhl -- Leibniz-Zentrum f{\"u}r Informatik},
  address   = {Dagstuhl, Germany},
  url       = {https://drops.dagstuhl.de/opus/volltexte/2021/13927},
  urn       = {urn:nbn:de:0030-drops-139273},
  doi       = {10.4230/LIPIcs.ITP.2021.32},
  hsconf    = {yes},
  plclub    = {yes}
}

@inproceedings{znmz12,
  author    = {Jianzhou Zhao and Santosh Nagarakatte and Milo M.~K.
               Martin and Steve Zdancewic},
  title     = {{Formalizing the LLVM Intermediate Representation for
               Verified Program Transformations}},
  doi       = {10.1145/2103621.2103709},
  booktitle = {Proc. of the {ACM} Symposium on Principles of Programming
               Languages (POPL)},
  year      = 2012
}

@inproceedings{znmz13,
  author    = {Jianzhou Zhao and Santosh Nagarakatte and Milo M. K.
               Martin and Steve Zdancewic},
  title     = {Formal Verification of {SSA}-Based Optimizations for
               {LLVM}},
  doi       = {10.1145/2499370.2462164},
  booktitle = {Proc. 2013 {ACM} SIGPLAN Conference on Programming
               Languages Design and Implementation (PLDI)},
  year      = 2013
}

@inproceedings{bourke20,
  title     = {Mechanized Semantics and Verified Compilation for a Dataflow Synchronous Language with Reset},
  booktitle = {Proceedings of the 47th {{ACM SIGPLAN Symposium}} on {{Principles}} of {{Programming Languages}}},
  author    = {Bourke, Timothy and Brun, Lélio and Pouzet, Marc},
  year      = 2020,
  volume    = {4},
  pagetotal = {29},
  publisher = {{Association for Computing Machinery}},
  location  = {{New Orleans, LA, USA}},
  doi       = {10.1145/3371112},
  url       = {https://www.leliobrun.net/publication/popl20/paper.pdf}
}

@inproceedings{ye2022formally,
  title        = {Formally verified animation for RoboChart using Interaction Trees},
  author       = {Ye, Kangfeng and Foster, Simon and Woodcock, Jim},
  booktitle    = {International Conference on Formal Engineering Methods},
  pages        = {404--420},
  year         = {2022},
  organization = {Springer}
}

@INPROCEEDINGS{Franchetti:09,
AUTHOR = {Franz Franchetti and Markus P{\"u}schel},
TITLE = {Generating High-Performance Pruned {FFT} Implementations},
BOOKTITLE = {International Conference on Acoustics, Speech, and Signal Processing (ICASSP)},
PAGES = {549--552},
YEAR = {2009}
}

@INPROCEEDINGS{Franchetti:02,
AUTHOR = {Franz Franchetti and Markus P{\"u}schel and Jos{\'e} M. F. Moura and Christoph W. Ueberhuber},
TITLE = {Short Vector {SIMD} Code Generation for {DSP} Algorithms},
BOOKTITLE = {High Performance Extreme Computing (HPEC)},
YEAR = {2002}
}

@INPROCEEDINGS{Moura:01,
AUTHOR = {Jos{\'e} M. F. Moura and Jeremy Johnson and Robert W. Johnson and David Padua and Viktor K. Prasanna and Markus P{\"u}schel and Bryan Singer and Manuela Veloso and Jianxin Xiong},
TITLE = {Generating Platform-Adapted {DSP} Libraries using {SPIRAL}},
BOOKTITLE = {High Performance Extreme Computing (HPEC)},
YEAR = {2001}
}

@article{kalman1960new,
  title={A new approach to linear filtering and prediction problems},
  author={Kalman, Rudolph Emil},
  year={1960}
}

@book{johnson2005pid,
  title={PID control},
  author={Johnson, Michael A and Moradi, Mohammad H},
  year={2005},
  publisher={Springer}
}

@article{macian2006dft,
  title={DFT-based controller for fuel injection unevenness correction in turbocharged diesel engines},
  author={Macian, Vicente and Lujan, Jose Manuel and Guardiola, Carlos and Yuste, Pedro},
  journal={IEEE transactions on control systems technology},
  volume={14},
  number={5},
  pages={819--827},
  year={2006},
  publisher={IEEE}
}

@article{twophase,
  author     = {Beck, Calvin and
                Yoon, Irene and
                Chen, Hanxi and
                Zakowski, Yannick and
                Zdancewic, Steve},
  title      = {A Two-Phase Infinite/Finite Low-Level Memory Model},
  year       = {2024},
  issue_date = {August 2024},
  publisher  = {Association for Computing Machinery},
  doi        = {10.1145/3674652},
  address    = {New York, NY, USA},
  number     = {ICFP},
  journal    = {Proc. ACM Program. Lang.},
  month      = {aug},
  notes      = {(To Appear)}
}

@phdthesis{yoon2023modular,
  title={Modular semantics and metatheory for {LLVM} {IR}},
  author={Yoon, Euisun},
  year={2023},
  school={University of Pennsylvania}
}

@phdthesis{zaliva2020helix,
  title={HELIX: From Math to Verified Code},
  author={Zaliva, Vadim},
  author+an = {1=highlight},
  year={2020},
  school={Carnegie Mellon University}
}

@inproceedings{DBLP:conf/pldi/BohrerTMMP18,
  author       = {Rose Bohrer and
                  Yong Kiam Tan and
                  Stefan Mitsch and
                  Magnus O. Myreen and
                  Andr{\'{e}} Platzer},
  editor       = {Jeffrey S. Foster and
                  Dan Grossman},
  title        = {VeriPhy: verified controller executables from verified cyber-physical
                  system models},
  booktitle    = {Proceedings of the 39th {ACM} {SIGPLAN} Conference on Programming
                  Language Design and Implementation, {PLDI} 2018, Philadelphia, PA,
                  USA, June 18-22, 2018},
  pages        = {617--630},
  publisher    = {{ACM}},
  year         = {2018},
  url          = {https://doi.org/10.1145/3192366.3192406},
  doi          = {10.1145/3192366.3192406},
  bibsource    = {dblp computer science bibliography, https://dblp.org/}
}

@inproceedings{DBLP:conf/rv/DesaiDS17,
  author       = {Ankush Desai and
                  Tommaso Dreossi and
                  Sanjit A. Seshia},
  editor       = {Shuvendu K. Lahiri and
                  Giles Reger},
  title        = {Combining Model Checking and Runtime Verification for Safe Robotics},
  booktitle    = {Runtime Verification - 17th International Conference, {RV} 2017, Seattle,
                  WA, USA, September 13-16, 2017, Proceedings},
  series       = {Lecture Notes in Computer Science},
  volume       = {10548},
  pages        = {172--189},
  publisher    = {Springer},
  year         = {2017},
  url          = {https://doi.org/10.1007/978-3-319-67531-2/_11},
  doi          = {10.1007/978-3-319-67531-2\_11},
  bibsource    = {dblp computer science bibliography, https://dblp.org/}
}

@inproceedings{DBLP:conf/itp/AnandK15,
  author       = {Abhishek Anand and
                  Ross A. Knepper},
  editor       = {Christian Urban and
                  Xingyuan Zhang},
  title        = {ROSCoq: Robots Powered by Constructive Reals},
  booktitle    = {Interactive Theorem Proving - 6th International Conference, {ITP}
                  2015, Nanjing, China, August 24-27, 2015, Proceedings},
  series       = {Lecture Notes in Computer Science},
  volume       = {9236},
  pages        = {34--50},
  publisher    = {Springer},
  year         = {2015},
  url          = {https://doi.org/10.1007/978-3-319-22102-1\_3},
  doi          = {10.1007/978-3-319-22102-1\_3},
  bibsource    = {dblp computer science bibliography, https://dblp.org}
}

@inproceedings{DBLP:conf/cpsweek/MalechaRAL16,
  author       = {Gregory Malecha and
                  Daniel Ricketts and
                  Mario M. Alvarez and
                  Sorin Lerner},
  title        = {Towards foundational verification of cyber-physical systems},
  booktitle    = {2016 Science of Security for Cyber-Physical Systems Workshop, SOSCYPS@CPSWeek
                  2016, Vienna, Austria, April 11, 2016},
  pages        = {1--5},
  publisher    = {{IEEE} Computer Society},
  year         = {2016},
  url          = {https://doi.org/10.1109/SOSCYPS.2016.7580000},
  doi          = {10.1109/SOSCYPS.2016.7580000},
  bibsource    = {dblp computer science bibliography, https://dblp.org}
}

\end{document}